\pdfoutput=1
\documentclass[binding=0.6cm, noexaminfo]{unifith}
\usepackage{natbib}
\usepackage{microtype}
\usepackage[english]{babel}
\usepackage[utf8]{inputenc}
\graphicspath{ {plot/} }
\usepackage{subcaption}
\usepackage{cancel}
\usepackage{setspace}
\usepackage{amssymb}
\usepackage{rotating}
\usepackage{amsmath}
\usepackage{wrapfig}
\usepackage{enumitem}
\usepackage[version=4]{mhchem}
\usepackage[printonlyused]{acronym}
\usepackage[acronym]{glossaries}
\newcommand{\araa}{ARA\&A} 
\newcommand{\mnras}{MNRAS}
\newcommand{\aap}{A\&A}
\newcommand{\apj}{ApJ}
\newcommand{\sovast}{SOVAST}
\newcommand{\nat}{Nature}
\newcommand{\apjl}{ApJL}
\newcommand{\apjs}{ApJS}
\newcommand{\prl}{PRL}
\newcommand{\prd}{PRD}
\newcommand{\na}{NA}
\newcommand{\apss}{Astrophysics and Space Science}
\newcommand{\aapr}{A\&AR}
\newcommand{\iaucirc}{IAU Circ.}
\usepackage{natbib} 

\newcommand{\corr}{\textcolor{black}}
\newcommand{\stick}{~}
\usepackage{comment}
\usepackage{enumitem}
\usepackage{ulem} 
\usepackage{hyperref}
\hypersetup{
    colorlinks=true,
    linkcolor=blue,   
    filecolor=magenta,      
    citecolor=blue,       
    urlcolor=magenta,     
    pdfpagemode=FullScreen,
    pdftitle={Modelling the Dynamics of Middle-Aged Pulsar Wind Nebulae in the Reverberation Phase},
    pdfauthor={Yuri Batini}
}

\title{Modelling the Dynamics of Middle-Aged Pulsar Wind Nebulae in the Reverberation Phase}
\author{Yuri Batini}
\IDnumber{7104205}
\course{Corso di Laurea Magistrale\\ in Scienze Fisiche e Astrofisiche}
\courseorganizer{Scuola di Scienze Matematiche,\\ Fisiche e Naturali}
\advisor{Prof. Niccolo' Bucciantini}
\AcademicYear{2024/2025}
\thesistype{Tesi di Laurea Magistrale}					   
\copyyear{2026}
\authoremail{yuri.batini@edu.unifi.it}

\usepackage{chngcntr}
\counterwithout{figure}{chapter} 
\usepackage{float}
\begin{document}

\frontmatter

\maketitle
\newgeometry{bottom=1cm}
\dedication{A nonna Gabri}
\begin{abstract}
\noindent
\textbf{\textit{Context.}} Pulsar Wind Nebulae (PWNe) are among the most important astrophysical sources emitting in the very-high-energy (VHE) $\gamma$-ray band. Predicting their long-term evolution is crucial in light of forthcoming high-energy observatories such as ASTRI and CTA.  
In this work, we investigate the dynamical evolution of \textit{middle-aged} PWNe -- which are major contributors to the TeV emission in the Galaxy -- and test the robustness of current approaches. \\\\
\textbf{\textit{This Work.}} To understand the diversity of these systems within the Galaxy, we derive the Pulsar (PSR) and Supernova Remnant (SNR) parameters governing PWNe evolution. Supernova kinetic energy, ejecta mass, and ambient density determine the SNR evolution, while PSR energy injection (depending on the surface magnetic field and initial period) powers the expansion of the PWN in the interior of the SNR. Adopting standard distributions, we first generate a synthetic PWN-SNR population, and we define a region of interest, encompassing the vast majority of these objects. 
Using a simple semi-analytical framework for the early evolution, we determine their physical properties as they interact with parent SNRs (the so-called \textit{reverberation phase}) and we follow the subsequent interaction with the SNR using a 1D Lagrangian code. 
In our region of interest, we find a large diversity in the late-evolution. Despite these differences, at late times all systems converge toward a relaxed state, consistent with the Sedov solution. Aware of 1D limitations, we perform a set of 2D simulations to study the growth of instabilities and long-term mixing. Employing a computational strategy -- designed to reduce computational cost, while preserving physical accuracy and consistency of the model -- we find that the growth of instabilities depends on initial perturbations but does not significantly alter global dynamics. Finally, while effective volume evolution is in good agreement with 1D predictions, multidimensional effects can increase the apparent size by up to 50\%. \\\\
\textbf{\textit{Conclusions.}} For the first time, we investigated in detail the multidimensional evolution of middle-aged PWNe in the reverberation phase. We characterized the late time evolution across the entire population, and our results confirm the robustness of 1D models, and that the predictions of current approaches, despite being based on strong assumptions, remain trustworthy.
\end{abstract}
\restoregeometry
\tableofcontents
\mainmatter
\chapter*{Introduction}
\addcontentsline{toc}{chapter}{Introduction}
\section*{The $\gamma$-ray Sky and its Observatories}
\addcontentsline{toc}{section}{The $\gamma$-ray Sky and its Observatories}
In astrophysics, $\gamma$-rays are defined as photons with energy above 0.5 MeV \citep{Degrange2015}. 
Based on their typical energy scale they are broadly
classified as MeV, GeV, TeV and PeV.
Among these, very-high-energy (VHE) $\gamma$-rays constitute a subset with energies in the range $[0.1 - 100]$ TeV \citep{VHE2024NEW}.
They often originate from non-thermal processes involving high-energy particles:
typically inverse Compton scattering (ICS) from electrons and 
positrons on background photons, or pion decay following proton-proton 
inelastic collisions.\\\\
The first source of VHE $\gamma$-rays to be detected was the Crab Nebula \citep{VHE1989}, but subsequently numerous other sources of VHE $\gamma$-rays were discovered. Amongst them, Supernova Remnants (\citealt{Tanimori98}, \citealt{Aharonian01}, \citealt{Enomoto02}, \citealt{Aharonian04}), Pulsar Wind Nebulae (\citealt{VHE1989}, \citealt{FANG2023}, \citealt{MACECRAB}) and Pulsars (\citealt{LINDEN2018}) are located within the Galaxy. 
Other Galactic sources include Microquasars \citep{RAMON06} and young stellar clusters \citep{BONOLLO26}.
VHE $\gamma$-rays are also observed from extragalactic sources coming from relativistic jets of active galactic nuclei -- the so-called AGN (\citealt{Punch92AGN}, \citealt{Quinn96AGN}) -- and from galaxies with high star formation rate \citep{Chen_2023}. Furthermore, VHE $\gamma$-rays have been detected in the afterglow emission from the extremely energetic events known as Gamma-Ray Bursts (GRBs, \citealt{MAGIC19GRBS}, \citealt{HESS21GRBS}).
In addition to these standard astrophysical sources, VHE $\gamma$-ray observations are also used to explore exotic physics, such as searching for signals from dark matter annihilation or decay, which could potentially explain observed excesses in TeV emission (\citealt{Aharonian06DM}, \citealt{Nardi09}).\\\\
Observations of $\gamma$-rays in the range GeV-PeV present unique challenges.
The Earth atmosphere is opaque, forcing us either to go into space or to use indirect techniques. On top of this, the flux of photons at these energies is extremely low, decreasing from GeV to TeV.
In the range $[0.1-100]$ GeV, the $\gamma$-ray flux is intense enough, and the energy low enough to be detectable using space-borne instruments.
Among them, the most important that is currently in use is Fermi-LAT (Large Area Telescope, \citealt{Atwood2009}), which operates as a pair-conversion detector.
An incident $\gamma$-ray interacts with the tracker, converting into an electron-positron pair, whose trajectories allow the original $\gamma$-ray direction to be reconstructed. 
The pair then enters a calorimeter and the total energy deposited provides a measure of the initial $\gamma$-ray energy.
The telescope has successfully explored the full $\gamma$-ray sky in the range $[0.1-100]$ GeV.\\\\
Observing the expected TeV signatures from Galactic sources requires sensitive instruments in the VHE $\gamma$-ray band.
The detection of VHE $\gamma$-rays is difficult for many reasons: the low photon flux and the limited size of space instruments like Fermi-LAT constrain the number of detectable $\gamma$-rays. 
Moreover, space instruments typically have limited angular resolution -- which prevents them from resolving neighbouring sources.\\\\
At TeV energies observations are made using Cherenkov imaging techniques.
When a VHE $\gamma$-ray reaches the top of the atmosphere, it produces an extensive air shower (EAS) of ultra-relativistic particles that in turn emit Cherenkov radiation, which can be detected by ground based Imaging Atmospheric Cherenkov Telescopes (IACTs).
To interpret the resulting images, the entire process is modelled using extensive Monte Carlo simulations of the EAS development, the absorption and scattering of Cherenkov photons in the atmosphere, and the detailed optical and electronic response of the telescope.
When Cherenkov radiation is detected, the results from these simulations are used to interpret the stereoscopic data provided by telescope arrays. Analyzing multiple images of the same EAS allows to achieve a high-precision reconstruction of energy and direction of the original $\gamma$-ray.\\\\
Recently, wide-field water-Cherenkov arrays like HAWC (High-Altitude Water Cherenkov \textit{Observatory}, \citealt{HAWC}), and LHAASO (Large High Altitude Air Shower Observatory, \citealt{LHAASONEW}) have extended coverage to the multi-TeV and PeV regime, identifying the first \textit{PeVatrons} - sources accelerating particles beyond $10^{15}$ eV.\\\\ 
Despite the many interesting results from existing IACTs like HESS (High Energy Stereoscopic System, \citealt{HESSCRAB}), MAGIC (Major Atmospheric Gamma Imaging Cherenkov \textit{Telescopes}, \citealt{MAGIC23}) and VERITAS (Very Energetic Radiation Imaging Telescope Array System, \citealt{AbeysekaraVERITAS_2018}) the interest in the VHE $\gamma$-ray sky 
has grown in recent years and new more powerful telescopes are being built and designed.\\\\
In order to advance VHE $\gamma$-ray astronomy, new IACTs are being 
developed particularly relevant for the Italian and European astrophysical 
community: the CTA (Cherenkov Telescope Array, \citealt{CTA2018}), and ASTRI (Astronomia a Specchi a Tecnologia Replicante Italiana, \citealt{ASTRI22}).\\\\
Individual CTA telescopes will have Cherenkov cameras with wide field of view, in order to improve the resolution of distant astrophysical sources. 
The angular resolution of CTA will approach 1 arc-minute at high energies -- the best resolution of any instrument operating in the $\gamma$-ray band -- allowing detailed imaging of a large number of $\gamma$-ray sources. CTA telescopes will collect data for VHE $\gamma$-rays in the range $[0.02-300]\stick$TeV.
With 99 telescopes at the southern site and 19 telescopes at the northern site, the observatory will provide full sky coverage.\\\\
CTA is not the only project developed in order to improve the detection of $\gamma$-rays: the ASTRI program was originally intended as a pathfinder of the CTA observatory at La Palma, but due to timeline issues it has been installed at the Teide Astronomical Observatory. This array consists of nine innovative IACTs and represents a particularly powerful instrument for deep observations up to 100 TeV \citep{ASTRI22}.
Its large field of view is also well-suited for observing multiple or extended sources, such as Pulsar Wind Nebulae, expected to emit TeV $\gamma$-rays during their late evolutionary phases.\\\\
VHE $\gamma$-rays bring us important information about the physical conditions, the emission mechanisms and the nature of Galactic and extragalactic high energy sources. VHE-ray emission can also be used to look for and identify sources of Cosmic Rays (CRs), that diffuse in the Galaxy, losing memory of their origin.
However, $\gamma$-ray sources form a complex zoo, particularly crowded in the Galactic plane, and $\gamma$-ray telescopes lack the resolution of optical or X-ray instruments. To make sense of our observations, to derive meaningful physical information about these sources, and to discriminate them, one needs good theoretical and phenomenological models for these objects that today are mostly lacking.
\section*{Pulsar Wind Nebulae in the $\gamma$-ray Sky}
\begin{figure}[h!]
    \centering
    \fbox{\includegraphics[width=1.0\textwidth]{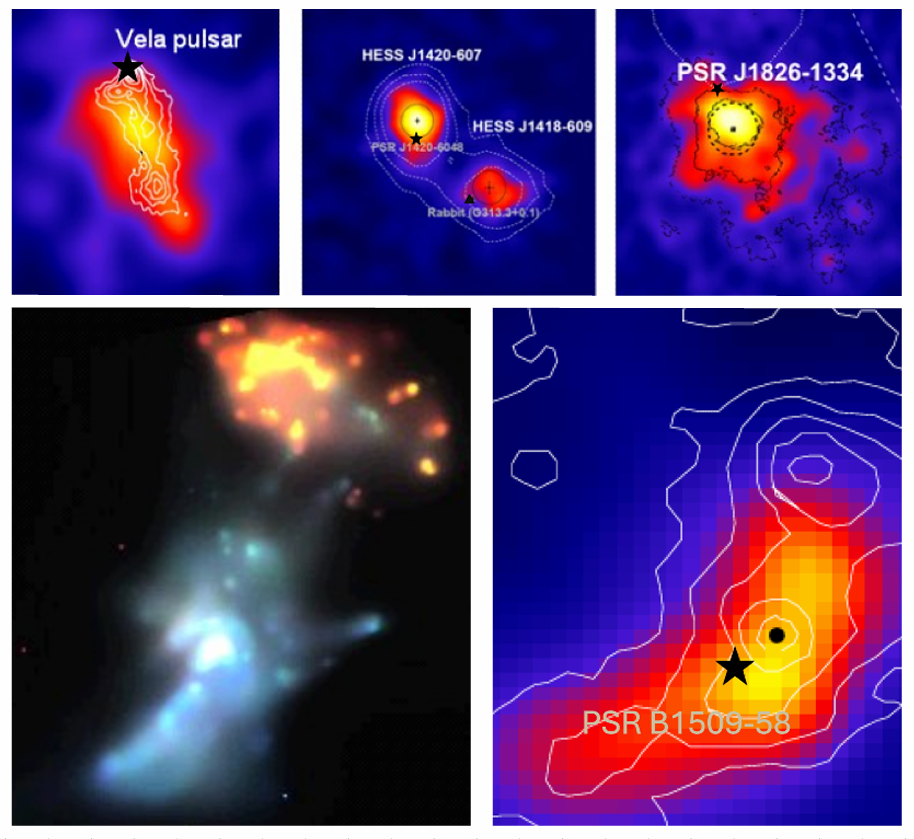}}
    \caption{\textit{Upper panels:} old PWNe observed by HESS in the TeV band. From left to right: Vela X, HESS J1420-607 \& HESS J1418-609, and HESS J1825–137. The color scale represents the excess of TeV emission with respect to the background in arbitrary units (\citealt{HESS_2006_VELAX}, \citealt{HESS_2006_RINGS}, \citealt{HESS_2006_THIRD_OLD_PSR}. \textit{Bottom panels:} SNR G320.4-1.2 with the inner, young PWN observed in the X-rays by Chandra (\textit{Left panel}, \citealt{SNR_OBSERVATION_YOUNG_PWN}) and in TeV band by HESS (\textit{Right panel}, \citealt{Aharonian_2005}). Here, the star symbol denotes the centroid of excess of the TeV emission. }
    \label{gamma_pwne}
\end{figure}

\addcontentsline{toc}{section}{Pulsar Wind Nebulae in the $\gamma$-ray Sky}
Among the most prominent Galactic sources in the $\gamma$-ray sky, two
play a special role in high-energy astrophysics and have drawn much
attention over the years: Supernova Remnants (SNRs), and Pulsar Wind Nebulae (PWNe).
Supernovae (SNe) are explosions of stars with masses above 9 $M_{\odot}$ \citep{Sukhbold+16}. 
Stars up to $\simeq 20 M_{\odot}$ experience core collapse leaving behind a rotating neutron star (NS) \citep{Zwicky34} and ejecting the stellar envelope in the surrounding medium -- interstellar medium (ISM) or circumstellar medium (CSM).
The fate of more massive stars is more uncertain: they might either undergo supernova explosion or collapse directly into a black hole (BH).\\\\
The matter ejected during the SN is usually referred to as \textit{ejecta}, and moves at typical velocities of thousands of km s$^{-1}$ \citep{Woltjer72}. When this gas begins to interact with the surrounding medium a shell preceded by a forward shock (FS) will 
form, together with a reverse shock (RS) propagating in the ejecta. 
These two shocks enclose a region of hot gas, which is itself divided by a contact discontinuity (CD) -- a boundary separating the shocked ejecta from the shocked ambient medium. This entire structure is generally referred to as a \textit{SNR shell}.\\\\
The NS instead, if strongly magnetised and rapidly rotating, can dissipate its rotational energy emitting a relativistic magnetized pair plasma \citep{Goldreich69} which powers the expansion of a 
relativistic, subsonic bubble in the SN ejecta, known as Pulsar Wind Nebula (PWN).  
Ultimately these NSs can be detected as Pulsars (PSRs).\\\\
PWNe can accelerate electrons and positrons very efficiently \citep{Baring2011}, and they are believed to be important contributors to the Galactic leptonic CR population.
The population of high-energy electrons and positrons injected into the nebula produces broadband non-thermal emission: synchrotron radiation due to the nebula's magnetic field, extending from radio to MeV and typically peaking between optical and X-ray energies, and ICS off ambient photons in the GeV-TeV range. Typical examples of background radiation are the cosmic microwave background radiation (CMB) and infrared photons (IR). Therefore, multiwavelength observations, particularly in X-rays (tracing synchrotron) and VHE $\gamma$-rays (tracing ICS), are crucial for determining key physical conditions of the nebula like the particle energy distribution and the magnetic field strength, thus contributing to the understanding of the PWNe.\\\\
Among the Galactic sources of VHE $\gamma$-rays, in this work we will focus on the PWNe. The interaction of the PWN with the surrounding SNR can be quite complex, with several evolutionary phases, each with its own different observational signatures.\\\\ At early stages, the expanding PWN drives a shock inside the cool ejecta, heating them and producing thermal emission. In this phase -- usually referred to as the \textit{free expansion phase} -- the nebular spectrum is non-thermal and peaks between optical and X-rays due to synchrotron emitting particles \citep{YoungPWN}.\\\\
As soon as the SNR blast wave has swept up an amount of material comparable to the original ejecta, the RS starts to propagate backward to the SNR centre \citep{Chevalier77}. In the presence of a PWN, the RS will interact with the nebula. 
This can cause the PWN to first compress and then \textit{reverberate} (or bounce) and re-expand, leading to a cycle of compressions and re-expansions that can last for thousands of years. 
This phase is usually referred to as the \textit{reverberation phase} and a PWN during this phase is called  \textit{middle-aged}. 
The PWN-SNR interface is Rayleigh-Taylor unstable and subject to the formation of filamentary structures, where the ejecta are mixed with the relativistic fluid inside the PWN \citep{Bucciantini_2004_RTI}.
During the reverberation phase the nebular magnetic field increases, resulting in enhanced synchrotron emission which burns off the highest energy particles.
Due to this, as PWNe age the ratio of ICS to synchrotron luminosity is expected to increase, to the point where most of the emission is in $\gamma$-rays \citep{Slane2017}.\\\\
Ultimately, the central PSR is expected to leave the SNR because of its initial kick velocity, most likely due to the asymmetries of the SN explosion \citep{BOMBACI04}. 
In the ISM the PSR motion is supersonic, and drives a shock, whose shape resembles that of a bow \citep{Gaensler_2006_bow_shock}. 
As a consequence, PWNe are expected to end their lives as bow shock nebulae (BSPWNe). 
Evolved pulsars are also associated with the formation of TeV halos \citep{OldPWNe}, due to high energy particles leaving the BSPWNe. 
Indeed, middle-aged and old PWNe are expected to dominate the TeV emission -- see Fig.$\stick$\ref{gamma_pwne} -- in the Galactic plane \citep{PWNETEV2013}.
\\\\
Predicting and interpreting the VHE emission expected from middle-aged PWNe during reverberation requires appropriate theoretical models for their evolution.
The simplest approach is the one-zone model in the thin shell approximation, which treats the PWN as a uniform relativistic bubble expanding in a spherically symmetric SNR (\citealt{Gunn_Ostriker_71}, \citealt{Chevalier77}, \citealt{Chevalier84}) and sweeping up the ambient material in a thin shell as it expands.
One-zone models are effective for semi-analytical descriptions of the early free expansion phase and provide good estimates for macroscopic quantities in young systems, like the particle energy distribution and the 
spectrum of the emitted radiation \citep{Pacini73}. 
Moreover, they are computationally very efficient given that the dynamics of the nebula is characterised just by its pressure and radius. This notwithstanding, they struggle to capture the complex, non-linear hydrodynamics (HD) during the reverberation phase, initiated when the RS of the SNR interacts with the nebula.
Accurately modelling the evolution of the PWN through reverberation necessitates a fully numerical HD approach, even in 1D. In order to capture both the free expansion phase and the complex dynamics of the interaction between the PWN and the RS a hybrid approach has been  recently developed by the Arcetri Group \citep{REV3}. 
This approach combines a thin shell formalism that describes the free 
expansion phase of a PWN in the ejecta, with an approach based on Lagrangian HD for the later PWN-SNR interaction. 
Like previous approaches, this new technique, despite its accuracy, treats the PWN-SNR system as purely 1D.
While 1D simulations capture the basic compression dynamics, real PWNe are intrinsically 3D systems, often exhibiting asymmetries and instabilities not present in 1D. 
The complex physics of the PWN reverberation phase, often simplified or ignored, is dominated by asymmetries and instabilities poorly described by simple models.
A primary source of asymmetry is the birth kick velocity of the PSR ($\sim 300 \text{ km s}^{-1}$ \citealt{Hansen_Phinney_71}), which displaces it from the SNR center, ensuring a geometrically asymmetric interaction with the RS. This, combined with HD Rayleigh-Taylor instabilities arising from the differences in the PWN-SNR densities (\citealt{BLONDIN01}, \citealt{Bucciantini_2004_RTI}), necessitates the use of detailed multi-dimensional HD simulations to properly model the system (\citealt{VANDER03}, \citealt{Temim_2015NEW}, \citealt{Kolb2017}).
Moving to higher dimensions -- 2D or 3D -- provides a more realistic description but significantly increases the computational cost, making extensive exploration of the parameter space challenging. Indeed, 2D and 3D simulations of PWNe have been done only for very young systems, and mostly focused just on the Crab Nebula (\citealt{Komissarov_2003_ultimate}, \citealt{DelZanna2004CRAB2D}, \citealt{Mignone2013CRAB3D}, \citealt{PORTH2014CRAB}).
\section*{This Work}
\addcontentsline{toc}{section}{This Work}
Making accurate predictions for these systems is crucial, as PWN-SNR complexes serve as valuable cosmic laboratories for studying particle acceleration, relativistic hydrodynamics (RHD) and magneto-hydrodynamics (RMHD), and non-thermal radiation processes, thereby testing fundamental physics in extreme environments.\\\\
This thesis investigates the dynamics of middle-aged PWNe interacting with their host SNRs, focusing on the reverberation phase, developing an approach that takes advantage of the efficiency of one-zone 1D models, to optimise the multidimensional study of  older systems. 
The primary aim is to characterize the physical conditions within these evolving nebulae in order to inform predictions about their late-time spectral and morphological evolution, particularly in the VHE band.\\\\
Individual PWN-SNR systems are defined by numerous physical parameters related to both the progenitor star, the SN explosion, the resulting NS, and the ISM/CSM. 
Exploring this extensive parameter space comprehensively through detailed numerical simulations (even in reduced dimensions) for every possible combination is computationally prohibitive. 
Neverthless, understanding the collective behavior and expected diversities across the entire population of PWN-SNR systems is crucial for interpreting observational surveys and understanding their Galactic contribution to the $\gamma$-ray emission.\\\\
Due to the vast parameter space governing the 
coupled evolution of PWNe and their host SNRs, we have initially characterised the physical conditions of these evolving nebulae, to guide our work on late-time evolution, by synthesising a population, based on the expected distributions of the various parameters that define the problem. 
The subsequent non-linear interaction during reverberation was then analysed using a detailed 1D Lagrangian numerical code, selecting typical cases that span the entire population, and provide a global overview of the expected behaviours.\\\\
Aware of the limitations of 1D, and aiming 
for a more realistic description while still managing computational cost, the results from the 1D evolution were then used as initial conditions for targeted 2D HD simulations of a PWN within its SNR. In this way we have been able to investigate for the first time, in a physically realistic setting, the multi-dimensional dynamical evolution of middle-aged PWNe, and to assess the robustness of previous 1D approaches.\\\\
By developing this multi-step strategy, we show that it is possible to combine several different approaches, and carefully select relevant cases, in order to optimise the study of old systems, over a vast parameter space, while retaining a high level of physical accuracy, necessary if one wants to derive meaningful predictions. 
\chapter{Supernova Remnants and Pulsar Wind Nebulae}
\label{SNE_SNR_CHAPTER1}
\section{Supernovae and Supernova Remnants}
\subsection{Introduction to Supernovae}
\label{SNE_SNRS}
Supernovae (SNe) are extremely luminous events observed in the sky throughout history. Given their immense brightness, they were historically interpreted as the ignition of a new star -- hence the term \textit{nova} -- appearing on the celestial vault. 
Later, to distinguish them from lower luminosity events, they were prefixed the term \textit{super} \citep{Zwicky34}.
Today it is well established that SNe are catastrophic events, where a star explodes driving an outflow expanding in the surrounding medium -- usually referred to as a Supernova Remnant (SNR) -- and possibly leaving behind a compact stellar remnant. 
SNe result from two distinct mechanisms.
Type Ia supernovae are explosions of white dwarfs -- a compact stellar remnant supported by electron degeneracy pressure \citep{Fowler26} -- exceeding the Chandrasekhar mass limit\footnote{This can happen either due to accretion (single degenerate scenario) or merger (double degenerate scenario).}. Type Ia SNe do not leave any stellar remnant, since a white dwarf (WD) exceeding the Chandrasekhar limit is entirely destroyed by thermonuclear instabilities. 
On the other hand, Type II, Type Ib and Type Ic SNe are the results of the collapse of the core in stars with masses above $\simeq 9 M_{\odot}$ (\citealt{Sukhbold+16}, \citealt{Alsabti17}) -- extending as low as $\simeq 8 M_{\odot}$ for Electron Capture Supernovae (\citealt{Nomoto1982}, \citealt{Nomoto1984}). For this reason, these types of SNe are usually referred to as \textit{Core Collapse Supernovae} (CCSNe). In this case, the collapse of the core can lead to the formation of a compact stellar remnant, a neutron star (NS) -- an even more compact remnant supported by neutron degeneracy pressure (\citealt{Tolman39}, \citealt{TOV39}) -- or a black hole (BH).\\\\
The mechanism driving the collapse of the core has been debated for decades. Today it is accepted that for stars with masses $8M_{\odot} \lesssim M \lesssim 20 M_{\odot}$ the collapse of the core can lead to the formation of a NS. In this case, once the stellar core exhausts all its nuclear fuel, it loses support against its own gravity. As a consequence, the core starts shrinking until its density becomes comparable to the nuclear density, eventually exceeding it ($\simeq 10^{15}\stick$g cm$^{-3}$). When this happens, the core stops collapsing and bounces, leading to the formation of a NS. This scenario has been supported by the detection of NSs inside the Crab Nebula (the remnant of SN 1054) and Vela SNR (\citealt{Staelin68}, \citealt{Large68}), and today the association of NSs with SNe is well established. At the same time, the shock wave resulting from the bounce of the core is expected to sweep up the infalling stellar envelope, driving the SN explosion.\\\\
The swept-up envelope expanding in the surrounding medium constitutes the \textit{ejecta} of the SN, and typically carries a kinetic energy $\simeq 10^{51}\stick$erg (\citealt{ColgateWhite66}, \citealt{Bethe1990}, \citealt{Sukhbold+16}). On the other hand, stars with higher masses are expected to collapse directly into a BH leaving no SNR.
\subsection{The Evolutionary Stages of Supernova Remnants}
\begin{figure}[h!]
	\begin{center}
    \includegraphics[width=1.0\textwidth]{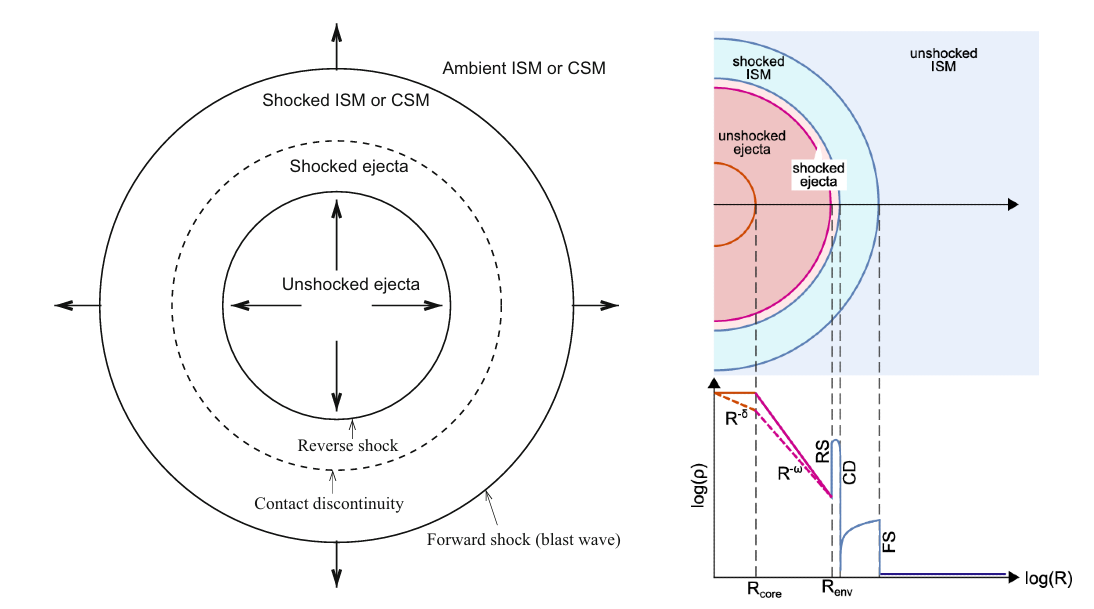}
    \caption{\textit{Left panel}: schematic representations of a spherically symmetric SNR by \cite{Reynolds2017}. \textit{Right panel}: density profile of a SNR expanding in the ISM by \cite{Bandiera+21}, characterized by a core ($R \leq R_{\rm core}$) and an envelope ($R_{\rm core} < R \leq R_{\rm env}$). We anticipate that the density of the cold ejecta scales as $r^{-\delta}$ -- with $\delta \simeq 0$ -- in the core, and as $r^{-\omega}$ -- with $\omega \simeq 12$ -- in the envelope \citep{MatznerMckee99}. Moving outward, the structures enclosed by the SNR shell: the reverse shock (RS), the contact discontinuity (CD) and the forward shock (FS).}
    \label{snr}
    \end{center}
\end{figure}

\begin{figure}[h!]
	\begin{center}
    \fbox{\includegraphics[width=1.0\textwidth]{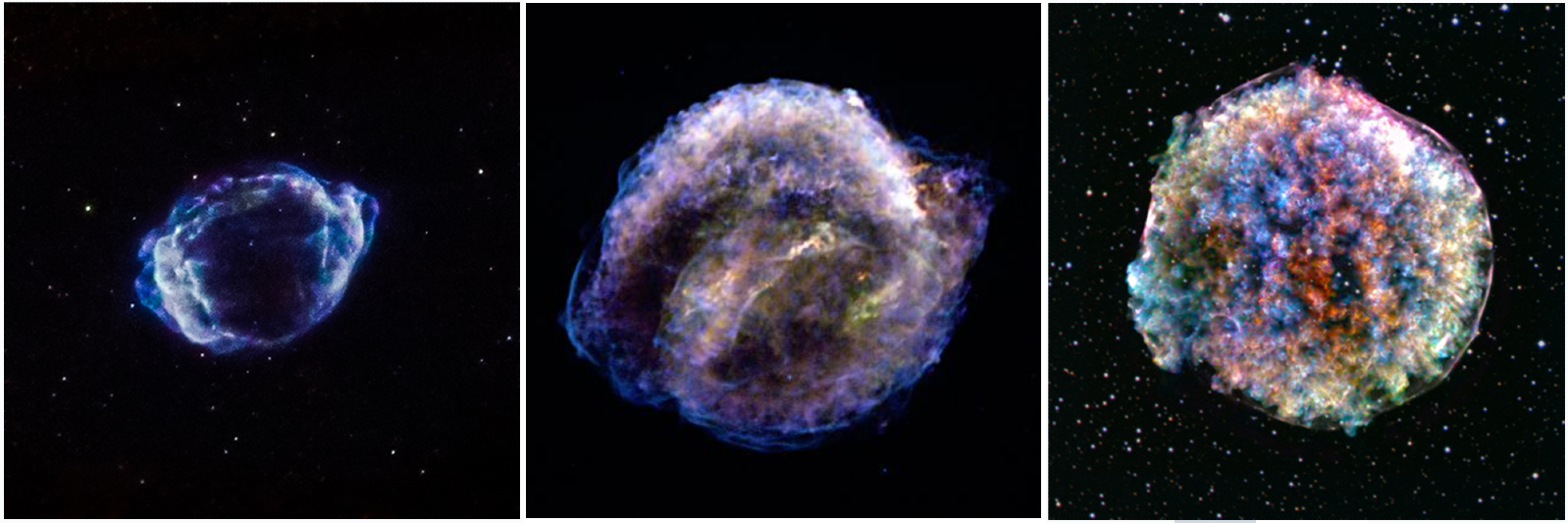}}
    \caption{SNRs observed in X-rays by Chandra. \textit{Left panel}: G1.9+0.3 (associated with a SN dating to $\simeq 1900$ AD). \textit{Center panel}: Kepler SNR G4.5+6.8 (associated with SN 1604). \textit{Right panel}: Tycho SNR G120.1+1.4 (associated with SN 1572).}
    \label{snr_observations}
    \end{center}
\end{figure}

In the simplest scenario, a SNR can be described in terms of spherically symmetric ejecta expanding in the medium surrounding the progenitor star, which in principle can be either the circumstellar medium (CSM) or the interstellar medium (ISM). 
As the SN is ignited, the shock wave resulting from the bounce of the core accelerates the ejecta up to $\simeq [10^3-10^4]\stick${km s$^{-1}$}. 
This SN blast wave -- usually referred to as \textit{forward shock} (FS) -- begins to propagate in the CSM/ISM, heating it up to X-ray temperatures. As the FS expands, it creates a shell of shocked CSM/ISM.
On the other hand, the rapid expansion of the ejecta in the SNR cools them to very low temperatures and rapidly lowers their sound speed. 
The FS decelerates as it sweeps up the ambient medium. 
As the freely-expanding ejecta reach the shocked decelerating ambient medium, since the relative velocity between the two is larger than the speed of sound in the ejecta, a \textit{reverse shock} (RS) forms and starts propagating back into the ejecta re-heating them.
These two shocks enclose an interface separating the shocked ejecta from the shocked CSM/ISM, across which the pressure is roughly constant, usually referred to as \textit{contact discontinuity} (CD) -- see Fig$\stick$\ref{snr}. The region enclosed by the two shock waves is usually referred to as \textit{SNR shell}.\\\\
The evolutionary stage when both FS and RS are present can last hundreds or thousands of years. In this phase the evolution of the SNR depends on the ejecta density profile, which in turn depends both on the density profile of the progenitor star and its interaction with the shock wave that disrupts the star itself. For this reason, this phase is usually referred to as the \textit{ejecta-dominated phase} \citep{Reynolds2017}.
A typical observational signature of this phase is the detection of X-ray radiation from metals -- see Fig.$\stick$\ref{snr_observations}. This emission is expected to arise both from the CSM/ISM -- heated by the FS -- and from the shocked ejecta -- heated by the RS.\\
\\
By the time the FS has swept up a significant amount of mass from the CSM/ISM, it starts decelerating. Numerical simulations show that the radius of the FS, after it has swept up about ten times the mass of the ejecta, approaches the self-similar Sedov solution $\propto t^{2/5}$ \citep{Sedov46}.
Once the RS has fully disappeared and all ejecta have been
reheated, SNRs experience the so-called \textit{Sedov phase}. This typically happens after thousands of years since the SN explosion. In this phase, the medium swept-up by the FS dominates the spectrum of emission. Observational signatures of this phase are X-rays and recombination continuum.\\\\
As the FS decelerates down to $\simeq 200\stick${km s$^{-1}$}, eventually the characteristic cooling time of the gas (which depends on its chemical composition and on its temperature, which is itself a function of the shock speed) becomes comparable with the age of the SNR itself. At this stage, the SNR experiences the so-called \textit{radiative phase}. Radiative cooling is typically dominated by UV, optical and near-infrared fine-structure transitions of C, O, and Fe. During this phase deceleration increases rapidly because of radiative losses, which reduce the thermal pressure driving the expansion of the SNR. This loss of thermal support can possibly lead the FS to decelerate down to the speed of sound of the surrounding medium, causing the dissolution of the SNR \citep{Reynolds2017}.  
\subsection{Dynamics and Structure of Supernova Remnants}
\label{snr_dynamics}
\begin{figure}[h!]
	\centering
    \includegraphics[width=1.0\textwidth]{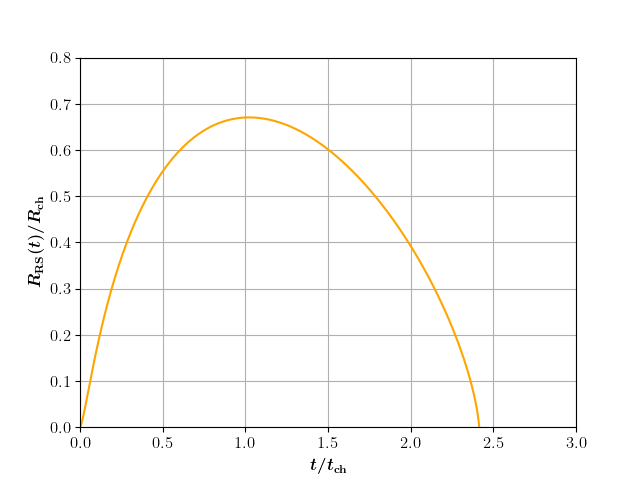}
    \caption{The evolutionary path of the RS inside a SNR for $\delta=0$ and $\omega=12$ \citep{Bandiera+21}.}
    \label{rs_image}
\end{figure}

In this section we will present a simplified model for the evolution of a young spherically symmetric SNR. As the ejecta expand in the surrounding medium, they are cooled down to very low temperatures, making their pressure negligible. In order to predict the structure of the ejecta, one must take into account the equations of mass and momentum conservation of a pressureless, non-relativistic fluid: 
\begin{align}
&\frac{\partial \rho_{\rm ej}}{\partial {t}}+\frac{1}{{r}^2}\frac{\partial}{\partial{r}}
\left[
{\rho_{\rm ej}{r}^2 v_{\rm ej}}
\right]=0,
\label{mass_cons_eje}\\
&\rho_{\rm ej}\frac{\partial v_{\rm ej}}{\partial{t}}+\rho_{\rm ej}v_{\rm ej}\frac{\partial v_{\rm ej}}{\partial{r}}=0,
\label{mom_cons_eje}
\end{align}
where $\rho_{\rm ej}$ and $v_{\rm eje}$ are the density and the velocity of the ejecta. For simplicity let us assume the solutions to have a form $\rho_{\rm ej}\propto r^{-\delta}t^{\xi}$ and $v_{\rm ej}\propto r^{\alpha}t^{\beta}$. Using these relations in Eq.$\stick$(\ref{mass_cons_eje}) and in Eq.$\stick$(\ref{mom_cons_eje}), the solution reads:
\begin{equation}
\xi = \delta-3\quad\quad
\alpha = 1 \quad\quad
\beta = -1.
\end{equation}
The simplest scenario corresponds to a spatially uniform density, implying $\delta=0$. Furthermore, assuming the SNR has not yet swept up a significant amount of mass from the ambient medium, the outer ejecta move at a constant velocity $V_{\rm out}$, and the radius of the remnant reads: $R_{\rm out} = V_{\rm out} t$. Under these assumptions, the density and the velocity of the ejecta scale as:
\begin{equation}
\rho_{\rm ej}\propto t^{-3}\quad\quad
v_{\rm ej}=\frac{r}{t}=V_{\rm out}\frac{r}{R_{\rm out}} .
\label{density_hydro}
\end{equation}
A more rigorous treatment, accounting for both the density of the envelope of the supergiant progenitor and the interaction of the stellar envelope with the shock wave generated by the bounce of the core, yields the following solution for $\rho_{\rm ej}$, derived by \cite{MatznerMckee99}:
\begin{equation}
\rho_{\rm ej}(r,t)\simeq 4\times 10^{-2}\rho_{\rm ch}
\left(
\frac{t}{t_{\rm ch}}
\right)^{-3}
\left[
\left(
\frac{r}{2V_{\rm ch}t}
\right)^{-\frac{\delta}{\alpha_{\rho}}}
+
\left(
\frac{r}{2V_{\rm ch}t}
\right)^{-\frac{\omega}{\alpha_{\rho}}}
\right]^{\alpha_{\rho}}.
\label{rho_eje_99}
\end{equation}
Here $\delta$ and $\omega$ represent the power-law indices characterizing the slope of the inner and outer density regions -- usually referred to as \textit{core} and \textit{envelope}. Specifically, for red supergiants, \cite{MatznerMckee99} find $\omega \simeq 12$, $\delta \simeq 0$ and $\alpha_{\rho} = -4.5$. As a consequence, the density of the ejecta consists of a core where it is approximately constant, and an envelope where it rapidly decreases. 
The quantities $\rho_{\rm ch}$, $t_{\rm ch}$ and $V_{\rm ch}$ are normalization variables representing the characteristic scales of density, time and velocity of the SNR, and their definitions read: 
\begin{align}
t_{\rm ch} &= E_{\rm sn}^{-\frac{1}{2}}M_{\rm ej}^{\frac{5}{6}}\rho_{0}^{-\frac{1}{3}} \simeq 3200
\left[\frac{E_{\rm sn}}{10^{51}\,\text{erg}}\right]^{-\frac{1}{2}}
\left[\frac{M_{\rm ej}}{10M_{\odot}}\right]^{\frac{5}{6}}
\left[\frac{m_0n_0}{{m_{\rm p}}{\text{cm}^{-3}}}\right]^{-\frac{1}{3}} \text{ yrs}, \label{t_ch}\\
V_{\rm ch} &= E_{\rm sn}^{\frac{1}{2}}M_{\rm ej}^{-\frac{1}{2}}\simeq 2200
\left[\frac{E_{\rm sn}}{10^{51}\,\text{erg}}\right]^{\frac{1}{2}}
\left[\frac{M_{\rm ej}}{10M_{\odot}}\right]^{-\frac{1}{2}} 
\text{km s}^{-1}, \label{v_ch}\\
\rho_{\rm ch} &= \rho_{0} = {m_{0} n_0}\label{rho_ch} ,
\end{align}
where $E_{\rm sn}$ represents the kinetic energy of the SN, $M_{\rm ej}$ the mass of the ejecta and $\rho_{0}$ the mass density of the ambient medium, which in turns depends on $n_{0}$ -- the number density -- and $m_0$ -- the mean mass of its particles. It is worth noting that different combinations of $E_{\rm sn}$, $M_{\rm ej}$ and $\rho_{0}$ can lead to equal results both for $t_{\rm ch}$ and $V_{\rm ch}$. As a consequence, SNRs differing in $E_{\rm sn}$, $M_{\rm ej}$ and $\rho_{0}$ can exhibit identical profiles of density, provided that they share the same characteristic scales of time and velocity. This leads their evolution to be self-similar \citep{TM99}.\\\\
However, Eq.$\stick$(\ref{t_ch}), Eq.$\stick$(\ref{v_ch}) and Eq.$\stick$(\ref{rho_ch}) do not represent the only useful physical scales for a SNR. In principle one can also define a \textit{characteristic scale} of length, luminosity and pressure:
\begin{align}
&R_{\rm ch}=M_{\rm ej}^{\frac{1}{3}}\rho^{-\frac{1}{3}}_{0}\simeq 7
\left[
\frac{M_{\rm ej}}{10 M_{\odot}}
\right]^{\frac{1}{3}}
\left[\frac{m_0n_0}{{m_{\rm p}}{\text{cm}^{-3}}}\right]^{-\frac{1}{3}}\text{pc},\label{r_ch}\\
&L_{\rm ch}=E_{\rm sn}t_{\rm ch}^{-1} \simeq 9.8\times 10^{39}
\left[
\frac{E_{\rm sn}}{10^{51}\stick{\text{erg}}}
\right]^{\frac{3}{2}}
\left[
\frac{M_{\rm ej}}{10M_{\odot}}
\right]^{-\frac{5}{6}}
\left[\frac{m_0n_0}{{m_{\rm p}}{\text{cm}^{-3}}}\right]^{\frac{1}{3}}\stick{\text{erg s}^{-1}},\label{L_ch}\\
&P_{\rm ch}=E_{\rm sn}R^{-3}_{\rm ch} \simeq 8.4\times 10^{-8}
\left[
\frac{E_{\rm sn}}{10^{51}\stick{\text{erg}}}
\right]
\left[
\frac{M_{\rm ej}}{10M_{\odot}}
\right]^{-1}
\left[\frac{m_0n_0}{{m_{\rm p}}\text{cm}^{-3}}\right]\text{ dyne cm}^{-2},\label{P_ch}
\end{align} 
and the variables defined in Eq.$\stick$(\ref{t_ch})--Eq.$\stick$(\ref{P_ch}), together with the kinetic energy of the explosion $E_{\rm sn}$ and the mass of the ejecta $M_{\rm ej}$ constitute a set of natural units for a SNR.
Although the solution in Eq.$\stick$(\ref{rho_eje_99}) offers a rigorous description of the ejecta density profile, it is convenient to approximate Eq.$\stick$(\ref{rho_eje_99}) using its asymptotic behavior. In the limit where $\alpha_{\rho}\rightarrow 0$, Eq$\stick$(\ref{rho_eje_99}) reduces to a broken power-law profile:
\begin{equation}
\rho_{\rm ej}(r,t)=
  \begin{cases}
  C_1 r^{-\delta}t^{\delta-3} & \text{for } r \leq R_{\rm c} \\
  C_2 r^{-\omega}t^{\omega-3} & \text{for } r > R_{\rm c},
  \end{cases}
\end{equation}
where $R_{\rm c}$ marks the radius of the transition between the inner flat core and the outer steep envelope. Imposing the continuity of the density at $r=R_{\rm c}$ one finds that $C_2=C_1R_{\rm c}^{\omega-\delta}t^{\delta-\omega}$. Furthermore, assuming the core expands with a characteristic velocity $v_{\rm c}$ -- such that $R_{\rm c} = v_{\rm c} t$ -- the density profile can be rewritten as:
\begin{equation}
\rho_{\rm ej}(r,t)=
  \begin{cases}
  A\left(\frac{r}{v_{\rm c}t}\right)^{-\delta}t^{-3}  & \text{for } r \leq v_{\rm c}t \\
  A\left(\frac{r}{v_{\rm c}t}\right)^{-\omega}t^{-3}& \text{for } r > v_{\rm c}t,
  \end{cases}
  \quad\text{where}\quad
  A=\frac{C_1}{v_{\rm c}^\delta}.
  \label{rho_piecewise}
\end{equation}
The constant $A$ and the velocity of the core $v_{\rm c}$ are not free parameters; they are strictly determined by the physical properties of the SN. Using the definitions of the mass of the ejecta and the kinetic energy of the SN, one finds that:
\begin{align}
&M_{\rm ej}=4\pi\int_{0}^{\infty}\rho_{\rm ej}r^2\text{d}r=\frac{4\pi A v_{\rm c}^{3}(\omega-\delta)}{(\omega-3)(3-\delta)},
\label{mass_ejecta}\\
&E_{\rm sn}=2\pi\int_{0}^{\infty}\rho_{\rm ej}v_{\rm ej}^2r^2\text{d}r=\frac{2\pi A v_{\rm c}^{5}(\omega-\delta)}{(\omega-5)(5-\delta)}.
\label{energy_sn}
\end{align}
It is evident that the slopes must satisfy $\delta < 3$ and $\omega > 5$.
Finally, by combining Eq.$\stick$(\ref{mass_ejecta}) and Eq.$\stick$(\ref{energy_sn}), we obtain the explicit expressions for the velocity of the core and the constant $A$ in terms of $E_{\rm sn}$ and $M_{\rm ej}$:
\begin{equation}
v_{\rm c}=
\sqrt{\frac{2(\omega-5)(5-\delta)E_{\rm sn}}{(\omega-3)(3-\delta)M_{\rm ej}}}
\quad{\text{and}}\quad
A = \frac{(\omega-5)(5-\delta)}{2\pi(\omega-\delta)}\frac{E_{\rm sn}}{v_{\rm c}^{5}}.
\label{vca}
\end{equation}
The interaction of the blast wave of the SN with the surrounding medium depends strictly on the ejecta density profile. Let us assume that at the early stages, the interaction takes place within the envelope of the ejecta and that the density of the surrounding medium scales as $\rho \propto r^{-s}$. In this case, under the hypotheses that $s < 3$, the evolution of the FS, CD and RS is self-similar and their radii scale as \citep{Chevalier82}: 
\begin{equation}
    R \propto t^{\frac{\omega-3}{\omega-s}}.
\end{equation}
Under these hypotheses, the spatial profiles of pressure, density, and velocity can be determined semi-analytically by numerically integrating the hydrodynamics equations in the region confined between the two shocks.
Focusing on the long term evolution of the RS, initially it expands 
outward behind the FS until it will eventually start to propagate backward in radius -- see Fig.$\stick$\ref{snr} and Fig.$\stick$\ref{rs_image}.
\\
Adopting the ejecta density profile of a broken power-law with indices $\delta=0$ and $\omega=12$, the temporal evolution of the RS is well approximated by the numerical solution derived by \citet{Bandiera+21}:
\begin{equation}
R_{\rm RS}(t)=R_{\rm ch}
\left[
\frac{a_1(1-a_2t/t_{\rm ch})^{\xi_1}t/t_{\rm ch}^{\xi_2}}{a_3+a_4(t/t_{\rm ch})+(t/t_{\rm ch})^2}
\right],
\label{r_rs_bandiera}
\end{equation}
where $a_1 = 7.90324$, $a_2=0.414355$, $a_3=0.611396$, $a_4=6.572$, $\xi_1=0.6824$ and $\xi_2=1.5548$. During the early expansion phase, the RS reaches a maximum radius, $R_{\rm max}$, given by:
\begin{equation}
R_{\rm max}\simeq 0.68 R_{\rm ch} \simeq 4.77\left[\frac{M_{\rm ej}}{10 M_{\odot}}
\right]^{\frac{1}{3}}
\left[\frac{m_0n_0}{{m_{\rm p}}\text{cm}^{-3}}\right]^{-\frac{1}{3}}\text{pc},
\end{equation}
at a time $t_{\rm max}$, corresponding to:
\begin{equation}
t_{\rm max} \simeq t_{\rm ch} \simeq 3200
\left[\frac{E_{\rm sn}}{10^{51}\,\text{erg}}\right]^{-\frac{1}{2}}
\left[\frac{M_{\rm ej}}{10M_{\odot}}\right]^{\frac{5}{6}}
\left[\frac{m_0n_0}{{m_{\rm p}}\text{cm}^{-3}}\right]^{-\frac{1}{3}} \text{ yrs}.
\end{equation}
Subsequently, as the mass of the swept-up medium becomes significant, the expansion decelerates. This causes the RS to propagate backward, reaching the center at a time $t_{\rm min}$, given by:
\begin{equation}
t_{\rm min} \simeq 2.5 t_{\rm ch} \simeq 8100
\left[\frac{E_{\rm sn}}{10^{51}\,\text{erg}}\right]^{-\frac{1}{2}}
\left[\frac{M_{\rm ej}}{10M_{\odot}}\right]^{\frac{5}{6}}
\left[\frac{m_0n_0}{{m_{\rm p}}\text{cm}^{-3}}\right]^{-\frac{1}{3}} \text{ yrs}.
\label{rs_inside}
\end{equation}
Once the RS has reached the center of the SNR, all the ejecta have been re-heated and the SNR relaxes into a fully self-similar state, where the radius of the FS evolves as $R_{\rm FS}\propto t^{2/5}$, according to the Sedov solution \citep{Sedov46}. 
%
%
%
%
%
%
%
%
%
%
%
%
%
%
%
%
%
%
%
%
%
%
%
%
\section{Pulsars and Neutron Stars}
\subsection{Discovery and Nature of Pulsars}
As discussed in the introduction, among the Galactic sources emitting in the VHE $\gamma$-ray band, Pulsars (PSRs) represent a fundamental class of objects.
The first PSR was originally detected in the Galaxy as a source of periodic radio pulses \citep{Hewish68}. The brief duration of the pulse at all frequencies ($\Delta t \simeq$ 0.016~s) constrained the size of its source to be smaller than $c \Delta t \sim $ 4.8 $\times 10^3$ km. Moreover, the high regularity of its pulse period ($P\simeq 1.337$~s) suggested initially a possible origin in terms of radial pulsations of an entire compact source. The pulsation period of stars with central densities in the range $[10^5 - 10^{19}]$~g~cm$^{-3}$ was computed by \cite{MeltzerThorne66}. 
The calculation first considered a very massive WD.
At the corresponding density ($10^7$~g~cm$^{-3}$) the fundamental mode has a period of about 8 s - much larger than the one detected. Despite this, the higher modes have periods similar to the emitted signal. 
The second case considered was a neutron star (NS). 
At its typical density -- $[10^{14}-10^{15}]$~g~cm$^{-3}$ -- the fundamental mode has a period of $\simeq 0.1-1$~ms, much smaller than the one from the PSR. As a consequence, the detected signal could be interpreted as a higher order mode of a WD but not as the fundamental mode of a NS.\\\\
As an alternative hypothesis, \cite{Gold68} proposed that these periods were fully consistent with the rotation of a NS. In this scenario, the period of the signal is not related to the pulsation of the source, but to its rotation period. Moreover, the observed signal should be a directional beam rotating like a \textit{lighthouse beacon}. Ultimately, the signal would be registered every time this rotating beam sweeps across the line of sight of the observer.\\\\
Subsequently, the detection of a PSR at the centre of the Crab Nebula (Crab PSR), inside the remnant of SN 1054 \citep{Staelin68} and the measurement of its period (${P} \simeq 33$~ms, \citealt{Lovelace69}) led to the conclusion that this PSR must be a NS, since a WD cannot pulsate this fast and, if rotating, it will exceed the mass shedding limit. 
Moreover, this PSR could not even be a pulsating NS since the period was too high \citep{MeltzerThorne66}. 
This PSR could only be a rotating NS. Thus, this observation provided the definitive proof, solidifying the now well-established consensus that PSRs are rotating NSs.\\\\
From 1969 to 1982 about 300 PSRs were identified in the Galaxy with periods in the range $[0.25-1]$~s  \citep{Manchester17}. Crucial exceptions were the Crab PSR, the Vela PSR (${P}\simeq$~89 ms, \citealt{Large68}) and the  Hulse-Taylor binary PSR (${P}\simeq$~59 ms, \citealt{HulseTaylor75}). 
This entire population of sources, from those with slow periods to the faster ones discovered up to that point, is usually referred to as \textit{canonical} PSRs.
Out of these discoveries emerged the idea that a rotating NS with a magnetic field would lose energy \citep{Pacini67} and emit radiation \citep{Gold68}.
Moreover, these discoveries confirmed the hypothesis that NSs could be born in SNe \citep{Zwicky34} with periods $[10-20]$~ms and gradually slowing down to
periods of order 1~s over millions of years (\citealt{Gold68}, \citealt{RICHARDS1969}, \citealt{RADHAKRISHNAN1969}, \citealt{HUNT1969}, \citealt{Manchester1972}). 
This cosy consensus was shaken by the discovery of a PSR with a rotation period of 1.6~ms \citep{Backer82}, the first \textit{millisecond} PSR (MSP), and to date, approximately 780 MSPs have been discovered.
\subsection{Pulsars Populations: Canonical and Millisecond} 
\label{can_mill}
The two families of PSRs show significantly different properties. 
Canonical PSRs have periods ${P}$ ranging from 30~ms to few seconds, period derivatives $\dot{{P}}$ in the range $[10^{-17}-10^{-12}]$~s~s$^{-1}$ and magnetic fields $\simeq 10^{12}$~G (\citealt{Dirson22}, \citealt{Vidana18}). On the contrary, MSPs have periods $\simeq\stick$ms and generally $\leq$~30~ms \citep{Lian25}, period derivatives $\dot{{P}}$ in the range $[10^{-18}-10^{-21}]$~s~s$^{-1}$ and magnetic fields $\simeq [10^{8}-10^{9}]$~G (\citealt{Manchester17}, \citealt{Vidana18}). The vast majority of the observed canonical PSRs have spin-down ages (${P}/2\dot{{P}}$, see Sect.$\stick$\ref{sec:NSelectrodynamics} for its definition) -- in the range $[1 - 100]$~Myr \citep{Dirson22}, while MSPs appear to be older than the Galactic disk, with typical spin-down ages $\geq$~Gyr \citep{Jiang13}. In order to explain the short rotational period of MSPs and longer age, it has been proposed that MSPs result from the recycling of an old, slowly rotating NS through the accretion from a low mass companion. Ultimately the accretion will lead to the spin-up of the PSR  (\citealt{Radhakrishnan82}, \citealt{Alpar82}) and the burial of the magnetic field \citep{Payne07}.\\\\ 
Currently, the various surveys have found 4303 PSRs in the Galaxy 
-- see the Australia Telescope National Facility (ATNF) Pulsar Catalogue\footnote{Credits: \url{https://www.atnf.csiro.au/research/pulsar/psrcat/catalogueHistory.html}} \citep{Manchester2005}. 
Despite the observation of more than 4000 PSRs, the total number of observable PSRs in the Galaxy is expected to be much higher ($\simeq 1.2 \times 10^{5}$ for canonical PSRs, \citealt{FaucherKaspi06} and $\simeq 3 \times 10^4$ for MSPs, \citealt{Lorimer13}).
The current census of PSRs is heavily concentrated in the Galactic disk, which hosts $\simeq$ 92\% of the total population.
The remaining 8\% 
reside in globular clusters \citep{Lian25}.
The two populations exhibit starkly different demographics. The Galactic disk is dominated by canonical PSRs ($\simeq$ 88\%), with MSPs making up the remaining 12\%.
The globular clusters, in contrast, are almost entirely composed of MSPs ($\simeq$ 93\%), with only 7\% being canonical PSRs.
\subsection{Neutron Stars: Basic Properties}
\label{sec:ns_m_r}
It is now well established that PSRs are NSs, compact objects with a typical radius of $\simeq 10\stick{\text{km}}$, a value originally derived from hydrostatic equilibrium in a degenerate gas of free neutrons (\citealt{Tolman39}, \citealt{TOV39}), and more recently precisely confirmed by observations in binary systems (\citealt{Ozel+09}, \citealt{Guver+10a}, \citealt{Ozel+12}, \citealt{Guver+13}) and masses in the range $[1.18-2.35]M_{\odot}$ (\citealt{Stairs+02}, \citealt{Kramer+06}, \citealt{Freire+11}, \citealt{Ferdman+14}, \citealt{Martinez+15}, \citealt{Ferdman+18}, \citealt{Wu25}).
As explained in the previous section, NSs are expected to be strongly magnetized. This was originally proposed by \cite{Ginzburg64} who argued that the magnetic flux $\Phi$ is conserved during the core collapse of a very massive star. As the radius of the core shrinks, this causes the magnetic field to increase dramatically: 
\begin{equation}
\Phi_{\rm core} \simeq B_{\rm core}R^2_{\rm core} = 
\Phi_{\rm ns} \simeq B_{\rm ns}R^2_{\rm ns} 
\Rightarrow
B_{\rm ns} = B_{\rm core}\frac{R^2_{\rm core}}{R^2_{\rm ns}}\quad , 
\label{B_flux_cons}
\end{equation}
where the subscripts \textit{core} and \textit{ns} denote the initial progenitor star core and final NS values for the magnetic field $B$ and radius $R$, respectively. Specifically, the core radius just before it collapses is $\simeq 10^{4}\stick{\text{km}}$ \citep{Sukhbold+16}.
As a consequence, if the magnetic flux is conserved during the core collapse, it leads the magnetic field of the NS to increase with respect to the core by an amount  $(R_{\rm core}/R_{\rm ns})^2 \simeq 10^{6}$.
This shows that any preexisting magnetic field can be amplified several orders of magnitudes. A simple argument states that, taking a star like the Sun, with a typical dipole magnetic field at the surface of $\simeq 1\stick{\text{G}}$, if one computes the dipolar field at the core, one finds $\simeq 10^6\stick{\text{G}}$. 
As a consequence, using Eq.\stick{(\ref{B_flux_cons})} the magnetic field at the surface of the NS is expected to increase easily up to $10^{12}$\stick{G} (\citealt{Ginzburg64}, \citealt{Gold68}).
This prediction is strongly supported by observations, which measure
the magnetic field at the surface of the NSs to be $\simeq 10^{12}$\stick{G} (\citealt{Pavlov01}, \citealt{Espinoza+2024}, \citealt{Liu_2025}) supporting Ginzburg's original idea.\\\\
In order to give an exhaustive description of the phenomenology of NSs it is necessary to discuss their typical range of masses. 
The first theoretical constraint was originally proposed by \cite{TOV39}, where the predicted upper limit for the mass of a non-rotating NS was 0.7$M_{\odot}$ assuming a gas of free degenerate neutrons.
This paradigm was significantly revised by \cite{Cameron59}, who introduced the crucial role of the nuclear interaction, ruling out the hypothesis of free neutrons. 
This inclusion shifted the predicted limit mass significantly upward, placing it at $\simeq 2M_{\odot}$ for a non-rotating NS. 
This prediction has been supported by observations of many binary NSs with masses in the range $[1.18-1.67] M_{\odot}$ over the years (\citealt{Stairs+02}, \citealt{Kramer+06}, \citealt{Freire+11}, \citealt{Ferdman+14}, \citealt{Martinez+15}, \citealt{Ferdman+18}). 
However, observations of massive NSs reaching up to $2.35 M_{\odot}$ (\citealt{Linares18}, \citealt{Wu25}) required an extension beyond the theory of Cameron. \cite{CookShapiro94} addressed this by modeling rotating NSs in general relativity, showing that rotation provides centrifugal support for NSs. As a consequence, the limit mass of NSs can be up to $20\%$ higher than the non-rotating mass limit. This provides a theoretical framework compatible with the existence of NS with masses $\simeq 2.35 M_{\odot}$. 
\subsection{Pulsar Electrodynamics: the Dipole Spin-Down Model}
\label{sec:NSelectrodynamics}
Having established that PSRs are rotating, highly magnetized NSs, we discuss now their electrodynamics. PSRs emit at most two radio pulses per rotation along the line of sight of an observer. This constrains the magnetic field at the surface to be dominated by the dipole term: 
\begin{equation}
{\boldsymbol{B}}_{\rm ns}(\boldsymbol{r})=
\frac{3(\boldsymbol{{\mu}}_{\rm ns}\cdot\mathbf{r})\mathbf{r}}{|\mathbf{r}|^5}-\frac{{\boldsymbol{\mu}}_{\rm ns}}{|\mathbf{r}|^3}, 
\label{eq_dipole}
\end{equation}
where ${\mu}_{\rm ns} = B_{\rm ns}R^3_{\rm ns}$ represents the magnetic dipole moment of the NS. 
It was \cite{Pacini67} who first proposed that rotation must power the energy loss of a NS through magnetic dipole emission in vacuum: 
\begin{align}
\frac{\text{d}E_{\rm rot}}{\text{d}t}
&=
-\frac{2}{3}\frac{|\ddot{\boldsymbol{\mu}}_{\rm ns}|^2}{c^3}.
\end{align}
where $E_{\rm rot}$ is the rotational energy. If the magnetic axis is inclined an angle $\chi$ with respect to the rotation axis, and the rotation rate is $\Omega_{\rm ns}$, the time derivative of the dipole moment will be $\dot{\boldsymbol{\mu}}_{\rm ns} = \boldsymbol{\Omega}_{\rm ns}\times\boldsymbol{\mu}_{\rm ns}$. One finds $\dot{\mu}_{\rm ns} = \Omega_{\rm ns}\mu_{\rm ns}\sin{\chi}$ and $\ddot{\mu}_{\rm ns} = \Omega^2_{\rm ns}\mu_{\rm ns}\sin{\chi}$. Moreover, given that the magnetic inclination is often unknown, one tipically uses the average of $\sin^2{\chi}$ on the solid angle, which is $\langle\sin^2{\chi}\rangle=2/3$. The NS will then lose rotational energy according to the spin-down equation:
\begin{align}
\frac{\text{d}E_{\rm rot}}{\text{d}t}
&=
\frac{\text{d}}{\text{d}t}
\left[
\frac{1}{2}I_{\rm ns}\Omega ^2 _{\rm ns}
\right]
=
-\frac{2}{3}\frac{|\ddot{\boldsymbol{\mu}}_{\rm ns}|^2}{c^3},
\label{spin_down_energy_magnetic_dipole}\\
&\Rightarrow \dot{\Omega}_{\rm ns} = -K_{\rm ns}\Omega^3_{\rm ns}\quad \text{with} \quad K_{\rm ns} = \frac{2 B^2_{\rm ns}R^6_{\rm ns}\langle\sin^2{\chi}\rangle}{3I_{\rm ns}c^3}.
\label{spin_down_equation_n3}
\end{align}
Measurements of masses and radii suggest that, for an average NS, $M_{\rm ns}\simeq 1.4M_{\odot}$ and $R_{\rm ns}\simeq 10$\stick{km}. The typical moment of inertia of a NS is $I_{\rm ns} \simeq M_{\rm ns}R^2_{\rm ns} \simeq 10^{45}$~g~cm$^{2}$. One can compute the derivative of the rotational period with respect to the time, and using Eq.\stick(\ref{spin_down_equation_n3}) one obtains:
\begin{equation}
P_{\rm ns}=
\frac{2\pi}{\Omega_{\rm ns}}\Rightarrow \dot{P}_{\rm ns}=-2\pi\frac{\dot{\Omega}_{\rm ns}}{\Omega^2_{\rm ns}}. 
\label{P_ns}
\end{equation}
One can use the expression for $K_{\rm ns}$ given by Eq.~{(\ref{spin_down_equation_n3})} in order to find an expression of the magnetic field $B_{\rm ns}$ as a function of $P_{\rm ns}$ and $\dot{P}_{\rm ns}$. Using the typical values of $R_{\rm ns}$, $M_{\rm ns}$, $I_{\rm ns}$ and the average $\langle\sin^2\chi\rangle$ one obtains:
\begin{equation}
B_{\rm ns}=\sqrt{\frac{3I_{\rm ns}c^3P_{\rm ns}\dot{P}_{\rm ns}}{8\pi^2 R^6_{\rm ns}\langle\sin^2\chi\rangle}}\simeq{10^{12}}
\left[\frac{P_{\rm ns}}{0.5\stick{\text{s}}}\right]^{\frac{1}{2}}\left[{\frac{\dot{P}_{\rm ns}}{10^{-15}\stick{\text{s s}^{-1}}}}\right]^{\frac{1}{2}}\stick{\text{G}}.
\label{B_ns_dipole}
\end{equation}
This implies that, from the measurements of $P_{\rm ns}$ and $\dot{P}_{\rm ns}$ one can infer the value of the magnetic field at the surface of the NS. Substituting typical values for $P$ and $\dot{P}$ into Eq.\stick(\ref{B_ns_dipole}) yields $B \simeq 10^{12}$ G for canonical PSRs and $B \simeq 10^{8}$ G for MSPs. These results support the idea that NSs are strongly magnetized.\\\\
In order to account for deviations from pure magnetic dipole emission, the spin-down Eq.$\stick$(\ref{spin_down_equation_n3}) can be generalized to:
\begin{equation}
\dot{\Omega}_{\rm ns}=-K \Omega^n_{\rm ns}, 
\label{general_spin_down_equation}
\end{equation}
where $n$ is called the \textit{braking index}. In the case of a pure magnetic dipole, $n=3$ and $K=K_{\rm ns}$. Performing the time derivative of Eq.\stick(\ref{general_spin_down_equation}) one finds that $n=\ddot{\Omega}_{\rm ns}\Omega_{\rm ns}/\dot{\Omega}^2_{\rm ns}$. Moreover, integrating Eq.\stick(\ref{general_spin_down_equation}) one can relate the age of the NS, $\tau_{\rm ns}$, to the observed period $P_{\rm ns}$, period derivative $\dot{P}_{\rm ns}$ and initial period $P_0 = 2\pi/\Omega_0$:
\begin{align}
\frac{\text{d}\Omega_{\rm ns}}{\text{d}t}
&=
-K \Omega^n_{\rm ns} \Rightarrow 
\tau_{\rm ns}=\frac{\Omega_{\rm ns}}{(1-n)\dot{\Omega}_{\rm ns}}
\left[
1-\left(\frac{\Omega_{\rm ns}}{\Omega_{0}}\right)^{n-1}
\right]\\
\label{fine}
&\Rightarrow \tau_{\rm ns}=\frac{P_{\rm ns}}{(n-1)\dot{P}_{\rm ns}}
\left[
1-\left(\frac{P_0}{P_{\rm ns}}\right)^{n-1}
\right]\\
&\Rightarrow \tau_{\rm ns}=\tau_{\rm c}\left[
1-\left(\frac{P_{0}}{P_{\rm ns}}\right)^{n-1}
\right]
\quad \text{with}
\quad
\tau_{\rm c}=\frac{P_{\rm ns}}{(n-1)\dot{P}_{\rm ns}}.
\end{align}
Assuming that the initial rotation of the NS was much faster than at present, one can neglect the ratio $P_0/P_{\rm ns}$. The remaining term $\tau_{\rm c}$ is usually referred to as the \textit{characteristic age}, and, if $n=3$, $\tau_{\rm c}$ is usually referred to as the \textit{characteristic dipole age}. The typical ages of canonical PSRs are $\simeq$ Myr, while for MSPs one finds typical ages comparable to the age of the universe ($\simeq\stick${Gyr}). This is a clear evidence that MSPs do not evolve with the same mechanism of the canonical ones.\\\\
However, measuring the braking index $n$ is notoriously challenging. The measurement of $\ddot{\Omega}$, given the low signal-to-noise ratio in the timing of old pulsars, requires phase-coherent timing analysis over long periods, exceeding several years. The large scatter in these values found on a large PSR sample by \cite{Parthasarathy+20} is evidence for such complexities. Moreover $n$ might not be constant in time as suggested by the case of PSR B0540-69 for which \citet{Yue+07} reported $n=2.41$; while recent results by \citet{Espinoza+2024}, show how it now approaches 1.5. Thereby, only a limited number of braking indices have been measured robustly.
In all systems for which we have solid measurements $n$ is always different from 3, and in general $n<3$. As a consequence, the canonical magnetic dipole radiation model does not adequately describe the rotational energy loss mechanism of the NS. 
In Table\stick\ref{tab_pulsar_indices}, we report a set of 11 PSRs with measured braking index, according to \citep{Espinoza+16}.
\begin{table}[h!]
\centering 
\begin{tabular}{ccccccc}
\hline
\hline
 PSR & Name &$\nu~[\rm{Hz}]$ & $\dot{\nu}~ [10^{-15}\text{ Hz$^2$}]$& $n$ & Notes & Ref.\\
\hline
\hline
B0531+21 & Crab &  29.9 & $-$377 000 & 2.342(1)& s & $1,2^*,5$\\
B0540-69 & Crab Twin &  19.7& $-$187 000& 2.13(1) & ig &$3, 4^*$,5\\
B0833-45 & Vela &  11.2 & $-$16 000&  1.7(2) & s &5\\
J0537-6910 & - & 62.0 & $-$199 000& 2.75(47)& ig & 5,12 \\
J1119-6127 & - &  2.44& $-$24 050 & 2.684(2)& ig& 5,6,7$^*$\\
B1509-58& - &  6.61 & $-$66 900& 2.832(3)& ig & 5,7$^*$,8\\
J1734-3333 & - &  0.85& $-$1670&0.9(2) & ig & 5,9$^*$\\
B1800-21 & - &   7.48& $-$7530 &  1.9(5) & s & 5,7$^*$\\
B1823-13 & - &   9.85& $-$7310 & 2.2(6) & s & 5,7$^*$\\
J1833-1034 & - & 16.2& $-$52 800 &1.857(1) & ig &5,10\\
J1846-0258&  Kes 75 &  3.06&  $-$66 600& 2.65(1) & ig &5,11\\
\hline
\hline
\end{tabular}
\caption{
The table reports PSRs with measured beaking indices, distinguished by two categories: inter-glitch (ig), where $n$ was calculated between two glitch events$^{\ref{fn:glitch}}$, and secular (s), where the measurement represents the long-term secular evolution. The number in parentheses following the value of $n$ denotes the uncertainty in the last significant digit. References labeled with $^{*}$ are taken from the ATNF catalogue$^{\ref{fn:atnfcat}}$, see also \cite{Manchester2005}. References: (1) \cite{Lyne+15}, (2) \cite{Lin+23}, (3) \cite{Fredman+15}, (4) \cite{Mignani+10}, (5) \cite{Espinoza+16}, (6) \cite{Weltevrede+11}, (7) \cite{Keith+24}, (8) \cite{Livingstone+11}, (9) \cite{Espinoza+11a}, (10) \cite{Camilo+06}, (11) \cite{Livingstone+07}, (12) \cite{Gügercinoğlu+25}.}
\label{tab_pulsar_indices}
\end{table}\\
\refstepcounter{footnote}
\label{fn:glitch}
\footnotetext[\value{footnote}]{A \textit{glitch} is a sudden, abrupt increase in the rotational frequency of a NS, interrupting its steady spin-down trend.}
\refstepcounter{footnote}
\label{fn:atnfcat} 
\footnotetext[\value{footnote}]{Credits: \url{https://www.atnf.csiro.au/research/pulsar/psrcat/}}
\noindent
Eq.$\stick$(\ref{general_spin_down_equation}) can also be used to compute $\Omega_{\rm ns}(t)$ and the resulting loss of rotational energy from the NS as a function of time $t$: 
\begin{align}
&\frac{\text{d}{\Omega}_{\rm ns}}{\text{d}t}=-K \Omega^n_{\rm ns}
\Rightarrow
\Omega_{\rm ns}(t)=\Omega_{0}
\left[
1+\frac{K t(n-1)}{\Omega^{1-n}_{0}}
\right]^{-\frac{1}{n-1}},\nonumber
\\
&\Rightarrow
\dot{E}_{\rm rot}=I_{\rm ns}\Omega_{\rm ns}\dot{\Omega}_{\rm ns}
=-I_{\rm ns}K \Omega^{n+1}_{\rm ns}
=-I_{\rm ns}K\Omega^{n+1}_{0}
\left[
1+\frac{\Omega^n_{0}K t(n-1)}{\Omega_{0}}
\right]^{-\frac{n+1}{n-1}}.
\label{per_trovare_la_spin_down_luminosity}
\end{align}
One can simplify Eq.\stick{(\ref{per_trovare_la_spin_down_luminosity}) using Eq.$\stick${(\ref{general_spin_down_equation})} and, using Eq. \stick{(\ref{P_ns})} for $\Omega$ and $\dot{\Omega}$, the resulting equation for $\dot{E}_{\rm rot}(t)$ reads:
\begin{align}
&\dot{E}_{\rm rot}(t)=I_{\rm ns}\Omega_{0}\dot{\Omega}_0
\left[
1-\frac{(n-1)t\dot{\Omega}_0}{\Omega_0}
\right]^{-\frac{n+1}{n-1}} = \dot{E}_0
\left(1+\frac{t}{\tau_0}\right)^{-\frac{n+1}{n-1}}, \label{general_spin_down_dissipation}\\
&\dot{E}_0 = I_{\rm ns}\Omega_{0}\dot{\Omega}_0 \quad \text{and} \quad \tau_0 = \frac{P_0}{(n-1)\dot{P}_0},
\label{general_E0_tau0}
\end{align}
where $\dot{E}_0$ represents the initial energy loss rate, and $\tau_0$ is defined as the initial spin-down time. 
As expected $\dot{E}_{\rm rot}$ is negative, since the neutron star is losing energy as it slows down. The rotational energy released by the NS powers the so-called spin-down luminosity $L_{\rm sd}(t)$. By convention, this is a positive quantity:
\begin{equation}
L_{\rm sd}(t) \equiv - \dot{E}_{\rm rot}(t).
\end{equation}
Applying this definition to Eq.\stick(\ref{general_spin_down_dissipation}) yields the luminosity evolution:
\begin{equation}
L_{\rm sd}(t)=L_{0} \left( 1+\frac{t}{\tau_{0}}\right)^{-\frac{n+1}{n-1}}
\quad \text{and} \quad L_0 = -I_{\rm ns}\Omega_{0}\dot{\Omega}_0.
\label{initial_sd_luminosity_and_time}
\end{equation}
In the magnetic dipole case ($n=3$), by substituting the definition of $K_{\rm ns}$ using Eq.$\stick$(\ref{spin_down_equation_n3}) and the definition of $\Omega_0$ (Eq.\stick{\ref{P_ns}}) into Eq.\stick{(\ref{initial_sd_luminosity_and_time})} and Eq.\stick{(\ref{general_E0_tau0})} one obtains:
\begin{align}
& L_0 =
\frac{32\pi^4B_{\rm ns}^{2}R_{\rm ns}^{6}\langle\sin^{2}\chi\rangle}{3c^{3}P^4_{0}}\quad \text{and} \quad
\tau_0 = \frac{3 I_{\rm ns}P_0^2 c^3}{16 \pi^2 B_{\rm ns}^2 R_{\rm ns}^6 \langle\sin^2\chi\rangle},\\
& L_0 \simeq 5\times 10^{38}
\left[
{\frac{B_{\rm ns}}{10^{12}\stick{\text{G}}}}
\right]^2
\left[
{\frac{R_{\rm ns}}{10\stick{\text{km}}}}
\right]^{6}
\left[
\frac{P_0}{15 \stick{\text{ms}}}
\right]^{-4}
\stick{\text{erg s}^{-1}},
\label{sd_lmdipole}\\
& \tau_0\simeq 5
\left[
\frac{B_{\rm ns}}{10^{12}\stick{\text{G}}}
\right]^{-2}
\left[
\frac{R_{\rm ns}}{10\stick{\text{km}}}
\right]^{-6}
\left[
\frac{P_{0}}{15\stick{\text{ms}}}
\right]^2
\left[
\frac{I_{\rm ns}}{10^{45}\stick{\text{g cm}^2}}
\right]
\stick{\text{kyr}}.
\label{initial_spin_down_luminosity_magnetic_dipole}
\end{align}
\subsection{Pulsar Electrodynamics: the Pulsar Wind}
\label{sec:psr_wind}
A complete description of the dynamics driving the loss of energy from a NS must also account for the structure of its magnetosphere.  
The first rigorous model addressing this was developed by \cite{Goldreich69}. They demonstrated that a rotating, strongly magnetized NS necessarily loses energy by pulling charges from its surface and launching a relativistic plasma outflow. This outflow, commonly referred to as \textit{PSR wind}, carries away the rotational energy of the NS.\\\\ One can assume that a NS is an ideal conductor and that the magnetic field just underneath the surface is dipolar. As a consequence, the electric field just below the surface will be:
\begin{align}
\boldsymbol{E}_{\rm ns}&=-\frac{\boldsymbol{v}_{\rm ns}}{c}\times{\boldsymbol{B}_{\rm ns}}\Rightarrow E_{\rm ns}
\simeq\frac{\Omega_{\rm ns}R_{\rm ns}}{c}B_{\rm ns}
\label{Electric_field_inside}
\\
&\Rightarrow E_{\rm ns}\simeq
4\times 10^8
\left[\frac{B_{\rm ns}}{10^{12}\stick{\text{G}}}\right]
\left[\frac{R_{\rm ns}}{10\stick{\text{km}}}\right]
\left[\frac{P_{\rm ns}}{0.5\stick{\text{s}^{-1}}}\right]^{-1}
\stick{\text{statV cm}^{-1}}.
\label{MHD_ideal_condition_equation}
\end{align}
It can be shown that, if the NS is surrounded by vacuum, the electric field outside its surface has a component $E_{\parallel}$ along the magnetic field, with a similar strength to Eq.$\stick$(\ref{MHD_ideal_condition_equation}). This component can pull charged particles from the surface of the NS out. These can be enforced to corotate with the NS only within a region where their angular velocity $|\boldsymbol{\Omega}_{\rm ns}\times\boldsymbol{r}|$ does not exceed the speed of light. The boundary of this region is known as the \textit{light cylinder} (LC):
\begin{equation}
|\boldsymbol{\Omega}_{\rm ns}\times\boldsymbol{r}|=c\Rightarrow r_{\rm LC}=\frac{c}{\Omega_{\rm ns}\sin\theta},
\label{light_cylinder_equation}
\end{equation}
expressed in spherical coordinates. Field lines remaining within this radius are closed and corotate with the NS. As a consequence, particles extracted by the electric field along these magnetic field lines can be set into corotational equilibrium. In contrast, particles extracted along field lines extending beyond the light cylinder cannot corotate, and force those field lines to open. Thus, these particles are able to escape the magnetosphere and fly away. The boundary of the open field line region on the surface of the NS can be estimated easily assuming the geometry of an aligned dipolar magnetic field. In this case, in polar coordinates, a field line is described by $r = A_0 \sin^2 \theta$. The constant $A_0$ is the maximum radial extent, which occurs at the equator ($\theta = \pi/2$). Since the last closed field line is defined as the line whose maximum extension is $R_{\rm LC} = c/\Omega_{\rm ns}$, its equation reads:
\begin{equation}
r = \frac{c}{\Omega_{\rm ns}}\sin^2\theta.
\label{Last_closed_line}
\end{equation}
The region around the magnetic pole where open magnetic field lines originates is known as \textit{polar cap} and its extent is:
\begin{equation}
\sin\theta_{\rm cap}\simeq\theta_{\rm cap}=
\sqrt{\frac{R_{\rm ns}}{R_{\rm LC}}}
\simeq 0.02
\left[
\frac{R_{\rm ns}}{10\stick{\text{km}}}
\right]^{\frac{1}{2}}
\left[
\frac{P_{\rm ns}}{0.5\stick{\text{s}}}
\right]^{-\frac{1}{2}}.
\label{polar_cap_equation}
\end{equation}
The particles extracted at the surface of the NS within $\theta < \theta_{\rm cap}$ are expected to leave the magnetosphere - see Fig.$\stick$\ref{open_field_lines} - constituting the so-called PSR wind.
\begin{figure}[h!]
	\begin{center}
    \fbox{\includegraphics[width=1.0\textwidth]{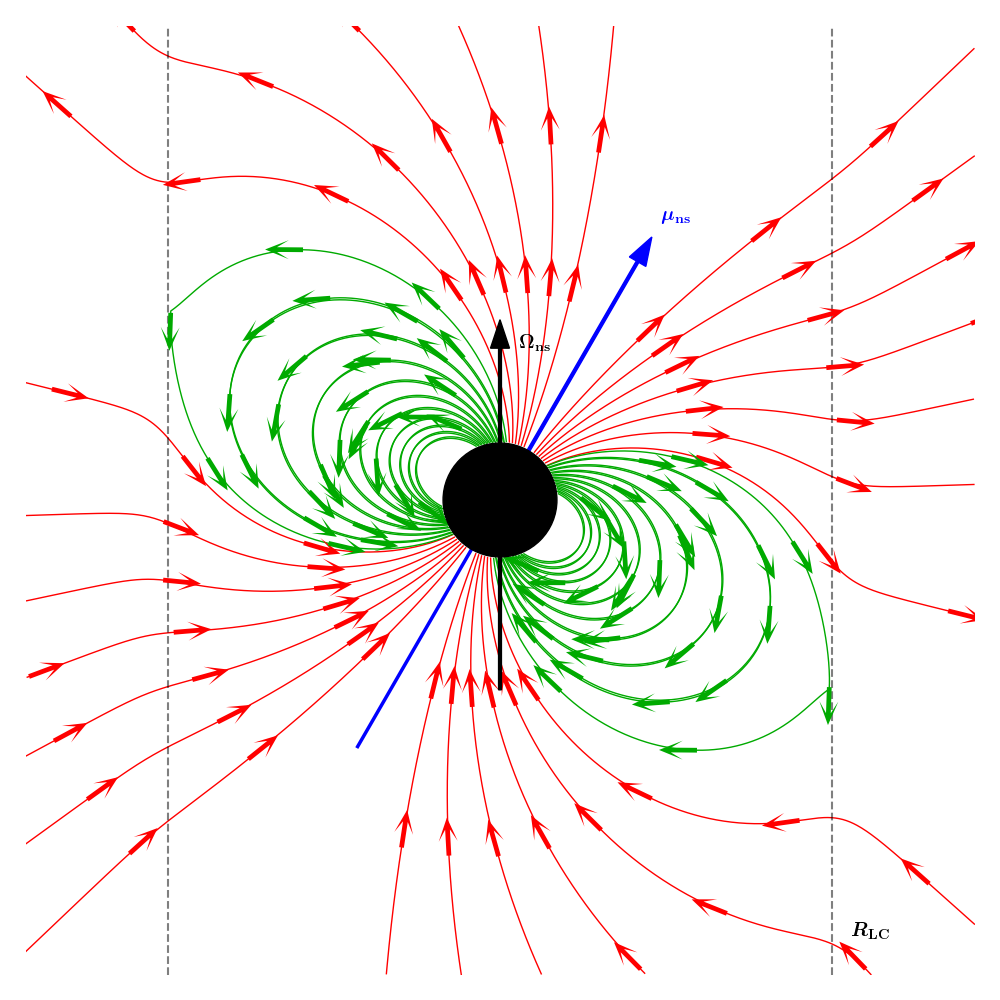}}
    \caption{Schematical representation of the magnetosphere of a NS. $\boldsymbol{\mu}_{\rm ns}$ represents the magnetic dipole moment of the NS and  $\boldsymbol{\Omega_{\rm ns}}$ its rotational axis. The vertical dashed lines mark the light cylinder radius $R_{\rm LC}$. Green lines represent the closed magnetic field lines, extended within $R_{\rm LC}$. On the contrary, red lines represent magnetic field lines extending beyond $R_{\rm LC}$.}
    \label{open_field_lines}
    \end{center}
\end{figure}
Taking into account the geometry of the dipolar magnetic field in Eq.$\stick${(\ref{Electric_field_inside})}, one finds the component of the outer electric field parallel to the magnetic field:
\begin{equation}
{E}_{\parallel}\simeq
10^{8}
\left[
\frac{B_{\rm ns}}{10^{12}\stick{G}}
\right]
\left[
\frac{R_{\rm ns}}{10\stick{\text{km}}}
\right]
\left[
\frac{P_{\rm ns}}{0.5\stick{\text{s}}}
\right]^{-1}
\theta^2
\stick{\text{statV cm}^{-1}},
\end{equation}
and a charged particle accelerated at the polar cap over a distance $\ell$ by ${E}_{\parallel}$ reaches very high Lorentz factor:
\begin{equation}
\gamma\simeq 6\times 10^{7}
\left[
\frac{B_{\rm ns}}{10^{12}\stick{\text{G}}}
\right]
\left[
\frac{R_{\rm ns}}{10\stick{\text{km}}}
\right]^2
\left[
\frac{P_{\rm ns}}{0.5\stick{\text{s}}}
\right]^{-2}
\left[
\frac{\ell}{1\stick{\text{km}}}
\right].
\end{equation} 
Such ultra-relativistic particles will emit curvature radiation along the curved magnetic field lines, with energy:
\begin{equation}
h\nu_{\rm curv} = h\frac{\gamma^3c}{R_{\rm curv}}\simeq 10^{7}\left[
\frac{B_{\rm ns}}{10^{12}\stick{\text{G}}}
\right]^3
\left[
\frac{R_{\rm ns}}{10\stick{\text{km}}}
\right]^5
\left[
\frac{P_{\rm ns}}{0.5\stick{\text{s}}}
\right]^{-6}
\left[
\frac{\ell}{1\stick{\text{km}}}
\right]^3\stick{\text{MeV}},
\label{curvature_radiation}
\end{equation} 
where $R_{\rm curv}$ is the curvature radius of the magnetic field line, and tipically $R_{\rm curv}\simeq R_{\rm ns}$. This radiation will be beamed along the magnetic field. This is the mechanism that is thought to be at the base of the collimated emission in the Lighthouse model.
However, particles lose energy so fast that they cannot really reach such high Lorentz factors, and the curvature radiation photons are typically produced at a much lower energy ($\simeq\stick$GeV).\\\\
Curvature photons, according to Quantum Electrodynamics (QED) are expected to produce virtual pairs $e^{-}$ $e^{+}$ if their energy $h\nu_{\rm curv}$ exceeds $2m_{\rm e}c^2$ in the regions where the magnetic field is very strong. The pair itself will produce curvature and synchrotron photons, giving rise to a \textit{pair production cascade} that continuously injects plasma along the open field lines, sustaining the PSR wind. 
Ultimately, the magnetosphere of the NS will be filled with a charge density screening the parallel electric field:
\begin{equation}
q_{\rm e}=\frac{\boldsymbol{\nabla}\cdot \boldsymbol{E}}{4\pi}\simeq\frac{{E}_{\parallel}}{4\pi R_{\rm ns}}.
\label{density_GJ}
\end{equation}
The presence of a charge density surrounding the NS shows that it cannot be emitting in vacuum - which was the fundamental assumption of the vacuum spin-down model.
A situation will arise where the net electromagnetic force acting on the surrounding plasma will be almost zero:
\begin{equation}
q_{\rm e}\boldsymbol{E}+\frac{\boldsymbol{J}}{c}\times{\boldsymbol{B}} \simeq 0\quad, 
\end{equation}
where $\boldsymbol{J}$ is the current density. This condition is known as \textit{force-free condition}, and it implies that $\boldsymbol{E}\cdot\boldsymbol{B}\simeq 0$, so the parallel electric field almost vanishes.  
In this case, one can determine expressions for $\boldsymbol{E}$ and $\boldsymbol{B}$ in the outer magnetosphere, and find a relation between the energy loss $\dot{E}_{\rm rot}$, the rotation rate $\Omega_{\rm ns}$ and the open magnetic flux $\Phi_{\rm open}$ (\citealt{Michel73}, \citealt{Bogovalov1999}):  
\begin{equation}
\dot{E}_{\rm rot}
=-\frac{2}{3}\frac{\Phi^2_{\rm open}\Omega^2_{\rm ns}}{c}
\simeq
-\frac{[B_{\rm ns}R^2_{\rm ns}\theta^2_{\rm cap}]^2\Omega^2_{\rm ns}}{c}=
-\frac{B^2_{\rm ns}R^6_{\rm ns}\Omega^4_{\rm ns}}{c^3}.
\label{split_monopole_rate}
\end{equation}
In conclusion, the rotational energy of a PSR is carried away by a relativistic MHD pair wind, since a pure vacuum environment is known to be physically inconsistent according to Eq$\stick$(\ref{density_GJ}). Nevertheless, the vacuum magnetic dipole formula is traditionally adopted to estimate this energy loss. This convention persists because both Eq.\stick(\ref{split_monopole_rate}) and Eq.\stick(\ref{spin_down_energy_magnetic_dipole}) yield expressions with the same functional structure. 
%
%
%
%
%
%
%
%
%
%
%
%
%
%
%
%
%
%
%
%
%
%
%
%
\section{Pulsar Wind Nebulae}
\subsection{The Crab Nebula and its Observations}
As we have seen previously, the explosion of a massive star, triggered by the collapse of its core, eventually leaves behind a compact stellar remnant: a NS. If rapidly rotating and strongly magnetized, a NS is expected to lose energy via a relativistic wind of particles, powering the expansion of a relativistic bubble inside the SNR, usually referred to as a Pulsar Wind Nebula (PWN).\\\\ At the early stages of their evolution, PWNe are confined within the parent SNR, but at later times the PSR eventually can leave the SNR, and the PWN will result from its interaction with the ambient medium.
The first evidence that PSRs transfer energy to the surrounding medium came from the Crab Nebula, which is historically considered the prototype of PWNe.
Spectral analysis reveals that while the outer region emits mostly optical thermal emission lines \citep{Minkowski1942}, the inner region is filled with a broadband non-thermal emission, extending from radio \citep{Greenstein1953} to X-ray \citep{Bowyer1964} -- see Fig.$\stick$\ref{crab_spectrum}. More recently emission have been detected at TeV energies \citep{VHE1989} up to PeV \citep{Peng22}. This observational evidence suggested the existence of a central engine powering the high energy emission from the nebula. 
Subsequently, the discovery of the Crab PSR \citep{Staelin68} solidified the idea that the emission from the nebula was related to the presence of a NS inside the SNR. As discussed in Sect.$\stick$\ref{sec:psr_wind}, \cite{Goldreich69} showed that a NS must lose its rotational energy by emitting a strongly magnetized wind of relativistic particles, which are expected to power the nebular emission through synchrotron (\citealt{Shklovskii57}, \citealt{Pacini73}, \citealt{ReesGunn74}) and inverse Compton scattering (ICS, \citealt{Gould65}, \citealt{VHE1989}).
Moreover, recent observations estimate the X-ray luminosity of the Crab Nebula to be $\simeq 10 \%$ of the spin-down luminosity of the PSR \citep{Dirson23}, confirming that the central PSR must power the continuum emission of the nebula itself. 
\begin{figure}[h!]
	\centering
	\includegraphics[width=0.82\textwidth]{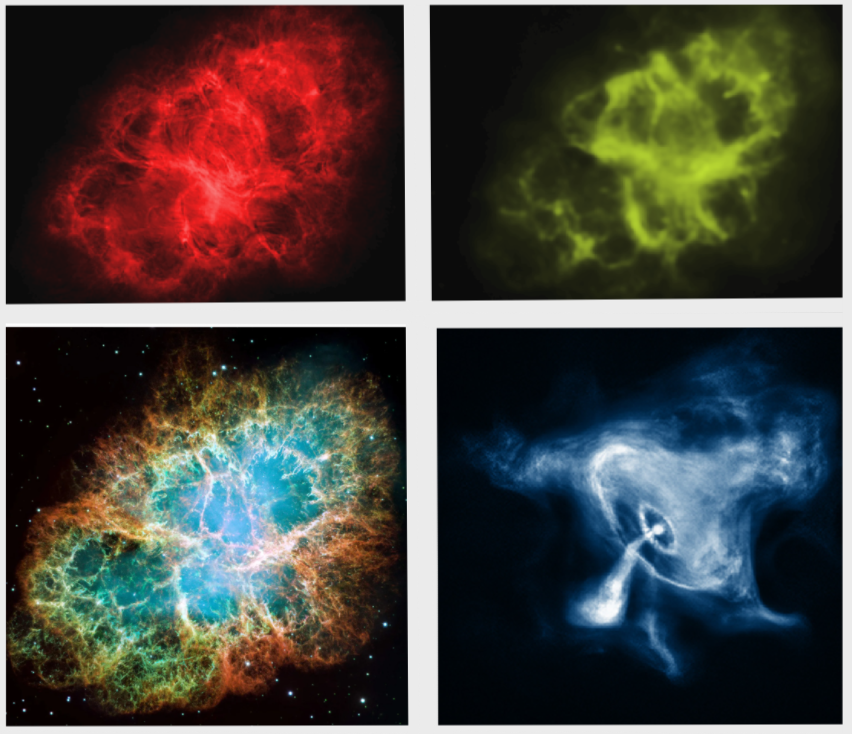}
    \caption{Images of the Crab Nebula in different bands. \textit{Upper-left panel}: Radio (VLA)$^{\ref{fn:crab_optical}}$. \textit{Upper-right panel}: Infrared (Spitzer)$^{\ref{fn:crab_optical}}$. \textit{Lower-left panel}: Optical (HST)$^{\ref{fn:crab_optical}}$. \textit{Lower-right panel}: X-rays (Chandra)$^{\ref{fn:crab_Xrays}}$. Note that panels are not in scale to each other; the radio, infrared and optical regions are significantly larger than the X-ray. }
    \label{crab_spectrum}
\end{figure}
\refstepcounter{footnote}
\label{fn:crab_optical} 
\footnotetext[\value{footnote}]{Credits: \url{https://esahubble.org/images/?search=crab+nebula}}
\refstepcounter{footnote}
\label{fn:crab_Xrays} 
\footnotetext[\value{footnote}]{Credits: \url{https://chandra.harvard.edu/photo/2018/crab/}}
\subsection{Models of Pulsar Wind Nebulae}
\label{ts_section}
In this section we will present a simplified model of the mechanism powering the expansion of the PWN inside the SNR ejecta. Upon leaving the PSR magnetosphere, the relativistic wind reaches the inner, freely expanding ejecta. This interaction leads to the formation of the so-called \textit{wind termination shock} (WTS), where eventually the kinetic energy of the cold, pressureless PSR wind is converted into random kinetic energy of non-thermal particles, resulting in a high-pressure flow downstream. Due to this, the wind downstream drives the expansion of a relativistic bubble -- the PWN -- inside the ejecta, sustained by the continuous injection from the PSR. 
Based on this picture, the location of the wind termination shock ($R_{\rm WTS}$) can be estimated equating the ram pressure of the PSR wind to the pressure inside the nebula:
\begin{equation}
\frac{L_{\rm sd}}{4\pi c R_{\rm WTS}^2}
=
\frac{E_{\rm pwn}}{4 \pi R^3_{\rm pwn}}
\simeq
\frac{L_{\rm sd}t_{\rm pwn}}{4 \pi R^3_{\rm pwn}}
\quad
\Rightarrow
\quad
R_{\rm WTS}\simeq\sqrt{\frac{V_{\rm pwn}}{c}}R_{\rm pwn},
\label{r_ts}
\end{equation}
where $E_{\rm pwn}$ is the total energy inside the nebula -- supplied by the spin-down luminosity of the PSR, ${L}_{\rm sd}$, over the age of the nebula $t_{\rm pwn}$. Furthermore, $R_{\rm pwn}$ and $V_{\rm pwn}$ denote the nebular radius and expansion velocity, respectively, and we have assumed a uniform radial motion law $R_{\rm pwn} = V_{\rm pwn}t_{\rm pwn}$. Note that Eq.$\stick$(\ref{r_ts}), defining the location of the WTS, depends on the expansion velocity of the nebula. As the PWN expands, it sweeps up the surrounding ejecta into a thin, massive shell. Initially, the expansion takes place within the core of the ejecta, containing most of the mass. Since the swept-up shell is thin, we can identify its position and velocity with the nebular radius and expansion speed, respectively. 
Given that the rotational energy of the central PSR powers the pressure of the nebula, driving the expansion of the shell in the ejecta, the kinetic energy of the swept-up ejecta is comparable to the internal energy of the PWN, and this yields:
\begin{align}
& E_{\rm rot}=\frac{I_{\rm ns}\Omega^2_{\rm ns}}{2}\simeq E_{\rm sh}\simeq\frac{M_{\rm ej}V^2_{\rm pwn}}{2},\\
& V_{\rm pwn}\simeq 1400
\left[
\frac{I_{\rm ns}}{10^{45}\text{ g cm}^2}
\right]^{\frac{1}{2}}
\left[
\frac{M_{\rm ej}}{10 M_{\odot}}
\right]^{-\frac{1}{2}}
\left[
\frac{P_{\rm ns}}{0.01\text{ s}}
\right]^{-1}\text{km s}^{-1}. 
\label{v_pwn}
\end{align}
Combining Eq.$\stick$(\ref{v_pwn}) and Eq.$\stick$(\ref{r_ts}) one gets $R_{\rm WTS}\simeq R_{\rm pwn}/20$. This implies that the WTS is not static; rather, as the PWN expands, it follows the expansion of the nebula itself. Specifically, X-ray observations of the Crab Nebula reveal a dark central region -- see Fig.$\stick$\ref{crab_spectrum} -- which does not contain any emitting material, and it is thought to represent the inner cavity occupied by the relativistic wind in the upstream region of the WTS.
\subsection{The Nebular Flow}
Modeling the post-shock flow requires solving the relativistic magneto-hydrodynamics (RMHD) jump conditions across the WTS. Far from the light cylinder and up to the WTS, the magnetic field carried by the PSR wind is expected to be almost purely toroidal. Assuming a purely radial outflow, the magnetic field is perpendicular to the wind velocity. Under these hypotheses, solving the RMHD equations across the WTS yields the downstream flow velocity, $v_{\rm d}$, valid for young PWNe confined within their parent SNR \citep{KennelCoroniti84a}:
\begin{equation}
\frac{v_{\rm d}}{c}=\frac{1+2\sigma+\sqrt{1+16\sigma+16\sigma^2}}{6(1+\sigma)}
\rightarrow
\begin{cases}
1 & \text{ if  }\sigma \gg 1\\
1/3 &\text{ if }\sigma \ll 1
\end{cases}
\quad
\text{where}
\quad
\sigma=\frac{B_{\rm u}^2}{\rho_{\rm u}\gamma_{\rm u}c^2}.
\end{equation}
The parameter $\sigma$ represents the so-called \textit{magnetization} of the PSR wind (the ratio of Poynting flux over kinetic energy flux), with the subscripts \textit{d} and \textit{u} denoting the regions downstream and upstream of the WTS, respectively. 
If the wind is strongly magnetized upstream, the shock is inefficient at decelerating the flow; conversely, low magnetization allows the flow to decelerate down to $c/3$. 
To determine the values of $\sigma$ compatible with the physical system, we must consider the evolution of the flow towards the boundary of the nebula. Assuming a radial flow from the WTS to the boundary of the nebula, the solution of the RMHD equations yields the expansion velocity:
\begin{equation}
V_{\rm pwn}\simeq
\begin{cases}
c & \text{ if } \sigma \gg 1 \\
\sigma c/3 & \text{ if } \sigma \ll 1 
\end{cases}. 
\end{equation}
It is evident that requiring the nebular flow to match the expansion velocity at the boundary of the PWN implies a weakly magnetized wind.
A strongly magnetized wind would result in a flow remaining relativistic up to the boundary, incompatible with the confinement of the PWN inside the SNR. This discrepancy between the high magnetization expected from the PSR wind theory and the low magnetization required to match the boundary conditions in PWNe is historically known as the \textit{$\sigma$-paradox}. Specifically, \cite{KennelCoroniti84a} derived a value of $\sigma \simeq 10^{-3}$.
To explain such low magnetizations, several solutions involving magnetic field dissipation have been proposed. Prominent among these is \textit{driven reconnection}, proposed by \cite{Lyubarsky2003} in a 1D framework and subsequently sustained by \cite{Sironi2012} via 2D and 3D simulations. Alternative scenarios invoked the development of the \textit{current-driven kink instability} in 3D \citep{Mizuno2011} -- allowing $\sigma > 1$ in the region of the PSR wind together with a small magnetization in the downstream region -- or the conversion of magnetic energy into downstream turbulence in 2D \citep{Lemoine2016}. 
Nevertheless, the $\sigma$-\text{paradox} remains a subject of debate, and the dominant mechanism responsible for the magnetic field dissipation is currently unknown.
\subsection{The Nebular Emission}
The radiation observed from PWNe can be broadly classified into two distinct components: the non-thermal emission originating from the population of relativistic particles filling the nebula, and the thermal radiation produced by the ejecta swept up by the nebula itself \citep{Slane2017}.\\\\
The non-thermal component is ultimately powered by the PSR wind. Upon crossing the WTS, the flow is decelerated and particles are eventually accelerated to ultra-relativistic Lorentz factors.
Electrons and positrons then produce the observed nebular spectrum 
via synchrotron radiation -- interacting with the nebular magnetic field -- and ICS off ambient photons. 
For electrons with energy $E_{\rm e}$, the characteristic energy of emitted synchrotron photons is:
\begin{equation}
h\nu_{\rm sync}\simeq 2.2
\left[
\frac{E_{\rm e}}{100 \text{ TeV}}
\right]^2
\left[
\frac{B_{\rm pwn}}{10\stick\mu\text{G}}
\right]
\stick
\text{keV},
\label{e_sync}
\end{equation}
where $B_{\rm pwn}$ is the nebular magnetic field, and the corresponding synchrotron cooling time-scale is:
\begin{equation}
\tau_{\rm sync}
\simeq
820
\left[
\frac{E_{\rm e}}{100\stick{\text{TeV}}}
\right]^{-1}
\left[
\frac{B_{\rm pwn}}{10\stick\mu\text{G}}
\right]^{-2}
\stick
\text{ yr.}
\label{t_sync}
\end{equation}
In contrast, IC scattering occurs when a relativistic electron or positron interacts with a low-energy ambient photon, typically from the 
Cosmic Microwave Background (CMB) or the infrared background. 
This process results in a significant boost to the photon energy, and the characteristic energy of the upscattered photon is given by:
\begin{equation}
h\nu_{\rm IC}\simeq
0.64
\left[
\frac{E_{\rm e}}{10\,\text{TeV}}
\right]^2
\left[
\frac{h\nu_{\rm bg}}{h\nu_{\rm cmb}}
\right]
\,\text{TeV},
\label{e_ic}
\end{equation}
where the subscript \textit{bg} denotes a general background photon, and $h\nu_{\rm cmb}\simeq 0.3\stick{\text{meV}}$. In the case of scattering off photons with energy density $U_{\gamma}$, the characteristic cooling time is:
\begin{equation}
\tau_{\rm IC}\simeq 200\
\left[
\frac{E_{\rm e}}{10\stick\text{TeV}}
\right]^{-1}
\left[
\frac{U_{\gamma}}{U_{\rm cmb}}
\right]^{-1}\stick\text{kyr,}
\label{t_ic}
\end{equation}
where $U_{\rm cmb}\simeq 0.26\stick\text{eV cm}^{-3}$. Comparing Eq.$\stick$(\ref{e_sync}) and Eq.$\stick$(\ref{e_ic}), it is evident that synchrotron radiation and IC scattering are responsible for the radio to X-ray and VHE $\gamma$-ray spectral components of the nebular spectrum, respectively -- see Fig.$\stick$\ref{crab_sync_ics}.
\begin{figure}[h!]
	\centering
   \includegraphics[width=1.0\textwidth]{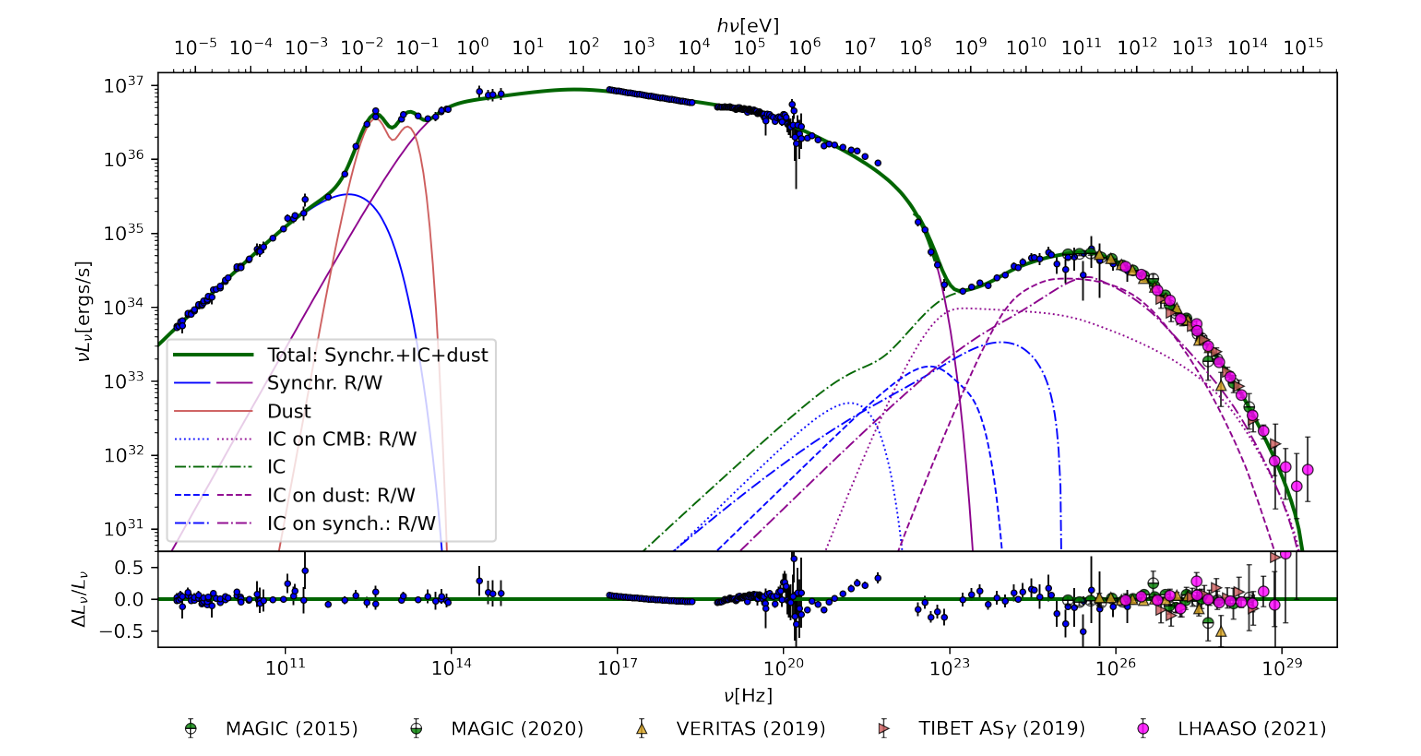}
    \caption{Spectral energy distribution of the Crab Nebula, comparing observational data with the best-fit model (green line, \citealt{Dirson23}). The model sums thermal dust emission with the synchrotron and IC radiation produced by two distinct electron populations: radio (R) and wind (W) electrons. The bottom panel shows the relative residuals.}
    \label{crab_sync_ics}
\end{figure}
Furthermore, the cooling timescales in Eq.$\stick$(\ref{t_sync}) and Eq.$\stick$(\ref{t_ic}) dictate the evolutionary path of the nebular emission. In young PWNe, radiative losses are dominated by synchrotron emission. Despite this, as PWNe age and the magnetic field decays, the synchrotron efficiency drops significantly, and evolved PWNe are expected to become increasingly dominated by IC emission.\\\\
While the emission inside the PWN is dominated by non-thermal processes, as the PWN expands in the ejecta, it drives a shock which compresses and heats the material. 
The resulting thermal radiation combines emission from hot, shocked ejecta with continuum emission from dust, condensed from the cold ejecta in the early expansion phase. 
The properties of the thermal emission depend on the velocity of the shock wave driven by the PWN in the ejecta, which is ultimately governed by the spin-down luminosity of the central PSR. Specifically, for slow shocks the emission may be observed in IR and optical bands -- as seen in the Crab Nebula -- whereas for faster shocks the emission may appear in the X-ray band, as observed in 3C 58 \citep{Slane2017}.  
\subsection{The Nebular Evolution}
The observational characteristics of PWNe, while dictated by the injection and cooling of relativistic particles and the nebular magnetic field, are strongly coupled to the dynamics of the parent SNR.\\\\
The evolution of a PWN is driven by the continuous injection of energy from the central PSR, which at the early stages is confined inside the SNR. The characteristic time-scale governing this phase is the initial spin-down time of the PSR, $\tau_0$. For a young PWN, with an age $t_{\rm pwn} \ll \tau_0$, the PSR spin-down luminosity $L_{\rm sd}(t)$ can be approximated as constant, so that $L_{\rm sd}(t)\simeq L_0$. During this phase, assuming a spherically symmetric PWN expanding in the ejecta, its radius evolves as \citep{Chevalier77}:
\begin{align}
R_{\rm pwn}(t)
&\simeq
1.5
\left(
\frac{L_0^2 E_{\rm sn}^3}{M_{\rm ej}^5}
\right)^{\frac{1}{10}}
t^{\frac{6}{5}}\\
&\simeq
1.1
\left[
\frac{L_0}{10^{38}\stick\text{erg s}^{-1}}
\right]^{\frac{1}{5}}
\left[
\frac{E_{\rm sn}}{10^{51}\stick\text{erg}}
\right]^{\frac{3}{10}}
\left[
\frac{M_{\rm ej}}{10 M_{\odot}}
\right]^{-\frac{1}{2}}
\left[
\frac{t}{\text{kyr}}
\right]^{\frac{6}{5}}
\text{pc},
\label{r_chev}
\end{align}
where we recall that $E_{\rm sn}$ is the kinetic energy of the SN and $M_{\rm ej}$ the mass of the ejecta. This phase is usually referred to as the \textit{free expansion phase}, and its observational signatures comprise synchrotron emission from the nebula and thermal emission from the shocked ejecta. Additionally, the PWN/ejecta interface is expected to exhibit Rayleigh-Taylor instabilities, driven by the strong density contrast between the light, relativistic plasma inside the nebula and the heavier, swept-up ejecta. These instabilities typically manifest themselves in the form of filaments observed both in optical and infrared bands \citep{Sankrit98}, resulting in a very complex morphology.
As discussed in Sect.$\stick$\ref{snr_dynamics}, as the blast wave of the SN sweeps up material from the ambient medium, it decelerates, driving a reverse shock (RS) back into the ejecta -- see Fig.$\stick$\ref{snr_pwn}.
\begin{figure}[h!]
	\centering
    \includegraphics[width=1.0\textwidth]{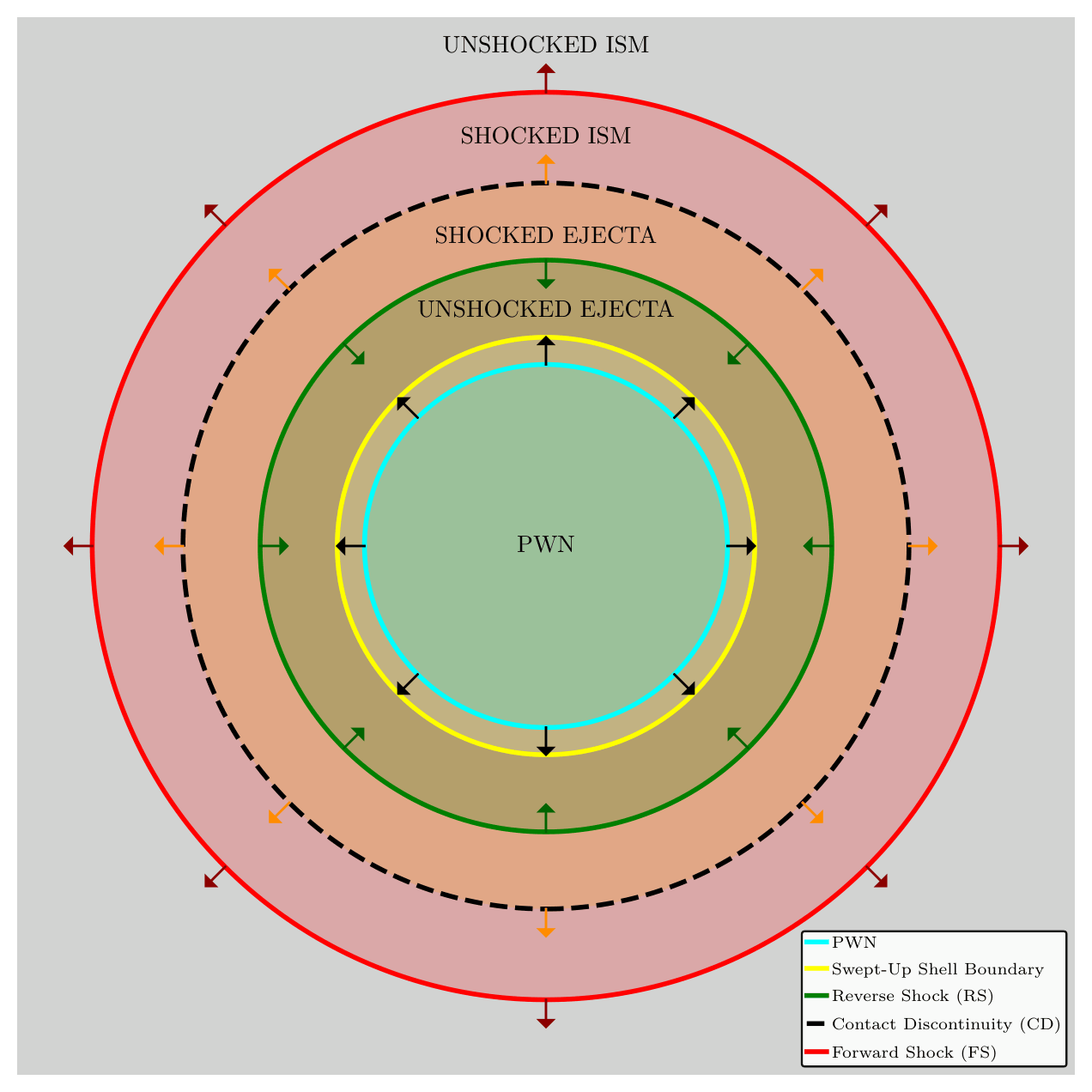}
    \caption{Schematic representation of a spherically symmetric PWN expanding in the parent SNR.}
    \label{snr_pwn}
\end{figure}
This shock eventually is expected to reach the center of the SNR on a time-scale $t_{\rm min}$, given by Eq.$\stick$(\ref{rs_inside}). In the presence of an expanding PWN, the inward-moving RS collides with the nebula itself after a time $t<t_{\rm min}$, typically a few thousand years. At this point the PWN experiences the so-called \textit{reverberation phase}, which leads to a cycle of compressions and re-expansions persisting for thousands of years. Crucially, the compression causes an increase in the nebular magnetic field, boosting the synchrotron emission of the nebula and rapidly depleting the highest energy particles \citep{Slane2017}. Concurrently, the RS/PWN interaction triggers Rayleigh-Taylor instabilities, leading to the formation of filamentary structures and mixing the ejecta material with the relativistic fluid filling the nebula. The RS/PWN interaction takes place on a timescale short compared to the lifetime of the SNR: for this reason, it is reasonable to assume that many observed PWNe are currently experiencing the reverberation phase. Specifically, remnants such as G328.4+0.2  \citep{Gelfand2007}, MSH 15-56 \citep{Temim_2013} and G327.1-1.1 \citep{Temim_2015NEW} exhibit complex structures, indicative of the RS/PWN interaction. Moreover, observations of extended sources in the TeV $\gamma$-ray band (\citealt{HESS_2006_VELAX}, \citealt{HESS_2006_RINGS}, \citealt{HESS_2006_THIRD_OLD_PSR}, \citealt{Lopez16}) indicate that many of these objects correspond to PWNe evolved beyond the reverberation phase.\\\\
As a PWN ages, the ratio of IC to synchrotron luminosity is expected to increase due to the declining nebular magnetic field. This results in a very extended and brighter TeV emission, with a fainter X-ray counterpart, as detected in HESS J1825-137 (\citealt{Slane2017}, \citealt{ABEYSEKARA2020102403}). In the end, since pulsars typically possess high birth kick velocities ($V_{\rm psr}\simeq 300\text{km s}^{-1}$, \citealt{Hansen_Phinney_71}) the central source is expected to eventually leave the parent SNR. Equating the displacement of the PSR with the radius of the SNR in the Sedov phase, yields the typical escape time $t_{\rm esc}$:
\begin{align}
&R_{\rm psr}=V_{\rm psr}t_{\rm esc}=R_{\rm snr}=
\left(
\frac{E_{\rm sn}}{\rho_{0}}
\right)^{\frac{1}{5}}t_{\rm esc}^{\frac{2}{5}},\\
& t_{\rm esc}\simeq 200
\left[
\frac{E_{\rm sn}}{10^{51}\stick\text{erg}}
\right]^{\frac{1}{3}}
\left[
\frac{V_{\rm psr}}{300\stick\text{km s}^{-1}}
\right]^{-\frac{5}{3}}
\left[
\frac{\rho_0}{m_{\rm p}\text{cm}^{-3}}
\right]^{-\frac{1}{3}}
\stick{\text{kyr}}.
\end{align}
As the PSR leaves the parent SNR, it starts propagating in the ambient medium, where its motion becomes supersonic. Moreover, the relative motion of the medium sweeps the PSR wind back, driving the formation of a so-called \textit{bow shock} structure. The interaction of the shock with the neutral hydrogen leads to the emission of H$\alpha$ radiation -- which is observed in a few systems, like the Guitar Nebula \citep{Chatterjee02}. Therefore, a sizeable fraction of PWNe is expected to evolve into bow shock nebulae (BSPWNe) in the late stages of their life. 

\subsection{Dynamical Evolution in the One-Zone Model}
\label{one_zone_model_early}
The simplest approach to describe the evolution of a PWN inside the SNR ejecta consists in assuming spherical symmetry and in neglecting any structure inside the nebula -- such as the wind region or the WTS. In this approximation -- usually referred to as the \textit{one-zone model} -- the PWN dynamics is fully characterised just by its radius and inner pressure as functions of time, $R_{\rm pwn}(t)$ and $P_{\rm pwn}(t)$ respectively. Moreover, as discussed before, during the free expansion phase, the nebula sweeps up a thin, massive shell inside the ejecta, whose thickness $\delta R_{\rm sh}$ is typically a few percent of $R_{\rm pwn}$ \citep{Jun_1998}. In this case one can approximate the outer radius of the shell as being equal to $R_{\rm pwn}$. For this reason, during the early expansion phase, we will work in the so-called \textit{thin-shell approximation}, neglecting any internal structure of the ejecta compressed by the nebula. The combination of these approaches allows one to efficiently characterize both the dynamics and the spectral evolution of young PWNe. 
Focusing on the dynamics of the nebula, the evolution of the swept-up shell is governed by the conservation of mass and momentum \citep{Rev1}: 
\begin{align}
&\frac{\text{d}M_{\rm sh}(t)}{\text{d}t}
=
4\pi R^{2}_{\rm pwn}(t)
\rho_{\rm ej}(R_{\rm pwn},t)
\left[
V_{\rm pwn}(t)
-
v_{\rm ej}(R_{\rm pwn},t)
\right],
\label{eq_ts_mass}\\
&M_{\rm sh}(t)\frac{\text{d}v_{\rm pwn}(t)}{\text{d}t}=F(t),
\label{eq_ts_momentum}
\end{align}
where $M_{\rm sh}(t)$ denotes the mass within the shell, $V_{\rm pwn}(t)$ the expansion velocity of the nebula, and $F(t)$ the force exerted on the shell, given by:
\begin{equation}
\begin{split}
F(t) &= 4\pi R^{2}_{\rm pwn}(t)[ P_{\rm pwn}(t) - P_{\rm ej}(R_{\rm pwn},t) \ +\\
&\qquad\qquad\qquad- \rho_{\rm ej}(R_{\rm pwn},t) \left(V_{\rm pwn}(t) - v_{\rm ej}(R_{\rm pwn},t)\right)^2 ].
\end{split}
\end{equation}
We recall that $v_{\rm ej}(R_{\rm pwn}, t)$, $\rho_{\rm ej}(R_{\rm pwn}, t)$ and $P_{\rm ej}(R_{\rm pwn}, t)$ represent the velocity, the density and the pressure of the ejecta computed at $R=R_{\rm pwn}$, respectively, while the term $\rho_{\rm ej}(V_{\rm pwn}-v_{\rm ej})^2$ represents the ram pressure exerted by the ejecta on the shell, and we have assumed $V_{\rm pwn} > v_{\rm ej}$. This arises from the relative velocity between the two components, as the shell acts as a moving piston sweeping through the surrounding medium. To complete the dynamical description, we must also account for the conservation of energy in the PWN:
\begin{equation}
L_{\rm sd}(t)
=
\frac{\text{d}E_{\rm pwn}(t)}{\text{d}t}
+
4\pi R^{2}_{\rm pwn}(t)V_{\rm pwn}(t)P_{\rm pwn}(t)
+
Q_{\rm rad}(t),
\label{eq_ts_energy}
\end{equation}
where $E_{\rm pwn}(t)$ denotes the internal energy of the nebula and $Q_{\rm rad}(t)$ the radiative losses term. The system is powered by the pulsar spin-down luminosity $L_{\rm sd}(t)$, given by Eq.~(\ref{initial_sd_luminosity_and_time}), which sustains the change in the internal energy of the PWN, the work done by the swept-up shell, and the emission of the system. 
In addition, since PWNe are filled with relativistic plasma, one might assume a relation between the total internal energy and the pressure, given by:
\begin{equation}
P_{\rm pwn}=\frac{E_{\rm pwn}}{4\pi R^3_{\rm pwn}}.
\label{rel_pwn}
\end{equation}
If one neglects radiative losses, Eqs.$\stick$(\ref{eq_ts_mass}), (\ref{eq_ts_momentum}) and (\ref{eq_ts_energy}) form a set of ordinary differential equation that can be applied to the free expansion phase, when a PWN expands in the unshocked, pressureless ejecta. 
In this case, the velocity and the density of the ejecta are given by Eq.$\stick$(\ref{density_hydro}) and Eq.$\stick$(\ref{rho_piecewise}), respectively. 
However, the mass, momentum and energy conservation equations become singular for $R\to 0$ and $t \to 0$; for this reason, in seeking a numerical solution, they must be initialized at a finite time.\\\\ 
It is possible to show that at the very early stages of the evolution -- when the spin-down luminosity is approximately constant -- the solutions are analytical. Dimensionality arguments suggest that the expansion of the PWN should be self-similar, and should follow a power-law in time, $R_{\rm pwn} \propto t^\alpha$. Inserting this relation into the system of equations for mass, momentum, and energy yields the following expression for the nebular radius:
\begin{equation}
R_{\rm pwn}(t)=C(\delta,\omega)
\left(\frac{{L^2_0} {E^{3-\delta }_{\rm sn}}}{{M^{5-\delta}_{\rm ej}}}\right)^{\frac{1}{2(5-\delta) }}t^{\frac{6-\delta}{5-\delta}}, 
\label{ini_rpwn}
\end{equation}
with the constant $C(\delta, \omega)$ corresponding to:
\begin{equation}
C(\delta,\omega)=
\left(
\frac{2^{\delta-3} [(2\delta-11)(2\delta-9)]^2 (\omega-5)^{\delta-3} (\omega-3)^{5-\delta}}{(\delta-3)^{\delta-3} (\delta-5)^{9-\delta} (\omega-\delta)^2}
\right)^{\frac{1}{2(\delta-5)}}.
\end{equation}
Here we recall that the density of the ejecta scales as $r^{-\delta}$ in the inner, flat core, and as $r^{-\omega}$ in the outer, steeper envelope, with $\delta\simeq 0$ and $\omega\simeq 12$  \citep{MatznerMckee99}. Notably, in this specific case, Eq.~(\ref{ini_rpwn}) has the same functional structure of the free-expansion solution described by Eq.~(\ref{r_chev}). Performing the time derivative of Eq.$\stick$(\ref{ini_rpwn}) yields the expression of the expansion speed:
\begin{align}
V_{\rm pwn}(t)&
\simeq 1.8
\left(
\frac{{L^2_0} {E^{3}_{\rm sn}}}{{M^{5}_{\rm ej}}}
\right)^{\frac{1}{10}}t^{\frac{1}{5}}\nonumber\\
& \simeq 1200
\left[
\frac{L_0}{10^{38}\stick{\text{erg s}^{-1}}}
\right]^{\frac{1}{5}}
\left[
\frac{E_{\rm sn}}{10^{51}\stick{\text{erg}}}
\right]^{\frac{3}{10}}
\left[
\frac{M_{\rm ej}}{10M_{\odot}}
\right]^{-\frac{1}{2}}
\left[
\frac{t}{\text{kyr}}
\right]^{\frac{1}{5}}\!\!\stick{\text{km s$^{-1}$}}.
\label{v_ts}
\end{align}
Furthermore, it is possible to show that the mass of the shell and the inner pressure are also self-similar and scale as a power-law in time. Inserting Eq.$\stick$(\ref{ini_rpwn}) in Eq.$\stick$(\ref{eq_ts_mass}) and Eq.$\stick$(\ref{eq_ts_energy}), and using Eq.$\stick$(\ref{rel_pwn}) one finds the swept-up mass:
\begin{align}
M_{\rm sh}(t)&\simeq0.6
\left(
\frac{L_0 t}{E_{\rm sn}}
\right)^{\frac{3}{5}}
 M_{\rm ej}\nonumber \\
&\simeq 0.2
\left[
\frac{L_0}{10^{38}\stick{\text{erg s}^{-1}}}
\right]^{\frac{3}{5}}
\left[
\frac{E_{\rm sn}}{10^{51}\stick{\text{erg}}}
\right]^{-\frac{3}{5}}
\left[
\frac{M_{\rm ej}}{10M_{\odot}}
\right]
\left[
\frac{t}{\text{kyr}}
\right]^{\frac{3}{5}}M_{\odot},
\label{m_ts}
\end{align}
and the nebular pressure:
\begin{equation}
\begin{split}
P_{\rm pwn}(t) &\simeq 1.1\times 10^{-2}
\left(
\frac{L_0^4 M_{\rm ej}^{15}}{E_{\rm sn}^9}
\right)^{\frac{1}{10}}
t^{-\frac{13}{5}} \\
&\simeq 3.3\times 10^{-9}
\left[
\frac{L_0}{10^{38}\,\text{erg s}^{-1}}
\right]^{\frac{2}{5}}
\left[
\frac{E_{\rm sn}}{10^{51}\,\text{erg}}
\right]^{-\frac{9}{10}}\times\\
&\qquad\qquad\qquad
\!\!\!\!\!\!\!\times
\left[
\frac{M_{\rm ej}}{10M_{\odot}}
\right]^{\frac{3}{2}}
\left[
\frac{t}{\text{kyr}}
\right]^{-\frac{13}{5}}
\,\text{dyne cm}^{-2}. 
\label{P_ts}
\end{split}
\end{equation}
Eqs.$\stick$(\ref{ini_rpwn}),$\stick$(\ref{v_ts}),$\stick$(\ref{m_ts}) and$\stick$(\ref{P_ts}) represent analytical solutions of mass, momentum and energy conservation equations, valid during the free expansion phase and $t\ll \tau_0$. Therefore, these solutions are essential to set the initial conditions on the radius, velocity, swept-up mass and nebular pressure at early times. This notwithstanding, these approximations break down once the spin-down luminosity begins to decay significantly, and the subsequent long-term evolution of the system must be computed numerically. 
\subsection{The Spectral Evolution in the One-Zone Model}
The one-zone thin-shell equations allow one to evolve young PWNe in a very efficient way. For this reason one-zone models are typically used to study the spectral evolution of both young and old PWNe (\citealt{TORRES201431}, \citealt{Tanaka2010}). To do this, one needs to evolve the particle distribution function $N(E,t)$. And this is typically done following the transport equation:
\begin{equation}
\frac{\partial N(E,t)}{\partial t}
+
\frac{\partial}{\partial E}
\left[\dot{E}(E,t)N(E,t)\right]
=
Q_{\rm inj}(E, t),
\label{transport_equation}
\end{equation}
where $E$ is the particle energy, $\dot{E}$ represents the energy loss rate and contains both adiabatic and radiation losses, while $Q_{\rm inj}(E, t)$ is the injection term, that represents particle acceleration at the WTS. Typically this is an input parameter, based on theoretical argument and/or observational constraints, and normalized to the PSR spin-down power. 
Adiabatic losses are just given by the expansion law of the PWN, $R_{\rm pwn}(t)$. ICS losses depend on the energy density of the background photon field, typically CMB and Galactic IR background, that are, again, input parameters. To compute synchrotron losses one needs to know the magnetic field inside the nebula (or alternatively the magnetic energy), which can be determined through different approaches. One can assume that the magnetic energy is a constant fraction of the internal energy of the PWN (e.g. equipartition, \citealt{Bucciantini2011}). Alternatively, one can add an equation for the magnetic energy, $E_{\rm B}$ as in \cite{Rev1}:
\begin{equation}
\frac{\text{d}E_{\rm B}(t)}{\text{d}t}+4\pi R^{2}_{\rm pwn}(t)V_{\rm pwn}(t)P_{\rm mag}(t)=\eta_{\rm B}L_{\rm sd}(t),
\end{equation}
where $P_{\rm mag}$ is the magnetic pressure and $\eta_{\rm B}$ is the fraction of the spin-down luminosity that goes into magnetic energy. In this case, instead of using Eq.$\stick$(\ref{rel_pwn}), it is customary to compute the total pressure as the sum of the magnetic one with the particle one, where the latter is obtained by directly integrating over the particle distribution function. Once the particle distribution function is known, one can easily compute the spectrum of the nebula. 

%
%
%
%
%
%
%
%
%
%
%
%
%
%
%
%
%
%
%
%
%
%
%
%
\chapter{Population Synthesis}
\label{CHAPTPOPSYN}
As discussed in the previous section, the one-zone model within the thin-shell approximation provides a robust framework for predicting the evolution of a young PWN. 
In this case, Eqs.$\stick$(\ref{eq_ts_mass}), (\ref{eq_ts_momentum}) and (\ref{eq_ts_energy}), together with the constraints given by Eqs.$\stick$(\ref{density_hydro}), (\ref{rho_piecewise}), (\ref{rel_pwn}) and the initial conditions set by Eqs.$\stick$(\ref{ini_rpwn}), (\ref{v_ts}), (\ref{m_ts}) and (\ref{P_ts}) determine the evolutionary path of a PWN inside the unshocked, pressureless ejecta of the parent SNR.
However, even neglecting radiative losses, the dynamics of a PWN-SNR system remains remarkably complex. This complexity arises from the vast parameter space required to describe the interaction between the PSR wind and the SNR, itself expanding into the surrounding medium.\\\\ 
Focusing on the central PSR, its spin down luminosity, $L_{\rm sd}(t)$, depends both on its initial value $L_0$, given by Eq.$\stick$(\ref{initial_sd_luminosity_and_time}), and on the initial spin-down time $\tau_0$, given by Eq.$\stick$(\ref{general_E0_tau0}). Additionally, the temporal evolution of the spin-down luminosity must also account for the mechanism driving the loss of rotational energy, in general described by Eq.$\stick$(\ref{general_spin_down_equation}), and depending on the value of the braking index $n$. \\\\
On the other hand, the evolution of a PWN also depends on the parameters of both the SNR and the ambient medium. Specifically, the ejecta density, $\rho_{\rm ej}$, depends both on the kinetic energy of the SN, $E_{\rm sn}$, and on the ejected mass, $M_{\rm ej}$. In addition, the radial scaling of the density depends on the profile of the ejecta density distribution -- the values of $\delta$ and $\omega$, defined in Sect.~\ref{SNE_SNRS} -- which in turn depends on the dynamics of the SN explosion. Finally, we have shown that the evolution of an expanding SNR depends also on the mass density of the ambient medium (CSM/ISM), $\rho_0$.\\\\ 
As a consequence, given the high dimensionality of the parameter space, studying the evolution of a single PWN-SNR system across all possible combinations becomes computationally prohibitive.
Therefore, our strategy shifts from studying a single system to a \textit{population synthesis} approach. By generating a synthetic population of PSRs and SNRs, we can statistically explore how the distributions of the initial parameters dictate the evolutionary trajectories of the resulting PWNe, \corr{which is the first step toward modelling their late time evolution and ultimately evaluating their contribution to the Galactic VHE $\gamma$-ray emission}.\\\\
Currently, the observed Galactic population of PSRs amounts to $\simeq 4300$ objects (see Sect.~\ref{can_mill}). However, here we will focus on the subset of PSRs energetic enough to power $\gamma$-ray emission. In this context, \corr{we recall that the VHE $\gamma$-ray emission in PWNe is mainly powered by ICS, typically persisting for $\simeq 100\stick$kyr -- see Eq.$\stick$(\ref{t_ic}). Furthermore, given that \corr{the rate of CCSN explosions} in the Galaxy is approximately $\simeq 1.6\stick$events per century \citep{Rozwadowska20}}, it is possible to estimate that there are $\simeq 1600$ PSRs -- \corr{and as a consequence $\simeq 1600$ PWNe} -- potentially contributing to the Galactic TeV emission. \corr{Based on these results}, we simulate a set of $10^4$ PWN-SNR systems, ensuring a statistically robust sample, and avoiding any under-sampling of their Galactic population.
\section{Constant Parameters}
To construct our synthetic population, we distinguish between parameters held constant and those characterized by statistical distributions. To maintain a manageable parameter space, several physical quantities are fixed to their canonical values, supported by theoretical results and observational evidence. Specifically, we hold constant: the braking index $n$, the mass $M_{\rm ns}$, radius $R_{\rm ns}$ and moment of inertia $I_{\rm ns}$ of the NS; and the indices $\delta$ and $\omega$ for the SNR ejecta density profile.\\\\
\corr{For the PSR parameters}, we adopt the canonical braking index $n=3$, consistent with the magnetic dipole radiation model. Although observed values for individual NSs show evident deviations \corr{-- as discussed in Sect.$\stick$\ref{sec:NSelectrodynamics}} -- we recall that this assumption remains robust for describing the loss of rotational energy for a strongly magnetized, rapidly rotating NS powering a relativistic wind of particles. This is \corr{because} both the energy loss via relativistic wind -- see Eq.$\stick$(\ref{split_monopole_rate}) -- and magnetic dipole emission -- see Eq.$\stick$(\ref{spin_down_equation_n3}) -- have the same functional structure. Moreover, under the hypothesis \corr{that losses follow} a pure magnetic dipole emission, the initial spin-down luminosity, $L_0$, and initial spin-down time, $\tau_0$, are given by Eq.$\stick$(\ref{sd_lmdipole}) and Eq.$\stick$(\ref{initial_spin_down_luminosity_magnetic_dipole}), respectively, providing simple closure relations. However, while the magnetic dipole model provides closure relations for $L_0$ and $\tau_0$, they remain dependent on the intrinsic properties of the NS, such as the magnetic field, $B_{\rm ns}$, radius, $R_{\rm ns}$, initial rotation period, $P_0$ and moment of inertia, $I_{\rm ns}$, which in turns depends on mass and radius of the NS. \\\\
In this regard, we recall that observations in binary systems indicate that NSs have small dispersions both in mass \corr{-- typically in the range $[1.15-2.35]M_{\odot}$} (\citealt{Stairs+02}, \citealt{Kramer+06}, \citealt{Freire+11}, \citealt{Ferdman+14}, \citealt{Martinez+15}, \citealt{Ferdman+18}, \citealt{Wu25}) and radius (\citealt{Ozel+09}, \citealt{Guver+10a}, \citealt{Ozel+12}, \citealt{Guver+13}). \corr{These constraints are derived from complementary observational channels. Masses are determined via the precise timing of observed pulsations in binary systems, which allows for orbital reconstruction and the subsequent determination of the component masses. Conversely, estimates of the radius rely often on spectroscopic modeling of thermal emission from the stellar surface -- for instance, during quiescent phases or thermonuclear bursts -- where the radius is inferred from the measured flux and the source distance}. In this context, by integrating constraints on the equation of state (EOS) for dense matter with observations, \cite{Biswas+24} demonstrated that the distribution of masses for NSs exhibits a prominent peak at approximately $1.33 M_{\odot}$, with a remarkably narrow spread \corr{of $\simeq 0.03 M_{\odot}$}. Based on this result, we adopt $M_{\rm ns} = 1.33 M_{\odot}$ as the representative mass for our population of NSs. Regarding the radius of the NSs, the \corr{aforementioned} measurements have suggested radii in the range $[9.8 - 11]\stick$km at a $\simeq 68\%$ confidence level \citep{Ozel+10}. We therefore use the mean of this interval, setting $R_{\rm ns} = 10.4\stick$km. Concerning the moment of inertia, $I_{\rm ns}$, \cite{Rezzolla+16} showed that it follows a universal relation depending on the compactness, $\xi \equiv M_{\rm ns}/R_{\rm ns}$, independently of the choice of the EOS. \corr{Adopting geometrized units -- $G=c=1$ -- such relation for a relativistic, rapidly rotating NS is}:
\begin{equation}
\frac{I_{\rm ns}}{M^3_{\rm ns}}=
\frac{\alpha_1}{\xi}
+
\frac{\alpha_2}{\xi^2}
+
\frac{\alpha_3}{\xi^3}
+
\frac{\alpha_4}{\xi^4},
\end{equation}
where $\alpha_1 = 8.134\times 10^{-1}$, $\alpha_2 = 2.101\times 10^{-1}$, $\alpha_3 = 3.175\times 10^{-3}$ and $\alpha_4 = -2.717\times 10^{-4}$. Applying this relation using our fiducial values for mass and radius, and converting to physical units, yields $I_{\rm ns} = 1.07\times 10^{45}\stick$g$\stick$cm$^2$, which we assume for every NS in the synthetic population.\\\\
Once we have fixed the constant parameters for the NS, we now turn to the hosting SNR. To characterize the structure of the SNR, we recall that the interaction of the SN blast wave with the stellar envelope leads to an ejecta density profile with a flat core and a steeper envelope. 
\corr{Anticipating that we will restrict our synthetic population to progenitors in the mass range $[9-22] M_{\odot}$, which are expected to explode as Red Supergiants (RSG), we adopt the corresponding canonical parameters ($\delta = 0$ and $\omega = 12$, \citealt{MatznerMckee99}). This choice is physically consistent with our mass selection, as opposed to the profile typical of Blue Supergiants (BSG) -- where $\delta = 0$ and $\omega = 10$ -- which would apply to significantly more massive stars ($\simeq [50-60] M_{\odot}$).}
By adopting these canonical values, we ensure a physically consistent ejecta density profile and reduce the dimensionality of the parameters space.
\section{Statistically Distributed Parameters}
While setting some parameters as constant is reasonable, due to the combination of observations and theoretical results, this approach is not suitable for the entire set of physical quantities governing the evolution of a PWN-SNR system. Key variables requiring statistical distributions include the initial period of rotation, $P_0$ -- together with its time derivative, $\dot{P}_0$ -- and the NS surface magnetic field, $B_{\rm ns}$. Similarly, we require statistical treatment for the parameters of the SNR, specifically the kinetic energy of the SN, $E_{\rm sn}$, the mass of the ejecta, $M_{\rm ej}$, as well as the density of the ambient medium, $\rho_0$.
\subsection{Distributions of Pulsars' Parameters}
\subsubsection{The Magnetic Field}
To model the magnetic field of the young PSRs population ($\tau_{\rm ns}<\stick$Myr), we adopt the statistical description provided by \cite{Watters+11}. 
\corr{This study is primarily constrained by the Galactic rate of SNe, which imposes a strict limit on the PSRs birthrate. By incorporating this physical constraint along with the properties of the detected young $\gamma$-ray PSRs, the authors determined the necessary distributions for initial spin periods and magnetic fields. This approach ensures that the synthetic population evolves exhibiting a distribution of spin and pulse properties in substantial agreement with the observed sample.} 
Focusing on young, energetic systems, it is reasonable to neglect field decay effects \citep{Payne07}, typical of older populations. Under this assumption, the birth magnetic field -- and thus the current field for young PSRs -- is well described by a log-normal distribution function. \corr{To} simplify the notation, we introduce the following variable: $\log{B}\equiv\log_{10}[B_{\rm ns}/\text{G}]$. \corr{Then}, the lognormal can be written as:
\begin{equation}
\sqrt{\frac{1}{2\pi \sigma^2_{\log B}}}
\exp\left\{ -\frac{(\log B - \mu_{\rm B})^2}{2\sigma_{ \log B}^2} \right\}\equiv
\sqrt{\frac{1}{2 \pi \sigma^2_{\log B}}}
f(\log B),
\end{equation}
with $\mu_{\rm B} = 12.65$ \corr{-- corresponding to a strength of $\simeq 4.5\times 10^{12}\stick$G --} and $\sigma_{\log B} = 0.3$. 
This parameterization is highly constrained, since even modest deviations from these values result in synthetic populations that fail to reproduce the observed Galactic distribution of young PSRs. 
We \corr{decided} to truncate the distribution within the range $[B_{\rm min}- B_{\rm max}]=[10^{11}-4.4\times 10^{13}]\stick$G. The choice of the lower limit serves to explicitly exclude Central Compact Objects (CCOs), a distinct class of young NSs characterized by weak magnetic fields ($[10^{10}-10^{11}]\stick$G) and the absence of PWNe (\citealt{Pavlov04}, \citealt{Ho_2012}). Conversely, the upper limit is set to the quantum critical field limit: this cutoff ensures the exclusion of the so-called \textit{Magnetars}, whose evolution is driven by the dissipation of their extremely high magnetic fields ($\simeq [10^{14}-10^{15}]\stick$G, \citealt{Vidana18}), rather than by the emission of a relativistic wind. 
\corr{Specifically, above this threshold, the photon splitting process becomes dominant \citep{Harding97} and suppresses the pair creation required to sustain the PSR wind itself}. Consistent with these physical constraints, the probability distribution function (PDF) of the magnetic field is given by:
\begin{equation}
\mathcal{P}(\log B) =
\begin{cases}
    C_{\rm B} f(\log B) & \text{if } \log B_{\rm min} \leq \log B \leq \log B_{\rm max} \\
    0 & \text{otherwise,}
\end{cases}
\label{LOGBDIST}
\end{equation}
\corr{with $\log B_{\rm min}\equiv\log_{10}[B_{\rm min}/\text{G}]$, and $\log B_{\rm max}\equiv\log_{10}[B_{\rm max}/\text{G}]$}. \corr{The normalization constant reads}:
\begin{equation}
C_{\rm B}
=
\sqrt{\frac{2}{\pi \sigma^2_{\log B}}}
\left[
\operatorname{erf}
\left(
\frac{\log B_{\rm max}-\mu_{\rm B}}{\sqrt{2 \sigma_{ \log B}^2}}
\right)
-
\operatorname{erf}
\left(
\frac{\log B_{\rm min}-\mu_{\rm B}}{\sqrt{2 \sigma_{ \log B}^2}}
\right)
\right]^{-1}.
\label{CBCTE}
\end{equation}
\subsubsection{The Initial Rotational Period}
The birth magnetic field is not the only parameter required to fully describe the complex physics of NSs. A complete characterization of their energy loss rate also requires knowledge of the initial rotation period, $P_0$. Focusing on the young population of PSRs, \cite{Watters+11} modeled the initial period distribution as a Gaussian:
\begin{equation}
\sqrt{\frac{1}{2 \pi \sigma^2_{P_0}}}\exp \left\lbrace - \frac{(P_0 - \bar{P}_0)^2}{2\sigma^2_{P_0}} \right\rbrace
\equiv
\sqrt{\frac{1}{2 \pi \sigma^2_{P_0}}}f(P_0),
\end{equation}
with $\bar{P}_0 = 50\,\text{ms}$ and $\sigma_{P_0} = \bar{P}_0/\sqrt{2}$. Following \cite{Watters+11}, we truncate the distribution within the interval $[P_{\min} - P_{\max}] = [10 - 500]\,\text{ms}$. 
\corr{Imposing a lower bound of $10\stick$ms is essential to prevent the overproduction of highly energetic, Crab-like PSRs in the synthetic population, consistent with the fact that only one object with that characteristics is observed in the entire Galaxy}. On the other hand, the choice of the upper limit is justified by both observational evidence and energetic constraints. Specifically, the population of observed young, high-energy $\gamma$-ray emitters is strictly confined to periods $\lesssim 500\text{ ms}$ \citep{Harding13}, beyond which it is dominated by older PSRs, typically emitting in the radio band, and magnetars. Following these prescriptions, yields the PDF of the initial period:
\begin{equation}
\mathcal{P}(P_0) =
\begin{cases}
    C_{P_0} f(P_0) & \text{if } P_{\rm min} \leq P_0 \leq P_{\rm max} \\
    0 & \text{otherwise,}
\end{cases}
\label{PDIST}
\end{equation}
with normalization constant given by:
\begin{equation}
C_{P_0}
=
\sqrt{\frac{2}{\pi \sigma^2_{P_0}}}
\left[
\operatorname{erf}
\left(
\frac{P_{\rm max}-\bar{P}_0}{\sqrt{2\sigma^2_{P_0}}}
\right)
-
\operatorname{erf}
\left(
\frac{P_{\rm min}-\bar{P}_0}{\sqrt{2\sigma^2_{P_0}}}
\right)
\right]^{-1}.
\label{CPCTE}
\end{equation}
We also recall that the PDF of the initial period strictly determines the distribution of its derivative with respect to time, $\dot{P}_0$, given by Eq.$\stick$(\ref{P_ns}).
The choices of the PDFs for both the magnetic field and the initial period of the NSs determine the distributions of both the initial spin-down luminosity and spin-down time, which we recall are given by Eq.$\stick$(\ref{sd_lmdipole}) and Eq.$\stick$(\ref{initial_spin_down_luminosity_magnetic_dipole}), respectively, and shown in Fig.$\stick$\ref{log_L0_tau0}.
\begin{figure}[h!]
	\centering
    \includegraphics[width=1.0\textwidth]{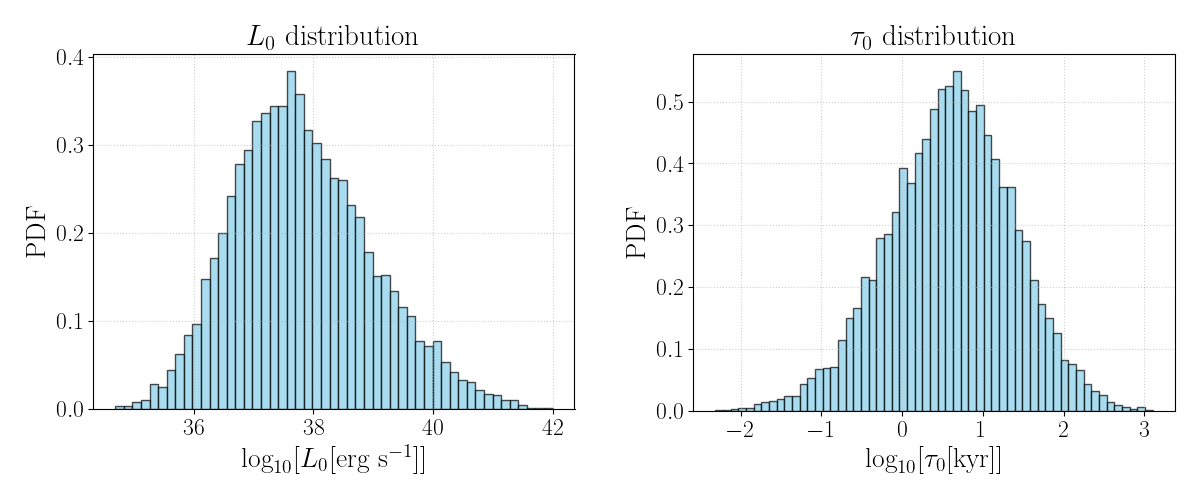}
    \caption{Normalized histograms of the distributions of $L_0$ and $\tau_0$ for a young population of $10^4$ PSRs.}
    \label{log_L0_tau0}
\end{figure}\\
Remarkably, the majority of the population exhibits initial spin-down luminosities $\simeq 10^{38}\stick$erg$\stick$s$^{-1}$ and initial spin-down timescales $\simeq 5\stick$kyr. In Fig.$\stick$\ref{PPDOT} we also report a comparison between the Galactic population of PSRs and the simulated sample.
\begin{figure}[H] 
	\centering
	 \includegraphics[width=0.85\linewidth]{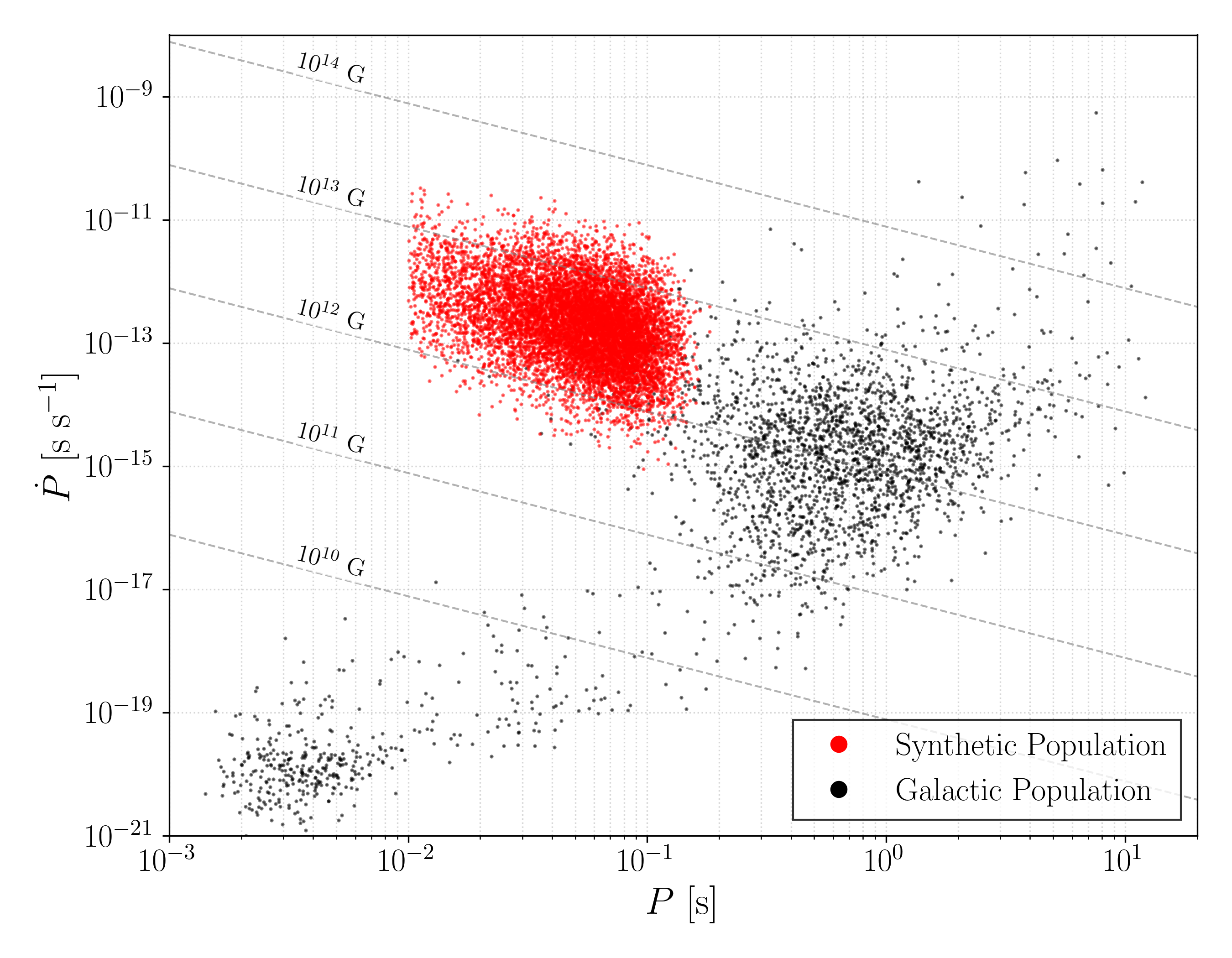}    
	
    \caption[Diagram of the period derivative vs period for both observed and synthetic PSRs.]{Diagram of the period derivative vs period for both observed and synthetic PSRs. For the synthetic population we report the initial period and its derivative, whereas for the observed Galactic population we plot the current values. The dashed diagonal lines represent configurations of constant magnetic field, calculated using Eq.$\stick$(\ref{B_ns_dipole}). \corr{Among the deteced PSRs we identify three main populations: canonical Radio PSRs (center; $P_0 \simeq 500\stick$ms, $B_{\rm ns}\simeq 10^{12}\stick$G), MSPs (lower left; $P_0\simeq [2-10]\stick$ms, $B_{\rm ns}\simeq 10^{8}\stick$G), and magnetars (upper right; $P_0\simeq 10\stick$s, $B_{\rm ns} \simeq [10^{14}-10^{15}]\stick$G). We draw the sample of the Galactic population from the ATNF catalogue\footnotemark \citep{Manchester2005}.}} 
    \label{PPDOT}
\end{figure}
\footnotetext{See also \url{https://www.atnf.csiro.au/research/pulsar/psrcat/}}
\noindent
The arrangement in the $P-\dot{P}$ diagram of sources we have generated reflects the specific choices made for the input PDFs. Specifically, the synthetic population is dominated by PSRs with characteristic magnetic fields of the order of $\sim 10^{12}\stick$G and initial periods of tens of milliseconds. \corr{These sources represent the progenitors that, upon evolution, are expected to reproduce the Galactic population of PSRs}. Crucially, the boundaries of the population demonstrate the effectiveness of our selection cuts: the distribution explicitly excludes low-field CCOs, radio-PSRs and Magnetars. For a more detailed description of the numerical procedure employed to generate the synthetic population of PSRs and SNRs, we refer the reader to Appendix~\ref{APPI}.
\subsection{Distributions of Supernova Remnants' Parameters}
\label{SNRSECTION}
Since we have established the properties of the population of NSs, we must now characterize the environment in which the PWNe evolve: the hosting SNR. The dynamical expansion of the nebula is, in fact, determined by the interplay between the energy injection of the NS and the confinement provided by the SNR ejecta. For this reason, we require PDFs for the three fundamental parameters governing the evolution of the SNR: the kinetic energy of the SN, $E_{\rm sn}$, the mass of the ejecta, $M_{\rm ej}$, and the density of the surrounding medium, $\rho_0$.
\subsubsection{The Mass of Progenitor Stars and Ejecta}
Focusing on the stars undergoing core collapse, \cite{Sukhbold+16} \corr{found} that those with masses in the range $[9-15] M_{\odot}$ successfully explode as CCSNe, leaving a remnant NS. On the other hand, the fate of more massive stars is less clear: stars with an initial mass in the range $[15 - 22] M_{\odot}$ can either explode as SNe or directly collapse into a BH. Conversely, beyond $22 M_{\odot}$, direct collapse becomes the dominant channel, with the exclusion of a narrow window in the range $[25-27] M_{\odot}$. \corr{As a consequence, the stellar population of CCSNe progenitors} is heavily peaked towards lower masses for which, according to observational evidence, the initial mass function (IMF) scales as $M^{-{\alpha}}$, where $M$ is the mass of the star and $\alpha=2.35$ (\citealt{Salpeter55}, \citealt{Selman+99}, \citealt{Elmegreen+00}).
The steepness of this power-law distribution implies that the number of stars drops rapidly with increasing mass, and the contribution of the narrow explodability island at $[25-27] M_{\odot}$ becomes statistically negligible, compared to the lower mass channels. Driven by these considerations, and to avoid the uncertainties associated with the direct collapse regime, we limit our population synthesis to the mass range $[M_{\rm min}- M_{\rm max}]=[9- 22] M_{\odot}$. Therefore, the PDF of the mass of the progenitor stars is given by:
\begin{equation}
\mathcal{P}(M)
=
\begin{cases}
C_{\rm M}M^{-{\alpha}} & \text{if } 9 M_{\odot} \leq M \leq 22 M_{\odot} \\
0 & \text{otherwise,}
\end{cases}
\label{Salpeter}
\end{equation}
where $C_{M}$ is the normalization factor:
\begin{equation}
C_{\rm M} = \frac{(\alpha-1) (M_{\rm min} M_{\rm max})^{\alpha-1}}{M_{\rm max}^{\alpha-1} - M_{\rm min}^{\alpha-1}}.
\label{CTEMASS}
\end{equation} 
In principle, a rigorous determination of the ejecta mass would require a detailed treatment of the mass losses throughout stellar evolution. Despite this, the physical mechanisms driving these losses -- such as the interplay between pulsations, dust formation in the outer envelopes, and convective instabilities -- remain theoretically challenging and difficult to constrain via observations \citep{VinkPuls2008}. Given the significant uncertainties affecting current numerical prescriptions for stellar mass losses, we define the mass of the ejecta simply as the difference between the initial mass of the progenitor star and the mass of the resulting NS. 
\corr{A more detailed treatment incorporating mass loss -- based on the results obtained by \cite{Sukhbold+16} -- is presented in Sect.$\stick$\ref{LOSS_MASS_ISM_SAMPLING}}.
\subsubsection{The Energy of Supernovae}
Turning to the kinetic energy of the SN, $E_{\rm sn}$, we adopted a uniform PDF covering the range observed in typical explosion models. Specifically, based on the SN explosions presented by \cite{Sukhbold+16}, we assumed $E_{\rm sn}$ to be uniformly distributed in the range $[E_{\rm min}- E_{\rm max}] =[0.5-2.0]\times 10^{51}\stick$erg. We remark that the choice of this range is consistent with the canonical explosion energy of $10^{51}\stick$erg established in pioneering studies (\citealt{ColgateWhite66}, \citealt{Bethe1990}), and the corresponding PDF is given by: 
\begin{equation}
\mathcal{P}(E_{\rm sn})
=
\begin{cases}
    \frac{1}{E_{\rm max}-E_{\rm min}} & \text{if } E_{\rm min} \leq E_{\rm sn} \leq E_{\rm max} \\
    0 & \text{otherwise.}
\end{cases}
\label{EDIST}
\end{equation}
\subsubsection{The Ambient Medium Density}
In principle the \corr{medium surrounding} the stars exploding as SNe, can either be  \corr{the} CSM or \corr{the} ISM. Due to the complexity and the uncertainties of the mass-loss mechanisms driving the formation of the CSM \citep{Sfaradi2024}, we have assumed SNe to take place in the ISM. This choice is justified, since a large fraction of massive stars is expected to leave their birth sites \citep{Castrillo2023}, ending their lives in the ISM \citep{Kim_2017}.
\corr{A more detailed account for these effects, following the results of \cite{Kim_2017}, weighting the contribution of low-density regions excavated by star formation and previous SNe, and including the distinction between clustered and runaway progenitors, will be discussed in Sect.$\stick$\ref{LOSS_MASS_ISM_SAMPLING} } Focusing on the density distribution of the ISM, we have followed the prescription of \cite{Elwood+18}, where through Monte Carlo simulations targeting extragalactic populations of SNRs -- such as those in M31 and M33 -- they established that the numerical density of the ISM follows a lognormal distribution. By introducing the following variable: $\log{n_0}\equiv\log_{10}[n_0/\text{cm}^{-3}]$, the distribution function of the density can be written as: 
\begin{equation}
\sqrt{\frac{1}{2 \pi \sigma^2_{\log n_0}}}\exp{\left\lbrace-\frac{(\log n_0-\mu_{0})^2}{2\sigma_{ \log{n_0}}^2}\right\rbrace}
\equiv
\sqrt{\frac{1}{2 \pi \sigma^2_{\log n_0}}}
f(\log n_0),
\end{equation} 
with $\mu_0 = -1.33\text{ and } \sigma_{\log{n_0}} = 0.7$. A proper modeling of the supernova environment requires accounting for the multiphase nature of the ISM, which is characterized by a complex structure consisting of distinct phases that span many orders of magnitude in density. These range from the tenuous, hot gas generated by shockwaves of SN explosions to the denser warm and cold neutral clouds. For this reason, we truncated the distribution of the ISM density by imposing a lower limit $n_{\rm min}=5 \times 10^{-3}\stick$cm$^{-3}$ -- a typical density of the hot, ionized medium -- and an upper limit $n_{\rm max}=2\stick$cm$^{-3}$ -- an appropriate value for the warm neutral medium \citep{Draine11}. Furthermore, the choice of these boundaries ensures that we cover the density range where CCSNe typically take place \citep{Kim_2017}. Performing this truncation to the distribution function yields the PDF of the ISM number density:
\begin{equation}
\mathcal{P}(\log n_0) =
\begin{cases}
    C_{0} f(\log n_0) & \text{if } \log n_{\min} \leq \log n_0 \leq \log n_{\max} \\
    0 & \text{otherwise,}
\end{cases}
\label{ISMDENSITYDIST}
\end{equation}
\corr{with $\log n_{\min} \equiv \log_{10}[n_{\rm min}/\text{cm}^{-3}]$ and $\log n_{\max} \equiv \log_{10}[n_{\rm max}/\text{cm}^{-3}]$}, where the normalization factor corresponds to:
\begin{equation}
C_{0}=
\sqrt{\frac{2}{\pi \sigma^2_{\log n_0}}}
\left[
\operatorname{erf}
\left(
\frac{\log n_{\rm max}-\mu_{0}}{\sqrt{2 \sigma_{ \log n_0}^2}}
\right)
-
\operatorname{erf}
\left(
\frac{\log n_{\rm min}-\mu_{0}}{\sqrt{2 \sigma_{ \log n_0}^2}}
\right)
\right]^{-1}.
\label{CNCTE}
\end{equation}
Finally, deriving the mass density distribution of the ISM requires its mean mass per particle, $m_0$, which in principle depends on the state of ionization of the ISM itself. 
This property varies significantly across its phases, sustained by diverse physical mechanisms ranging from shock heating in the hot phase, to UV photoionization in warm regions; and explicitly tracking the contribution of ionization for each phase introduces complications that are beyond the scope of this work. Therefore, for simplicity, we neglected these effects and adopted a constant mean mass per particle corresponding to the neutral gas composition, setting $m_0 = 1.3 \, m_{\rm p}$. This value assumes a standard cosmic chemical composition dominated by hydrogen and helium and neglects the electron contribution from ionization.\\\\
The definition of the PDFs for the kinetic energy of the SNe, $E_{\rm sn}$, mass of the ejecta, $M_{\rm ej}$, and density of the ambient medium, $\rho_{0}$, provides the necessary framework to synthesize a population of SNRs. By sampling from these distributions, we assign a specific set of physical parameters to each simulated SNR. Therefore, we can combine these physical quantities to obtain the characteristic scales of luminosity, time and radius -- $L_{\rm ch}, t_{\rm ch}$ and $R_{\rm ch}$ -- given by Eq.$\stick$(\ref{L_ch}), (\ref{t_ch}) and (\ref{r_ch}) respectively -- see Fig.$\stick$\ref{tch_Rch_Lch}.
\begin{figure}[h!]
    \centering 
        \begin{minipage}{0.90 \textwidth} 
            \centering 

            \includegraphics[width=1.0\linewidth]{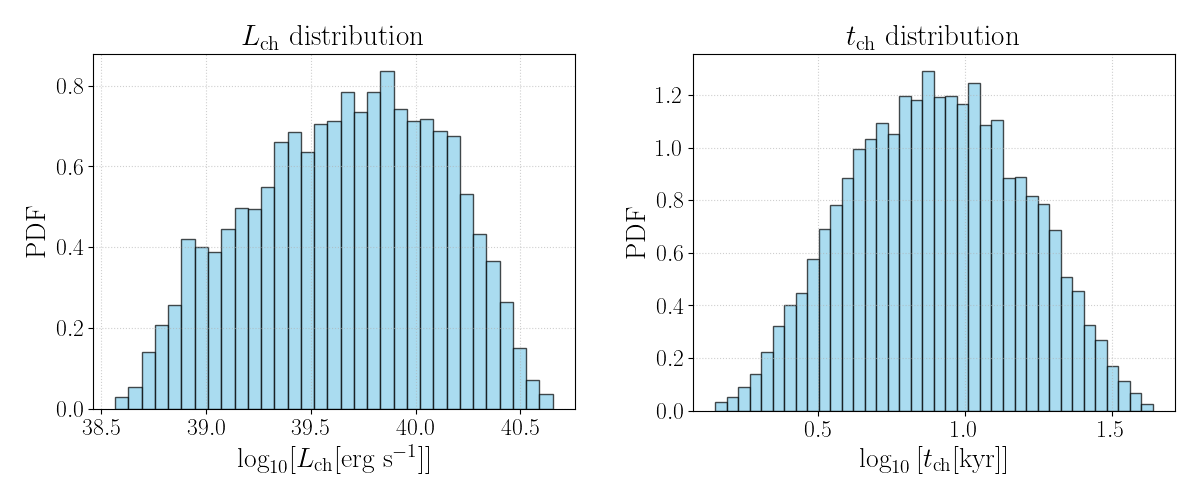}
            


            \includegraphics[width=0.5\linewidth]{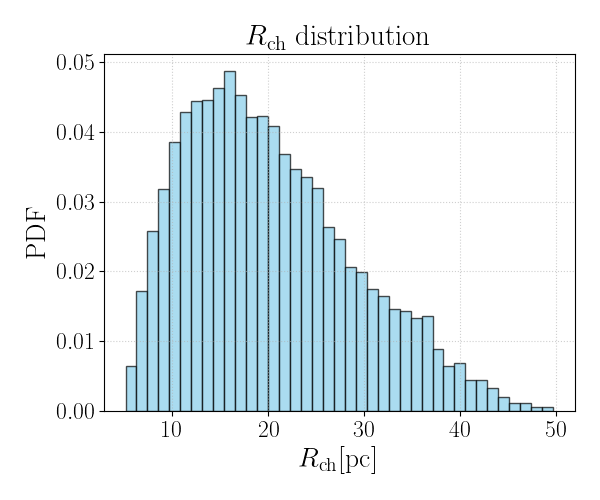}
        \end{minipage}
    \caption{Normalized histograms of the distributions of $L_{\rm ch}$, $t_{\rm ch}$ and $R_{\rm ch}$ for a population of $10^4$ SNRs. }
    \label{tch_Rch_Lch}
\end{figure}
\noindent
It is evident that the distributions of the characteristic scales of both luminosity and time inside SNRs are significantly narrower, spanning fewer decades compared to the scales of the PSRs, exhibiting distinct peaks at $L_{\rm ch} \simeq 10^{40}\stick$erg$\stick$s$^{-1}$ and $t_{\rm ch} \simeq 10\stick$kyr. Regarding the characteristic radius, its distribution is confined in the range $[5-50]\stick$pc and peaks at $R_{\rm ch}\simeq 15$pc.
We recall that these scales, together with the characteristic velocity and pressure -- $V_{\rm ch}$ and $P_{\rm ch}$ -- given by Eq.$\stick$(\ref{v_ch}) and Eq.$\stick$(\ref{P_ch}), govern the dynamical evolution of SNRs. We remark that a single characteristic scale does not uniquely identify a single progenitor: in principle, distinct SNRs with different combinations of explosion energy, mass of the ejecta, and density of the ISM can yield identical characteristic quantities. However, adopting a normalized evolution turns this degeneracy into a significant advantage. 
\section{The Synthetic Population and the Characteristic Plane}
\label{L0LCHT0TCH}
To analyze the evolutionary properties of the synthetic population of PWN-SNR systems in a generalized framework, we introduce the dimensionless \textit{characteristic plane}, defined by the coordinates $\log_{10}[L_0/L_{\rm ch}]$ and $\log_{10}[\tau_0/t_{\rm ch}]$ and shown in Fig.$\stick$\ref{ROI95}.
\begin{figure}[h!]
	\centering
	 \includegraphics[width=0.9\linewidth]{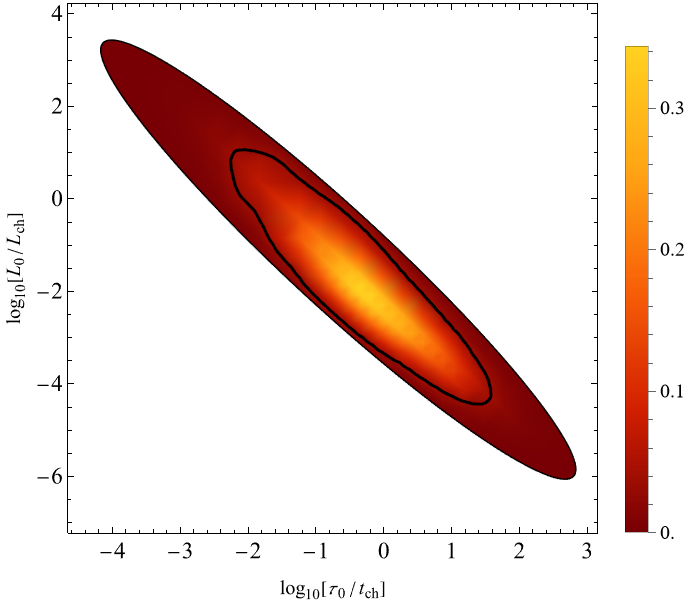}    
	\caption{PDF of the PWN-SNR synthetic population in the characteristic plane, with the colorbar indicating its values. While the external, thin ellipse encloses the entire sample, the inner, thick boundary encloses 95 \% of the population, and defines our ROI.}  
    \label{ROI95}
\end{figure}\\
In the characteristic plane our systems are fully confined in an elliptical region. Within this larger region we define the area encompassing $95\%$ of the systems, the one enclosed by the inner boundary in Fig.$\stick$\ref{ROI95}, as our \textit{Region of Interest} (ROI). Crucially, it is within this region that we will evaluate the evolution of the population up to the reverberation phase.
It is evident that the initial spin-down times of the PSRs, $\tau_0$, typically exhibit values of the same order of magnitude as the characteristic time of the hosting remnants, $t_{\rm ch}$. On the other hand, the initial spin-down luminosities, $L_0$, are much smaller compared to the luminosity scale of the SNR, $L_{\rm ch}$. Specifically, the ROI is confined to the region where $\log_{10}[\tau_0/t_{\rm ch}]$ spans the interval $[-2 \text{ --} \ 1.5]$ and $\log_{10}[L_0/L_{\rm ch}]$ lies within $[-5 \text{ --} \ 1]$, with the distribution peaking at $\log_{10}[\tau_0/t_{\rm ch}] \simeq 0$ and $\log_{10}[L_0/L_{\rm ch}]\simeq -2$.
In this representation, the intrinsic physical scales of each NS are normalized with respect to the luminosity and time scales of the hosting remnant. Therefore, each point on this plane no longer identifies a single PWN-SNR system, but a class of equivalent objects.  
It is interesting to note that we can integrate this normalization strategy into the one-zone model described in Sect.$\stick$\ref{one_zone_model_early}. Indeed, by normalizing the dynamical variables -- specifically the nebular radius, $R_{\rm pwn}/R_{\rm ch}$, and expansion velocity, $V_{\rm pwn}/V_{\rm ch}$, along with the mass of the swept-up shell, $M_{\rm sh}/M_{\rm ej}$, and the internal pressure of the nebula, $P_{\rm pwn}/P_{\rm ch}$ -- the resulting evolution \corr{becomes universally valid for all equivalent systems}. 
\section{The Free-Expansion Phase}
\label{LOSEMASSINCLUSION}
Building upon the normalization strategy discussed above, we can use the one-zone model in the thin-shell approximation to describe the evolution of a population of PWNe inside their hosting SNRs during the initial free-expansion phase. This evolutionary stage proceeds until the shell driven by the expansion of the PWN interacts with the SNR via the collision with the RS.\\\\ We recall that the model is governed by the conservation of mass, momentum, and energy as described in Eqs.~(\ref{eq_ts_mass}), (\ref{eq_ts_momentum}), and (\ref{eq_ts_energy}), subject to the constraints given by Eqs.~(\ref{density_hydro}), (\ref{rho_piecewise}), (\ref{rel_pwn}), and (\ref{initial_sd_luminosity_and_time}), with the initial conditions defined in Eqs.~(\ref{ini_rpwn})--(\ref{P_ts}). By initializing the simulations, according to our population synthesis, at an early age of a few years, the subsequent dynamical evolution is determined numerically for the entire population, until the expanding PWN impacts the RS -- described by Eq.$\stick$(\ref{r_rs_bandiera}) -- at the onset of the reverberation phase.\\\\
We can therefore characterize how different families of equivalent objects within the ROI enter the reverberation phase by analyzing the values of their physical variables at the moment of impact. In the following, we present the distributions of the normalized quantities, specifically the  time setting the beginning of the reverberation phase, $t_{\rm rev}/t_{\rm ch}$ -- see Fig.$\stick$\ref{trev_dist} -- the nebular radius, $R_{\rm pwn}(t_{\rm rev}/t_{\rm ch})/R_{\rm ch}$, the expansion velocity, $V_{\rm pwn}(t_{\rm rev}/t_{\rm ch})/V_{\rm ch}$, the total swept-up mass, $M_{\rm sh}(t_{\rm rev}/t_{\rm ch})/M_{\rm ej}$ and internal pressure $P_{\rm pwn}(t_{\rm rev}/t_{\rm ch})/P_{\rm ch}$ -- see Fig.$\stick$\ref{RPMV}.
\begin{figure}[h!]
	\centering
	 \includegraphics[width=0.8\linewidth]{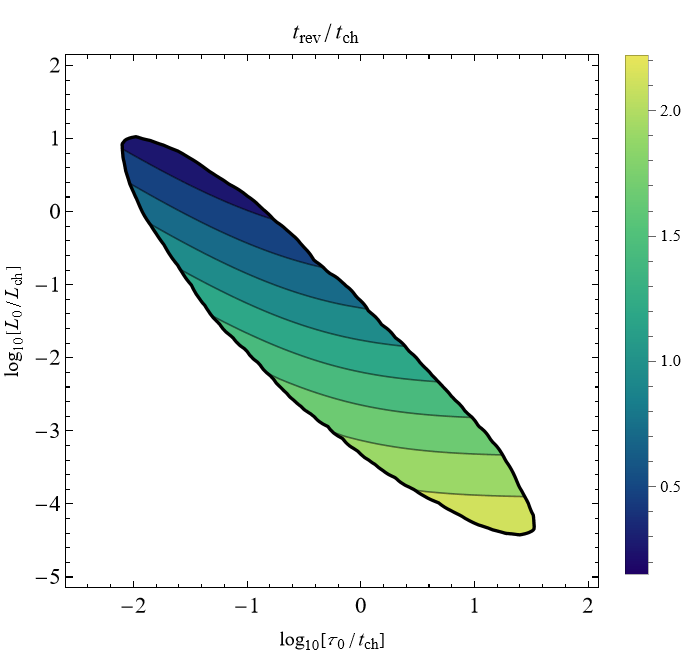}    
	\caption{Distribution of $t_{\rm rev}/t_{\rm ch}$ within the ROI. The color gradient indicates the magnitude of the variable. The smooth internal contours trace the isolines of the quantity, \corr{uniformly spaced from its minimum to its maximum value.}} 
    \label{trev_dist}
\end{figure}
\noindent
It is evident that, across the population, classes of PWNe characterized by higher initial spin-down luminosity and lower initial spin-down time -- compared to the physical scales of the remnant -- encounter the RS significantly earlier. Specifically, while high-luminosity PWNe enter the reverberation phase at $t_{\rm rev} \simeq 0.5\, t_{\rm ch}$, the systems at the peak of the populations -- located at the center of the ROI -- typically reach the RS around $t_{\rm rev} \simeq t_{\rm ch}$, whereas fainter PWNe impact the RS at $t_{\rm rev} \simeq 2\, t_{\rm ch}$. A similar variation aligned with the major axis of the ROI also characterizes the distributions of the swept-up mass and expansion velocity, \corr{while the nebular radius exhibits a slightly different trend}. Conversely, the internal pressure at reverberation exhibits a significant variation along the minor axis of the ROI. 
Therefore, these results suggest that the dynamical state of the system at reverberation varies significantly along these two orthogonal directions -- see Fig.$\stick$\ref{RPMV}.
\begin{figure}[h!]
	\centering
	 \includegraphics[width=0.8\linewidth]{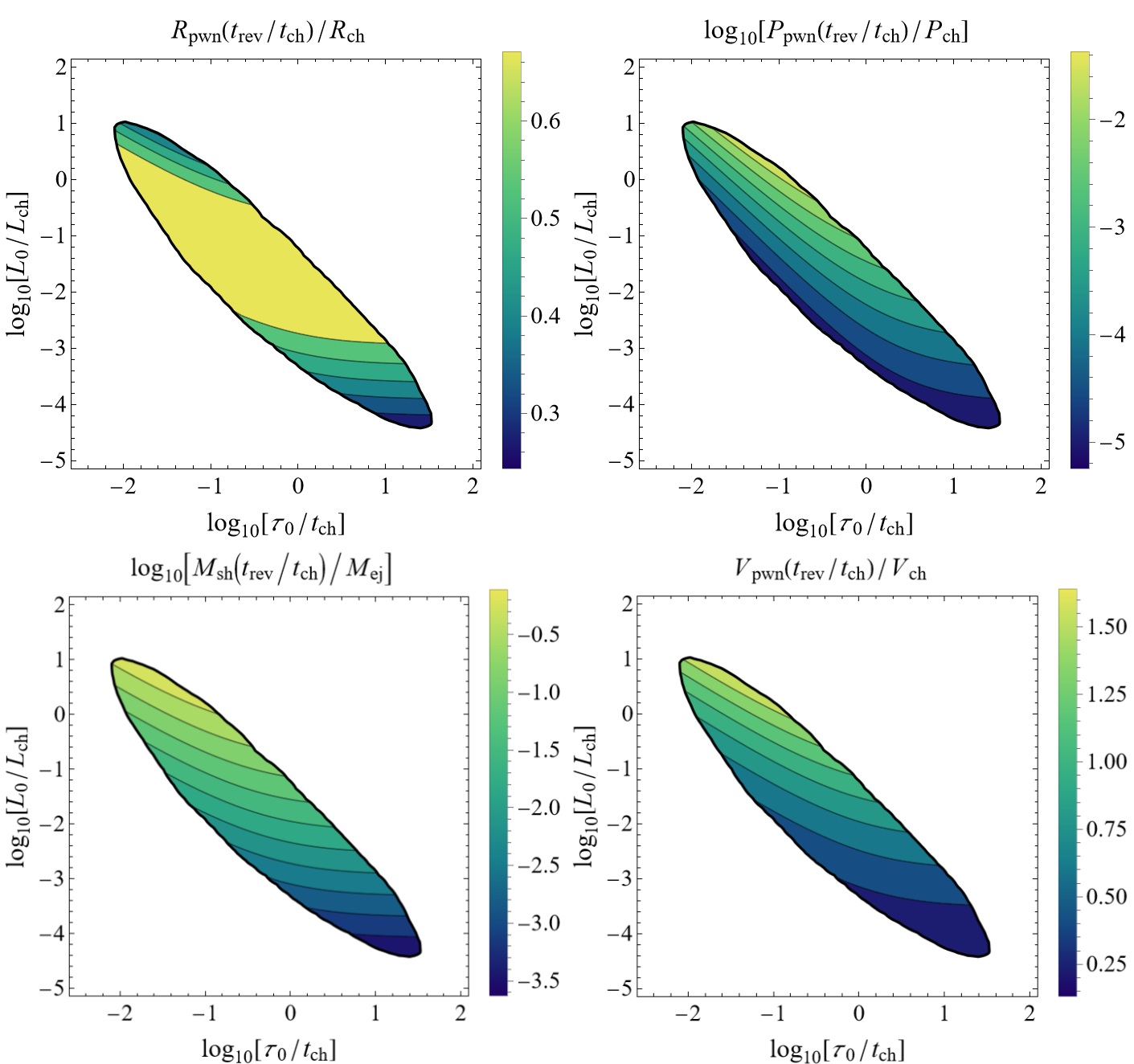}
	\caption{Distributions of $R_{\rm pwn}(t_{\rm rev}/t_{\rm ch})/R_{\rm ch}$, $P_{\rm pwn}(t_{\rm rev}/t_{\rm ch})/P_{\rm ch}$, $M_{\rm sh}(t_{\rm rev}/t_{\rm ch})/M_{\rm ej}$ and $V_{\rm pwn}(t_{\rm rev}/t_{\rm ch})/V_{\rm ch}$ within the ROI. The smooth internal contours trace the isolines uniformly spaced from the minimum to the maximum value.}  
    \label{RPMV}
\end{figure}
\noindent
The above picture clearly shows that systems where the spin-down luminosity of the PSR is large relative to the characteristic value of the SNR encounter the RS earlier, with higher velocities -- typically $\gtrsim V_{\rm ch}$ -- larger swept-up masses $(\gtrsim 0.3 M_{\rm ej})$ -- and higher internal pressure $(\gtrsim 10^{-2}P_{\rm ch})$ than the fainter ones, despite exhibiting similar radii, $\simeq [0.3-0.4] R_{\rm ch}$. 
On the other hand, weaker PWNe reach the RS with typical velocities $\lesssim 0.5 V_{\rm ch}$, swept-up mass $\simeq 10^{-3} M_{\rm ej}$ and internal pressure $\simeq 10^{-5}P_{\rm ch}$.
In comparison, the most populous systems reach the reverberation phase with a radius of $\simeq 0.7 R_{\rm ch}$ and an expansion velocity $\simeq V_{\rm ch}$. At this stage, they possess an intermediate internal pressure of $\simeq 10^{-3} P_{\rm ch}$ and have swept up approximately $10^{-2} M_{\rm ej}$. \\\\
However, these quantities alone do not provide a complete description of a PWN expanding within its hosting SNR. In principle, a full characterization requires taking into account the normalized energetic budget of the system, specifically: the kinetic energy of the shell, $E_{\rm sh}(t_{\rm rev}/t_{\rm ch})/E_{\rm sn}$, the internal energy of the nebula, $E_{\rm pwn}(t_{\rm rev}/t_{\rm ch})/E_{\rm sn}$, the total energy injected by the PSR, $E_{\rm inj}(t_{\rm rev}/t_{\rm ch})/E_{\rm sn}$, and the energy available to sustain the nebular expansion after the collision with the RS, $E_{\rm ava}(t_{\rm rev}/t_{\rm ch})/E_{\rm sn}$ -- see Fig.$\stick$\ref{EEEE}. 
\corr{We remark that, defining the total spin-down energetic budget as:
\begin{equation}
E_{\rm sd} \equiv \int_{0}^{\infty}L_{\rm sd}(t)\text{d}t,
\end{equation}
we identify:
\begin{equation}
E_{\rm inj}(t_{\rm rev})\equiv \int_{0}^{t_{\rm rev}}L_{\rm sd}(t)\text{d}t
\quad \text{and} \quad
E_{\rm ava}(t_{\rm rev})\equiv \int_{t_{\rm rev}}^{\infty}L_{\rm sd}(t)\text{d}t.
\end{equation}
Note that $E_{\rm inj}(t_{\rm rev}) + E_{\rm ava}(t_{\rm rev}) = E_{\rm sd}$}.\\\\ 
\begin{figure}[h!]
	\centering
	 \includegraphics[width=0.8\linewidth]{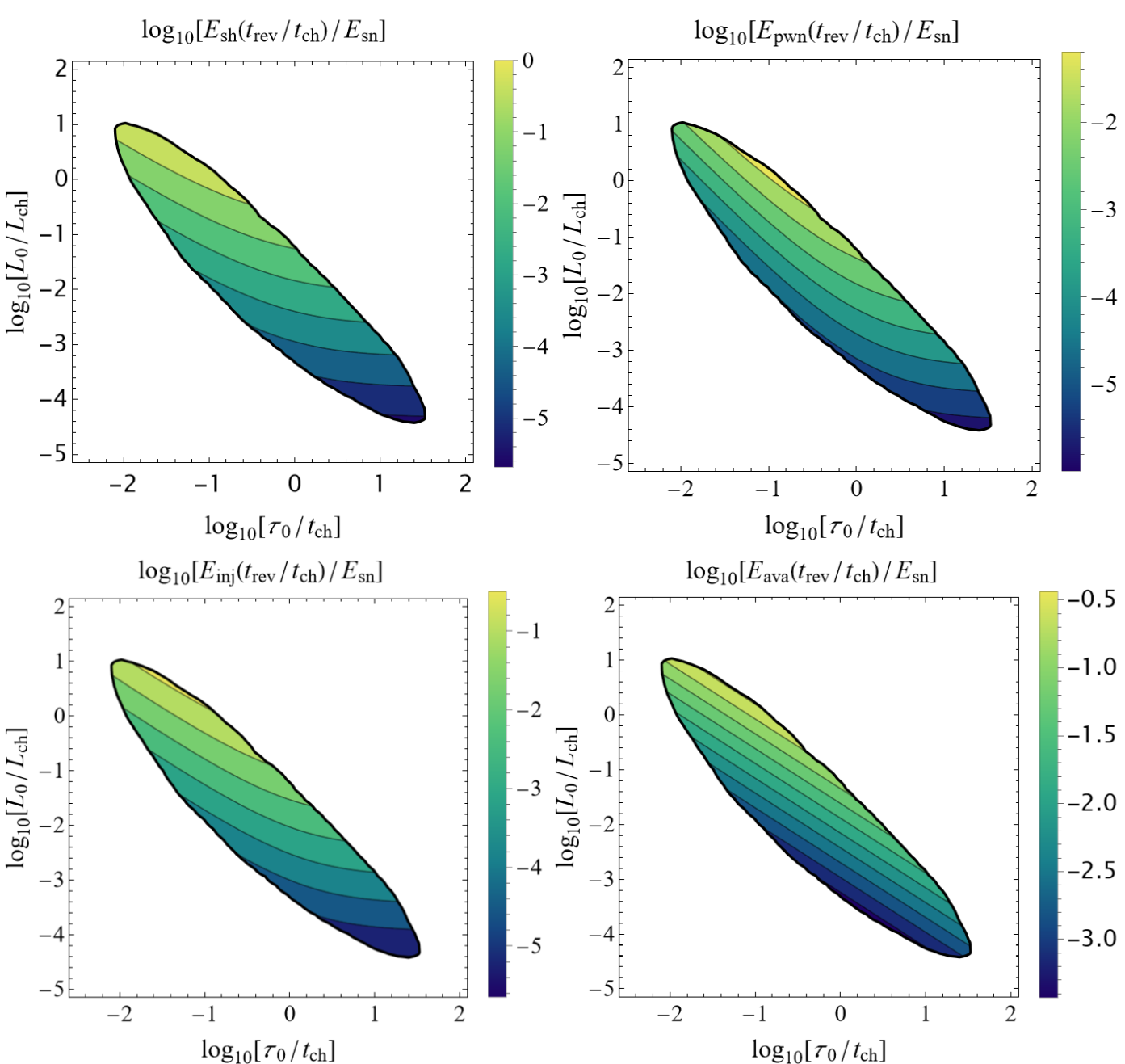}   
	\caption{Distributions of $E_{\rm sh}(t_{\rm rev}/t_{\rm ch})/E_{\rm sn}$, $E_{\rm pwn}(t_{\rm rev}/t_{\rm ch})/E_{\rm sn}$, $E_{\rm inj}(t_{\rm rev}/t_{\rm ch})/E_{\rm sn}$ and $E_{\rm ava}(t_{\rm rev}/t_{\rm ch})/E_{\rm sn}$ within the ROI.  The smooth internal contours trace the isolines uniformly spaced from the minimum to the maximum value.}  
    \label{EEEE}
\end{figure}\\
Evidently, the kinetic energy of the shell dominates over both the internal energy of the nebula and the energy injected by the PSR. Specifically, systems where the initial spin-down luminosity is comparable or higher than the characteristic value of the SNR, feature highly energetic swept-up shells, eventually reaching $E_{\rm sh} \simeq E_{\rm sn}$, sustained by a strong injection from the PSR -- $E_{\rm inj} \simeq 0.1 E_{\rm sn}$ -- compared to fainter nebulae. Conversely, PWNe with lower relative initial spin-down luminosities possess shells with small kinetic energy -- $E_{\rm sh} \simeq 10^{-4} E_{\rm sn}$ -- supported by minimal energy injection -- $E_{\rm inj} \simeq 10^{-5} E_{\rm sn}$. Regarding the internal energy, although more powerful systems exhibit higher values at the beginning of the reverberation phase, this component remains significantly subdominant to the shell kinetic energy, typically amounting to $\simeq 0.01 E_{\rm sn}$. Weaker PWNe show even lower internal energies $\simeq [10^{-5}-10^{-6}]E_{\rm sn}$. Furthermore, more powerful PWNe retain a large fraction of available spin-down energy to sustain the reverberation -- $E_{\rm ava} \simeq 0.2 E_{\rm sn}$ -- whereas fainter systems reach only $E_{\rm ava} \simeq 10^{-3} E_{\rm sn}$. In comparison, the majority of the systems reach the reverberation phase with intermediate values of energy. Specifically, for these systems we find $E_{\rm sh} \simeq [10^{-2}-10^{-3}] E_{\rm sn}$, $E_{\rm pwn} \simeq [10^{-3}-10^{-4}] E_{\rm sn}$, $E_{\rm inj} \simeq [10^{-3}-10^{-4}] E_{\rm sn}$, and $E_{\rm ava} \simeq 10^{-3} E_{\rm sn}$. Finally, while the shell kinetic energy and the injected energy vary strongly along the major axis of the ROI, the nebular internal energy and the available energy exhibit a strong gradient along the minor axis.\\\\
\corr{At this point we have fully characterized the physical and energetic properties of PWN-SNR equivalent systems at the onset of the reverberation phase. The analysis reveals significant gradients aligned with both the major and minor axes of the ROI. These variations provide the crucial initial conditions for the subsequent interaction between the PWN and the SNR. Thereby, this suggests that the subsequent evolution will follow distinct evolutionary paths, depending on the specific location of the system within the characteristic plane.} 
\section{Effects of Pre-SN Mass Loss and ISM Diversity}
\label{LOSS_MASS_ISM_SAMPLING}
Before performing the numerical evolution to characterize the reverberation phase, we introduce two critical refinements to our initial conditions, accounting for pre-SN mass losses and more complex ISM distributions.
First, we account for losses of mass by adopting the final stellar mass before the SN explosion, rather than the initial one of the progenitor. This choice is based on the evolutionary models of \cite{Sukhbold+16}, which include stellar winds and the complex interaction between late-stage burning shells and the outer convective envelope. Operationally, we first generated the population of progenitors by sampling their initial masses from Eq.$\stick$(\ref{Salpeter}), and subsequently derived their final masses by applying the mass-loss prescriptions provided by \cite{Sukhbold+16} -- see Fig.$\stick$\ref{zams_vs_presn}.
\begin{figure}[h!]
	\centering
	 \includegraphics[width=0.65\linewidth]{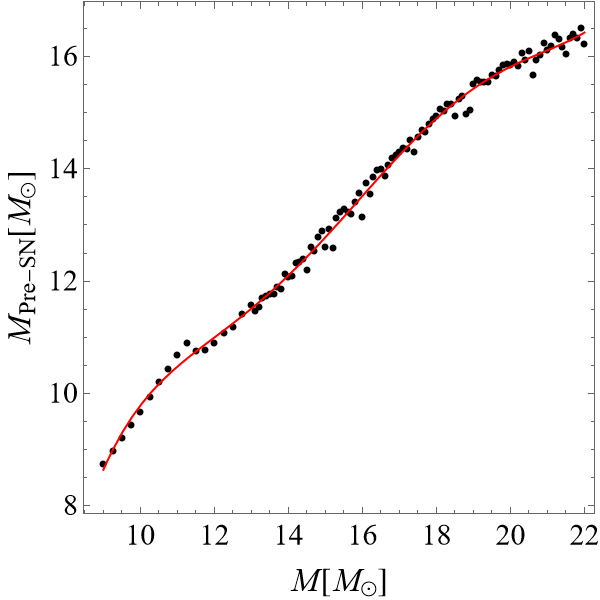}    
	\caption{Relation between the zero age main sequence mass of the progenitor -- $M$ -- and the stellar mass just before the SN explosion occurs -- $M_{\rm Pre-SN}$. While the black dots represents the results of \cite{Sukhbold+16}, the red, solid line represents the best-fit function, obtained using a 5th-order polynomial. } 
    \label{zams_vs_presn}
\end{figure}
\noindent
Second, we address the environmental diversity of the explosion sites. While massive stars predominantly form in associations, a significant fraction leave their birth sites due to the weak gravitational binding of these systems, ultimately exploding as isolated SNe. \cite{Castrillo2023} estimate that up to $\sim 25\%$ of massive stars are runaways, a fraction that is expected to increase when accounting for mechanisms of expulsion from clusters. Considering also that $\sim 10\%$ of the O-B stars are likely formed in isolation \citep{Chu2008}, this population adds to the runaway component. Crucially, both subgroups sample the ambient ISM density in an equivalent manner.\\\\
\corr{To quantify this variability in the density of the explosion site, we adopt the results of \cite{Kim_2017}, whose high-resolution MHD simulations naturally account for the environmental diversity of SN progenitors. Their models resolve the gravitational collapse driving star formation and explicitly compute SN explosions, which actively shape the density distribution of the surrounding medium. By tracking the specific location and time of each event, they determine the ambient density encountered by both runaway stars and those still near the original star formation site (cluster), revealing a clear dichotomy between the two populations -- see Fig.$\stick$\ref{SNe_ism}. 
\noindent
Specifically, for our analysis we adopt the distributions derived from their run with the highest spatial resolution ($2\stick$pc)}.
\begin{figure}[h!]
	\centering
	 \includegraphics[width=0.70\linewidth]{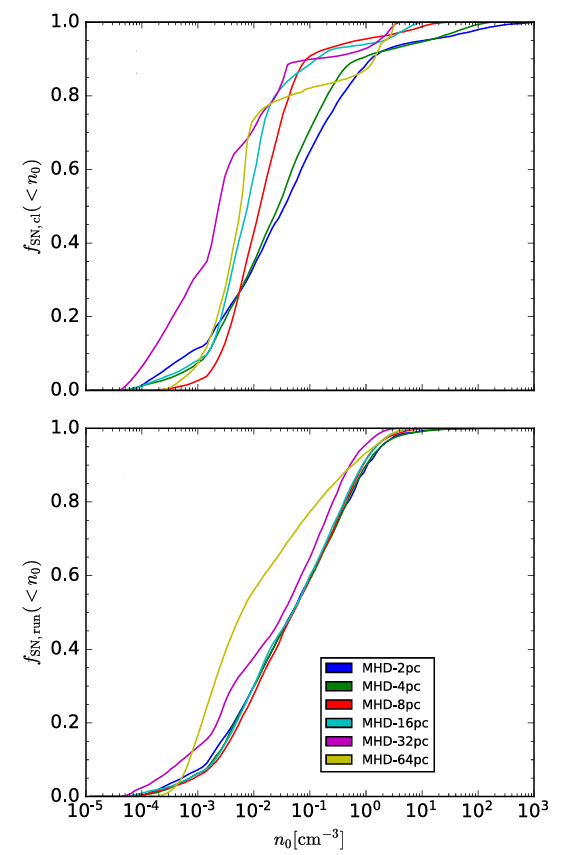}    
	\caption{Cumulative distribution function of the density at the site of the SN explosion for varying numerical resolution ($f_{\rm SN}$ representing cumulative fraction of SNe in regions where the ambient density is smaller than $n_0$ \citep{Kim_2017}). \textit{Upper panel}: SNe explosions occurring in clusters (subsctipt \textit{cl}). \textit{Lower panel}: SNe explosions of runaway stars (subscript \textit{run}).} 
    \label{SNe_ism}
\end{figure}\\\\
\noindent
Therefore, we assume that \corr{the same number of} SNe occur in cluster environments as those in isolation, but we apply distinct density cuts. Following the prescription of \cite{Kim_2017}, for SNe in clusters we sample the ambient density setting a lower limit of $10^{-4}\stick$cm$^{-3}$, which is the minimal density for the hot ionized medium, and an upper limit of $10^{3}\stick$cm$^{-3}$, corresponding to the density of molecular clouds. Conversely, for the isolated/runaway population, while we retain the same lower limit, we truncate the distribution to the typical density of the cold neutral medium, $10\stick$cm$^{-3}$. This approach allows us to investigate if more realistic variations recipes for both the mass of the progenitor star and the ambient density affect the statistical distribution of PWN-SNR systems in the ROI -- see Fig.$\stick${\ref{roi_modified}}.
\begin{figure}[h!]
	\centering
	 \includegraphics[width=0.72\linewidth]{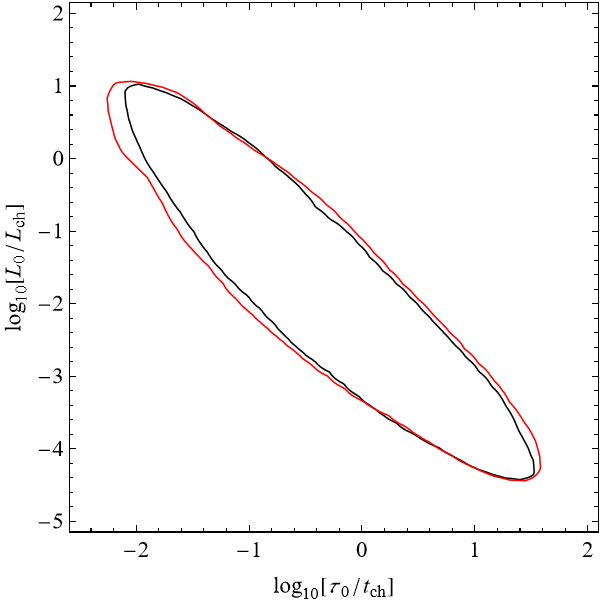}    
	\caption{Comparison between the canonical ROI, derived in Sect.$\stick${\ref{L0LCHT0TCH}} (black) and the region obtained by accounting for the mass loss of the progenitor \citep{Sukhbold+16} and the sampling of the ISM given by \cite{Kim_2017} (red).} 
    \label{roi_modified}
\end{figure}\\\\
Remarkably, these changes do not significantly modify the distribution of PWN-SNR equivalent systems in the characteristic plane. The corresponding new ROI does not exhibit substantial deviations from the reference case, preserving both the region of the peak of the distribution and its overall extent. This evidence confirms that the previously defined ROI is statistically robust, thereby justifying its adoption for the subsequent analysis.
\chapter{The Reverberation Phase in 1D}
\label{CHAPREV1D}
Once the RS reaches the boundary of the PWN, the assumptions valid during the free-expansion phase break down, as the nebula no longer expands into the unshocked, cold ejecta. Due to this, simple analytical prescriptions for the conditions in the SNR shell become inapplicable. This interaction profoundly alters the shell dynamics, leading to a severe compression and the potential onset of Rayleigh-Taylor instabilities, which arise when the heavier shocked ejecta push against the lighter relativistic fluid of the PWN. For these reasons, the thin-shell approximation is inadequate for describing the reverberation phase. Therefore, to address the complete evolutionary history, it is necessary to employ an approach capable of tracking the evolution of the PWN from its very birth up to the late stages. At the same time, to achieve good computational performances, this method should retain the simplifications of the one-zone model, thereby avoiding resolving the PWN internal dynamics.
\section{Lagrangian Hydrodynamics}
To track the evolution of the coupled system from the early beginning until the reverberation phase, we adopt a fully numerical approach using a Lagrangian hydrodynamical scheme. This allows one to maintain high resolution on the structures of the fluid as they evolve and expand in time, to track contact discontinuities, and to follow the various components, specifically the PWN, the SNR and the ISM. Assuming spherical symmetry, the evolution is governed by the fundamental equations of Lagrangian hydrodynamics, where the mass coordinate $m$ -- defined as the mass enclosed within a radius $r$ -- serves as the independent Lagrangian variable. The code solves the hydrodynamic equations formulated in Lagrangian mass coordinates. In this formulation, the continuity equation is expressed using the relation between mass and radius, together with momentum and energy conservation \citep{Morozova2015}:
\begin{align}
\label{LAGM}
\frac{\partial r}{\partial m}&=\frac{1}{4\pi r^2 \rho},
\\
\label{LAGMOM}
\frac{\text{D}v}{\text{D}t}&= -4\pi r^2 \frac{\partial P}{\partial m}  -4\pi \frac{\partial(r^2 Q)}{\partial m}.,
\\
\label{LAGENE}
\frac{\text{D}\epsilon}{\text{D}t}&= -P \frac{\text{D}}{\text{D}t}\left(\frac{1}{\rho}\right) - 2\pi r^2 Q \frac{\partial v}{\partial m}
\end{align}
Here, $\text{D}/\text{D}t \equiv \partial/\partial t + \boldsymbol{v}\cdot\boldsymbol{\nabla}$ is the Lagrangian time derivative at constant mass. The physical quantities $v$, $P$, $\rho$, $Q$ and $\epsilon$ represent the fluid radial velocity, pressure, mass density, the bulk viscosity and specific internal energy, respectively. We recall that, in the case of a non-relativistic fluid, one has:
\begin{equation}
\epsilon = \frac{3}{2}\frac{P}{\rho}. 
\end{equation}
Following the recipes described in \cite{Mezzacappa1993}, shock formations are handled through the implementation of standard von Neumann-Richtmyer viscosity \citep{VonNeumann1950}, $Q$, defined as:
\begin{equation}
Q \equiv
\begin{cases}
2\rho \left(\frac{\partial v}{\partial k} \right)^2 & \text{if } \frac{\partial v}{\partial k} < 0 \\
0 & \text{otherwise},
\end{cases}
\end{equation}
where $\partial / \partial k$ denotes the derivative with respect to an integer Lagrangian coordinate $k$, specifically, the mass of the cell. For a detailed description of the discretized equations implemented in the code, we refer the reader to Appendix~\ref{APPII}.\\\\
The Lagrangian time derivative, $\text{D}/\text{D}t$, physically shifts the description to a comoving reference frame. In this setup, the non-linear advection term, $\boldsymbol{v}\cdot\boldsymbol{\nabla}$, is naturally absorbed into the grid dynamics: the spatial configuration is not fixed a priori, but is effectively selected by the flow itself. The radial coordinate, $r$, ceases to be an independent integration variable and becomes a time-dependent quantity, whose evolution is explicitly updated at each timestep to follow the motion of the fluid. On the other hand, the mass enclosed within each computational shell remains invariant, whereas the cell volume evolves dynamically in response to the local expansion or compression. By physically anchoring the grid to the fluid elements, this approach allows the grid to naturally adapt to the flow, thereby ensuring high spatial resolution and preserving the sharpness of contact discontinuities.
\section{Numerical Setup and Initial Conditions}
\label{SETUP_1D_SECTION}
The necessity to treat the PWN and SNR as a coupled system determines our numerical strategy. To effectively capture both the free expansion phase and the complex dynamics of the interaction between the PWN and the RS, we adopt the hybrid approach recently developed by the Arcetri Group \citep{REV3}. This framework couples the one-zone model -- describing the global evolution of the PWN -- with a fully Lagrangian hydrodynamical scheme, that handles the evolution of the SNR. Consistent with the standard one-zone approximation, the PWN is treated as a uniform bubble of relativistic fluid, and we recall that its internal pressure is given by:
\begin{equation}
P_{\rm pwn} = \frac{E_{\rm pwn}}{4 \pi R^3_{\rm pwn}},
\end{equation}
where $E_{\rm pwn}$ denotes the internal energy, $R_{\rm pwn}$ the nebular radius, and the PWN pressure is evolved according to Eq.$\stick$(\ref{eq_ts_energy}). Conversely, the Lagrangian approach is specifically employed to spatially resolve the evolution of the SNR and the surrounding medium. In this configuration, the nebula acts as a spherical piston, effectively governing the dynamics of the surrounding remnant.\\\\
The physical setup is initialized with a standard set of parameters: an explosion energy $E_{\rm sn} = 10^{51}\stick$erg, an ejecta mass $M_{\rm ej} = 10 \, M_{\odot}$, and an ambient number density $n_{0} = 0.125\stick$cm$^{-3}$, itself characterized by an average mass per particle $m_0=m_{\rm p}$. We recall that the ejecta density profile assumes a flat core ($\delta = 0$) and a steeper envelope ($\omega = 12$), according to Eq.$\stick$(\ref{rho_piecewise}).
Regarding the initialization of the system, the starting time, $t_{\rm ini}$, is set in the range $[3-5]$ years, a value significantly smaller than the typical PSR spin-down time scale, $\tau_0$.
The initial radius of the nebula -- $R_{\rm ini}$ -- and the internal pressure are given by Eq.$\stick$(\ref{r_chev}) and Eq.$\stick$(\ref{P_ts}), respectively, evaluated at $t=t_{\rm ini}$. 
The ejecta velocity, $v_{\rm ej}$, is initialized assuming homologous expansion -- as described in Eq.$\stick$(\ref{density_hydro}). Furthermore, consistent with the assumption of cold expanding ejecta, their pressure, $P_{\rm ej}$, is set to a negligible value compared to the characteristic pressure of the SNR, specifically $P_{\rm ej} = 10^{-12} P_{\rm ch}$. Recalling the definition of the initial ejecta core radius -- $R_{\rm core}(t_{\rm ini}) \equiv v_{\rm c}t_{\rm ini}$, with $v_{\rm c}$ following Eq.\stick(\ref{vca}) -- the outer boundary of the SNR is initially truncated at $2R_{\rm core}(t_{\rm ini})$: this radius is sufficiently large to enclose 99.8\% of the ejecta mass, given the selected density profile. With the chosen physical parameters of the ejecta, the characteristic time-scale of the remnant, $t_{\rm ch}$, is $6489\stick$yr, whereas the characteristic length-scale, $R_{\rm ch}$, amounts to $14.84\stick$pc. 
To capture the long-term evolution of the system during the reverberation phase, we set the total duration of the simulation to $10t_{\rm ch}$, corresponding to $\simeq 65\stick$kyr.
Although we present our results for this specific set of SNR values, we recall that the hydrodynamical evolution can be rescaled to any specific source by normalizing against the characteristic variables, defined in Eqs.$\stick$(\ref{v_ch})$-$(\ref{P_ch}), and the evolution becomes universal.
For this reason our models will be referred to as classes of equivalent systems (systems for brevity).\\\\
Consistent with the one-zone approximation, which treats the nebula as a uniform bubble, the inner boundary of the first mass shell effectively acts as the driving interface. The domain of the simulation extends up to a radius of 64$\stick$pc. This, given the mass of the ejecta and the density of the ambient medium, corresponds to $\simeq 6.8 R_{\rm ch}$, ensuring that the FS remains well within the grid boundaries throughout the entire evolution. Regarding the computational grid, it consists of 5800 mass shells, extending from $R_{\rm pwn}(t_{\rm ini})$ to the outer boundary. The mesh is structurally divided to capture the relevant scales: the inner 2000 shells cover uniformly the SN cold ejecta -- extending from the PWN to the edge of the envelope up to $2R_{\rm core}(t_{\rm ini})$ --  while the remaining 3800 shells cover uniformly the ISM.
\section{The Evolution Beyond the Free-Expansion Phase}
\label{sec:beyond_rev}
In this framework, we now proceed to investigate the complex interaction between the nebula and the remnant.
We recall that within the characteristic plane, systems display extremely diverse dynamic properties in the ROI when they enter the reverberation phase -- see Figs.~\ref{trev_dist}, \ref{RPMV} and \ref{EEEE}. Specifically, we recall that the variations are highly pronounced along the major and minor axes of the ROI. Motivated by these findings, and with the aim of probing systems that are both statistically representative and exhibit distinct evolutionary histories, we selected specific groups of models to be investigated during the reverberation phase. While the major axis effectively captures the dominant evolutionary trend, simply sampling along this axis would yield an incomplete picture. To investigate the evolution along "orthogonal directions", we sampled the parameter space by selecting systems sharing identical intrinsic properties. Specifically, we identified groups distributed along the major axis, on regions of constant normalized reverberation time, $t_{\rm rev}/t_{\rm ch}$, initial spin-down luminosity $L_0/L_{\rm ch}$, and initial spin-down time $\tau_0/t_{\rm ch}$ -- see Fig.$\stick$\ref{SELECTEDSYSTEMSROI} and Table$\stick$(\ref{COES_compact}). The latter also includes the compression factor (CF) -- defined in Eq.~(\ref{EQCF}) as the ratio between the maximum and minimum nebular radius during the first compression cycle -- for each system.
\begin{figure}[h!]
    \centering
    \includegraphics[width=0.37\linewidth]{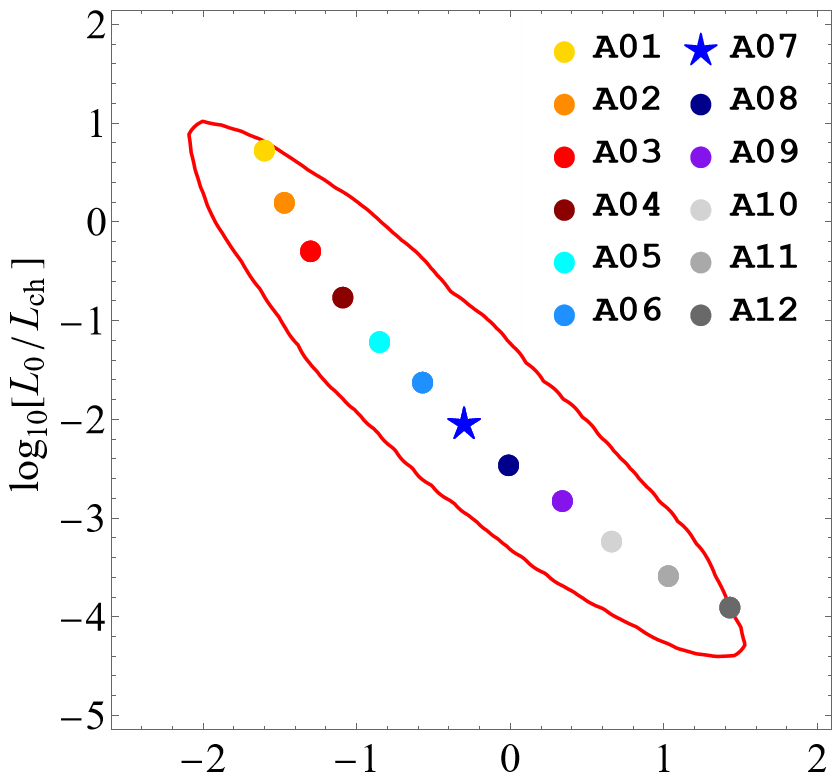}%
    \hspace{0.02\linewidth}%
        \includegraphics[width=0.35\linewidth]{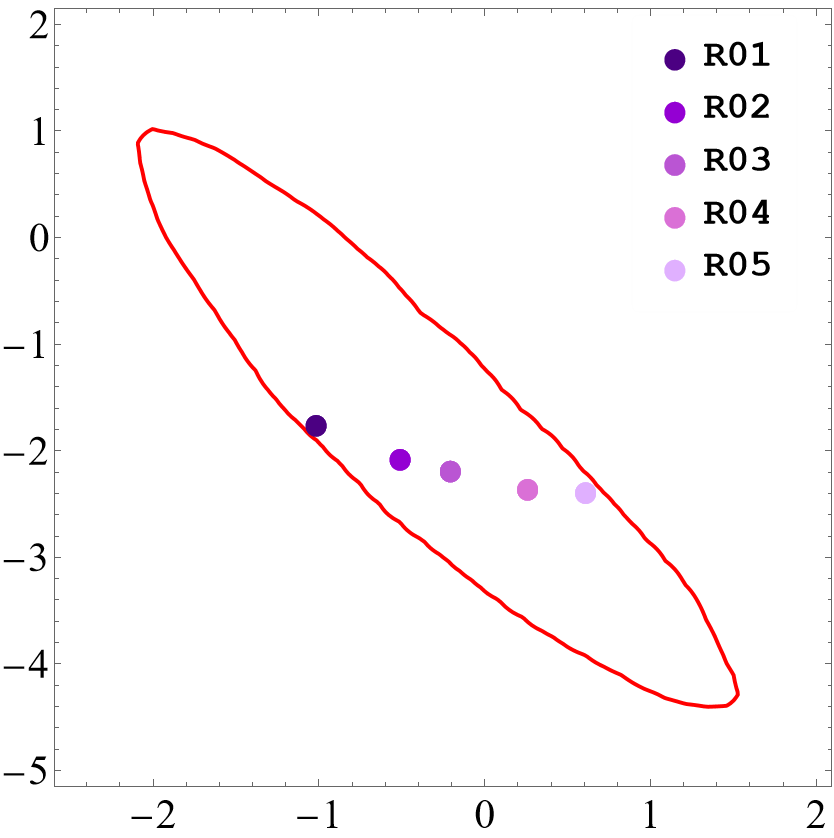}%
    \vspace{0.2cm} %
    
    \includegraphics[width=0.37\linewidth]{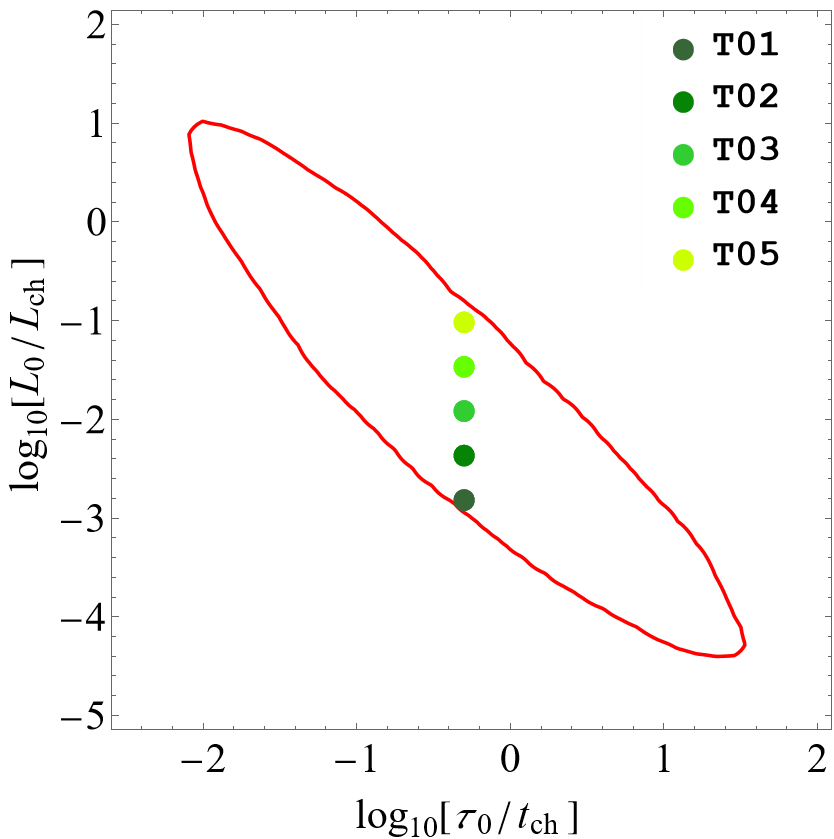}%
    \hspace{0.02\linewidth}%
    \includegraphics[width=0.35\linewidth]{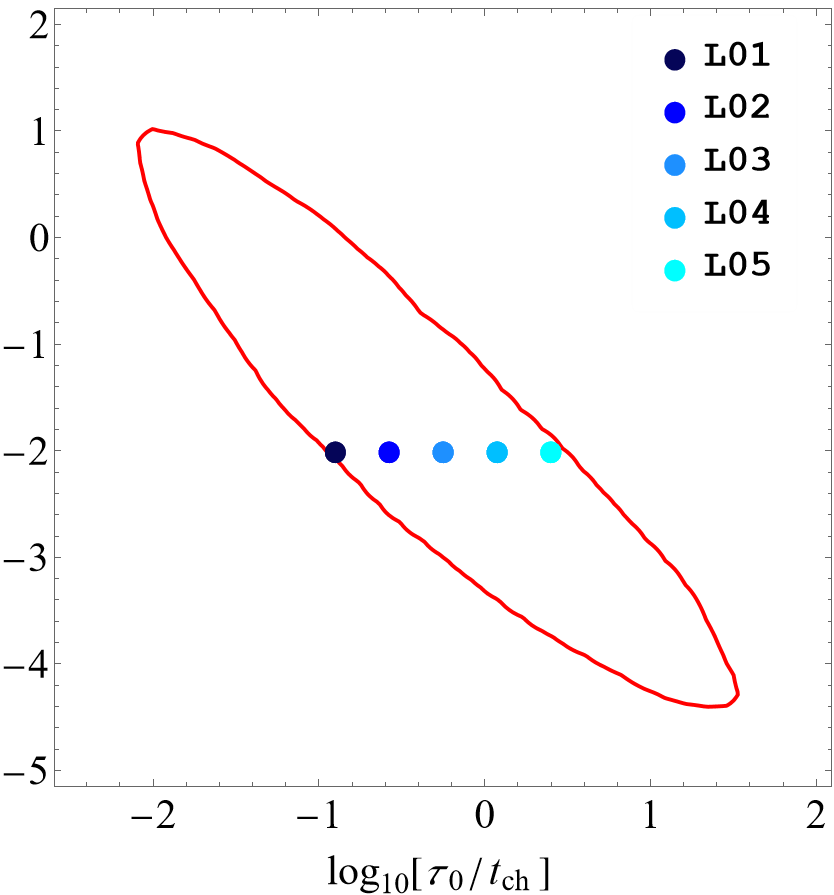}
    
    \caption{Groups of systems sampled across the ROI (red solid line), along the major axis (\texttt{A}, \textit{Upper-left panel}), in the same isolevel of $t_{\rm rev}/t_{\rm ch}$ (\texttt{R}, \textit{Upper-right panel}), at fixed $\tau_0/t_{\rm ch}$ (\texttt{T}, \textit{Lower-left panel}) and $L_0/L_{\rm ch}$ (\texttt{L}, \textit{Lower-right panel}), see also Table$\stick$(\ref{COES_compact}).}
    \label{SELECTEDSYSTEMSROI}
\end{figure}
\begin{table}[h!]
\centering
\setlength{\tabcolsep}{4pt} 
\begin{tabular}{cccc | cccc} 
\hline
\hline
System & $\log_{10}[\tau_0/t_{\rm ch}]$ & $\log_{10}[L_0/L_{\rm ch}]$ & \text{CF} & System & $\log_{10}[\tau_0/t_{\rm ch}]$ & $\log_{10}[L_0/L_{\rm ch}]$ & \text{CF} \\
\hline
\hline
\texttt{A01} & $-1.59$ & \stick$\,\,0.74$ & $1.81$ & 
\texttt{R01} & $-1.02$ & $-1.75$ & $11.1$\\

\texttt{A02} & $-1.47$ & \stick$\,\,0.21$ & $2.27$ &
\texttt{R02} & $-0.51$ & $-2.07$ & $5.46$\\

\texttt{A03} & $-1.30$ & $-0.28$ & $2.34$ &
\texttt{R03} & $-0.21$ & $-2.18$ & $3.72$\\

\texttt{A04} & $-1.09$ & $-0.75$ & $2.85$ &
\texttt{R04} & $\,\,\,\,0.26$ & $-2.35$ & $2.51$\\

\texttt{A05} & $-0.85$ & $-1.20$ & $3.61$ &
\texttt{R05} & $\,\,\,\,0.61$ & $-2.38$ & $2.08$\\

\cline{5-8} 

\texttt{A06} & $-0.57$ & $-1.61$ & $3.55$ &
\texttt{T01} & $-0.30$ & $-2.80$ & $9.27$\\

\texttt{A07} & $-0.30$ & $-2.05$ & $3.70$ &
\texttt{T02} & $-0.30$ & $-2.35$ & $5.27$\\

\texttt{A08} & $-0.01$ & $-2.45$ & $3.81$ &
\texttt{T03} & $-0.30$ & $-1.95$ & $3.14$\\

\texttt{A09} & \stick$\,\,0.34$ & $-2.81$ & $3.75$ &
\texttt{T04} & $-0.30$ & $-1.45$ & $1.94$\\

\texttt{A10} & \stick$\,\,0.66$ & $-3.22$ & $4.51$ &
\texttt{T05} & $-0.30$ & $-1.00$ & $1.18$\\

\cline{5-8} 

\texttt{A11} & \stick$\,\,1.03$ & $-3.57$ & $5.53$ &
\texttt{L01} & $-0.90$ & $-2.00$ & $12.0$\\

\texttt{A12} & \stick$\,\,1.43$ & $-3.89$ & $7.23$ &
\texttt{L02} & $-0.57$ & $-2.00$ & $5.76$\\

\cline{1-4} 

 & & & & 
\texttt{L03} & $-0.25$ & $-2.00$ & $3.20$\\

 & & & & 
\texttt{L04} & $\,\,\,\,0.08$ & $-2.00$ & $3.31$\\

 & & & & 
\texttt{L05} & $\,\,\,\,0.40$ & $-2.00$ & $1.70$\\

\hline
\hline
\end{tabular}
\caption{Parameters $\log_{10}[\tau_0/t_{\rm ch}]$, $\log_{10}[L_0/L_{\rm ch}]$ and Compression Factors (CF) -- defined in Eq.$\stick$(\ref{EQCF}) -- for the systems illustrated in Fig. \ref{SELECTEDSYSTEMSROI}.}
\label{COES_compact}
\end{table}
\noindent
However, before discussing these cumulative results across the ROI, we first focus on a specific system, \texttt{A07}, which corresponds to the peak of the distribution -- see Fig.$\stick$\ref{ROI95}. This will serve to introduce key aspects and properties of the PWN-SNR dynamics, to discuss the properties of the various phases, as well as to demonstrate the capability of our Lagrangian code to evolve PWN-SNR coupled systems and accurately capture the formation of contact discontinuities -- see Fig.$\stick$\ref{SNAPSHOTS}.
\begin{figure}[h!]
	\centering
        \begin{minipage}{1.0\textwidth}
            \centering
            \includegraphics[width=1.0\linewidth]{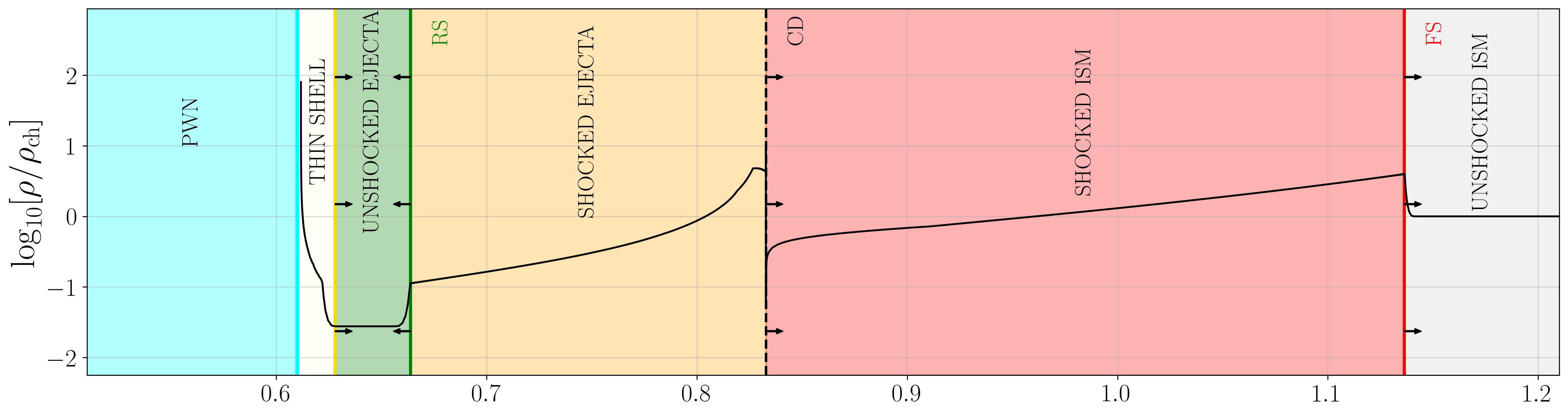}
            \includegraphics[width=1.0\linewidth]{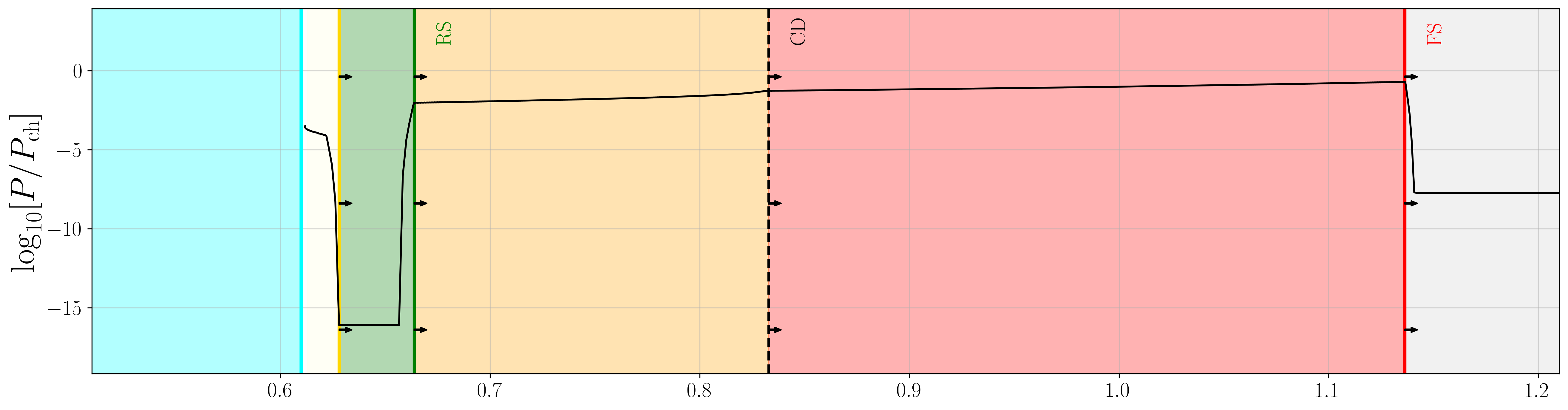}
            \includegraphics[width=1.0\linewidth]{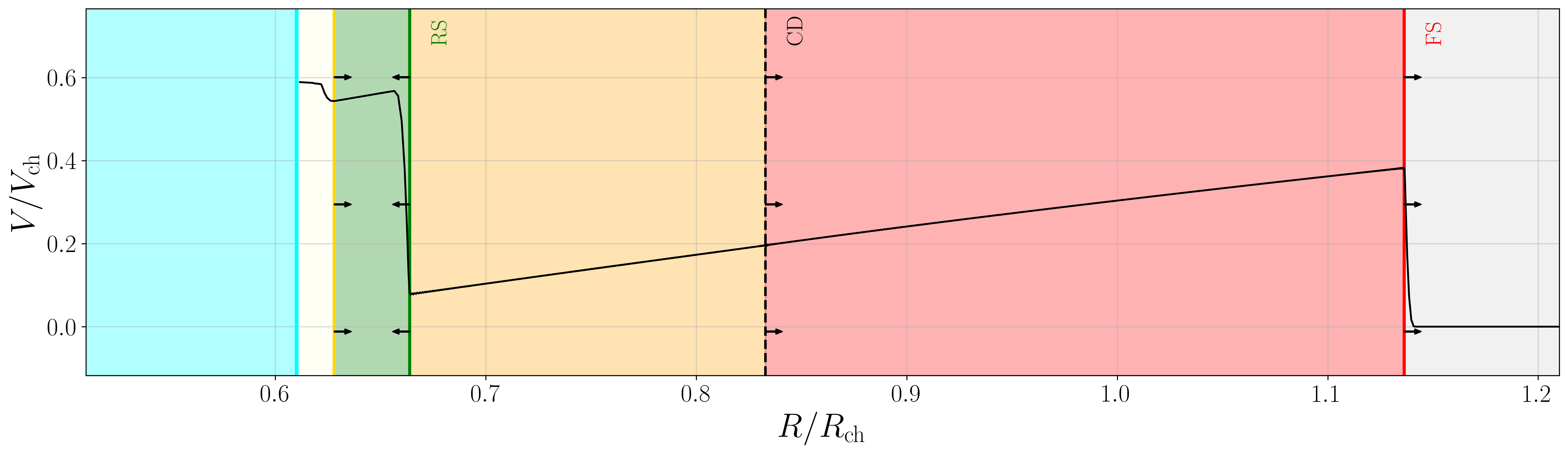}
        \end{minipage}  
	\caption{Profiles of normalized density (\textit{Upper panel}), pressure (\textit{Center panel}) and velocity (\textit{Lower panel}), extending from the boundary of the PWN to the ISM for the class of equivalent systems \texttt{A07}, evaluated at $t = 1.1 t_{\rm ch}$, reproducing the schematical representation in Fig.$\stick$\ref{snr_pwn}.} 
    \label{SNAPSHOTS}
\end{figure}\\
Starting from the inner region, the expanding nebula sweeps up a remarkably dense thin shell. This structure is characterized by a peak density of $\simeq 10^2  \rho_{\rm ch}$ at $\simeq 0.6 R_{\rm ch}$, and maintains a high internal pressure of $\simeq 10^{-3}  P_{\rm ch}$, in contrast to the negligible pressure of the unshocked ejecta ($\simeq 10^{-16}  P_{\rm ch}$). Expanding with a velocity of $\simeq 0.6  V_{\rm ch}$, the shell exhibits a steep internal density gradient, with values dropping to $\simeq 10^{-1} \rho_{\rm ch}$ at $\simeq 0.62 R_{\rm ch}$. Here, we can identify the cold, unshocked ejecta, where the density declines to $\simeq 10^{-2}  \rho_{\rm ch}$ and the velocity profile follows the homologous expansion described in Eq.$\stick$(\ref{density_hydro}). At $\simeq 1.1 t_{\rm ch}$, the domain of the unshocked ejecta extends outward from $\simeq 0.62  R_{\rm ch}$ to $\simeq 0.66  R_{\rm ch}$, which denote the outer radius of the swept-up shell and the location of the RS, respectively. The RS propagates into the cold ejecta, thermalizing the flow and delimiting the region of the shocked ejecta. Across the RS, the velocity drops from $0.6 V_{\rm ch}$ to $0.1 V_{\rm ch}$, while the pressure rises sharply to $\simeq 10^{-2} P_{\rm ch}$. Concurrently, the density increases from $\simeq 2.8\times 10^{-2} \rho_{\rm ch}$ immediately behind the RS, to $\simeq 0.11 \rho_{\rm ch}$, as expected for a strong shock. In this region, up to the CD, the density increases up to $\simeq 2$ orders of magnitude at the outer boundary, while the pressure remains approximately constant at $\simeq 10^{-2}P_{\rm ch}$, with the velocity rising to $\simeq 0.2 V_{\rm ch}$.\\\\
The outer edge of the shocked ejecta is defined by the CD, located at $\simeq  0.83 R_{\rm ch}$, which separates the ejected stellar material from the shocked ISM. While exhibiting continuity in pressure, the CD is characterized by a sharp discontinuity in density of approximately two orders of magnitude. Moving outward into the shocked ISM, both density and velocity increase radially, reaching peak values of $\simeq 4 \rho_{\rm ch}$ and $\simeq 0.4  V_{\rm ch}$, respectively, immediately behind the FS, located at $\simeq 1.15  R_{\rm ch}$ and separating the shocked ISM from the unshocked medium. Conversely, the pressure profile remains approximately constant at $\simeq 10^{-2}  P_{\rm ch}$ up to the FS.\\\\
Over a total duration of $10 t_{\rm ch}$, in order to follow the evolution with a fine sampling, we chose to output system snapshots approximately every $0.015 t_{\rm ch}$ -- yielding a total of 648 outputs -- recording density, pressure and velocity values across the entire computational grid -- see Fig.$\stick$\ref{SNAPSHOTS}.
Combining such snapshots, we can reconstruct the full temporal evolution of the \texttt{A07} system over a duration of $10 t_{\rm ch}$, as visualized in Fig.$\stick$\ref{SNAPSHOTS_TOGETHER}.
Crucially, it is important to recall that while the \texttt{A07} system serves as an illustrative example, it does not capture the full dynamical diversity of our population. Specifically, this system is characterized by an intermediate compression regime $(\text{CF}= 3.70)$ and for a more exhaustive analysis, covering scenarios that exhibit both weak (case \texttt{L05}, $\text{CF}=1.70$) and strong compressions (case \texttt{T01}, $\text{CF} = 9.27$) we refer the reader to Appendix \ref{APPII}.
\begin{figure}[h!]
    \centering 
     \begin{minipage}{1.0 \textwidth}
     	\includegraphics[width=0.99\linewidth]{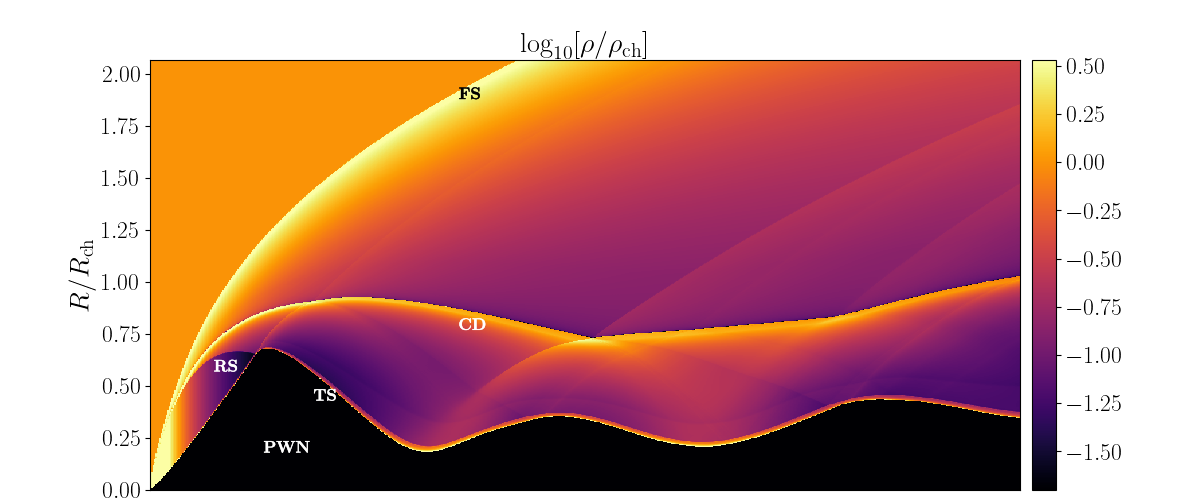}
     	\includegraphics[width=0.99\linewidth]{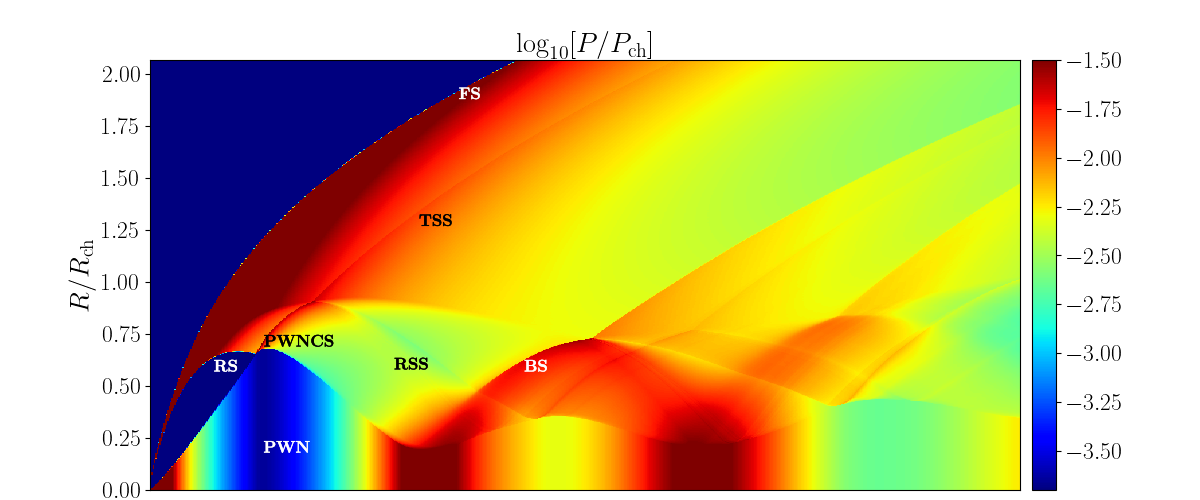}
    		\includegraphics[width=0.99\linewidth]{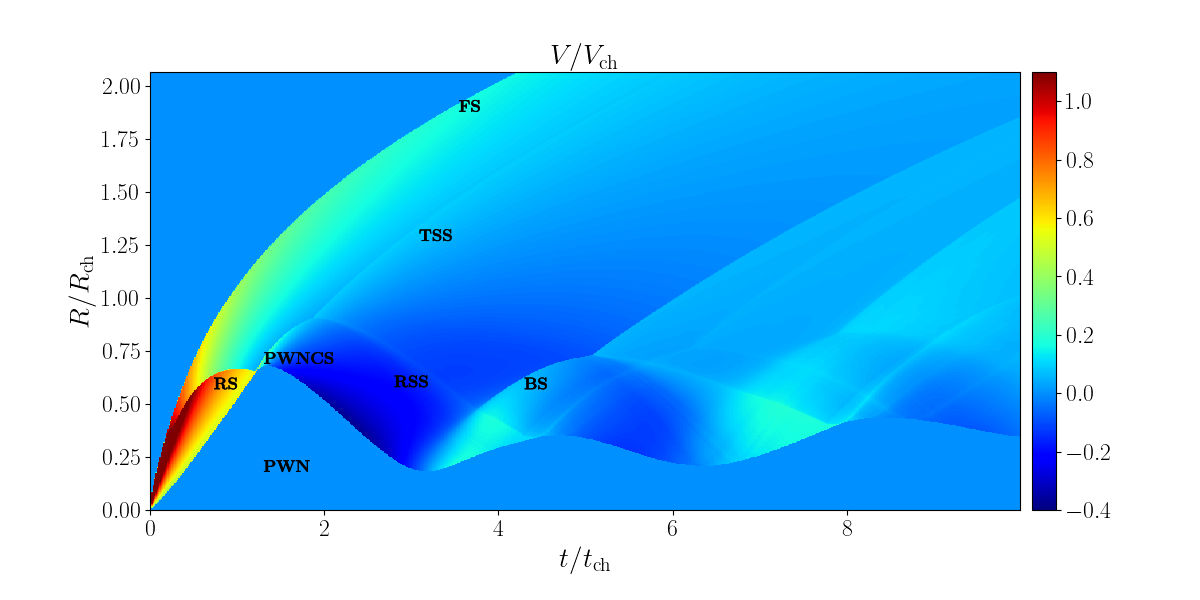}
    	\end{minipage}
    \caption{Evolution of the normalized density (\textit{Upper panel}), pressure (\textit{Center panel}), and velocity (\textit{Lower panel}) for the case \texttt{A07}. It is possible to distinguish the PWN (black area in the density graphics), the RS, the CD, the RS and the Thin-Shell (TS). Several shocks are produced as contact discontinuities interact. Among these, we recognize the PWN compression shock (PWNCS), the Transmitted Secondary Shock (TSS), the Reflected Secondary Shock (RSS) and the Bounce Shock (BS). }  
    \label{SNAPSHOTS_TOGETHER}
\end{figure}
Focusing on the density panel of Fig.$\stick$\ref{SNAPSHOTS_TOGETHER}, the PWN can be clearly identified as the low density (dark) central region. Consistent with the one-zone model, its internal pressure is uniform and its velocity is set to zero.
Since in the one-zone model the mass density of the PWN is not defined, this is set to an arbitrary low value of $10^{-7}\rho_{\rm ch}$ for visualization purposes.
The free-expansion phase lasts until $\simeq 1.2 t_{\rm ch}$, when the PWN has reached the RS at a radius $\simeq 0.65 R_{\rm ch}$. In the meantime, the FS has reached a radius of $\simeq 1.1 R_{\rm ch}$, while the CD, separating the shocked ISM from the shocked ejecta, is distinguishable at $\simeq 0.8 R_{\rm ch}$. While initially very close to each other, the RS, the CD and the FS become clearly distinct as the system evolves. 
While the nebula sweeps-up a thin shell (TS) during the free-expansion phase, its thickness -- $ 0.01 R_{\rm ch}$ -- is so small that is not clearly visible, and it becomes readily identifiable only after the interaction with the RS, when its thickness grows up to $0.03 R_{\rm ch}$ during the compression phase.\\\\
The evolution following the reverberation phase is quite complex. Recalling that the interaction with the RS occurs at $\simeq 1.2 t_{\rm ch}$, just after this time, the PWN keeps expanding because of the inertia of the swept-up shell, up to a maximum of $\simeq 0.68 R_{\rm ch}$, reached at $\simeq 1.3 t_{\rm ch}$. This overshooting launches a secondary shock front, the so-called PWN compressive shock (PWNCS), that propagates within the shocked ejecta region, between the RS and the CD, clearly visible in the velocity panel. Subsequently, the PWNCS reaches the CD at $\simeq 1.9 t_{\rm ch}$, when its extension is $\simeq 0.9 R_{\rm ch}$. This interaction generates a transmitted secondary shock (TSS), that propagates outward in the shocked ISM region of the SNR shell, and a reflected secondary shock (RSS), that propagates inward. In the meantime, the nebula undergoes compression, until it reaches a minimum radius of $\simeq 0.2 R_{\rm ch}$ at a time $\simeq 3.2 t_{\rm ch}$. During the compression, the pressure in the PWN rises, and at $\simeq 2.3 t_{\rm ch}$, when the extension of the nebula is $\simeq 0.4 R_{\rm ch}$, it begins to exceed the one in the surrounding SNR. 
As a consequence, after $\simeq 3.2 t_{\rm ch}$, the PWN re-expands, and in doing so a bounce shock (BS) is generated, that propagates in the shocked ejecta of the SNR shell. When the BS reaches the CD, at $\simeq 5 t_{\rm ch}$, it also splits in a transmitted and a reflected shock.
Following the first compression, a re-expansion regime begins, that lasts until $\simeq 4.8 t_{\rm ch}$, when the nebular radius measures $\simeq 0.35 R_{\rm ch}$. This phase is driven both by the overpressure reached at the minimum of the radius and by the continuous injection of the PSR spin-down energy. Subsequently, the interaction with the RSS triggers a second compression cycle, proceeding up to $\simeq 6 t_{\rm ch}$. Following this contraction, the nebula re-expands until it intercepts the reflected BS at $\simeq 8 t_{\rm ch}$, initiating a further compression cycle. In the meantime a complex network of secondary shocks forms and propagates in the SNR shell, and that drives the SNR interior toward the Sedov solution, characteristic of very late times. 
\subsection{Evolution Across the Major Axis of the ROI}
Shifting our focus exclusively to the nebular dynamics, we now examine the 
evolutionary history of the group \texttt{A}. Recalling that these systems populate the major axis of the ROI, they possess widely diverse evolutionary characteristics as they enter the reverberation phase -- see Figs.~\ref{trev_dist}, \ref{RPMV} and \ref{EEEE}. As expected, the simulations reveal different evolutions during the reverberation phase -- see Fig.$\stick$\ref{PWNRADPRSA}.
\begin{figure}[h!]
    \centering 
     	\includegraphics[width=0.95\linewidth]{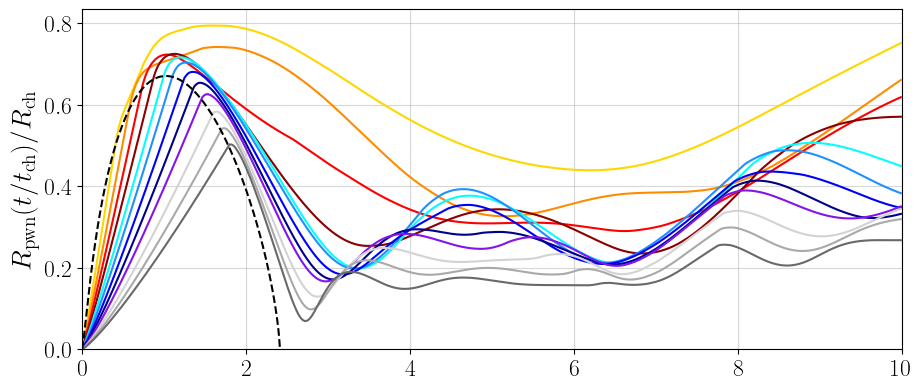}
     	\includegraphics[width=1.0\linewidth]{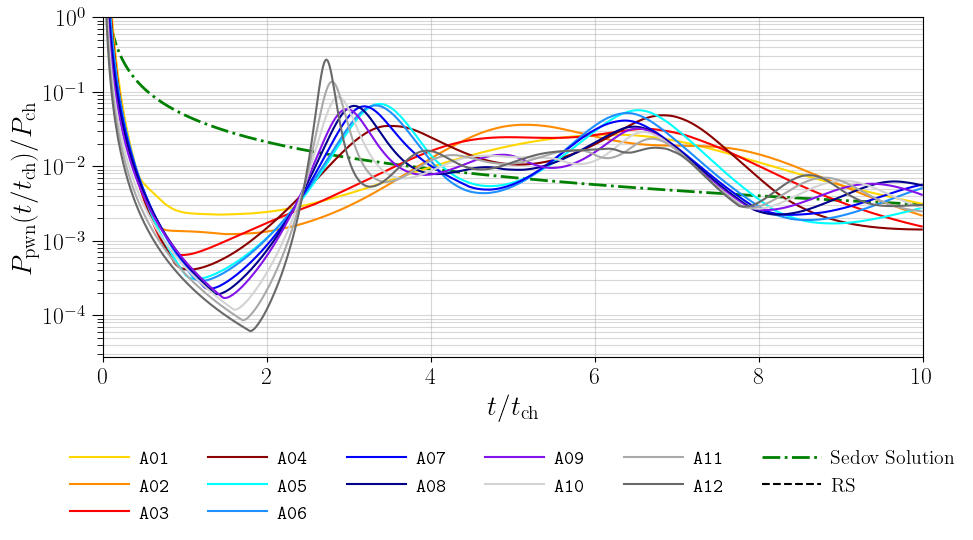}
    \caption{Evolution of the normalized nebular radius (\textit{Upper panel}) and pressure (\textit{Lower panel}) for the PWN-SNR systems sampled along the major axis of the ROI (group \texttt{A}).}  
    \label{PWNRADPRSA}
\end{figure}\\\\
Inspection of the radial evolution reveals that the majority of PWNe
interact with the RS when the nebular radius is $\simeq 0.6 R_{\rm ch}$. 
This interaction is followed by a cycle of compressions and expansions that 
persists significantly longer than the initial free-expansion phase. 
As a consequence, the observational probability of detecting a PWN that has 
already undergone the collision with the RS is considerably higher than 
finding one in the free-expansion stage, provided its residual luminosity remains above the instrumental limits in every observational band.
To quantify the subsequent contraction of the nebula, we define the compression factor, CF, as:
\begin{equation}
    \text{CF} \equiv \frac{R_{\rm pwn}(t_{\rm max})}{R_{\rm pwn}(t_{\rm min})},
    \label{EQCF}
\end{equation}
where $t_{\rm max}$ and $t_{\rm min}$ denote the times of maximum expansion 
and subsequent (first) maximum compression following the RS interaction, respectively. The CF is a central quantity in our analysis, acting as a discriminant for the nebular response to the RS, determining the subsequent spectral evolution. The results indicate different evolutionary paths based on energetics. The most luminous PWNe, characterized by short spin-down times -- $\tau_0 \simeq 0.05 t_{\rm ch}$ -- and high initial spin-down luminosities -- $L_0 \simeq L_{\rm ch}$ -- exhibit strong resistance 
to compression, with $\text{CF} \leq 2$ -- see Table (\ref{COES_compact}) -- displaying at most a single compression cycle over long time-scales.
These systems eventually re-expand to radii exceeding their size during the free-expansion phase. 
Conversely, fainter systems -- $\tau_0 \gtrsim 10 t_{\rm ch}$ and
$L_0 \lesssim 10^{-3} L_{\rm ch}$ -- suffer severe compressions, reaching 
$\text{CF} \simeq 7$. For these objects, after the first cycle of contraction, the radius tends to saturate, subsequently showing only weak oscillations.
Intermediate cases -- $\tau_0 \simeq t_{\rm ch}$ and
$L_0 \simeq 0.01 L_{\rm ch}$ -- display moderate compression factor values around $\simeq 3.5$. However, they exhibit the most interesting and dynamical behavior, characterized by distinct, multiple cycles of expansion and compression. It is crucial to highlight the energetic context: luminous PWNe possess high energy reservoirs within the shell and the nebula, alongside a substantial remaining budget in the PSR. In contrast, faint systems -- subject to the highest compression -- are characterized by low energies -- specifically, $\sim 10^{-3}$ times lower than most powerful PWNe. This suggests that the susceptibility to compression is strictly governed by the energy scales of the nebula. Furthermore, it is evident that high-luminosity PWNe complete the first compression cycle over extended timescales ($\simeq 6\,t_{\rm ch}$), whereas intermediate systems conclude it within $\simeq 3\,t_{\rm ch}$, with fainter systems requiring even less time ($\lesssim 3\,t_{\rm ch}$).\\\\
Turning to the normalized nebular pressure, $P_{\rm pwn}/P_{\rm ch}$, we 
observe a distinct transition between the last part of the free-expansion phase -- characterized by low pressure due to the expansion into cold ejecta -- and the reverberation phase, where compression drives a significant increase, consistent with Eq.~(\ref{rel_pwn}). 
The magnitude of this pressure growth is strictly connected to the compression efficiency. High-luminosity systems reach only moderate peaks during the first contraction ($\simeq 10^{-2} P_{\rm ch}$), while intermediate systems cluster around $\simeq 6 \times 10^{-2} P_{\rm ch}$. In sharp contrast, faint systems, driven by extreme compression, reach higher pressures $(\simeq 0.3 P_{\rm ch})$. Following this peak, the pressure evolution mirrors the dynamical behavior of the nebular radius. Consistent with their efficient re-expansion, luminous systems show a marked decrease in pressure immediately after reaching the minimum radius. Intermediate systems, characterized by a highly dynamic evolution, exhibit strong pressure oscillations that frequently approach values comparable to the first compression peak. Conversely, faint systems display a flatter profile; consistent with the lack of significant radial oscillations after the maximum compression, their pressure remains relatively stable without exhibiting large fluctuations.
Ultimately, at the end of the simulation -- $t = 10 t_{\rm ch}$ -- the systems converge to a narrow pressure range of 
$\simeq [1-7] \times 10^{-3} P_{\rm ch}$, indicating that the final 
pressure dispersion is remarkably contained.
It is instructive to compare the evolution of the nebular pressure with the theoretical expectation for the SNR interior during the Sedov-Taylor phase, where the expansion follows a self-similar solution. In this regime, assuming a non-relativistic adiabatic index $\Gamma = 5/3$ in the remnant, the pressure of the center of the SNR is given by \citep{Shu92}:
\begin{equation}
P_{\rm SNR} \simeq 0.306 \left(\frac{2}{\Gamma + 1} \right)\rho_0 V_{\rm FS}^2,
\end{equation}
where $\rho_0$ denotes the unshocked ISM density and $V_{\rm FS}$ represents the velocity of the FS, which is propagating in the ISM. The evolution of the FS radius, $R_{\rm FS}$, follows the standard Sedov scaling (\citealt{Sedov46}, \citealt{TM99}): 
\begin{equation}
R_{\rm FS} \simeq 1.15\left(\frac{E_{\rm sn}}{\rho_0}\right)^{1/5} t^{2/5},
\end{equation}
where $E_{\rm sn}$ is the kinetic energy of the SN. By deriving the corresponding expansion velocity and normalizing the resulting pressure in terms of characteristic units, we obtain the analytical profile for the pressure:
\begin{equation}
P_{\rm SNR} \simeq 4.8\times 10^{-2} P_{\rm ch }\left(\frac{t}{t_{\rm ch}}\right)^{-6/5}.
\label{Sedov_Pressure_SNR}
\end{equation}
Remarkably, while this solution is unsuitable for describing the pressure at the remnant center during the complex reverberation phase, it captures the asymptotic behavior of the system. At late times $(\gtrsim 8 t_{\rm ch})$ the nebular pressure converges to this analytical prediction, indicating that both the PWN and the host SNR relax toward a self-similar state, in agreement with the evolution presented in Fig.~\ref{SNAPSHOTS_TOGETHER}.
\subsection{Evolution at Constant $t_{\rm rev}/t_{\rm ch}$}
Proceeding with the analysis, we now focus on the evolutionary outcomes of systems with the same normalized reverberation time (group \texttt{R}), specifically selecting the value $t_{\rm rev}/t_{\rm ch} = 1.3$ -- see Fig.$\stick$\ref{TREVLAG}.
\begin{figure}[h!]
    \centering 
     	\includegraphics[width=0.95\linewidth]{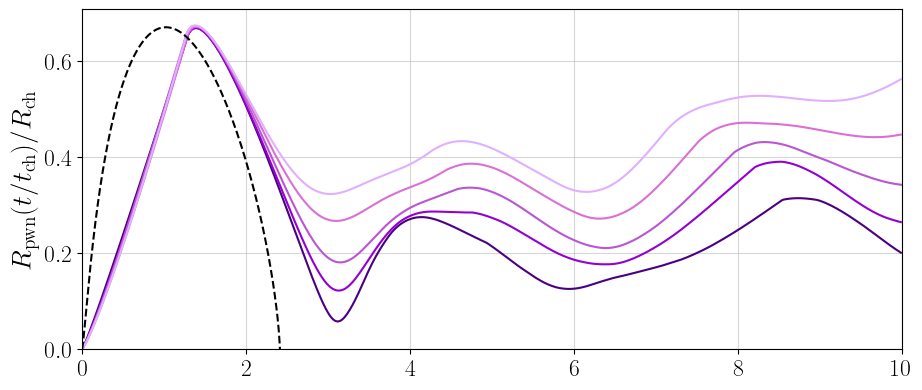}
     	\includegraphics[width=1.0\linewidth]{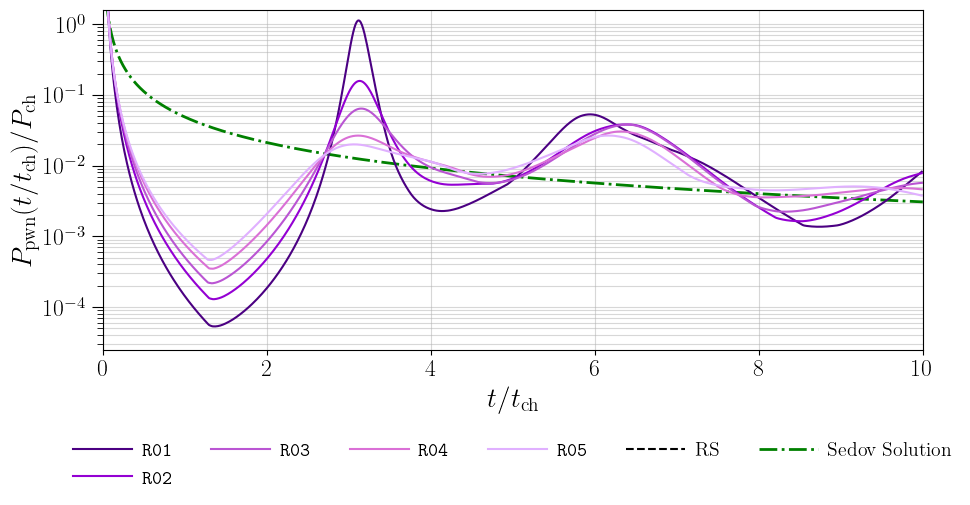}
    \caption{Evolution of the normalized nebular radius (\textit{Upper panel}) and pressure (\textit{Lower panel}) for the PWN-SNR systems with the same normalized reverberation time (group \texttt{R}).}  
    \label{TREVLAG}
\end{figure}\\\\
These models are characterized by a high uniformity at the moment of the interaction: they encounter the RS with nearly identical radii and velocities. Furthermore, as illustrated in Fig.$\stick$\ref{EEEE}, they possess similar energy contents within the swept-up shell, alongside comparable energy injections from the PSR. More pronounced discrepancies emerge regarding the internal nebular energy and the remaining energy reservoir available to sustain the reverberation phase. Crucially, these quantities increase with increasing initial spin-down time. The evolutionary trajectories display marked differences in terms of radial profile. It is noteworthy, however, that all systems reach the minimum radii simultaneously at $\simeq 3 t_{\rm ch}$ and at $\simeq 6 t_{\rm ch}$. In agreement with our previous findings, systems possessing larger energy reserves exhibit lower compression factors. 
Quantitatively, the most luminous systems -- characterized by short spin-down times and then faster dissipation of their spin-down energy -- experience strong compression, reaching CF $\simeq 11$. On the other hand, systems with higher spin-down timescales show a better resistance, with CF $\simeq 2$, along with intermediate cases showing values of CF $\simeq 3.5$.
It is important to notice that this group of systems, which spans less than a decade in luminosity, reveals a behavior distinct from that of group \texttt{A}. Unlike the previous case, all systems in this sample exhibit an evident cycle of compression and re-expansion. In this context, the most luminous systems exhibit a lower resistance to compression, a direct consequence of their depleted energy budget.
Conversely, the fainter systems are characterized by very high initial spin-down times. Crucially, while both subsets reach the reverberation phase with a comparable amount of injected energy, the remaining energy budget of the fainter systems is $\simeq 100$ times larger than that of the luminous ones. This vast reservoir allows the PWNe to re-expand efficiently, and eventually recover radii comparable to their extent at the end of the free-expansion phase. Finally, despite these different radial evolutionary histories, at the end of the simulation, we found pressures similar to the previous case, converging to $\simeq [4-9] \times 10^{-3} P_{\rm ch}$. 
It is worth noting that, in analogy with the behavior observed in group \texttt{A}, the Sedov solution accurately describes the internal pressure evolution for group \texttt{R} systems only at late times $(\gtrsim 8 t_{\rm ch})$ confirming that the system is relaxing toward a self-similar state. This behavior appears evident focusing on the most powerful systems, while the fainter ones exhibit more significant deviations. 
\subsection{Evolution at Constant $\tau_0/t_{\rm ch}$}
Turning our attention to the systems sampled at a constant normalized initial spin-down time (group \texttt{T}), we observe evolutionary trajectories that differ markedly from group \texttt{R} -- see Fig.$\stick$\ref{T0SDLAG}.
\begin{figure}[h!]
    \centering 
     	\includegraphics[width=0.95\linewidth]{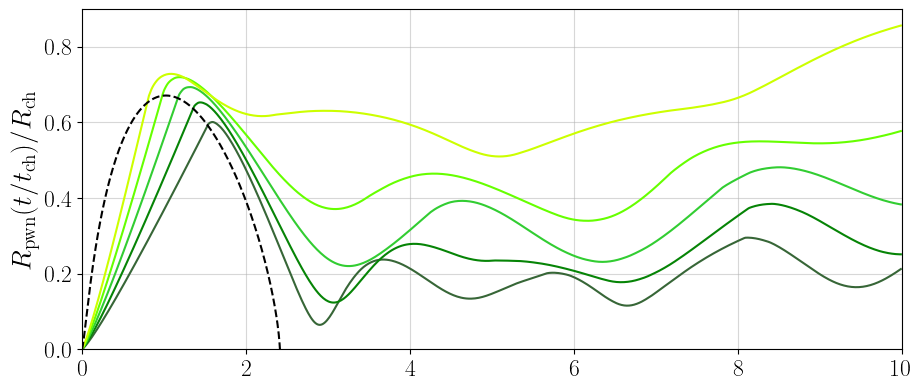}
     	\includegraphics[width=1.0\linewidth]{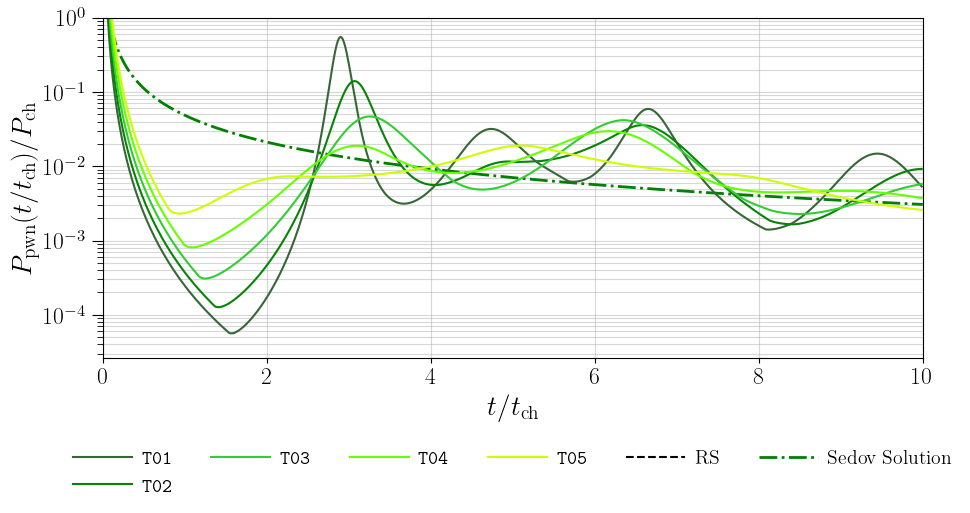}
    \caption{Evolution of the normalized nebular radius (\textit{Upper panel}) and pressure (\textit{Lower panel}) for the PWN-SNR systems with identical initial spin-down time (group \texttt{T}).}  
    \label{T0SDLAG}
\end{figure}\\\\
Given the substantial variability of the dynamical state across this specific sequence in the parameter space -- as evidenced in Fig.$\stick$\ref{RPMV} and Fig.$\stick$\ref{EEEE} -- these evolutional differences in pressure and radius were expected. A detailed inspection reveals that, while systems with higher luminosities encounter the RS at radii comparable to their fainter counterparts $(\simeq 0.6 R_{\rm ch})$, they enter the reverberation phase exhibiting significantly higher velocities. 
It is evident that there is a drastic inversion of the energy balance compared to the previous case. Here, the luminous systems possess an accumulated energy injection and a remaining reservoir for reverberation that are nearly two orders of magnitude higher than their fainter ones.
This energetic difference directly governs the dynamical response: luminous systems exhibit minimal compression, with $\text{CF} \simeq 1.2$. On the other hand, fainter systems undergo severe contraction, reaching $\text{CF} \simeq 9.3$, with intermediate cases falling between these extremes, typically settling at $\text{CF} \simeq 3.1$.
Furthermore, in contrast to the previous case, the evolutionary trajectories do not exhibit any similarity; notably, the compression and subsequent re-expansion phases occur asynchronously across the sample, rather than concentrating around a specific time. Specifically, the most luminous systems exhibit a single compression cycle followed by a strong re-expansion. As the luminosity decreases, the behavior remains highly dynamic, consistent with the fact that we are sampling a region of the parameter space including the intermediate systems of groups \texttt{R} and \texttt{A}. Despite this temporal and dynamical heterogeneity, the systems once again exhibit a remarkable convergence in their final states of pressure. By the end of the simulation, the dispersion of the pressure is minimal, settling within the narrow range of $[3-10] \times 10^{-3} P_{\rm ch}$. While the tendency toward the Sedov solution described in Eq.~(\ref{Sedov_Pressure_SNR}) is less evident here, the internal pressure still appears to approach this regime at late times ($\gtrsim 10t_{\rm ch}$) for the most luminous systems, while the fainter show more evident deviations, similarly to group \texttt{R}.
\subsection{Evolution at Constant $L_0/L_{\rm ch}$}
Shifting our focus to systems characterized by fixed normalized luminosity (group \texttt{L}), we observe, once again, distinct evolutionary paths -- see Fig.$\stick$\ref{L0SDLAG}.
\begin{figure}[h!]
    \centering 
     	\includegraphics[width=0.95\linewidth]{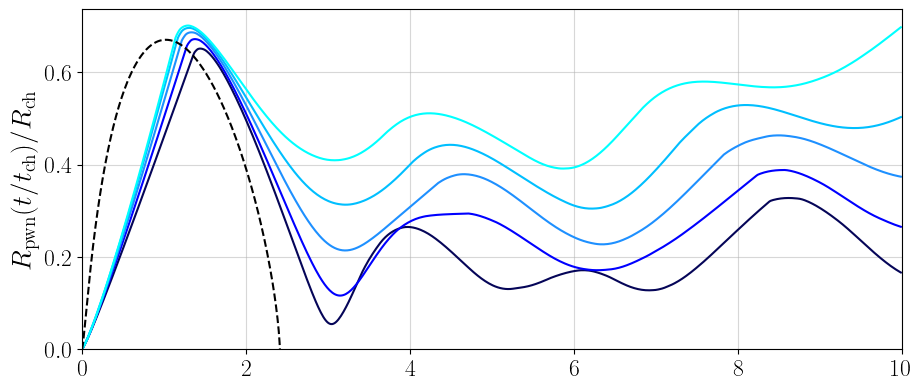}
     	\includegraphics[width=1.00\linewidth]{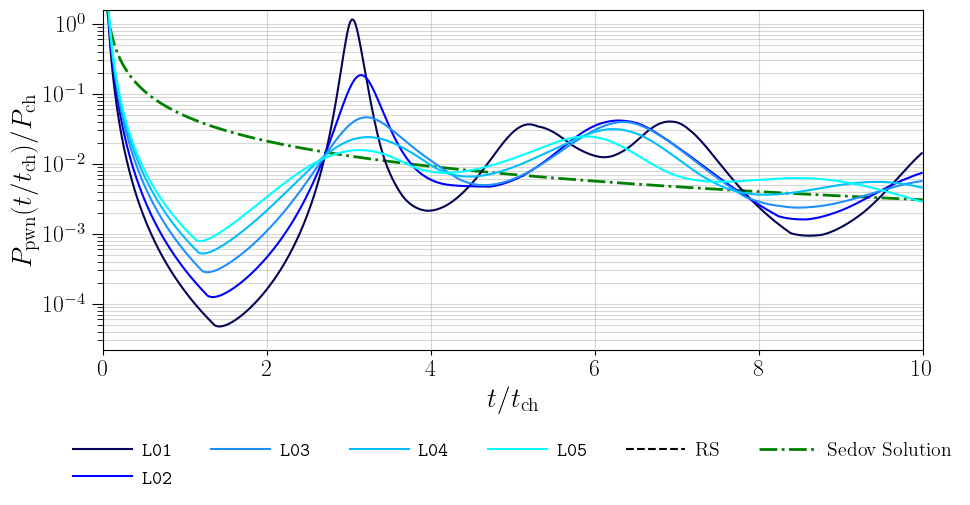}
    \caption{Evolution of the normalized nebular radius (\textit{Upper panel}) and pressure (\textit{Lower panel}) for the PWN-SNR systems with identical initial spin-down luminosity (group \texttt{L}).}  
    \label{L0SDLAG}
\end{figure}\\\\
Recalling the findings presented in Sect.$\stick$\ref{LOSEMASSINCLUSION} and Figs.$\stick$\ref{RPMV}--$\stick$\ref{EEEE}, although these systems arrive at the reverberation phase with a comparable energy injection from the PSR, they possess vastly different energy budgets available to sustain the subsequent evolution. Specifically, while the total injected energy remains approximately constant, the remaining reservoir available to power reverberation increases with the initial spin-down time, up to two orders of magnitude. Therefore, the analysis highlights that compressive phases are significantly more effective in regimes characterized by rapid spin-down energy dissipation. Indeed, systems with low initial spin-down time suffer strong compressions, evidenced by a compression factor $\text{CF}\simeq 12$. As $\tau_0/t_{\rm ch}$ increases, the resistance grows, leading up to CF $\simeq 3.2$, for intermediate systems, eventually reaching values of $\text{CF}\simeq 1.7$ (system \texttt{L05}). It is crucial to remark that the evolution of these systems is fully consistent with that of group \texttt{R}, a coherence anticipated by their proximity in the parameter space. Ultimately, and consistent with the behavior identified in the preceding cases, the systems converge to a final pressure state within the range of $\simeq [3-13]\times 10^{-3} P_{\rm ch}$. Again, for group \texttt{L}, although the pressure approaches the Sedov limit, this trend is clearly distinguishable in luminous systems, whereas faint ones diverge more markedly from the self-similar solution.
\section{Discussion}
The extensive analysis conducted on the evolutionary paths of groups \texttt{A}, \texttt{R}, \texttt{T}, and \texttt{L} leads to the robust conclusion that the radial history of PWNe changes drastically across the characteristic plane. We recall that systems exhibiting the most complex dynamics -- characterized by multiple cycles of compression and re-expansion -- are those most commonly found within the ROI. In contrast, the most luminous systems generally undergo a single compression event followed by a vigorous re-expansion. 
A central quantity in interpreting these diverse behaviors is the CF. Systems with low energies offer minimal resistance to the RS, suffering extreme contractions $(\text{CF} \gtrsim 7)$, whereas energetic pulsars limit the compression to modest factors, exhibiting $\text{CF} \simeq 2$, with intermediate systems showing $\text{CF}\simeq 3.5$. Crucially, as discussed in \cite{Bandiera2023}, the magnitude of the CF determines the subsequent spectral evolution of the nebula. High compression drives drastic magnetic field amplification and adiabatic heating of the particles. This can trigger a \textit{super-efficiency} (\citealt{Rev1}, \citealt{REV3}) in synchrotron emission -- see Eq.$\stick$(\ref{t_sync}) -- leading to a rapid \textit{burn-off} of the particle population. Remarkably, despite these profound disparities in radial evolution, all systems exhibit a striking convergence in pressure at late times $(\gtrsim 5 t_{\rm ch})$ as shown in Fig.$\stick$\ref{PRSAVG}.
\begin{figure}[h!]
    \centering 
     	\includegraphics[width=1.0\linewidth]{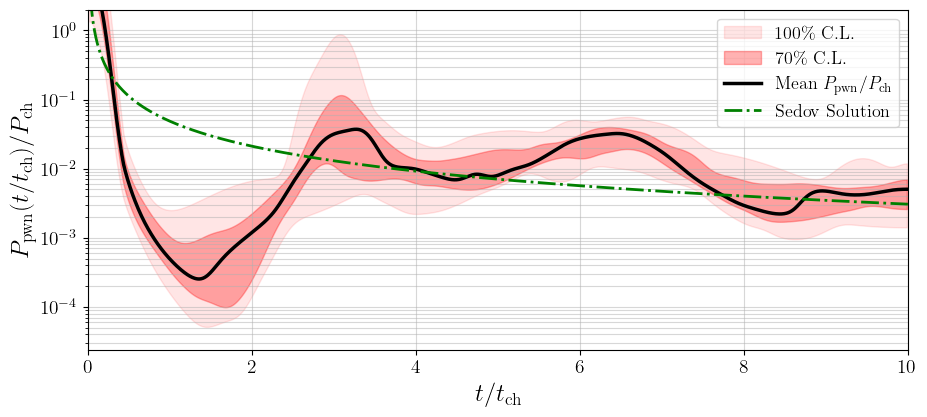}
    \caption{Mean pressure (black solid line) obtained by the evolution of the models reported in Table~\ref{COES_compact}, compared to the Sedov solution (green dashed line), while coloured areas represent the confidence bands. Notably, the 70\% C.L. band encompasses the pressure profiles of 19 out of the 27 selected systems within the ROI.}  
    \label{PRSAVG}
\end{figure}\\\\
It is important to clarify that the mean pressure trend displayed here is not weighted over the entire population, but represents a purely numerical average over all the systems in Table~\ref{COES_compact}. 
In this sense, the confidence bands serve as an indicator of the expected range in which the final pressure is likely to settle. 
However, many of the systems overlap in the central region of the ROI, corresponding to the most populated area -- see also Fig.$\stick$\ref{ROI95}. As a consequence, the resulting mean pressure weights more systems that are intrinsically more probable. It is evident that the pressure profiles, initially distinct, settle into a narrow range, approaching the Sedov solution at late times, typically $\gtrsim 8 t_{\rm ch}$. 

\chapter{The Reverberation Phase in 2D}
\label{CHAPREV2D}
The Lagrangian approach that we have adopted to evolve a PWN-SNR system, and that was illustrated in the previous chapter, efficiently captures the dynamics of the compression of the PWN and the formation of shocks and contact discontinuities as it expands and contracts in the hosting SNR. This notwithstanding, this scheme treats the system as purely 1D, while PWN-SNR systems are intrinsically 3D, often exhibiting asymmetries and instabilities absent in 1D models. For these reasons, one expects that increasing the dimensionality of the system should lead to a more realistic description. Specifically, Rayleigh-Taylor instabilities \citep{Hester1996} are expected to arise from the interaction between the low-density, relativistic PWN plasma and the hot, denser SN shocked ejecta. Despite the importance of these effects, the dynamics in multiple dimensions, during the compression phase, remains a largely unexplored problem \citep{Kolb2017}. Investigating this regime is crucial, as it provides a necessary test for the robustness of the results derived from 1D Lagrangian simulations which, at present, are the only models capable of making predictions about the spectral evolution of these systems. 
\section{The PLUTO Code}
To achieve this and capture the full evolution of PWN-SNR systems, it is necessary to use a multi-dimensional code for fluid dynamics. For this purpose, we employ the PLUTO\footnote{See also: \url{https://plutocode.ph.unito.it/}}  code \citep{Pluto_Static}. PLUTO is a finite-volume / finite-difference shock-capturing code, designed to integrate the hydrodynamic conservation laws of mass, momentum and energy:
\begin{equation}
	\label{pluto_mass}
    \frac{\partial \rho}{\partial t} + \boldsymbol{\nabla}\cdot(\rho\boldsymbol{v})=0,
\end{equation}
\begin{equation}
\frac{\partial (\rho\boldsymbol{v})}{\partial t} + \boldsymbol{\nabla}\cdot\left[\rho\boldsymbol{v}\boldsymbol{v}+P\boldsymbol{I}\right]=0,
\label{pluto_momentum}
\end{equation}
\begin{equation}
\frac{\partial}{\partial t}\left[\rho\epsilon + \frac{\rho \boldsymbol{v}^2}{2} \right] + \boldsymbol{\nabla}\cdot\left[\left(\rho\epsilon + \frac{\rho \boldsymbol{v}^2}{2}+P\right)\boldsymbol{v}\right] = 0,
\label{pluto_energy}
\end{equation}
where $\rho$ denotes the mass density, $\boldsymbol{v}$ the velocity, $P$ the pressure, $\epsilon$ the specific internal energy and $\boldsymbol{I}$ the metric matrix. We recall that, in the case of an ideal non relativistic fluid, the specific internal energy is given by:
\begin{equation}
\epsilon = \frac{3P}{2\rho}.
\end{equation}
PLUTO is an Eulerian code, and, as a consequence, the spatial resolution is set \textit{a priori}, potentially leading to a loss of accuracy on sharp structures compared to the naturally adaptive grid of a Lagrangian approach. To mitigate this drawback, PLUTO employs an Adaptive Mesh Refinement (AMR) module \citep{Pluto_AMR}. Instead of using a uniform grid with high-resolution -- which would be computationally prohibitive in multiple dimensions -- this module allows the code to automatically increase the spatial resolution only where complex flow patterns occur. 
In this way, the local resolution can be progressively enhanced up to a maximum number of refinement levels, $l_{\rm max}$, arbitrarily chosen by the user, ensuring high precision in capturing sharp structures and instabilities.\\\\
Building on this framework, we perform our simulations in a 2D domain, specifically adopting a spherical geometry in the $(r, \theta)-$plane, where $r$ denotes the radial coordinate and $\theta$ the polar angle.
The selection of the $(r, \theta)$ plane over $(r, \phi)$ -- where $\phi$ is the azimuthal angle -- is dictated by the technical constraints of the Adaptive Mesh Refinement module, which natively supports only the first domain in a 2D spherical setup. 
Transitioning to multi-D implies a drastic increase in computational time, and this makes a parameter study in 3D computationally very expensive. For this reason, we have decided to carry out our study in 2D as a necessary methodological preliminary. Indeed, 2D can already reveal inconsistencies with respect to 1D models, and serves as a starting point for more demanding 3D runs.
In any case, replicating the full set of 1D simulations presented in Sect.$\stick$\ref{sec:beyond_rev} is prohibitive also in 2D. We therefore focus on evolving only one of the most probable systems within the ROI, namely the \texttt{A07} system.
\section{Numerical Strategy for 2D Simulations}
One of the objectives of this work is not merely to simulate the full evolution of a system from its initial conditions, but -- perhaps even more important -- to identify an efficient computational strategy to investigate a multidimensional physical problem while optimizing computational times. The initial expansion phase involves extremely small scales that would require high computational cost to be accurately resolved in multi-D, together with the scales characteristic of later phases. 
For this reason, we chose not to evolve the system in 2D from the very beginning. Instead, we used the results of 1D Lagrangian simulations as input for the 2D simulations in PLUTO. Recalling that the Lagrangian code outputs comprehensive system snapshots -- recording density, pressure, and velocity of the SNR across the grid -- we initialize the 2D simulation by directly mapping the 1D hydrodynamic profiles of the SNR from the snapshot extracted just before the onset of the reverberation phase. For the PWN the one-zone Lagrangian simulation only provides the pressure, which we assume uniform, leaving both velocity and density undefined, to be chosen arbitrarily -- see Sect.$\stick$\ref{NRPWN}. By exploiting this feature, we bypass the free-expansion phase -- which we already fully characterized in Sect.$\stick$\ref{LOSEMASSINCLUSION}.\\\\ 
Despite the computational efficiency provided by the AMR module and the significant advantage of bypassing the free-expansion phase, transitioning to 2D simulations in PLUTO introduces significant numerical and physical challenges. Primarily, the code cannot handle the spatial coexistence of two thermodynamically distinct fluids: the relativistic PWN, characterized by an adiabatic index $\Gamma = 4/3$, and the non-relativistic SNR ejecta, where $\Gamma = 5/3$; and even more advanced techniques fail in the presence of mixing \citep{Bucciantini2002}. Furthermore, continuously tracking the energy injection from the central PSR during the complex reverberation phase within a multidimensional grid becomes computationally hard to handle, unless one resolves the PSR wind region and the wind termination shock. 
The wind termination shock radius ($R_{\rm WTS}$) is determined by the condition that the ram pressure of the PSR wind equals the internal pressure of the nebula, as given by Eq.~(\ref{r_ts}). By normalizing this expression with respect to the characteristic variables of the SNR, we obtain the dimensionless radius of the wind termination shock:
\begin{equation}
    \label{eq:termination_shock_radius_normalized}
    \frac{R_{\rm WTS}(t/t_{\rm ch})}{R_{\rm ch}} = 
        \left[ \frac{P_{\rm pwn}(t/t_{\rm ch})}{P_{\rm ch}} \right]^{-\frac{1}{2}} 
        \left[ \frac{L_0}{L_{\rm ch}} \right]^{\frac{1}{2}} 
        \left[ 1 + \frac{t}{t_{\rm ch}}\frac{t_{\rm ch}}{\tau_0} \right]^{-1} 
        \left[ \frac{V_{\rm ch}}{c} \right]^{-\frac{1}{2}},
\end{equation}
where $P_{\rm pwn}$ is the inner pressure of the nebula, $L_0$ the initial spin-down luminosity of the PSR, and $\tau_0$ the initial spin-down time.
In addition, the evolution of the wind termination shock radius is determined by the properties of the SNR, depending on the ratio $V_{\rm ch}/c$ -- with $V_{\rm ch}$ given by Eq.$\stick$(\ref{v_ch}).
Given the physical properties of the SNR, that we selected in Sect.$\stick$\ref{SETUP_1D_SECTION}, the characteristic velocity amounts to $ \simeq 2200 \, {\rm km \, s^{-1}}$, yielding a ratio $V_{\rm ch}/c \simeq 7.3 \times 10^{-3}$. 
Choosing the case \texttt{A07} -- identified by $\tau_0/t_{\rm ch} \simeq 0.50$ and $L_0/L_{\rm ch} \simeq 8.9 \times 10^{-3}$, as summarized in Table~\ref{COES_compact} -- and using its pressure profile -- see Fig.~\ref{PWNRADPRSA} -- we computed the radius of the wind termination shock for this specific case, as reported in Fig.~\ref{terminationA07}.
\begin{figure}[h!]
\centering
\includegraphics[width=1.0\linewidth]{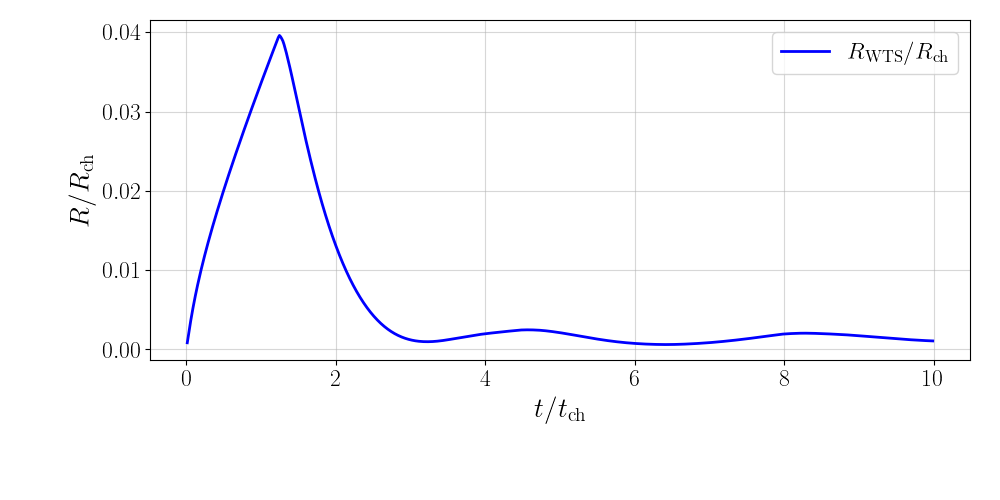}
\caption{The termination shock of the PSR wind for the system \texttt{A07}. For this specific case, $\tau_0/t_{\rm ch} \simeq 0.50$, $L_0/L_{\rm ch} \simeq 8.9 \times 10^{-3}$, and $V_{\rm ch}/c \simeq 7.3 \times 10^{-3}$.}
\label{terminationA07}
\end{figure}\\
Comparing Fig.$\stick$\ref{terminationA07} and Fig.$\stick$\ref{PWNRADPRSA} it is evident that the radius of the wind termination shock remains much smaller than the minimum radius reached by the nebula during its evolution. For this reason we decided to neglect the effects of the inner region occupied by the PSR wind in our simulations.
In addition, at late times, due to spin-down, the injection of energy from the PSR can become quite small, causing the radius of the wind termination shock to recede even further, as shown in Fig.~\ref{terminationA07}.\\\\
To overcome these multidimensional limitations while preserving the physical accuracy of the pre-reverberation phase, we adopt a hybrid strategy, transforming the PWN into a non-relativistic fluid ($\Gamma = 5/3$) -- thereby allowing the 2D code to evolve a single fluid -- and simultaneously switching off the continuous energy injection from the PSR.\\\\
We recall that for a fluid undergoing adiabatic expansion / compression, the relation $P \propto V^{-\Gamma}$ holds, where $P$ is the pressure, $V$ the volume and $\Gamma$ the adiabatic index. Due to this, the pressure of a non-relativistic, spherical bubble scales as $P \propto R^{-5}$, where $R$ denotes its radius, whereas the pressure of a relativistic PWN should scale as $P \propto R^{-4}$. By altering the equation of state, the steeper $R^{-5}$ scaling during the compression phase causes the pressure to rise more rapidly than it would in the relativistic scenario. Qualitatively, this artificially enhanced pressure during compression is expected to counterbalance the lack of pressure support that would otherwise be provided by the PSR spin-down energy. To assess the validity and the robustness of this combined choice, we performed a 1D comparative simulation between the system, \texttt{A07} in the correct regime, and the modified, non-relativistic case, \texttt{A07-NR}. In the latter, the energy injection from the PSR is switched off just before the interaction with the RS, and the PWN is subsequently treated as a non-relativistic gas with its internal pressure scaling adiabatically as $R^{-5}$. The results of this simulation are shown in Fig.~\ref{REL_NON_REL}.
\begin{figure}[h!]
    \centering 
     	\includegraphics[width=0.93\linewidth]{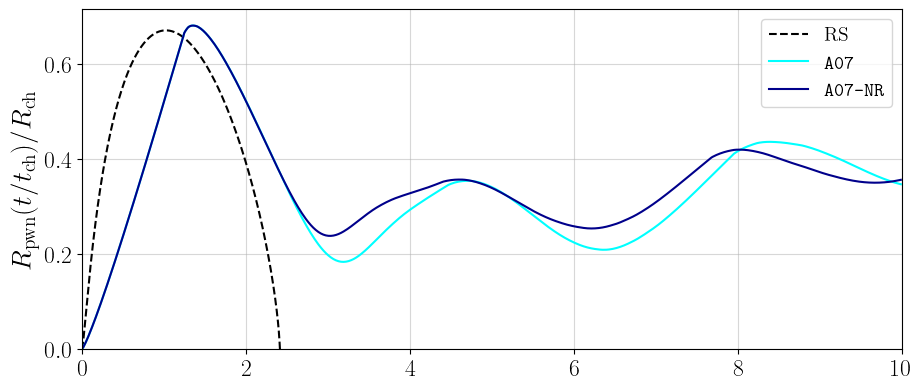}
     \centering
     \hspace*{-4mm}
     	\includegraphics[width=0.95\linewidth]{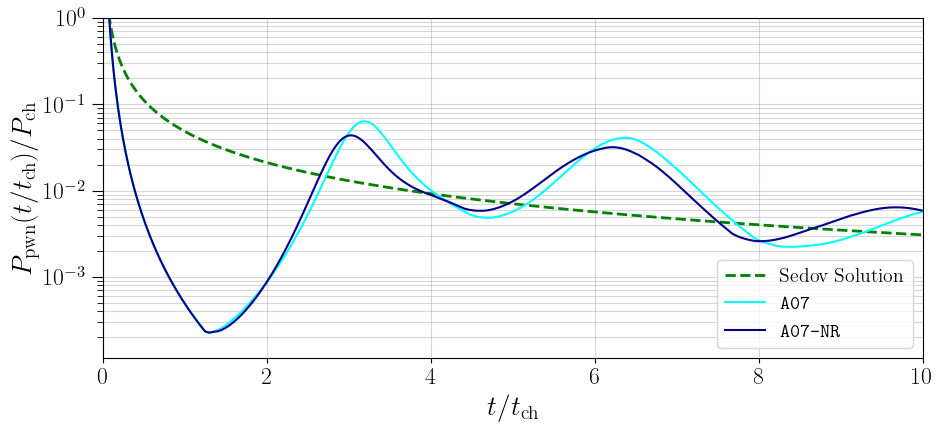}
    \caption{Evolution of the normalized nebular radius (\textit{Upper panel}) and pressure (\textit{Lower panel}) for the PWN-SNR systems \texttt{A07} and its non-relativistic counterpart, without energy injection from the central PSR (\texttt{A07-NR}).}  
    \label{REL_NON_REL}
\end{figure}
\noindent
\\The two systems obviously share identical profiles both of radius and pressure during the free-expansion phase up to the compression, and their subsequent evolution is quite similar, with minor quantitative changes present, as we anticipated. Notably, the system \texttt{A07-NR} undergoes less compression, reaching a minimum radius of $\simeq 0.25R_{\rm ch}$, whereas the \texttt{A07} case is compressed up to $\simeq 0.20R_{\rm ch}$. \\\\
Regarding the pressure, this increases more rapidly in the \texttt{A07-NR} case during compression, followed by a more pronounced decrease during the subsequent re-expansion. Despite this faster growth, at the peak of the compression, the pressure of the \texttt{A07-NR} system is $\simeq 4\times 10^{-2}P_{\rm ch}$, while in the case \texttt{A07} it reaches $\simeq 7\times 10^{-2}P_{\rm ch}$. 
Throughout the entire evolution, the two pressures generally differ by no more than a factor $\simeq 2$, where the pressure of the system \texttt{A07-NR} remains higher during the compression cycles while it is generally lower during the re-expansion phases. Ultimately, at late times -- $ \gtrsim 8 \, t_{\rm ch}$ -- the pressure of the \texttt{A07-NR} model asymptotically approaches the Sedov solution, following the overall behavior of the canonical case, despite some minor differences.\\\\ 
Coherently with the steeper radial scaling of pressure, during the re-expansion cycles, the \texttt{A07-NR} system generally reaches somewhat larger radii, typically exceeding the \texttt{A07} case by no more than $\simeq 0.05 R_{\rm ch}$. Regarding the timing of these cycles, the first compression in the \texttt{A07} system occurs at $\simeq 3.2 \, t_{\rm ch}$, whereas it is slightly shifted earlier to $\simeq 3 \, t_{\rm ch}$ in the \texttt{A07-NR} case. This temporal delay of $\simeq 0.2 \, t_{\rm ch}$ persists throughout the subsequent compression phases, which are shifted earlier in the non-relativistic case. A similar trend is seen for the re-expansion phases, with the \texttt{A07-NR} system anticipating the subsequent re-expansion cycles compared to the \texttt{A07} case.  \\\\
Despite these minor discrepancies, the comparison shows good qualitative agreement between the evolutions of the two systems. Aiming for a more complete analysis, we also show the overall hydrodynamic evolution of the \texttt{A07-NR} system in time -- see Fig.$\stick$\ref{NON_REL_OVERALL}.
\begin{figure}[h!]
    \centering 
     \begin{minipage}{1.0 \textwidth}
     	\includegraphics[width=0.99\linewidth]{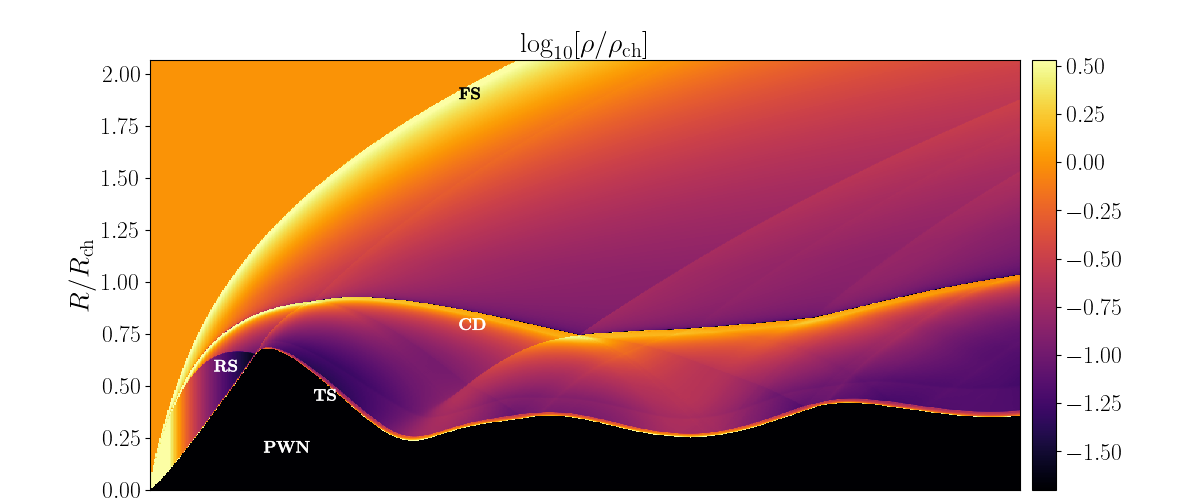}
     	\includegraphics[width=0.99\linewidth]{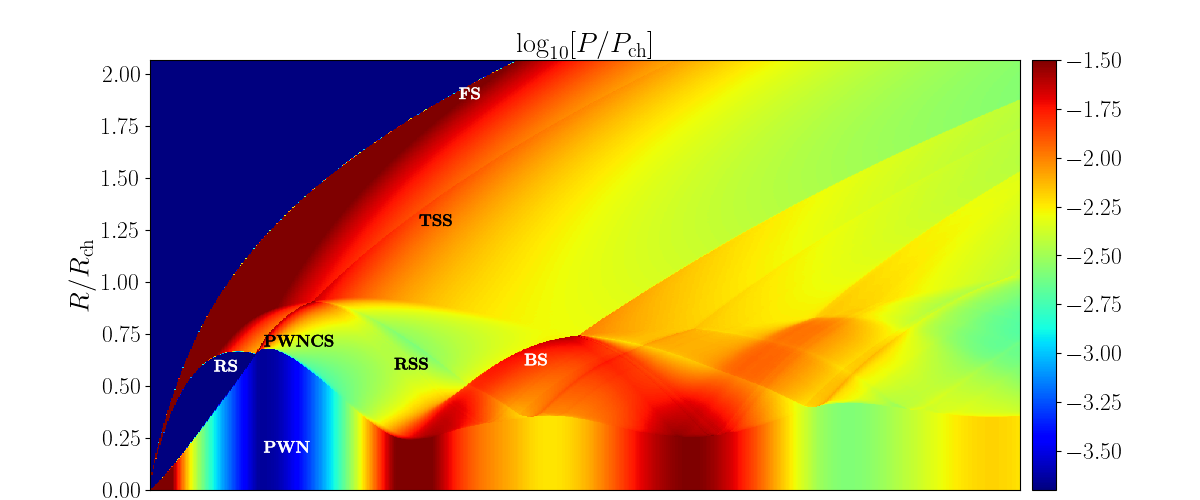}
    		\includegraphics[width=0.99\linewidth]{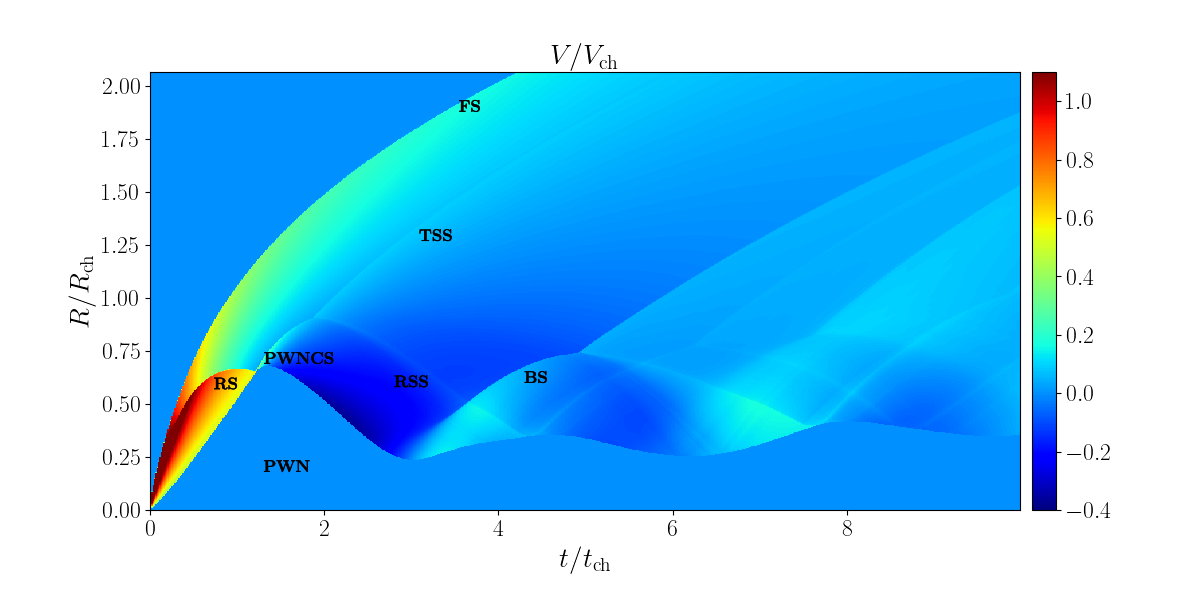}
    	\end{minipage}
    \caption{Evolution of the normalized density (\textit{Upper panel}), pressure (\textit{Center panel}), and velocity (\textit{Lower panel}) for the system \texttt{A07-NR}. It is possible to distinguish the PWN (black area in the density plot), the RS, the CD, the FS and the Thin-Shell (TS). Several shocks are produced as contact discontinuities interact. Among these, we recognize the PWN compression shock (PWNCS), the Transmitted Secondary Shock (TSS), the Reflected Secondary Shock (RSS) and the Bounce Shock (BS).}  
    \label{NON_REL_OVERALL}
\end{figure}
\noindent
Comparing the evolution shown in Figs.~\ref{NON_REL_OVERALL} and \ref{SNAPSHOTS_TOGETHER}, further evidence of the good agreement between the models emerges. Coherently with the fact that the compression of \texttt{A07-NR} is anticipated with respect to the original \texttt{A07} case, the BS generated during this phase interacts with the CD earlier. Specifically, this interaction occurs at $\simeq 4.9 \, t_{\rm ch}$ for \texttt{A07-NR}, while it is delayed until $\simeq 5.1 \, t_{\rm ch}$ in the \texttt{A07} system. As a consequence, this interaction triggers the formation of several secondary shocks earlier in the \texttt{A07-NR} case. It is worth remarking that, despite this generally anticipated dynamics, the overall agreement between the two cases for what concerns the dynamical structures remains remarkably good. 
To ensure that this consistency is not merely an isolated feature of the \texttt{A07} system, we expanded our comparative analysis to encompass all the cases selected within the ROI -- and given in Table$\stick$\ref{COES_compact_53}. 
\begin{table}[h!]
\centering
\setlength{\tabcolsep}{4pt} 
\begin{tabular}{ccc | ccc} 
\hline
\hline
System & \text{CF} & \text{CF(\texttt{NR})} & System & \text{CF} & \text{CF(\texttt{NR})} \\
\hline
\hline
\texttt{A01} & $1.81$ & $1.62$ & 
\texttt{R01} & $11.1$ & $5.29$ \\

\texttt{A02} & $2.27$ & $1.89$ &
\texttt{R02} & $5.46$ & $3.59$ \\

\texttt{A03} & $2.34$ & $1.97$ &
\texttt{R03} & $3.72$ & $2.87$ \\

\texttt{A04} & $2.85$ & $2.33$ &
\texttt{R04} & $2.51$ & $2.37$ \\

\texttt{A05} & $3.61$ & $2.75$ &
\texttt{R05} & $2.08$ & $2.17$ \\

\cline{4-6} 

\texttt{A06} & $3.55$ & $2.71$ &
\texttt{T01} & $9.27$ & $4.84$ \\

\texttt{A07} & $3.70$ & $2.85$ &
\texttt{T02} & $5.27$ & $3.52$ \\

\texttt{A08} & $3.81$ & $3.00$ &
\texttt{T03} & $3.14$ & $2.58$ \\

\texttt{A09} & $3.75$ & $3.02$ &
\texttt{T04} & $1.94$ & $1.90$ \\

\texttt{A10} & $4.51$ & $3.39$ &
\texttt{T05} & $1.18$ & $1.25$ \\

\cline{4-6} 

\texttt{A11} & $5.53$ & $3.86$ &
\texttt{L01} & $12.0$ & $5.43$ \\

\texttt{A12} & $7.23$ & $4.49$ &
\texttt{L02} & $5.76$ & $3.64$ \\

\cline{1-3} 

 & & & 
\texttt{L03} & $3.20$ & $2.62$ \\

 & & & 
\texttt{L04} & $3.31$ & $2.14$ \\

 & & & 
\texttt{L05} & $1.70$ & $1.85$ \\

\hline
\hline
\end{tabular}
\caption{Compression Factors (CF) for the systems sampled within the ROI -- see also Fig. \ref{SELECTEDSYSTEMSROI} and Table \ref{COES_compact} -- with CF(\texttt{NR}) corresponding to systems with adiabatic index 5/3 and with PSRs turned off during reverberation.}
\label{COES_compact_53}
\end{table}\\\\
By examining the compression factors of the non-relativistic PWNe -- CF(\texttt{NR}) -- it becomes evident that, although the overall compression is reduced, the general trend of the canonical compression factor (CF) is preserved across the ROI. It is worth noting that CF(\texttt{NR}) generally exhibits values comparable to those of the canonical case, with a maximum decrease by a factor $\simeq 2$ observed exclusively in systems \texttt{T01}, \texttt{R01}, and \texttt{L01}. These specifically have very low initial spin-down times or initial spin-down luminosities, implying that the internal pressure of the nebula driven by such a limited energy injection is weaker, compared to the other systems. Due to this, in these specific cases, modelling the PWN as a non-relativistic gas alters the final compression factor more significantly than in the other systems. These results further justify the adoption of our strategy of modelling the PWN as a non-relativistic fluid and simultaneously turning off the energy injection by the PSR during the reverberation for the subsequent 2D simulations.
\section{Physical Setup of the Pulsar Wind Nebula}
\label{NRPWN}
Transitioning from a 1D simulation performed with the Lagrangian code to a 2D framework in PLUTO presents a challenge, since the one-zone Lagrangian code computes the PWN evolution just in terms of its radius and internal pressure, leaving its internal velocity field and mass density undefined, while the PLUTO setup requires an explicit initialization of density, pressure, and velocity across the entire multidimensional grid. In addition, recalling that we decided to perform our simulations within the framework of non-relativistic hydrodynamics, it is crucial that the PWN retains the dynamical response of a relativistic plasma. For this, we must define a physically motivated effective density and velocity field for the nebula. For a non-relativistic ideal gas, the sound speed is given by:
\begin{equation}
c_{\rm s} = \sqrt{\frac{5P}{3\rho}},
\end{equation}
where $P$ is the pressure and $\rho$ the mass density. By contrast, a realistic PWN is a relativistic subsonic bubble, characterized by a sound speed $c_{\rm s} = c/\sqrt{3}$. To ensure that the non-relativistic fluid in PLUTO responds exactly as a relativistic plasma would, we equate the two sound speeds, yielding the following prescription for the initial PWN mass density:
\begin{equation}
\rho_{\rm pwn} = \frac{5P_{\rm pwn}}{c^2},
\label{rho_right}
\end{equation}
where $P_{\rm pwn}$ is the nebular pressure. By initializing the density in this manner, we guarantee that the nebula correctly replicates the properties of a relativistic plasma, despite the non-relativistic setup.
However, while this density prescription is formally correct, we can still increase it to significantly optimize computational times without altering the global dynamics of the system. To strictly preserve causality, the time integration is governed by the Courant-Friedrichs-Lewy (CFL) condition \citep{Lewy1928}. This condition states that the time-step, $\Delta t$, must be smaller than the characteristic time required for a sound-wave -- more generally a signal -- to traverse the numerical cell. Specifically:
\begin{equation}
\Delta t = C_{\rm CFL} \min_{i} \left\lbrace \frac{\Delta l_{i}}{| \lambda_{i}^{\rm max}|} \right\rbrace,
\end{equation}
where $C_{\rm CFL}$ must be $<1$, but in PLUTO it is set to 0.4 to ensure numerical stability, and $\Delta l_{i}$ and $\lambda_{i}^{\rm max}$ denote the spatial extension of the cell $i$ and the maximum physical velocity of the signal in the cell, respectively.\\\\ 
In our setup, this condition is dominated by the angular extent of the cells -- $\delta\theta$ -- at the minimum radius -- $R_{\rm min}$ -- and by the sound speed of the nebula, $c_{\rm s}$, leading to $\Delta t \simeq R_{\rm min} \delta\theta / c_{\rm s}$. Since $c_{\rm s} \propto \rho^{-1/2}$, it follows that $\Delta t \propto \rho^{1/2}$. Thereby, artificially increasing the initial density of the PWN allows larger time steps, drastically reducing the overall computational cost.
This notwithstanding, this density increase cannot be arbitrary; one needs to make sure that the sound speed within the nebula remains significantly larger than the characteristic velocity of the remnant, $V_{\rm ch}$ -- given by Eq.$\stick$(\ref{v_ch}). Using the kinetic energy of the SN and ejected mass adopted in 1D Lagrangian simulations, yields $V_{\rm ch} \simeq 2200\stick$km$\stick\text{s}^{-1}$.
Based on this, we choose an initial PWN density ten times higher than the value given by Eq.$\stick$(\ref{rho_right}), by imposing:
\begin{equation}
\rho_{\rm pwn} = \frac{50 P_{\rm pwn}}{c^{2}}.
\label{rho_used}
\end{equation}
With this assumption, the sound speed in the nebula is $\simeq 55.000\stick{\text{km s}^{-1}}\simeq 25 V_{\rm ch}$, safely satisfying the condition $c_{\rm s} \gg V_{\rm ch}$. This ensures that the nebula continues to respond dynamically to compressions as a relativistic gas would, allowing to increase the integration time-step by a factor of $\sqrt{10} \simeq 3$, and making the simulation three times faster. To verify that this choice preserves the physical accuracy of the simulation, we performed a preliminary 1D test in PLUTO comparing the cases with $\rho_{\rm pwn} = 5 P_{\rm pwn} c^{-2}$ and $\rho_{\rm pwn} = 50 P_{\rm pwn} c^{-2}$. \\\\
Regarding the velocity field inside the PWN, we initialize it with a simple linear profile:
\begin{equation}
v_{\rm pwn} = V_{\rm sh} \frac{r}{R_{\rm sh}},
\label{v_pwn_NR}
\end{equation}
where $V_{\rm sh}$ is the velocity of the inner boundary of the thin shell swept-up by the nebula at the initial time. This allows us to match the correct condition at the center, where the velocity goes to zero, without introducing unphysical velocity jumps at the PWN boundary, that might lead to spurious numerical relaxation effects. For the aforementioned test, both setups used the pressure derived from the Lagrangian code and this linear velocity profile. The results of the test are shown in Fig.$\stick$\ref{5_VS_50}.
\begin{figure}[h!]
\centering
\includegraphics[width=0.95\textwidth]{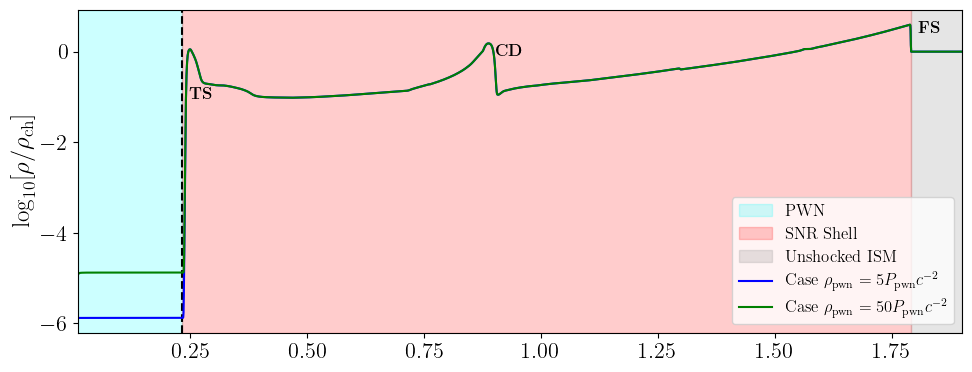}
\hspace*{-4mm}
\centering
\includegraphics[width=0.965\textwidth]{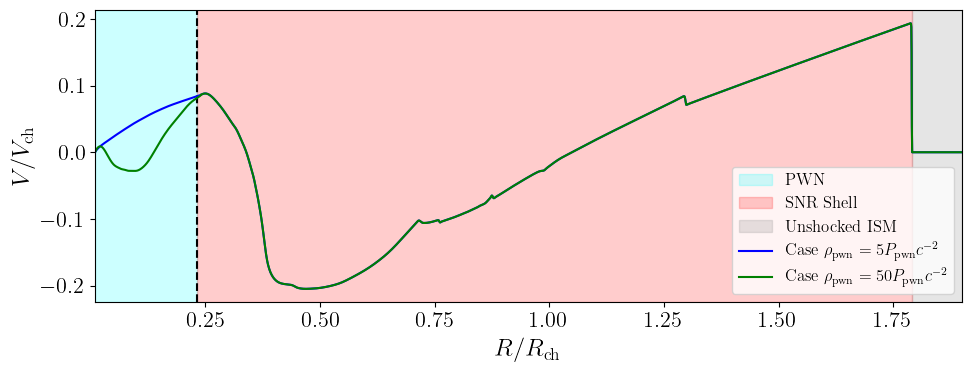}
\caption{Profiles of density (\textit{Upper panel}) and velocity (\textit{Lower panel}) in the case $\rho_{\rm pwn} = 5 P_{\rm pwn}c^{-2}$ and $\rho_{\rm pwn} = 50 P_{\rm pwn}c^{-2}$ at $\simeq 3 t_{\rm ch}$, during maximum compression. While the RS has been dissolved by the interaction with the Thin Swept-Up Shell (TS) -- here located between $\simeq 0.23R_{\rm ch}$ and $0.27 R_{\rm ch}$ -- it is still possible to identify the CD and the FS, located at $\simeq 0.9 R_{\rm ch}$ and $\simeq 1.8 R_{\rm ch}$, respectively.}
\label{5_VS_50}
\end{figure}
\noindent
While the internal density and velocity profiles of the nebula naturally differ at the time of maximum compression -- occurring at $\simeq 3 t_{\rm ch}$ -- the overall dynamical evolution of the remnant remains identical and this holds at every time step of the simulation. This justifies the use of the fictitious higher density in the nebula -- $\rho_{\rm pwn} = 50 P_{\rm pwn} c^{-2}$ -- as a physically robust strategy to reduce computational time.
\section{Numerical Setup}
\subsection{Spatial Grid}
We adopt a spherical coordinate system $(r, \theta)$, assuming azimuthal symmetry. The basic computational domain extends radially from an inner boundary $R_{\rm min} = 0.01 R_{\rm ch}$ to an outer boundary $R_{\rm max} = 3.80 R_{\rm ch}$. In the angular direction, the grid covers a wedge across the equatorial plane from $\theta_{\rm min} = \pi/2 - \Delta\theta\stick$ to $\theta_{\rm max} = \pi/2 + \Delta\theta\stick$, with $\Delta\theta = 0.2\stick$. To dynamically track sharp features such as shock waves and contact discontinuities without a prohibitive computational cost, we employ the Chombo Adaptive Mesh Refinement module \citep{Martin2025}. We configure the grid to allow up to 4 additional levels of refinement. A refinement ratio of 2 is adopted across all levels -- a standard default choice meaning that the spatial resolution doubles at each consecutive nested level. The level 0 grid has 400 uniformly spaced cells in the radial direction, with a fundamental spatial resolution of $9.475\times 10^{-3}R_{\rm ch}$, and 16 uniformly spaced cells in the polar direction, yielding basic angular resolution of $0.025\stick$radians. At the highest resolution, the level 4 corresponds to 6400 effective cells in the radial direction and 256 cells in the angular direction, with a resolution of $5.94\times 10^{-4}R_{\rm ch}$ and $1.56\times 10^{-3}\stick$radians, respectively.
\subsection{Numerical Reconstruction and Hydrodynamic Solvers}
PLUTO employs a shock-capturing methodology. The spatial reconstruction of the left and right states at the cell interfaces is performed using a \texttt{\textit{LINEAR}} Total Variation Diminishing (\texttt{\textit{TVD}}, \citealt{Harten1983}) reconstruction applied to the density, pressure, and velocity. This method operates on a 3-point stencil, reconstructing the physical quantities within a given cell $i$ by evaluating the data in the adjacent cells $i-1$ and $i+1$. To compute the slopes of these piecewise linear distributions, we employ the Monotonized Central (\texttt{\textit{MC}}, \citealt{VANLEER1977}) limiter, which is the default configuration in PLUTO. This specific slope limiter is highly effective in capturing shock waves and contact discontinuities, while maintaining second-order spatial accuracy in regions where density, pressure and velocity exhibit smooth spatial behavior.
To compute the inter-cell fluxes, we use the Harten-Lax-van Leer-Contact solver (\texttt{\textit{HLLC}}, \citealt{ToroSpruceSpeares94}), which is robust and highly accurate for resolving isolated shocks waves and contact discontinuities. We selected this solver because it offers an optimal trade-off between computational cost and physical reliability; while marginally more diffusive than exact solvers, it is exceptionally stable when dealing with very low-density plasmas, such as PWNe. Finally, the system is advanced in time using a second-order Runge-Kutta (\texttt{\textit{RK2}}, \citealt{Numerical_Recipies}) scheme, which evaluates fluxes at a half-time step to compute the full-step update.
\subsection{Initial and Boundary Conditions}
\subsubsection{Initial Conditions}
The initial conditions for the density, pressure, and velocity profiles within the SNR, alongside the internal pressure of the PWN, were extracted from the 1D Lagrangian simulation at a time $t = 1.23t_{\rm ch}$. At this evolutionary stage, the profiles of these quantities as initial conditions exhibit a structure analogous to the one reported in Fig.~\ref{SNAPSHOTS}. Recalling that the PLUTO code requires explicit initial values for density and velocity also within the nebula itself, we prescribed these quantities according to Eqs.~(\ref{rho_used}) and (\ref{v_pwn_NR}), respectively.\\\\ 
We know, from observational evidence in the Crab Nebula, that the swept-up shell is Rayleigh-Taylor unstable, as revealed by the presence of thermally emitting filamentary structures \citep{Sankrit98}. This instability arises from the interaction between the expanding PWN and the shocked ejecta, and leads to the fragmentation of the swept-up shell. While we cannot emulate in our conditions the complex structure of a Rayleigh-Taylor unstable layer, we can model the associated density fluctuations by modifying the angular density distribution in the shell. Moreover, we limit our analysis to single mode perturbations by introducing an initial density fluctuation within the shell: 
\begin{equation}
{\delta\rho(r, \theta)} = \eta {\rho_{\rm 0}(r)} \cos\left[\frac{n_{\theta}\pi}{\Delta\theta}\left(\theta-\frac{\pi}{2}\right)\right]
\text{ for }
R_{\rm sh}^{(\rm in)}\leq R \leq R_{\rm sh}^{(\rm out)},
\end{equation}
where $\rho_{\rm 0}$ represents the unperturbed mass density in the shell, $\eta$ the perturbation amplitude relative to the unperturbed shell density profile, with $n_{\theta}$ the angular wavenumber of the perturbation, whereas $R_{\rm sh}^{(\rm in)}$ and $R_{\rm sh}^{(\rm out)}$ denote the inner and outer boundary of the shell, respectively. Observationally, these filaments develop on angular scales ranging from $0.03$ to $0.1$ radians \citep{Hester1996}. Given our chosen angular domain of $2\Delta\theta = 0.4$ radians, this physical scale would correspond to a range of $4$ to $12$ wavelengths. However, shorter wavelengths require a significantly higher spatial resolution. Therefore, to optimize computational time while still capturing the core of the physical mechanisms, we restricted our investigation to the modes $n_{\theta} = 1, 2$, and $4$. Additionally, we explored different perturbation amplitudes by setting the parameter $\eta$ to $0.05$, $0.10$, $0.20$, and $0.30$. The resulting 2D density initialization is illustrated in Fig.$\stick$\ref{fig:initial_2d_density}.
\begin{figure}[h!]
\centering
\includegraphics[width=1.0\linewidth]{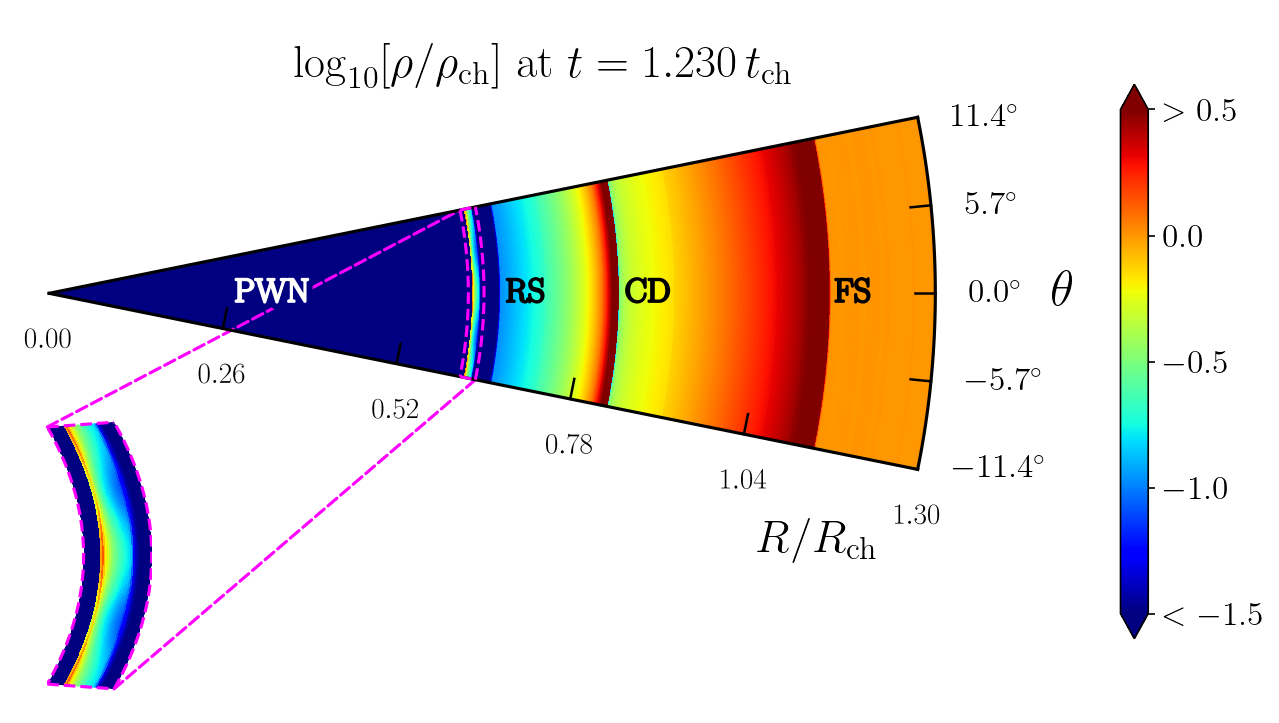}
\caption{Initial 2D density map of the \texttt{A07-NR} PWN-SNR system, with amplitude perturbation $\eta=0.30$ and wavenumber $n_{\theta}=2$. From the center outward: the PWN (blue area), the thin perturbed swept-up shell, unshocked ejecta, RS, shocked ejecta, CD, shocked ISM, FS, followed by the unshocked ISM.}
\label{fig:initial_2d_density}
\end{figure} 
\subsubsection{Radial Boundary Conditions}
At the inner boundary -- $R_{\rm min}$ -- we impose a \textit{radial reflective condition}. In this regime, pressure, density and angular velocity are extrapolated at the 0-th order, whereas the radial velocity is inverted across the boundary. Specifically,
\begin{equation}
\rho\rightarrow\rho,\quad
P\rightarrow P ,\quad
v_{\rm r} \rightarrow -v_{\rm r},\quad
v_{\theta} \rightarrow v_{\theta},
\end{equation}
where $\rho$ is the density, $P$ the pressure, $v_{\rm r}$ the radial velocity and $v_{\theta}$ the angular velocity. Conversely, at the outer boundary -- $R_{\rm max}$ -- we apply a \textit{radial outflow condition}, which enforces a zero-gradient across the boundary by copying the values of the last active interior cell into the outer zones. This choice is physically consistent with our assumption of a uniformly distributed ISM surrounding the expanding SNR. 
\subsubsection{Angular Boundary Conditions}
Regarding the angular boundaries at $\theta_{\rm min}$ and $\theta_{\rm max}$, we opted for a \textit{reflective condition} instead of \textit{periodic} or \textit{outflow} for two reasons. A periodic boundary would be physically and mathematically incorrect in spherical coordinates, since the spherical volume element depends explicitly on the polar angle. As a consequence, applying periodicity would mean artificially attaching cells that are fundamentally different in physical space, forcing a mapping between regions with completely mismatched interface areas and volumes. Conversely, employing an outflow condition in a highly subsonic regime -- such as the interior of the PWN -- leads to a severe loss of control over the incoming and outgoing fluxes, potentially causing catastrophic and unphysical emptying or filling of the domain. Under the chosen \textit{angular reflective conditions}, scalar quantities such as pressure and density, along with the radial velocity are extrapolated at 0-th order, whereas the angular velocity is inverted across both boundaries. Specifically, 
\begin{equation}
\rho\rightarrow\rho,\quad
P\rightarrow P ,\quad
v_{\rm r} \rightarrow v_{\rm r},\quad
v_{\theta} \rightarrow -v_{\theta}.
\end{equation}
By setting a reflective boundary, we prevent the artificial injection or escape of physical quantities. Subsequent verification of our results confirms that our boundary conditions produce very minor boundary effects and mostly at very late times.
\section{The Compression Phase}
Having defined the initial setup and the boundary conditions of the simulation domain, we now examine the subsequent dynamical evolution of the system. Following the initial expansion, the PWN reaches its maximum radial extent and enters a compression phase. The overall evolution of the density structure during this phase is illustrated in Fig.~\ref{fig:compression_evolution}, for the case $\eta=0.30$ and $n_{\theta}=4$.
\begin{figure}[h!]
\centering
\includegraphics[width=0.7\linewidth]{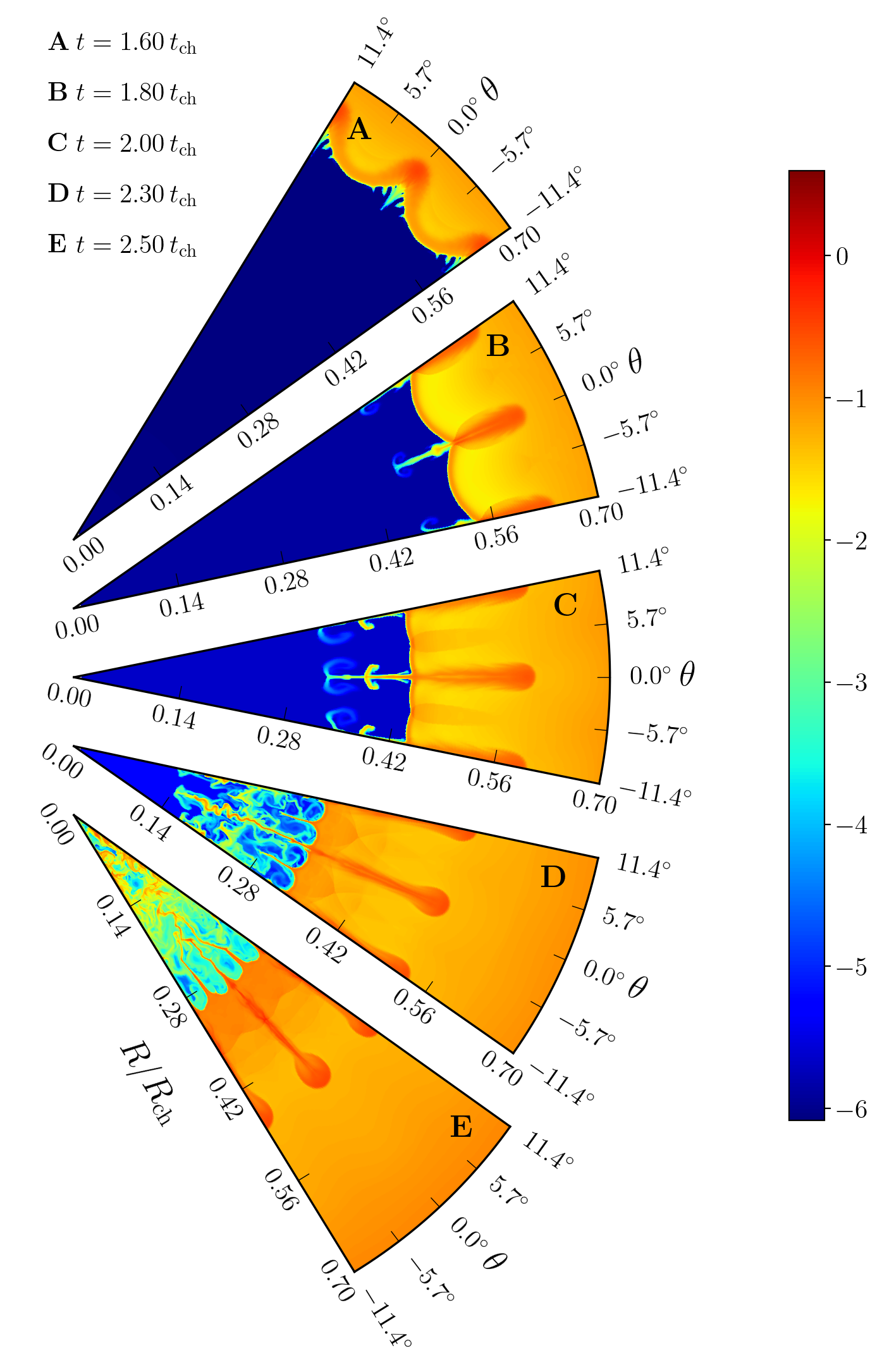}
\caption{Time evolution of $\log_{10}[\rho/\rho_{\rm ch}]$ during the compression phase of the PWN, with an initial perturbation amplitude $\eta = 0.30$ and an angular wavenumber $n_{\theta} = 2$.}
\label{fig:compression_evolution}
\end{figure}\\\\
\noindent
As the PWN begins to contract, the portions of the swept-up shell that are over-dense tend to lag behind those that are under-dense -- see panel A of Fig.$\stick$\ref{fig:compression_evolution}, corresponding to $t = 1.6 t_{\rm ch}$. This is mainly due to the fact that the initial compression phase is primarily regulated by the pressure in the SNR shell -- between the RS and the CD -- and the inertia in the shell. This lagging leads to a deformation of the interface between the shell and the PWN boundary. This deformation causes vorticity to develop in the PWN, close to the boundary itself, with convective motions converging toward the regions with over-density. This vorticity drives the growth of the Kelvin-Helmholtz instability at the interface that, together with shear instabilities, causes matter to be stripped by these convective motions from the shell and dragged into the PWN. This process leads to the formation of the so-called \textit{penetration fingers}, which can be clearly observed in panel B of Fig.$\stick$\ref{fig:compression_evolution} at $t = 1.8 t_{\rm ch}$.\\\\
As these fingers advance further inward, the localized vorticity surrounding their tips curls the entrained material, causing them to develop the so-called mushroom-like \textit{heads}. These structures originate from the initially over-dense regions of the shell, as illustrated in panel C of Fig.~\ref{fig:compression_evolution}, corresponding to $t = 2.0 t_{\rm ch}$. As the compression proceeds, it becomes evident that these structures recede toward the center of the nebula at a velocity of $\simeq 0.4 V_{\rm ch}$, while the over-dense shell material from which they originate retreats at $\simeq 0.1 V_{\rm ch}$. Furthermore, the evolution is consistent across the angular domain: the structures developing at the boundaries are similar to those emerging in the central region.\\\\
As the compression enters into its final stages -- see panel D of Fig.~\ref{fig:compression_evolution}, evaluated at $t = 2.3 t_{\rm ch}$ -- the advancing \textit{mushroom-like heads} start to interact, making the nebula heavily mixed with the shocked ejecta and turbulent. The turbulence in turn drives vorticity, and the resulting shear suppresses the growth of high wavelength modes, dampening the further development of the Rayleigh-Taylor instabilities.
Ultimately, as shown in panel E of Fig.~\ref{fig:compression_evolution} -- corresponding to $t = 2.5 t_{\rm ch}$-- as the heads of the mushroom-like structures reach the PWN interior, the nebula becomes fully mixed and the bounce shock is launched, marking the definitive end of the compression phase and the onset of the re-expansion of the nebula.
It is worth noting that, in this case, the \text{mushroom-like structures} reach the core of the SNR on a timescale comparable to that required for the RS to converge at the center -- see Eq.~(\ref{rs_inside}).\\\\
While the sequence described above provides a comprehensive view of the compression dynamics for the specific case with perturbation amplitude $\eta=0.30$ and wavenumber $n_{\theta}=2$, the overall evolution of the system is highly sensitive to the properties of the initial density perturbation. In general, the dynamics of the compression and of the instability depends on both $\eta$ and $n_{\theta}$. To characterize these dependencies more accurately and to isolate their respective physical effects, we performed a comparative analysis. Specifically, we characterised the compression dynamics across different models both by holding the perturbation amplitude constant to investigate the role of varying the wavenumber, and by fixing the wavenumber to explore the impact of different perturbation amplitudes.
We first examine a scenario with a fixed amplitude of the perturbation of mass density -- $\eta=0.30$ -- to explore how varying the wavenumber $n_{\theta}$ influences the overall evolution of the instability as shown in Fig.~\ref{compression_fixed_eta_one}. In order to make a comparison, we will evaluate these dynamics at a subset of the times previously established in Fig.~\ref{fig:compression_evolution}, specifically $t = 1.6, \ 2.0,$ and $2.5\ t_{\rm ch}$. We begin this analysis from Fig.~\ref{compression_fixed_eta_one}, by describing the early compression phase at $t = 1.6\ t_{\rm ch}$.
\begin{figure}[h!]
\centering
\includegraphics[width=0.7\linewidth]{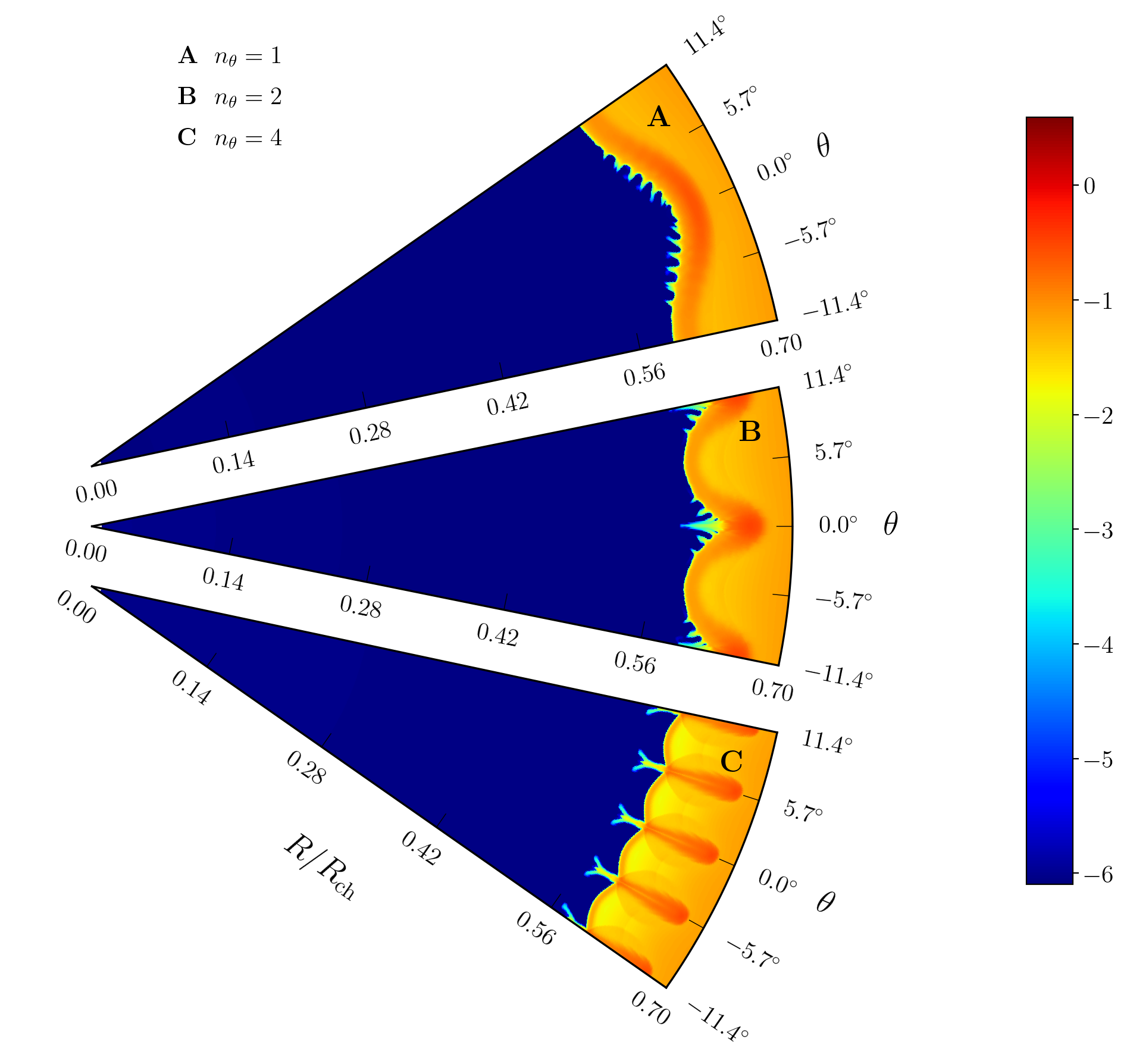}
\caption{$\log_{10}[\rho/\rho_{\rm ch}]$ during the compression phase of the PWN, at $\eta$ = 0.3 and $t = 1.6 t_{\rm ch}$.}
\label{compression_fixed_eta_one}
\end{figure}\\\\
By maintaining a fixed amplitude of the perturbation  -- $\eta = 0.30$ -- it becomes evident that increasing the angular wavenumber $n_{\theta}$ significantly anticipates the development of the instabilities. Notably, while the trailing over-dense regions maintain a characteristic mass density of $\simeq 0.1 \rho_{\rm ch}$ across all analyzed models, the density within the penetrating structures varies significantly.\\\\ Starting with the $n_{\theta} = 1$ case, the \text{penetration fingers} have not yet emerged at this evolutionary stage, and the instability has not yet developed. In this regime, the over-dense material merely occupies a narrow region between $\simeq 0.60 R_{\rm ch}$ and $\simeq 0.63 R_{\rm ch}$. 
Advancing to the $n_{\theta} = 2$ case, the \text{penetration fingers} have a more defined shape. In this intermediate case, the fingers extend from $\simeq 0.60 R_{\rm ch}$ to $0.63 R_{\rm ch}$ and exhibit a mass density of $\simeq 10^{-4}\rho_{\rm ch}$. Here, the over-dense regions have a length roughly comparable to that of the underlying \text{penetration fingers} and extend from $\simeq 0.62 R_{\rm ch}$ to $\simeq 0.66 R_{\rm ch}$. 
Shifting to the $n_{\theta} = 4$ case, the \text{fingers} are already well-defined, extending radially from $\simeq 0.57 R_{\rm ch}$ to $\simeq 0.60 R_{\rm ch}$ with an internal density of $\simeq 10^{-2} \rho_{\rm ch}$, two orders of magnitude higher than the $n_{\theta} = 2$ case. The corresponding over-dense regions span from $0.60 R_{\rm ch}$ out to $\simeq 0.68 R_{\rm ch}$, roughly doubling their radial extent compared to the $n_{\theta} = 2$ case. This pronounced elongation shows that these structures recede inward more rapidly as the angular wavenumber increases. Remarkably, the \text{fingers} exhibit a bifurcation at their tips. This is a consequence of the enhanced localized vorticity generated by the faster growing instabilities, which will lead to the anticipated formation of the aforementioned \text{mushroom-like} structures shown in panel C of Fig.~\ref{fig:compression_evolution}. Crucially, while these fingers exhibit a length similar to those observed in the $n_{\theta} = 2$ model, they have advanced into a more internal region of the domain, showing that a higher wavenumber anticipates the growth and the dynamics of the instabilities.
A similar trend regarding the delayed onset of the instability is observed holding the wavenumber constant while reducing the initial perturbation amplitude at $t = 1.6 t_{\rm ch}$ -- see Fig.~\ref{compression_fixed_n_one}.
\begin{figure}[h!]
\centering
\includegraphics[width=0.75\linewidth]{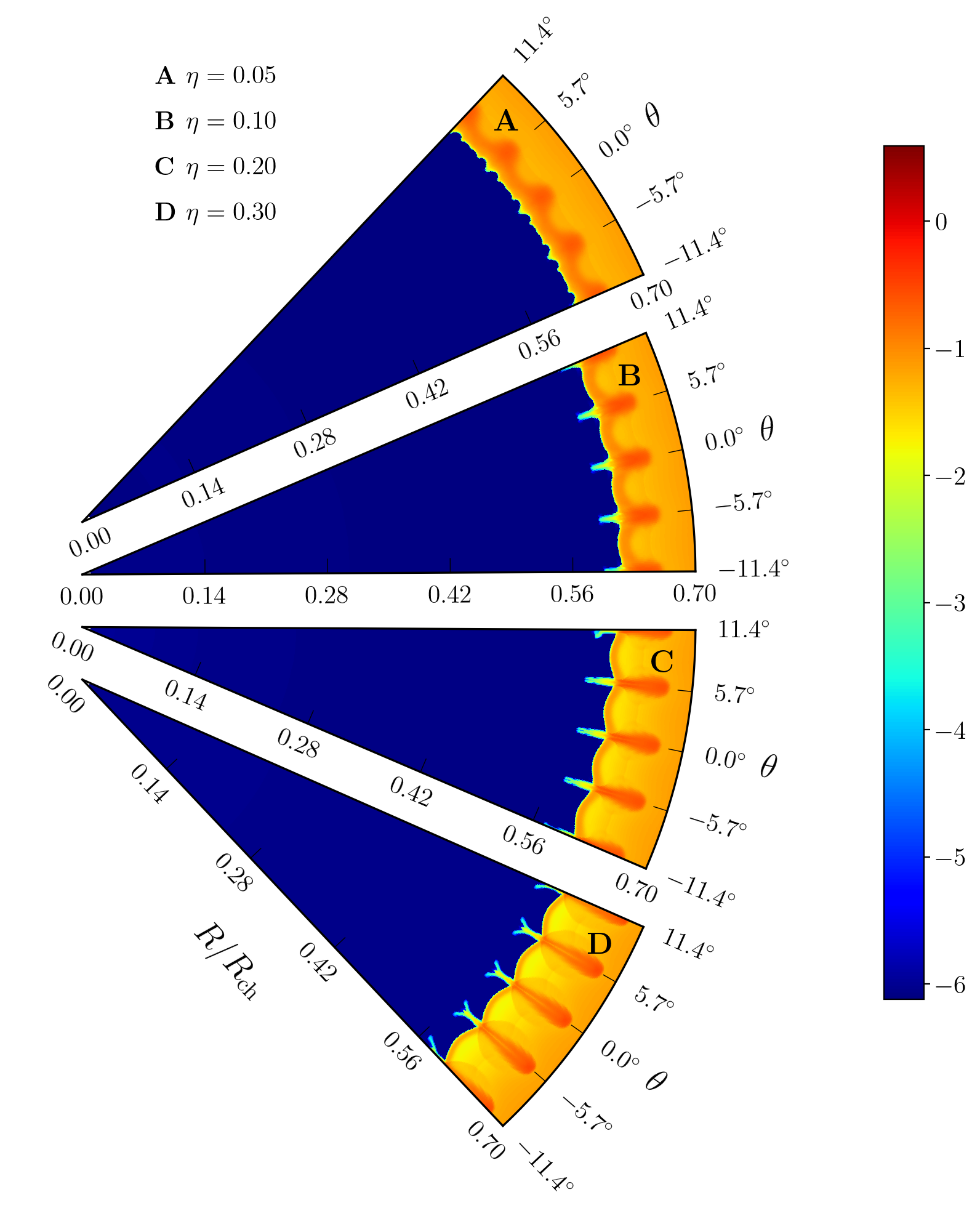}
\caption{$\log_{10}[\rho/\rho_{\rm ch}]$ during the compression phase of the PWN, at $n_{\theta}=4$ and $t = 1.6 t_{\rm ch}$.}
\label{compression_fixed_n_one}
\end{figure}
\noindent
In the case with the weakest perturbation -- $\eta = 0.05$ -- the evolution is notably delayed, with only the trailing over-dense regions between $0.60 R_{\rm ch}$ and $0.65 R_{\rm ch}$. As the amplitude is increased to $\eta = 0.10$, the \text{penetration fingers} begin to emerge, approximately extending from $\simeq 0.60 R_{\rm ch}$ to $\simeq 0.62 R_{\rm ch}$ and exhibiting an internal mass density of $\simeq 10^{-4} \rho_{\rm ch}$. \\\\
Progressing further to the $\eta = 0.20$ and $\eta = 0.30$ cases, the fingers exhibit a similar extension between $0.56 R_{\rm ch}$ and $0.59 R_{\rm ch}$, while their internal density rises by two orders of magnitude, up to $\simeq 10^{-2} \rho_{\rm ch}$. It is evident that in the highest $\eta$ scenario, the tips of the fingers display a distinct bifurcation. As established earlier, this splitting is an evidence of an advanced dynamical stage, marking the onset of formation of the \text{mushroom-like structures} displayed in panel C of Fig.~\ref{fig:compression_evolution}. In addition, the trailing over-dense regions directly above these structures maintain a characteristic mass density of $\simeq 0.1 \rho_{\rm ch}$. Notably, these over-dense zones extend from $\simeq 0.60 R_{\rm ch}$ to $\simeq 0.64 R_{\rm ch}$ in the $\eta= 0.20$ case, whereas they advance further outward to $\simeq 0.67 R_{\rm ch}$ in the $\eta = 0.30$ model. This broader radial expansion highlights that the growth of the instabilities is anticipated as the perturbation increases. \\\\
Aiming to trace the dynamical evolution into the compression phase, our analysis now shifts to the development of the \text{mushroom-like structures}. We begin by returning to the constant-amplitude regime, fixing $\eta = 0.30$ to evaluate the wavenumber dependence on the dynamics both of the instability and of the compression at $t = 2.0 t_{\rm ch}$ -- as illustrated in Fig.~\ref{compression_fixed_eta_two}.
\begin{figure}[h!]
\centering
\includegraphics[width=0.7\linewidth]{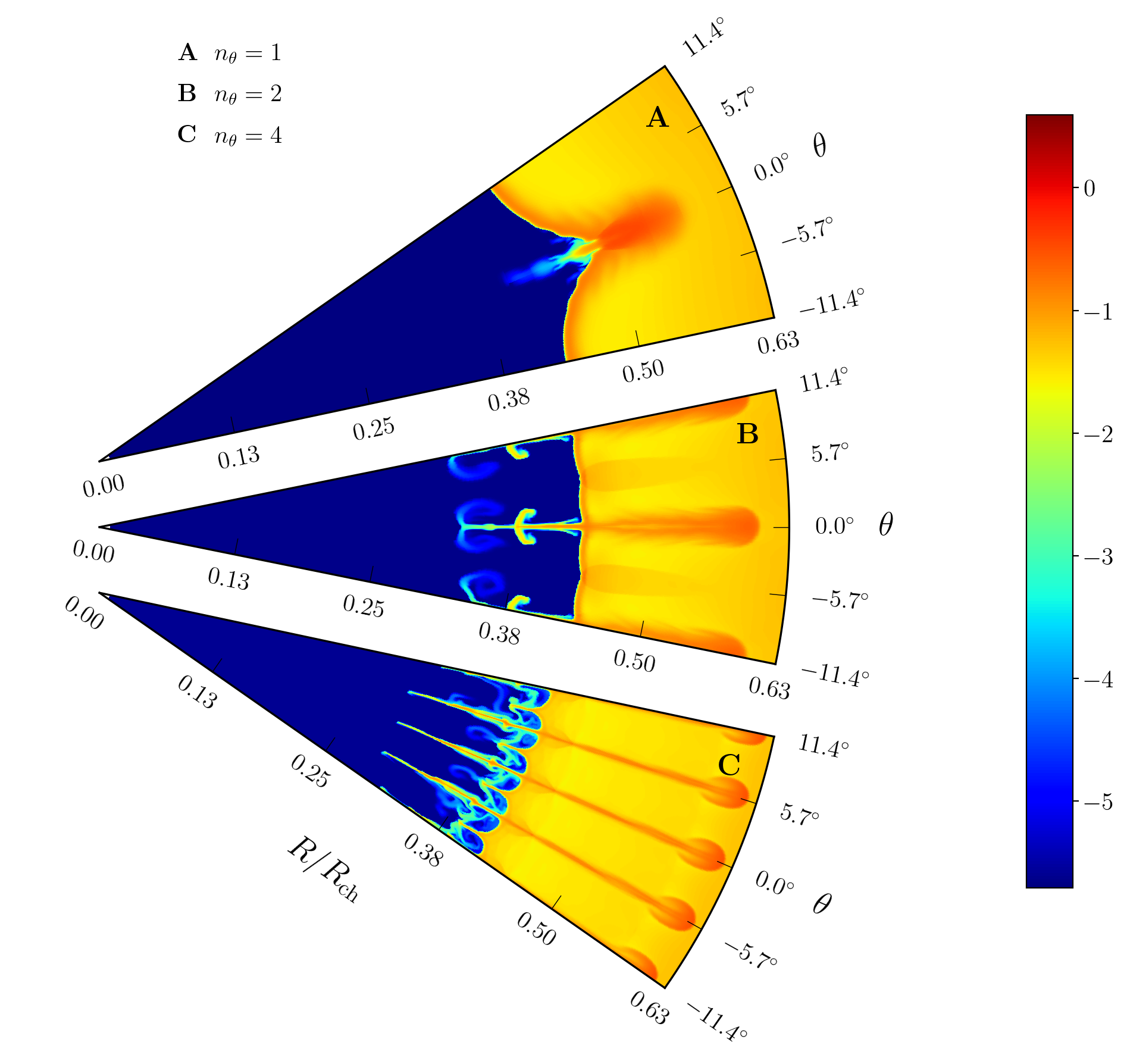}
\caption{$\log_{10}[\rho/\rho_{\rm ch}]$ during the compression phase of the PWN, at $\eta=0.30$ and $t = 2.0 t_{\rm ch}$.}
\label{compression_fixed_eta_two}
\end{figure}\\
\noindent
The trend that we have previously established regarding the faster onset of the instability at higher wavenumbers -- observed in Fig.~\ref{compression_fixed_eta_one} -- is maintained at this advanced evolutionary stage. The \text{penetration finger} in the $n_{\theta} = 1$ model now becomes clearly visible, spanning radially from $\simeq 0.38 R_{\rm ch}$ to $\simeq 0.42 R_{\rm ch}$, exhibiting a density of $\simeq 10^{-4} \rho_{\rm ch}$, while the density increases toward its base. This structure drags material from the trailing over-dense region -- which extends for $\simeq 0.08 R_{\rm ch}$ behind the base of the \text{finger}. Notably, these trailing regions in the $n_{\theta} = 1$ case reach a density peaked up to $\simeq \rho_{\rm ch}$ and a similar behavior is observed in the $n_{\theta} = 2$ and $n_{\theta} = 4$ models.\\\\
Moving to the $n_{\theta} = 2$ case, the formation of \text{mushroom-like structures}, as a result of the enhanced localized vorticity triggered by the increased wavenumber, is evident. These structures advance toward the core, detaching from the trailing over-dense zones much more efficiently than in the $n_{\theta} = 1$ case. The primary mushroom formations extend from $\simeq 0.38 R_{\rm ch}$ to $\simeq 0.44 R_{\rm ch}$, generating secondary fluid structures from their tips that plunge deeper down to $\simeq 0.34 R_{\rm ch}$. The mass density within these structures displays a distinct gradient: the main stem maintains a density of $\simeq 0.1 \rho_{\rm ch}$, which drops to $\simeq 10^{-2} \rho_{\rm ch}$ in the \text{mushroom heads}, and further decreases to $\simeq 10^{-3} \rho_{\rm ch}$ within the secondary fingers, extending for $\simeq 0.05 R_{\rm ch}$ starting from the tip of the central structure.\\\\
The dynamics of the instability becomes significantly more complex in the $n_{\theta} = 4$ model. Here, multiple \text{mushroom-like} structures are interacting, driving a turbulent mixing within the nebula, and, as the compression proceeds, the shocked ejecta penetrate the lower-density structures. Although the main stems of the \text{mushrooms-like structures} are partially disrupted by this interaction, they remain distinguishable between $\simeq 0.38 R_{\rm ch}$ and $\simeq 0.42 R_{\rm ch}$, together with a penetration layer of shocked ejecta of similar length. Below these stems, highly elongated secondary \text{fingers} extend deeply from $\simeq 0.28 R_{\rm ch}$ up to $\simeq 0.38 R_{\rm ch}$, doubling their radial length compared to the $n_{\theta} = 2$ case. The outer boundaries of these secondary structures are characterized by a lower density of $\simeq 10^{-3} \rho_{\rm ch}$, while their interior core density rises to $\simeq 0.1 \rho_{\rm ch}$, similarly to the $n_{\theta}=2$ case. Crucially, at these early times, the system with $n_{\theta} = 4$ is already undergoing an intense mixing phase that lower wavenumber models have yet to experience. 
Furthermore, the nebula in the $n_{\theta} = 4$ model appears slightly more compressed than in the $n_{\theta} = 1$ and $n_{\theta} = 2$ scenarios. This suggests that increasing the angular wavenumber not only makes the growth of the instabilities more rapid, but can also anticipates the overall compression. This complexity is similarly observed when holding the wavenumber constant and varying the initial perturbation amplitude at $t = 2.0 t_{\rm ch}$ -- see Fig.~\ref{compression_fixed_n_two}.
\begin{figure}[h!]
\centering
\includegraphics[width=0.75\linewidth]{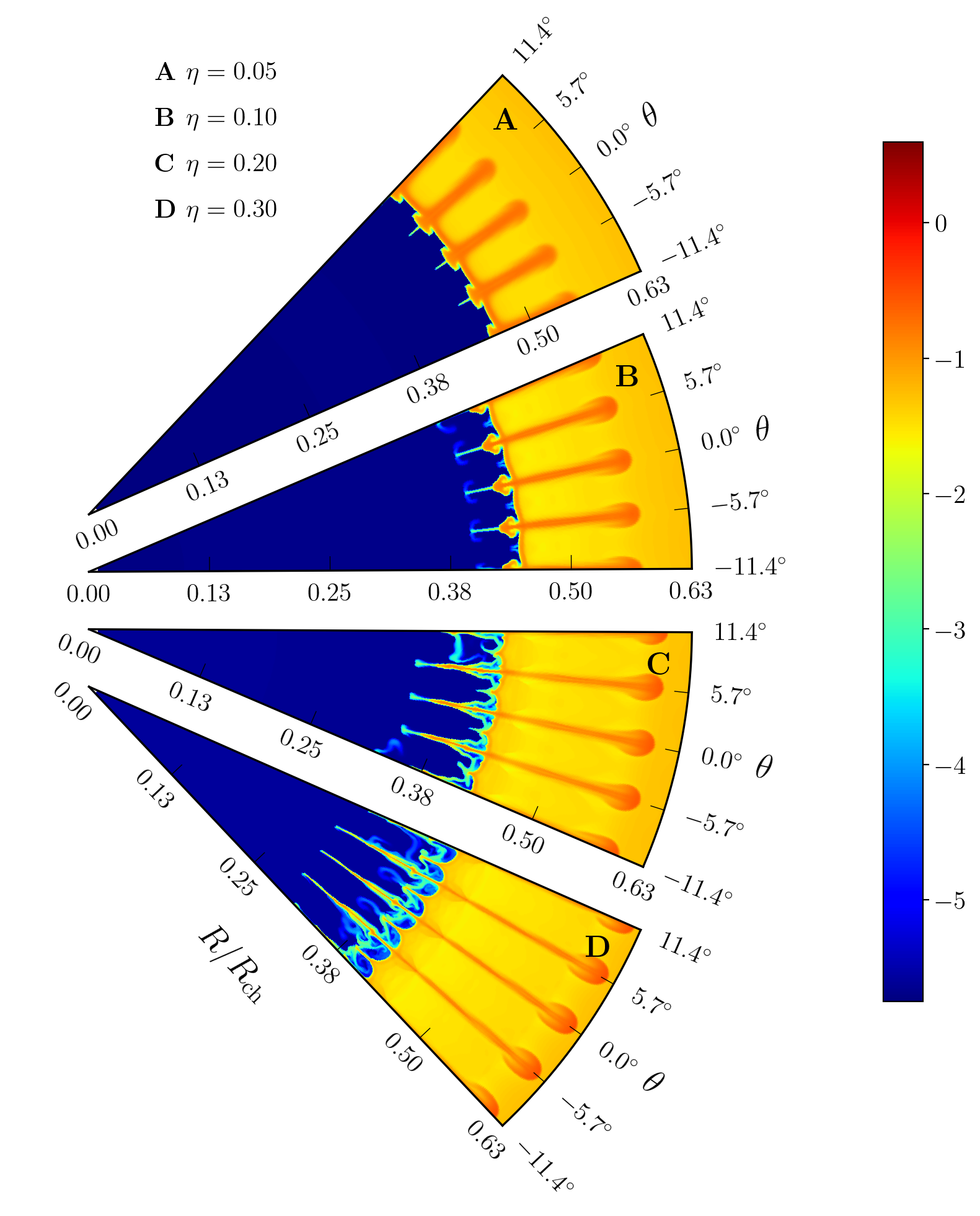}
\caption{$\log_{10}[\rho/\rho_{\rm ch}]$ during the compression phase of the PWN, at $n_{\theta}=4$ and $t = 2.0 t_{\rm ch}$.}
\label{compression_fixed_n_two}
\end{figure}
\noindent
It is evident that in the lowest amplitude case -- $\eta = 0.05$ -- the mushroom-like structures have not yet developed. Despite this, the formation of \text{penetration fingers} -- with a characteristic density of $\simeq 10^{-3}\rho_{\rm ch}$ -- is evident, anchored by a dense base -- where $\rho \simeq 0.1 \rho_{\rm ch}$. Combined, these \text{fingers} and their base extend from $\simeq 0.44 R_{\rm ch}$ to $\simeq 0.46 R_{\rm ch}$. The trailing over-densities -- which maintain a characteristic density of $\simeq 0.1 \rho_{\rm ch}$ across all panels, though exhibiting a slight decrease at higher amplitudes -- are much elongated, up to $\simeq 0.54 R_{\rm ch}$.\\\\ 
A similar overall behavior characterizes the $\eta = 0.10$ case; however, the mushroom-like structures are now distinguishable, driven by the enhanced localized vorticity associated with the larger initial perturbation. These structures span from $\simeq 0.42 R_{\rm ch}$ to $\simeq 0.44 R_{\rm ch}$, with secondary \text{fingers} already emerging between $\simeq 0.40 R_{\rm ch}$ and $\simeq 0.42 R_{\rm ch}$, whose mass density is $\simeq 10^{-3}\rho_{\rm ch}$. As a consequence, the structure including both the primary \text{mushrooms} and their secondary fingers effectively doubles in radial length compared to the features observed in the $\eta = 0.05$ model.
Furthermore, the high-density structures extend outward from the \text{mushroom} bases for $\simeq 0.10 R_{\rm ch}$, effectively moving outward while the nebula continues its contraction.\\\\ 
The dynamical evolution is further anticipated at higher amplitudes -- $\eta = 0.20$ and $\eta = 0.30$. In these regimes, the \text{penetration fingers} are about twice as long as those observed in the $\eta = 0.10$ case and two orders of magnitude denser, with the turbulent mixing in the nebula becoming increasingly pronounced. This enhanced mixing is particularly evident in the highest amplitude scenario -- see panel D of Fig.~\ref{compression_fixed_n_two} -- where shocked ejecta clearly precipitate deeply into the low-density regions of the shell.\\\\ 
At this stage, simulations suggest that increasing either the perturbation amplitude or the angular wavenumber anticipates both the growth of the instabilities and the dynamics of the compression of the nebula: indeed, while the PWN radius is $\simeq 0.45 R_{\rm ch}$ for $\eta = 0.05$, it has already contracted to $\simeq 0.41 R_{\rm ch}$ for $\eta = 0.30$. However, it should be emphasized that, although the onset of compression appears to be advanced, this effect is considerably less pronounced than the accelerated dynamics of the instabilities.\\\\ 
To complete our analysis of the compression phase, we now focus on its final stages. By returning to the constant-amplitude regime -- $\eta = 0.30$ -- and varying the wavenumber $n_{\theta}$, we explore the evolution characterizing the final phases of the compression of the system, as illustrated in Fig.~\ref{compression_fixed_eta_three}.
\begin{figure}[h!]
\centering
\includegraphics[width=0.7\linewidth]{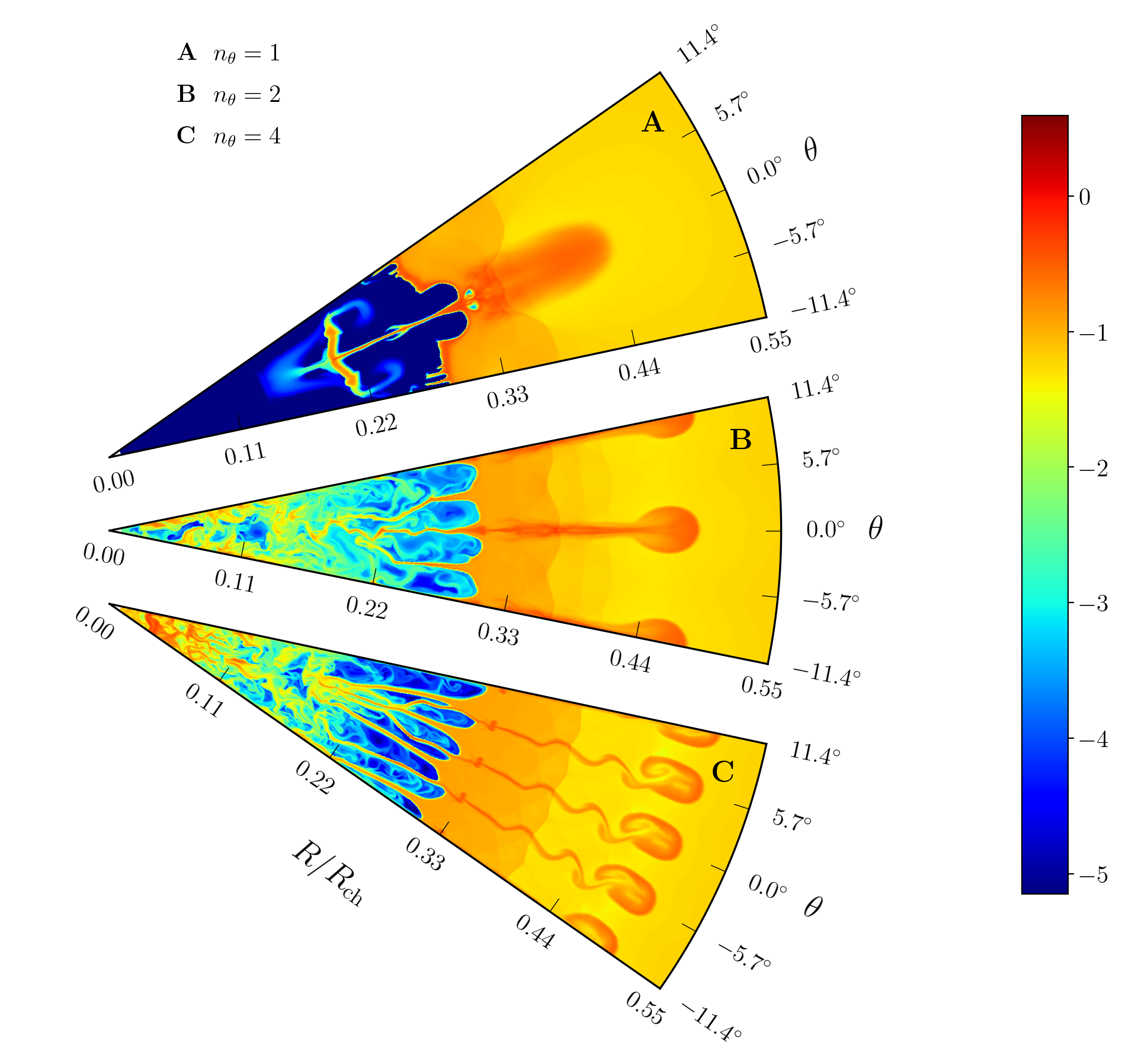}
\caption{$\log_{10}[\rho/\rho_{\rm ch}]$ during the compression phase of the PWN, at $\eta=0.30$ and $t = 2.5 t_{\rm ch}$.}
\label{compression_fixed_eta_three}
\end{figure}
\noindent
At $t = 2.5 t_{\rm ch}$, it becomes evident that the differences across the models are even more pronounced. Starting with the $n_{\theta} = 1$ case, the trailing over-dense region now extends from $\simeq 0.30 R_{\rm ch}$ to $\simeq 0.44 R_{\rm ch}$, maintaining a characteristic mass density of $\simeq 0.1 \rho_{\rm ch}$. At this stage, the \text{mushroom-like structure} is well-defined, occupying the region between $\simeq 0.20 R_{\rm ch}$ and $0.30 R_{\rm ch}$, accompanied by a secondary \text{finger} that penetrates down to $\simeq 0.15 R_{\rm ch}$ as the nebula recedes, with the bounce shock located at $\simeq 0.37 R_{\rm ch}$. Nevertheless, it is crucial to note that the interior of the nebula has not yet undergone any significant mixing. \\\\
The dynamics becomes more complex as the wavenumber increases and in the $n_{\theta} = 2$ scenario, the turbulent mixing is more pronounced: the dense heads of the \text{mushroom structures} have already interacted and merged, resulting in a visible mixing of the material between the origin and $\simeq 0.11 R_{\rm ch}$. Additionally, the trailing over-densities stretch outward from the boundary of the nebula -- located at $\simeq 0.3 R_{\rm ch}$ -- up to $\simeq 0.5 R_{\rm ch}$. Similarly to the $n_{\theta} = 1$ model, the bounce shock is observed at $\simeq 0.37 R_{\rm ch}$.\\\\
Finally, the $n_{\theta} = 4$ model shows the compression in a far more advanced state and in this latter scenario, the bounce shock is slightly advanced compared to the previous cases. In this case, the bounce shock is destined to propagate through a region where the over-densities are significantly more frayed -- although more extended in comparison with the $n_{\theta} = 1$ and $n_{\theta} = 2$ models, where these structures are more compact, and this fragmentation directly influences the overall shape of the bounce shock itself. \\\\
Ultimately, within this high-perturbation regime, the mixing washes out the growth of the Rayleigh-Taylor instabilities. To further explore these late-stage dynamics, we now shift our focus back to analysing cases with different values of $\eta$, examining the final stages of the compression while holding the wavenumber constant, at $t = 2.5 t_{\rm ch}$.
\begin{figure}[h!]
\centering
\includegraphics[width=0.75\linewidth]{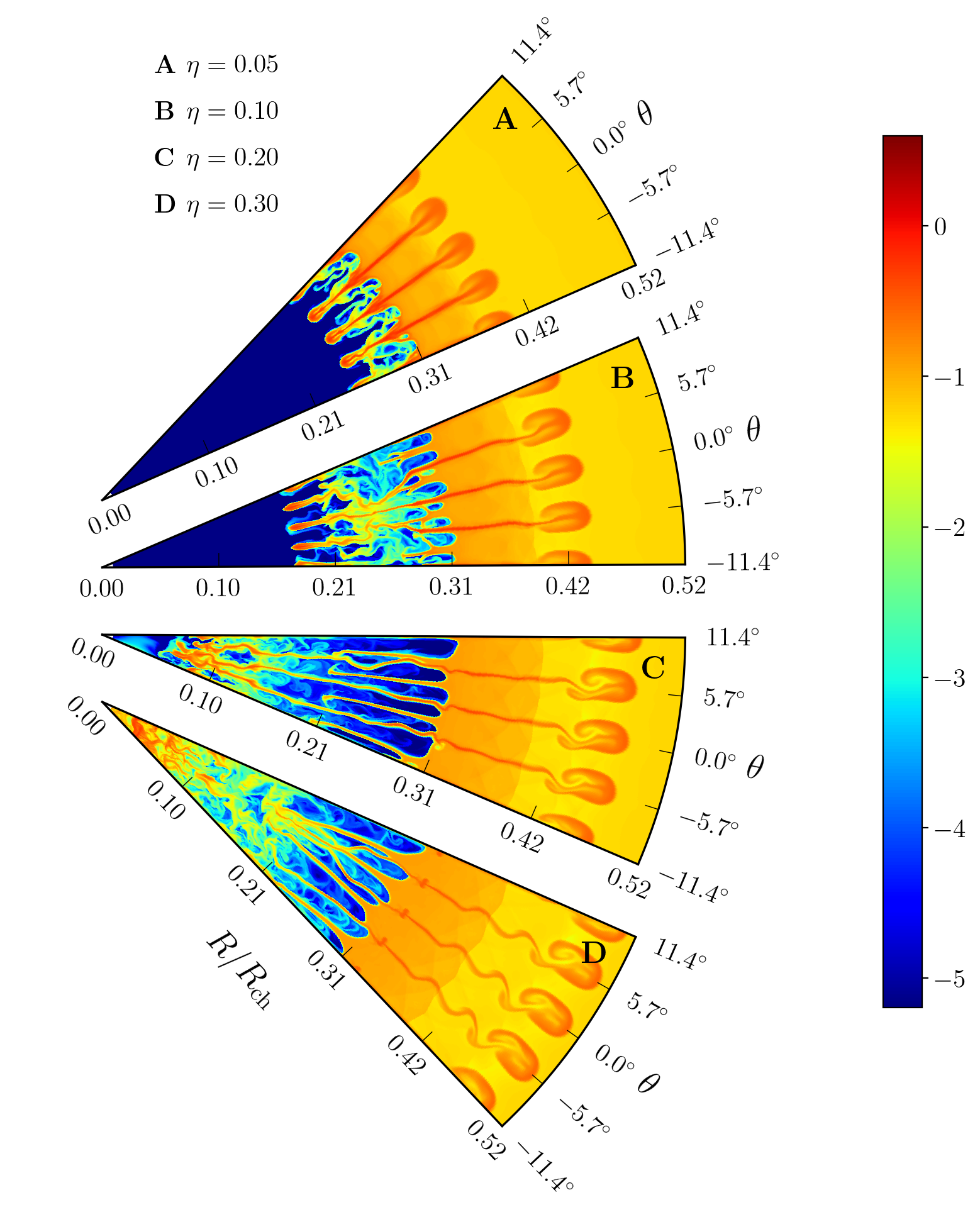}
\caption{$\log_{10}[\rho/\rho_{\rm ch}]$ during the compression phase of the PWN, at $n_{\theta}=4$ and $t = 2.5 t_{\rm ch}$.}
\label{compression_fixed_n_three}
\end{figure}\\
\noindent
As illustrated in Fig.~\ref{compression_fixed_n_three}, for the $\eta = 0.05$ case, the penetration of dense \text{fingers} -- extending from $\simeq 0.23 R_{\rm ch}$ to $\simeq 0.25 R_{\rm ch}$ -- triggers the emergence of \textit{plume-like structures}. 
Notably, these \text{plumes} emerge with a well defined shape exclusively in the case with the weakest perturbation, extending from $\simeq 0.25 R_{\rm ch}$ to $\simeq 0.31 R_{\rm ch}$ and they exhibit some turbulent mixing. 
Furthermore, the over-dense regions above the nebula extend from $\simeq 0.30 R_{\rm ch}$ to $\simeq 0.40 R_{\rm ch}$ both in the $\eta=0.05$ and $\eta=0.10$ case, their global density is $\simeq 0.1 \rho_{\rm ch}$ and their structure is well-defined. 
Notably, the internal unmixed structure of the nebula is significantly broader for $\eta = 0.05$ compared to the $\eta = 0.10$ case, extending up to $\simeq 0.25 R_{\rm ch}$, whereas for $\eta = 0.10$ the unmixed PWN only reaches $\simeq 0.17 R_{\rm ch}$. 
Rayleigh-Taylor \text{filaments} are evident only in the $\eta = 0.10$ case, located between $\simeq 0.25 R_{\rm ch}$ and $\simeq 0.31 R_{\rm ch}$, and characterized by turbulent mixing.\\\\
For the higher perturbation amplitudes $\eta = 0.20$ and $\eta = 0.30$, it is evident that the Rayleigh-Taylor instabilities have not formed and the heads of the mushroom-like structures have almost reached the center of the nebula. 
In these regimes, the over-dense regions extend up to $\simeq 0.50 R_{\rm ch}$ and are generally more frayed. This effect is even more pronounced for $\eta = 0.30$, where the nebula exhibits a substantially higher mixing, indicating a much more advanced dynamical stage of the instability. \\\\
In conclusion, it must be emphasized that despite these differences in the growth rate of the instability, the nebulae exhibit similar overall radii at this evolutionary stage, $\simeq 0.3 R_{\rm ch}$, which is consistent with the 1D case \texttt{A07-NR} -- see Fig.~\ref{NON_REL_OVERALL}. Subsequent analyses have shown that the bounce shock is launched at $2.8 t_{\rm ch}$ in all the 2D simulations, anticipating the corresponding 1D case \texttt{A07-NR} -- see Fig.~\ref{NON_REL_OVERALL} -- by $0.2 t_{\rm ch}$ and the ordinary \texttt{A07} case by $0.4 t_{\rm ch}$ -- see Fig.~\ref{SNAPSHOTS_TOGETHER}. As these analyses suggest, although increasing $\eta$ and $n_{\theta}$ initially slightly advances the onset of compression, the details of the perturbations in the swept-up shell do not appear to influence the long-term compression dynamics, as all nebulae reach the final stages in a similar way. However, it remains evident that increasing $\eta$ and $n_{\theta}$ anticipates and amplifies the dynamical evolution of the instabilities.\\\\
Furthermore, when compared to the 1D case in Fig.~\ref{NON_REL_OVERALL}, it is clear that at maximum compression, the denser shell surrounding the PWN broadens as the perturbation is amplified. Only in the scenario with a  minimal perturbation -- with $\eta = 0.05$ and $n_{\theta} = 4$ -- the thickness of the shell remains comparable to that of the corresponding 1D model, which measures $\simeq 0.03 R_{\rm ch}$. Specifically, in the $\eta = 0.05$ case, the shell extends between $\simeq 0.30 R_{\rm ch}$ and $\simeq 0.35 R_{\rm ch}$, whereas it expands up to $\simeq 0.40 R_{\rm ch}$ in the case with $\eta = 0.30$ and $n_{\theta}=4$. In addition, as shown in Fig.~\ref{compression_fixed_eta_three}, increasing the wavenumber generally amplifies the shell broadening.
\section{The Late Evolution}
In this section, we present the late evolution of the system following the compression phase -- see Fig.$\stick$\ref{fig_late_evolution}. As established in the preceding analysis, increasing both the perturbation amplitude, $\eta$, and its wavenumber, $n_{\theta}$, progressively anticipates and amplifies the growth of instabilities, whereas the compression dynamics is preserved. For this reason, anticipating that the late-stage evolution is not dominated by the previous growth of instabilities in any of the cases we have examined, we have chosen to present a single representative case, with $\eta = 0.30$ and $n_{\theta} = 4$, and to compare the evolution of the system directly with the corresponding 1D model -- see Fig.~\ref{NON_REL_OVERALL}.
\begin{figure}[h!]
\centering
\includegraphics[width=0.65\linewidth]{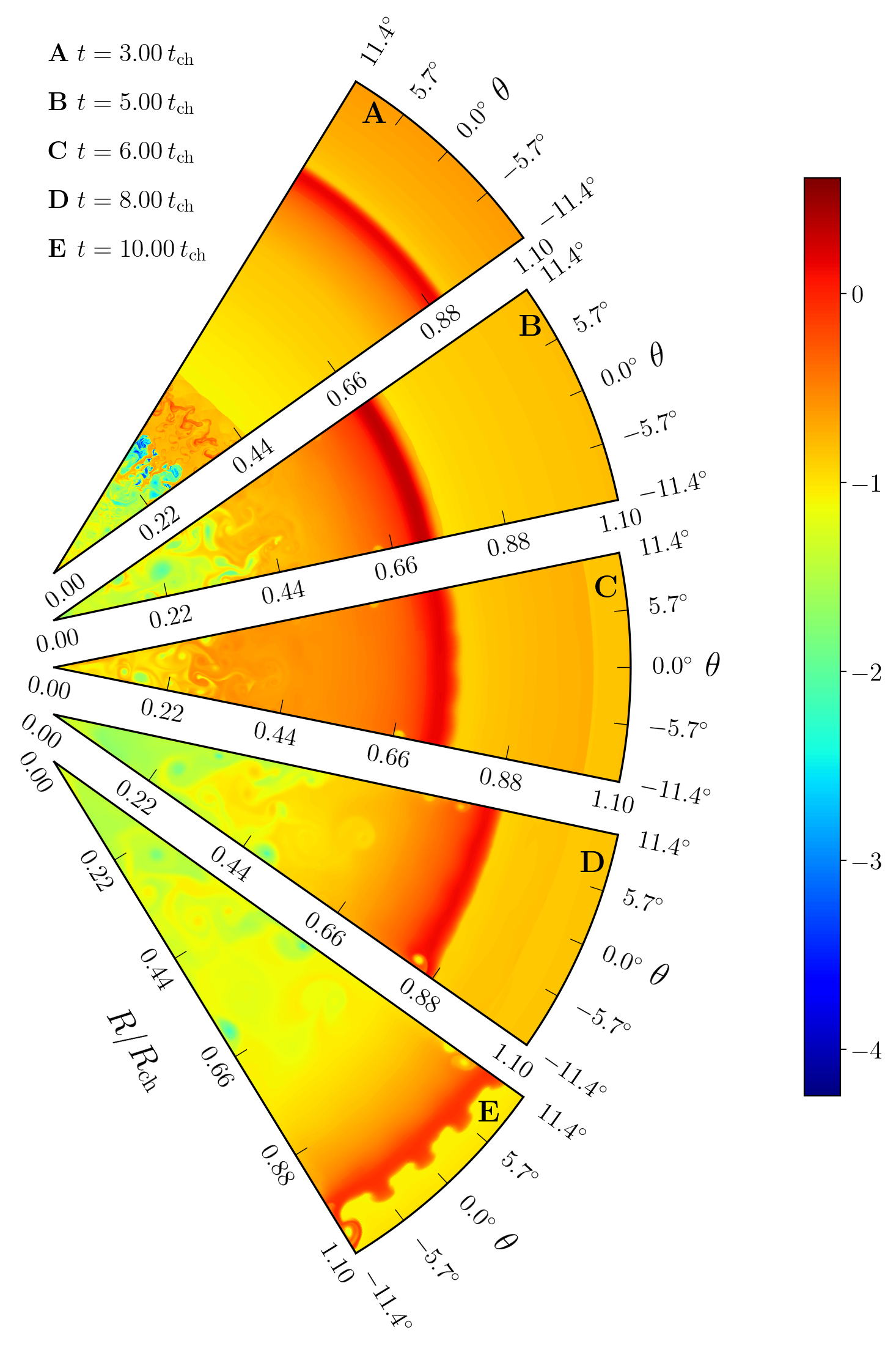}
\caption{Time evolution of $\log_{10}[\rho/\rho_{\rm ch}]$ after the compression, with an initial perturbation amplitude $\eta = 0.30$ and an angular wavenumber $n_{\theta} = 4$.}
\label{fig_late_evolution}
\end{figure}\\
\noindent
Examining panel A, at $3 t_{\rm ch}$ the PWN is still clearly distinguishable, extending up to $\simeq 0.30 R_{\rm ch}$. Further outward, the bounce shock is located at $\simeq 0.44 R_{\rm ch}$. Having propagated through the overdense perturbations, the bounce shock carries the signature of these density fluctuations, with the CD identifiable at $\simeq 0.90 R_{\rm ch}$.\\\\
As the nebula re-expands, reaches a maximum radius of $\simeq 0.40 R_{\rm ch}$ at $5 t_{\rm ch}$ -- see panel B. By this time, the bounce shock has reached the CD at $\simeq 0.75 R_{\rm ch}$. This interaction generates both a reflected shock and a transmitted shock, the latter of which carries outward the perturbation initially imprinted in the bounce shock. At this stage, the nebula undergoes a second compression cycle driven by its interaction with the reflected shock. While this shock is not visible in the density profile -- though it could be more easily traced in the velocity field, as demonstrated by its 1D counterpart in Fig.~\ref{NON_REL_OVERALL} -- its impact is evident in the dynamics of the system.\\\\ 
By $6 t_{\rm ch}$ -- see panel C -- the PWN is compressed to a radius of $\simeq 0.30 R_{\rm ch}$. Meanwhile, the CD is situated at $\simeq 0.75 R_{\rm ch}$, and its front is undergoing further deformation due to the preceding collision with the bounce shock. Additionally, the transmitted shock resulting from this collision is now distinguishable at $\simeq 1 R_{\rm ch}$.\\\\
Moving to panel D, at $8 t_{\rm ch}$, the nebula has re-expanded, and now appears more deformed, making it challenging to define its extension. Nevertheless, it is roughly still traceable up to $\simeq 0.40 R_{\rm ch}$, while very mixed and fragmented. In the subsequent phase, consistent with the 1D counterpart shown in Fig.~\ref{NON_REL_OVERALL}, the following compression is difficult to track, given the deformation of the nebula. However, the PWN exhibits a radius consistent with the 1D model -- approximately $0.40 R_{\rm ch}$ prior to the next full compression -- while the CD advances to $\simeq 0.90 R_{\rm ch}$. The nebula then enters a further compression cycle, triggered by the collision with the reflected shock, itself generated from the interaction of the bounce shock with the CD. Finally, at $10 t_{\rm ch}$ -- see panel E -- the nebula retreats to $\simeq 0.30 R_{\rm ch}$, while the CD advances to $\simeq 1 R_{\rm ch}$ and begins to exhibit pronounced instability.\\\\
In conclusion, the 2D analysis demonstrates good agreement with the corresponding 1D model -- Fig.~\ref{NON_REL_OVERALL} -- regarding the overall reverberation dynamics, since the temporal evolution of the PWN, the CD, and of the other shock waves is consistent across both cases. This confirms that the results derived from 1D Lagrangian simulations are robust for capturing the evolution of the system.
\section{Discussion}
While the analysis of the overall dynamics revealed that the reverberation phase in 2D is generally consistent with the 1D simulations, we require a more in-depth comparison between the evolution calculated using the Lagrangian code and PLUTO. Primarily, we are interested in assessing the robustness of the assumption of uniform pressure inside the PWN adopted in the one-zone model, as well as on the derived nebular size.
Relying on the results of the 2D simulations, we sampled the pressure at three different radii -- $0.07 R_{\rm ch}$, $0.14 R_{\rm ch}$, and $0.21 R_{\rm ch}$ -- given that these points remain within the low-density bubble that defines the PWN at the center of the remnant. This sampling was performed at every output time, and for each radius, the pressure was extracted at three different angles: $-5^\circ$, $0^\circ$, and $5^\circ$. This analysis was performed for each of the $n_\theta = 4$ cases across varying values of $\eta$, since variations of $n_{\theta}$ at fixed $\eta$ yielded no substantial differences in the pressure profiles across the selected points. From these points, we derived an average pressure profile to compare against our 1D reference case, \texttt{A07-NR}, and the comparison is shown in Fig.~\ref{meanprs2d}.
\begin{figure}[h!]
\centering
\includegraphics[width=1.0\linewidth]{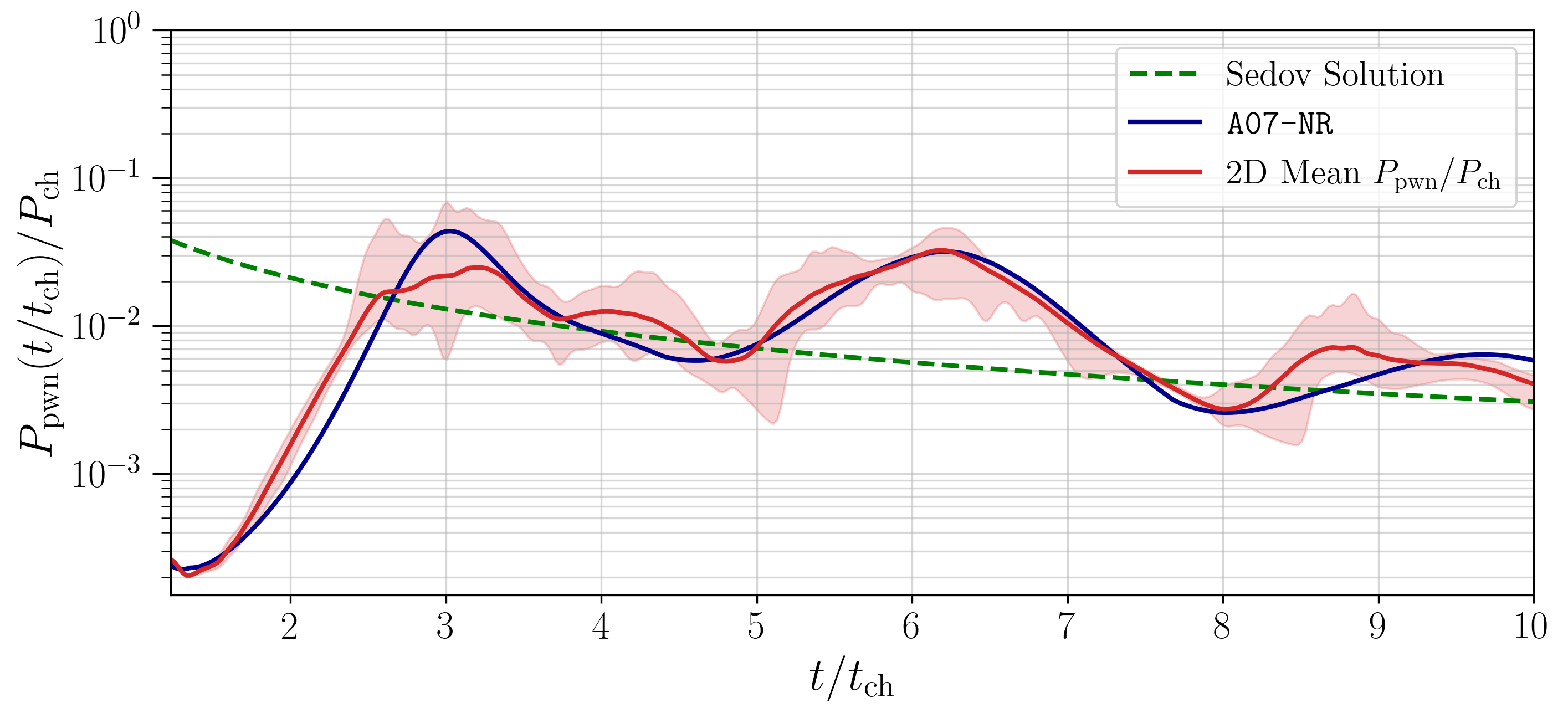}
\caption{Evolution of the internal pressure of the PWN. The thick, blue solid line represents the 1D non-relativistic case \texttt{A07-NR}, while the red line indicates the average pressure derived from the profiles sampled at various points within the PWN, with the shaded transparent band encompassing all the pressure profiles extracted across the selected points. The green, dashed line representing the Sedov solution. }
\label{meanprs2d}
\end{figure}\\\\
As indicated by the broad spread of the shaded band, which captures the variation of the pressure among the different sampling points, the 2D models reveal a gradient of internal pressure, with inner radii experiencing pressures up to an order of magnitude greater than the outer radii during the first compression. This notwithstanding, the average pressure profile does not differ by more than a factor of $2-3$ from the uniform pressure of the \texttt{A07-NR} case. This indicates that the Lagrangian approach yields robust results regarding the internal pressure.\\\\
Regarding the radius of the nebula, we found that the initial perturbation in the shell triggers a turbulent mixing between the PWN and the ejecta as the nebula interacts with the RS, and the hypothesis of a spherically symmetric PWN breaks down. As a consequence, defining a radius to describe the characteristic dimension of the nebula in 2D becomes non-trivial. However, recalling the adiabatic relation $PV^\Gamma = \text{constant}$ -- where $P$ is pressure, $V$ the volume occupied by the material and $\Gamma = 5/3$ -- it is possible to define the \textit{effective radius} of the nebula, $R^{\rm (eff)}_{\rm pwn}$, in the phase following the interaction with the RS:
\begin{equation}
\frac{R^{\rm (eff)}_{\rm pwn}(t/t_{\rm ch})}{R_{\rm ch}} = \frac{R_{\rm pwn}(t_{\rm rev}/t_{\rm ch})}{R_{\rm ch}}\left[ \frac{P_{\rm pwn}(t/t_{\rm ch})}{P_{\rm pwn}(t_{\rm rev}/t_{\rm ch})} \right]^{-\frac{1}{5}},
\label{EQRADEFF}
\end{equation}
where $t_{\rm rev}$ denotes the beginning of the \text{reverberation} phase. The \textit{effective radius} is the relevant quantity for determining the adiabatic losses of the system. In spite of this, during the reverberation phase, this is not the only relevant dimension of the nebula. As observed in the analysis of the compression in 2D, the mixing between the PWN and the shocked ejecta of the SNR makes the localization of the material of the nebula less clear. To address this, we employed a tracer\footnote{In PLUTO, a tracer is a dimensionless scalar quantity advected with the fluid; in our setup, it was initialized to $1$ inside the PWN at the beginning of the simulation, and $0$ outside.} to follow the material of the nebula as it expands and compresses -- see Fig.~\ref{TRACER_IMAGE}.
\begin{figure}[h!]
\centering
\includegraphics[width=1.0\linewidth]{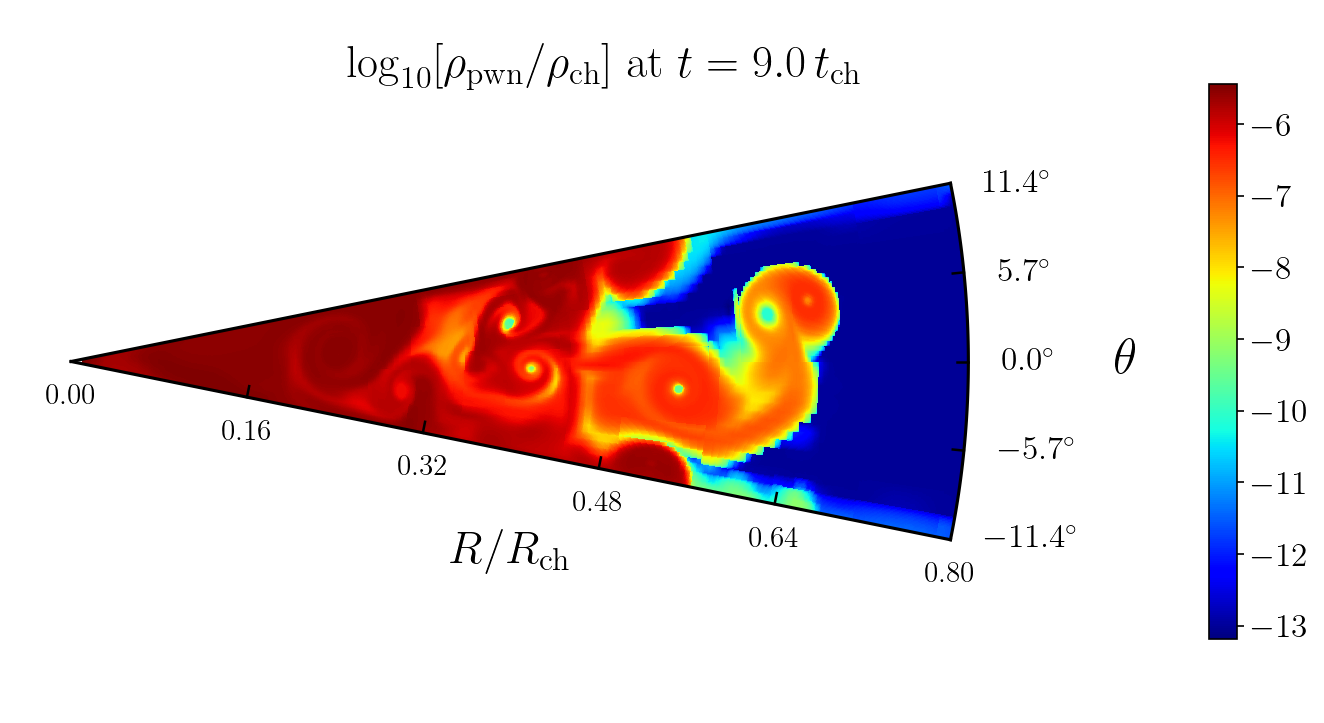}
\caption{Density of the PWN material at $t = 9.0\,t_{\rm ch}$ in the case with amplitude $\eta = 0.30$ and wavenumber $n_\theta = 4$.}
\label{TRACER_IMAGE}
\end{figure}\\
Up to $\simeq 0.30 R_{\rm ch}$, the density of the PWN material looks pretty uniform in distribution. Beyond this radius, mixing is evident. At radii $\gtrsim 0.40 R_{\rm ch}$ \textit{bubbles} of PWN material embedded into the SNR material are present, extending up to $\simeq 0.70 R_{\rm ch}$. There is some indication that \text{bubbles} present at the edge around $\simeq 0.56 R_{\rm ch}$ might somehow be enhanced by boundary effects, at very late times. 
Given that the PWN material is scattered up to $\simeq 0.70 R_{\rm ch}$, it is useful to define an \textit{apparent radius}, $R^{\rm (app)}_{\rm pwn}$. To do so, we first compute the average (normalized) density of the nebula -- $\langle{\rho_{\rm pwn}}/{\rho_{\rm ch}}\rangle_{\rm c}$ -- in the innermost region:
\begin{equation}
\langle \frac{\rho_{\rm pwn}}{\rho_{\rm ch}} \rangle_{\rm c} =
\frac{1}{\rho_{\rm ch}}
\frac{\int_{\theta_{\rm min}}^{\theta_{\rm max}} \int_{r_{\rm min}}^{0.20 R_{\rm ch}} \rho_{\rm pwn} r^2 \sin\theta  \text{d}r\text{d}\theta}{\int_{\theta_{\rm min}}^{\theta_{\rm max}} \int_{r_{\rm min}}^{0.20 R_{\rm ch}} r^2 \sin\theta \text{d}r \text{d}\theta },
\end{equation}
and we define the \text{apparent radius} as the radius where the local density of the nebula drops to $1/10$ of this averaged value. This is the quantity relevant for the apparent observable size of these systems. Since we observed the nebula exhibiting material at low-density extending beyond its denser central region -- see Fig.~\ref{TRACER_IMAGE} -- we generally expect the \text{apparent radius} of the PWN to be larger than the \text{effective radius}. A comparison between the mean \text{apparent radius}, the mean \text{effective radius} -- averaged across cases with varying values of $\eta$ and fixing $n_\theta = 4$ -- and the radius of the 1D non-relativistic case (\texttt{A07-NR}) is presented in Fig.~\ref{NORM_EFF_APP}.
\begin{figure}[h!]
\centering
\includegraphics[width=1.0\linewidth]{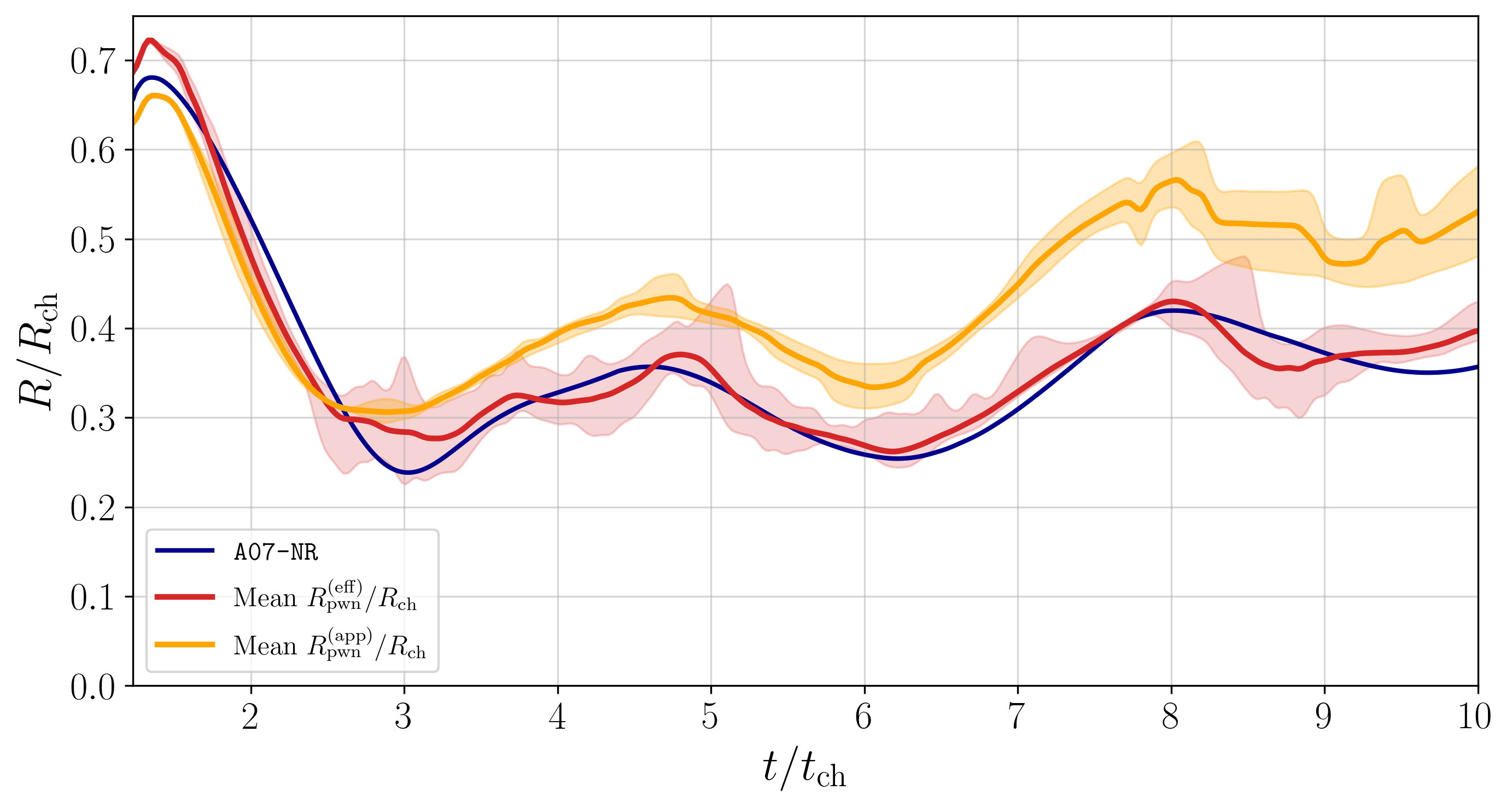}
\caption{Comparison among the radius of the 1D case \texttt{A07-NR} (blue solid line), the mean effective radius (red solid line), and the mean apparent radius (yellow solid line). The yellow and red shaded transparent bands encompass all the \text{effective} and \text{apparent} radial profiles in the cases with $n_{\theta} = 4$, with varying $\eta$.}
\label{NORM_EFF_APP}
\end{figure}\\\\
Consistent with the findings from the average pressure of the nebula, the effective radius $R^{(\rm eff)}_{\rm pwn}$ shows a good agreement with the non-relativistic 1D profile. This serves as a further consistency check, confirming the robustness of the results obtained from 1D simulations for what concerns the net volume. Conversely, the apparent radius is generally larger: following the mixing of the PWN with the shocked ejecta, at late times ($\gtrsim 8 t_{\rm ch}$) it is possible to find the material of the nebula up to $\simeq 0.5 R_{\rm ch}$, well beyond the effective radius.
\chapter{Conclusions and Future Outlook}
\section{Conclusions}
In this work, we have analyzed the dynamics of Pulsar Wind Nebulae (PWNe) expanding in the hosting Supernova Remnants (SNRs), investigating both their \textit{free-expansion} and \textit{reverberation} phases. Given the vast parameter space that characterizes a single PWN-SNR system, in terms of the properties of the Pulsar (PSR), the remnant itself, and the Interstellar Medium (ISM), we investigated all the possible combinations adopting a \textit{population synthesis} approach. 
This was done, in Chapter$\stick$\ref{SNE_SNR_CHAPTER1}, by considering the Galactic distribution of the main properties that define the role of both PSRs, like their initial spin period, and magnetic field, and SNRs, like ejecta mass, supernova energy and ambient density.
By generating a synthetic population of young PSRs and SNRs, we have statistically explored how they are distributed. We have found that in the so-called \textit{characteristic plane} -- that allows us to reduce the dimensionality of the problem by rescaling several quantities -- they occupy a well defined Region of Interest (ROI), encompassing the $95\%$ of the synthetic population. Then one can correlate the distributions of the initial parameters to the subsequent evolution of the resulting PWNe.
In Chapter~\ref{CHAPTPOPSYN} we characterized the \text{free-expansion phase} by using the approach based on the one-zone model within the thin-shell approximation. Using this simplified, semi-analytical framework, we successfully characterized the distribution of the physical variables of the PWNe across the entire ROI at the onset of the \text{reverberation phase}.
We found significant variations across the ROI, suggesting that, over our population, one is likely to have very different evolutionary histories, in the late phases. \\\\ 
However, the interaction between the expanding PWN and the Reverse Shock (RS) of the SNR profoundly alters the dynamics of the nebula. Therefore, simple analytical prescriptions for the conditions within the SNR shell -- that are a mandatory ingredient for the one-zone model within the thin-shell approximation --  become inapplicable, and the evolution during the \text{reverberation phase} must be computed purely numerically. To probe this phase, in Chapter$\stick$\ref{CHAPREV1D} we sampled representative systems across the ROI itself along various directions, investigating the subsequent evolution using a Lagrangian approach. The results revealed that systems hosting a PSR with a high initial spin-down luminosity undergo a weak compression ($\text{Compression Factor} \lesssim 2$), followed by a vigorous re-expansion. Conversely, systems hosting a weaker PSR experience severe compression ($\text{Compression Factor} \gtrsim 7$) followed by subsequent weak oscillations. Ultimately, intermediate systems undergo moderate compression ($\text{Compression Factor} \simeq 3$) and exhibit a more complex dynamics, featuring multiple cycles of expansion and compression driven by the interaction of the nebula with secondary shocks, themselves generated during the compression phase. The analysis revealed that, while exhibiting different dynamical behaviors -- dictated by the initial PWN energetics -- at late times ($\gtrsim 8 t_{\rm ch}$) all systems exhibit a convergence to a relaxed state, where the pressure asymptotically approaches the Sedov solution.\\\\
While the 1D Lagrangian approach captures the dynamics of the compression of the nebula and the formation of shocks and contact discontinuities, real PWN-SNR systems are intrinsically 3D and exhibit morphological asymmetries and instabilities that 1D models cannot reproduce. Therefore, extending the analysis to higher dimensions is essential for a more realistic description. Due to the high computational cost of 3D simulations, we opted to limit our analysis to 2D, and in Chapter$\stick$\ref{CHAPREV2D} we have focused on a single representative system -- corresponding to the most probable PWN-SNR system in the ROI -- using a hybrid strategy. For our aims we have used the PLUTO code, and to bypass the computationally expensive \text{free-expansion} phase, we initialized PLUTO by mapping in 2D the profiles of density, pressure and velocity of both of the PWN and the SNR, from our Lagrangian 1D solution, just before the onset of the \text{reverberation} phase. Furthermore, because PLUTO can neither handle mixing between thermodynamically distinct fluids nor continuously track the energy injection from the PSR without resolving the inner region of the PSR wind, we adopted a physical workaround. Immediately before the interaction with the RS, we switched off the injection from the PSR and modelled the PWN as a non-relativistic gas. 
Performing a preliminary 1D comparative test, we showed that these two choices combined do not lead to significant changes in the dynamics. This result justifies the use of the non-relativistic approximation for our 2D simulations. In addition, aiming to investigate the growth of instabilities, we seeded our initial condition with a single-mode perturbation of the density in the swept-up shell just before the interaction with the RS.\\\\ 
Our findings indicate that higher amplitudes of the initial perturbation  accelerate both the development of instabilities and the subsequent mixing between the shocked SNR ejecta and the material of the PWN. However, the overall dynamics, from a more qualitative point of view, does not vary much among various cases: the time at which the nebula reaches its maximum compression does not exhibit significant differences by varying the initial perturbation in the shell, nor does the subsequent mixing. To further verify the robustness of our 1D models when transitioning to higher dimensions, we also evaluated the assumption of uniform pressure within the PWN. The 2D results confirm that this assumption remains generally valid. We then introduced two new quantities, an effective radius and an apparent one, related to the volume and the extension of the PWN, respectively, and found that in term of effective volume the PWN follows the predictions of 1D simulations, while the apparent size can be up to $50\%$ larger. 
\section{Future Outlook}
While the consistency observed between 1D Lagrangian models and our 2D simulations establishes a robust baseline for characterizing the \text{reverberation} phase, there are several possible extensions and improvements of this work that could be pursued in the future. \\\\
Primarily, the models presented in this thesis are purely hydrodynamic and adiabatic. Regarding the first point, we know that the inclusion of magnetic fields is necessary to give a more realistic description, since the tension exerted by the magnetic field lines might strongly affect and even suppress the growth of fluid instabilities at the PWN-SNR interface. Concerning the second, PWNe are subject to radiative cooling through synchrotron emission and inverse Compton scattering, and incorporating these processes is necessary to characterize these systems in a more realistic way.\\\\
Future multi-dimensional simulations should also sample a broader range of systems to capture different evolutionary histories, from weak to severe compression, since in this work we restricted our study to a single representative model in 2D, selected at the peak of the distribution in our region of interest. Additionally, the fluid instabilities we explored in 2D were triggered by a simple single-mode perturbation of the density in the shell swept-up by the nebula. Implementing more realistic perturbations -- such as clumps or a multi-mode spectrum as observed in objects such as the Crab Nebula -- could provide a more realistic description. The present work also assumed that the SNR expands directly into the ISM, while progenitor stars modify their surroundings through stellar winds, creating the so-called Circumstellar Medium. Future studies should account for this interaction, as it is expected to influence the early expansion both of the SNR and of the PWN. \\\\
While the most natural extension of this work would be increasing the dimensionality of the system, moving to 3D, our 2D runs constitute a necessary methodological preliminary due to the prohibitive computational costs of 3D simulations, establishing a robust agreement between 1D and 2D models. Advancing to 3D, which reflects the real nature of PWN-SNR systems, could help us to make a better assessment of the properties of instabilities and mixing, to be directly compared with observations. Moreover the inclusion of magnetic field must necessarily be done in 3D, given that it responds differently to parallel or perpendicular perturbations.\\\\
We recall that ultimately, this work aims at providing a more robust description of the evolution of middle-aged PWN-SNR systems. This framework can then be used to model their emission, especially in the TeV-PeV band, in the light of present and future $\gamma$-ray observatories such as ASTRI and CTA. So, an obvious follow-up would be to use our findings to inform current spectral evolutionary codes, in order to make more reliable phenomenological descriptions.
\appendix
\renewcommand{\chaptername}{Appendix}
\renewcommand{\thechapter}{\Alph{chapter}}
\chapter{The Population of Pulsars and Supernova Remnants}
\label{APPI}
In this Appendix we detail the scheme adopted to synthesize the population of PWN-SNR systems. Specifically, the parameters for the population of PSRs, that are statistically distributed, are the logarithm of the magnetic field, $\log B$, and the initial period, $P_0$, which follow the PDFs defined in Eqs.$\stick$(\ref{LOGBDIST}) and (\ref{PDIST}), respectively. Regarding the SNRs, the distributions for the mass of progenitor stars, the energy of the SN, and the number density of the ISM are governed by Eqs.$\stick$(\ref{Salpeter}), (\ref{EDIST}), and (\ref{ISMDENSITYDIST}).
From these PDFs, we have derived the corresponding Cumulative Distribution Functions (CDFs). Letting $x$ denote the variables $\log B$, $P_0$, and $\log n_0$, their CDFs can be expressed in a compact common form:
\begin{equation}
\mathcal{C}(x) = 
\begin{cases}
0 & \text{if } x < x_{\min}\\
\frac{
\operatorname{erf}
\left(
\frac{x-\mu_x}{\sigma_x\sqrt{2}}
\right)
-
\operatorname{erf}
\left(
\frac{x_{\rm min}-\mu_x}{\sigma_x\sqrt{2}}
\right)
}
{
\operatorname{erf}
\left(
\frac{x_{\rm max}-\mu_x}{\sigma_x\sqrt{2}}
\right)
-
\operatorname{erf}
\left(
\frac{x_{\rm min}-\mu_x}{\sigma_x\sqrt{2}}
\right)
}
& \text{if } x_{\min} \leq x \leq x_{\max}\\
1 & \text{if } x > x_{\max}, \\
\end{cases}
\label{eq:general_CDF}
\end{equation}
where $\mathcal{C}$ denotes the cumulative function, and 
the symbols $\mu_x$ and $\sigma_x$ identify the characteristic parameters assigned in the corresponding PDFs -- namely $\mu_{\rm B}$, $\bar{P}_0$ and $\mu_0$ and the associated values $\sigma_{\log B}$, $\sigma_{P_0}$ and $\sigma_{\log n_0}$ -- while $x_{\rm min}$ and $x_{\rm max}$ stand for the lower and upper limits. 
Furthermore, the progenitor mass distribution yields the following CDF:
\begin{equation}
\mathcal{C}(M) = 
\begin{cases}
	0 & \text{if } M < M_{\min} \\
	\frac{ M_{\rm max}^{\alpha - 1}(M^{\alpha - 1} - M_{\rm min}^{\alpha - 1} )}{M^{\alpha - 1}(M_{\rm max}^{\alpha - 1} - M_{\rm min}^{\alpha - 1} )} & \text{if } M_{\min} \leq M \leq M_{\max}\\
	1 & \text{if } M > M_{\max}, \\
\end{cases}
\end{equation}
where $M$ denotes the mass of the progenitor star. We recall that $\alpha$ is the spectral index of the IMF, with $M_{\rm min}$ and $M_{\rm max}$ accounting for the lower and upper limits.
Conversely, the PDF of the kinetic energy of the SNe, $E_{\rm sn}$, leads to a linear CDF of the form:
\begin{equation}
\mathcal{C}(E_{\rm sn}) 
= 
\begin{cases}
   0 & \text{if } E_{\rm sn} < E_{\min}\\
   \frac{E_{\rm sn}-E_{\rm min}}{E_{\rm max}-E_{\rm min}} & \text{if } E_{\min}\leq E_{\rm sn}\leq E_{\max}\\
   1 & \text{if } E_{\rm sn} > E_{\max},\\
\end{cases}
\end{equation}
where we recall that $E_{\rm min}$ and $E_{\rm max}$ define the upper and lower limits, respectively. 
After defining the analytic CDFs and applying the truncations detailed in Sect.~\ref{SNRSECTION}, each function was inverted against a set of random numbers uniformly distributed in the interval $[0, 1]$. This procedure allows us to generate a set of physical variables consistent with their respective PDFs -- see Fig.$\stick$\ref{appendix_graphics} -- ensuring a statistically robust synthetic population of PWN-SNR systems.
\begin{figure}[h!]
            \centering 
            \includegraphics[width=1.0\linewidth]{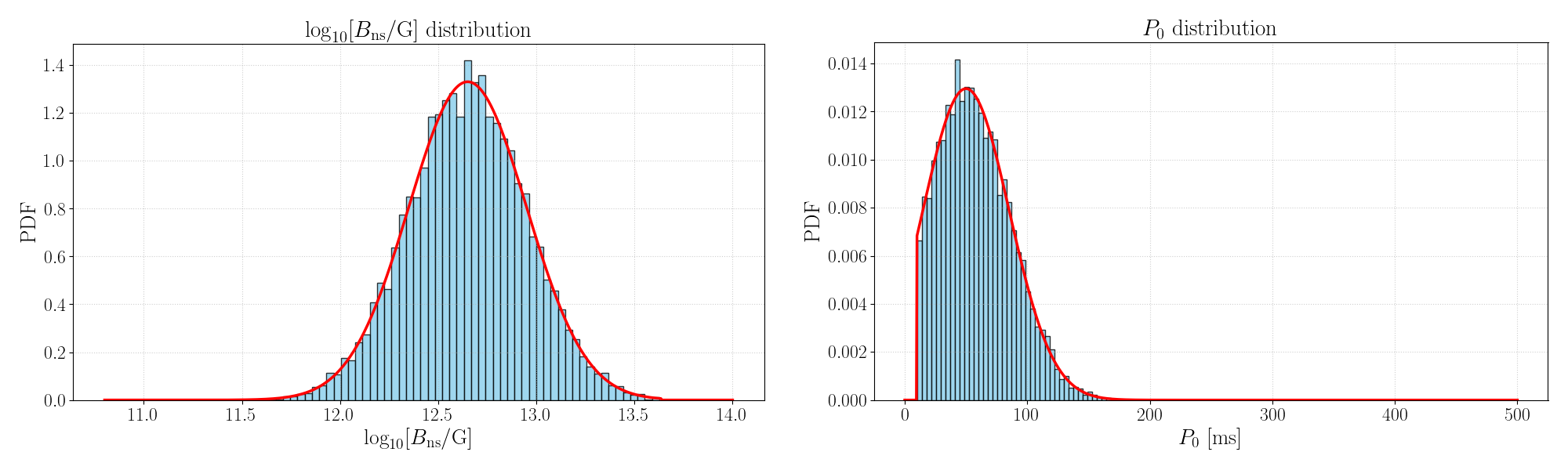}
			\centering
            \includegraphics[width=1.0\linewidth]{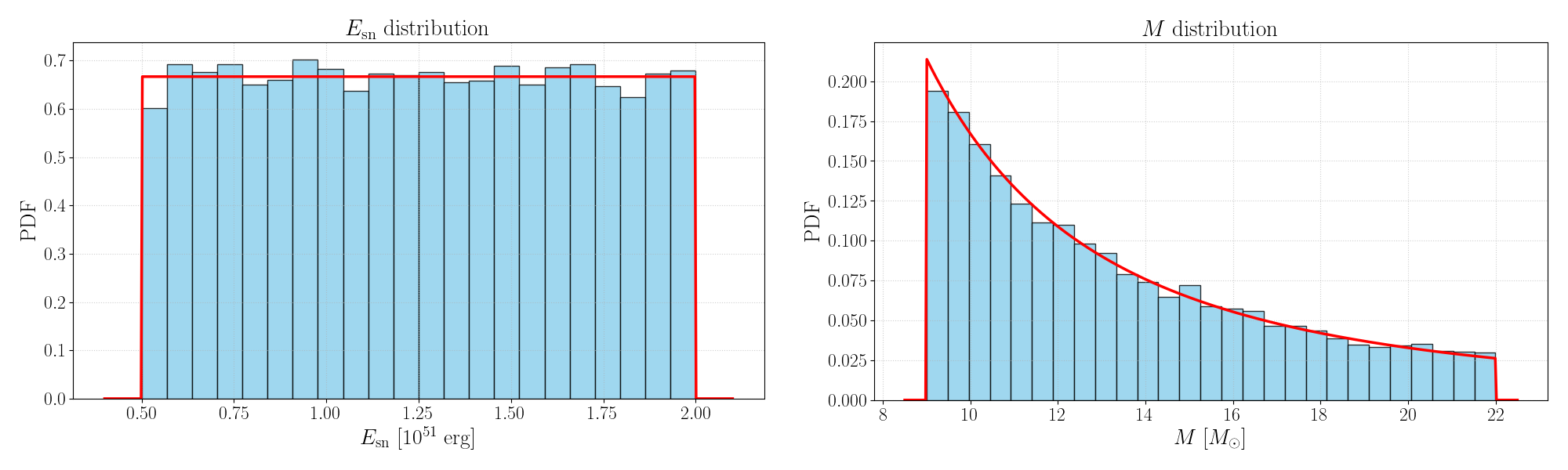}
            \centering
            \includegraphics[width=0.5\linewidth]{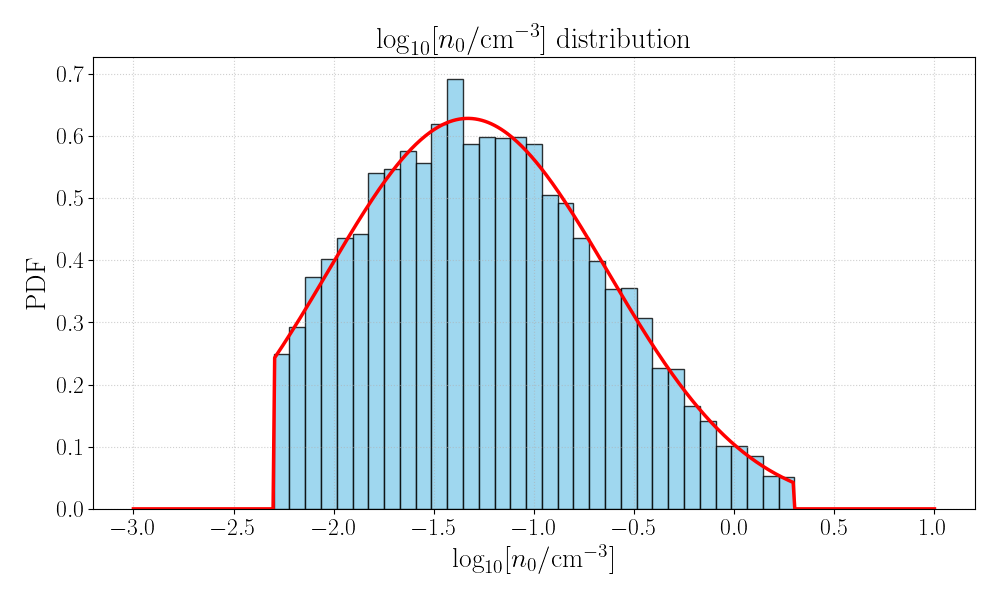}
    \caption{Simulated data (blue) and associated PDFs (red lines) for the physical quantities discussed above.}
    \label{appendix_graphics}
\end{figure}
%
%
%
%
%
%
%
%
%
%
%
%
%
%
%
%
%
%
%
%
%
%
%
%
\renewcommand{\chaptername}{Appendix}
\renewcommand{\thechapter}{\Alph{chapter}}
\chapter{Discretized Equations of Lagrangian Hydrodynamics}
\label{APPII}
In this section we present the discretized equations of the hydrodynamic Lagrangian code that
we have used to model numerically the evolution of a PWN coupled to its SNR. Here we briefly summarize the equations that are solved for the radius ($r$) velocity ($v$) and acceleration ($a$), where the subscripts $i+1/2$ indicate the interface between the cells $i$ and $i+1$, while the index $n$ accounts for the time step of the simulation \citep{Bandiera+21}: 
\begin{align}
\label{RADLAGFIRST}
    r_{i+1/2}^{n+1}
     &= r_{i+1/2}^{n} + 4\pi\Delta t v_{i+1/2}^{n} + 2\pi(\Delta t)^{2}a_{i+1/2}^{n}, \\
    v_{i+1/2}^{n+1}
     &= v_{i+1/2}^{n} + 4\pi\Delta t a_{i+1/2}^{n}, \\
     \label{ACCLAG}
    a_{i+1/2}^{n}
     &=-\left\lbrace(r_{i+1/2}^{n})^{2}(P_{i+1}^{n}-P_{i}^{n})+\left[(r_{i+1}^{n})^{2}Q_{i+1}^{n}-(r_{i}^{n})^{2}Q_{i}^{n}\right]\right\rbrace / {\Delta m_{i+1/2}},
\end{align}
with $\Delta t = t^{n+1}-t^{n}$. To strictly preserve causality, the time integration is governed by the Courant-Friedrichs-Lewy (CFL) condition \citep{Lewy1928}. This condition mandates that the time-step, $\Delta t$, must be smaller than the characteristic time required for a sound wave to traverse the numerical cell. Specifically, for the $i$-th shell:
\begin{equation}\Delta t < C_{\rm CFL} \frac{r_{i+1/2}-r_{i-1/2}}{|v_{i+1/2}-v_{i-1/2}| + 2c_{\text{s},i}},
\end{equation}
where $C_{\rm CFL}$ is set to 0.2 to ensure numerical stability, and $c_{\text{s},i}$ denotes the sound speed in the cell \textit{i}. The latter, assuming an ideal non-relativistic gas, is defined as:
\begin{equation}
c_{\text{s},i}\equiv
\sqrt{\frac{5}{3}\frac{P_i}{\rho_i}},
\end{equation}
with $P_i$ and $\rho_i$ pressure and mass density inside the $i$-th cell. Furthermore, to prevent the overlapping of the interfaces, we require
\begin{equation}
\Delta t < \frac{r_{i+1/2}-r_{i-1/2}}{2|v_{i+1/2}-v_{i-1/2}|},
\end{equation} 
thereby ensuring that the radial contraction of any shell during a single time step is limited to at most half of its current width. The artificial viscosity, $Q$, the radius of the barycenter of the shell, $r$, its mass density, $\rho$, and internal energy per unit mass, $\epsilon$, are given by:
\begin{align}
     Q_{i}^{n}
     &= \eta_{\rm vnr}\rho_{i}^{n}(v_{i+1/2}^{n}-v_{i-1/2}^{n})^{2}\Theta[v_{i-1/2}^{n}-v_{i+1/2}^{n}], \\
     r_{i}^{n}
    &= \left[\frac{(r_{i+1/2}^{n})^{3}+(r_{i-1/2}^{n})^{3}}{2}\right]^{1/3}, \\
    \label{RHOLAG}
     \rho_{i}^{n}
     &=\frac{3\Delta m_{i}}{4\pi[(r_{i+1/2}^{n})^{3}-(r_{i-1/2}^{n})^{3}]},\\
     \label{ENELAG}
     \begin{split}
\epsilon_{i}^{n+1} &= \epsilon_{i}^{n} - \frac{P_{i}^{n+1} + P_{i}^{n}}{2} \left[ \frac{1}{\rho_{i}^{n+1}} - \frac{1}{\rho_{i}^{n}} \right] + \\
&\quad - 2\pi\Delta t \left[ r_{i}^{n+1} + r_{i}^{n} \right]^2 Q_{i}^{n} \frac{v_{i+1/2}^{n} - v_{i-1/2}^{n}}{\Delta m_{i}},
\end{split}
\end{align}
where $\epsilon = 3P/2\rho$ in the case of a non-relativistic fluid and $\Theta[\cdot]$ is the Heaviside function. It is evident that, in contrast to Eqs.(\ref{RADLAGFIRST})$-$(\ref{RHOLAG}), Eq.$\stick$(\ref{ENELAG}) is implicit because the pressure term $P_{i}^{n+1}$ is itself a function of the unknown internal energy $\epsilon_{i}^{n+1}$, through the EOS, thus requiring an iterative solution. We choose $\eta_{\rm vnr} = 4$, larger than the canonical value 2 \citep{VonNeumann1950}, in order to suppress numerical noise arising from strong shocks. Regarding the boundary conditions, we set the artificial viscosity at the inner boundary of the innermost mass shell to zero, and the pressure equal to the nebular one. In addition, $\Delta m_i$ and $\Delta m_{i+1/2}$ denote the mass at the center of each shell $i$ and the mass within its interface, respectively:
\begin{align}
    \Delta m_i
     &= m_{i+1/2} - m_{i-1/2},  \\
    \Delta m_{i+1/2}
     &= [\Delta m_i + \Delta m_{i+1}]/2,
\end{align}
while $m$ denotes the Lagrangian mass coordinate, defined as the cumulative mass enclosed from the center up to the interface $\textit{i}+1/2$. It is evident that Eqs.$\stick$(\ref{ACCLAG}), (\ref{RHOLAG}) and (\ref{ENELAG}) have the same functional structure of Eqs.$\stick$(\ref{LAGMOM}), (\ref{LAGM}) and (\ref{LAGENE}), respectively.
Kinematic quantities -- specifically radius, velocity, and acceleration -- are defined at the cell interfaces, which represent the physical boundaries of the Lagrangian fluid elements. The dynamical evolution of these boundaries is governed by the mass at the interfaces, which provides the local inertial term required to solve the momentum equation. Conversely, dynamic variables -- including density, pressure, and specific internal energy -- are volume-averaged quantities defined at the cell centers.\\\\
The interplay between quantities defined at cell centers and interfaces is governed by a bidirectional coupling: the quantities at the cell centers drive the dynamics by generating pressure gradients that accelerate the interfaces. Reciprocally, the interfaces dictate the dynamical evolution, as their displacement redefines the shell volume, updating density and internal energy.\\\\
As anticipated in Sect. \ref{sec:beyond_rev}, we present two distinct regimes, examining both a highly compressive scenario (system \texttt{T01}, $\text{CF} = 9.27$) and a regime of weak compression (system \texttt{L05}, $\text{CF} = 1.70$). The resulting profiles for these extreme cases are illustrated in Figs. \ref{high_cf} and \ref{low_cf}, which are analogous to Fig$\stick$\ref{SNAPSHOTS_TOGETHER}. 
\begin{figure}[h!]
    \centering 
     	\includegraphics[width=0.98\linewidth]{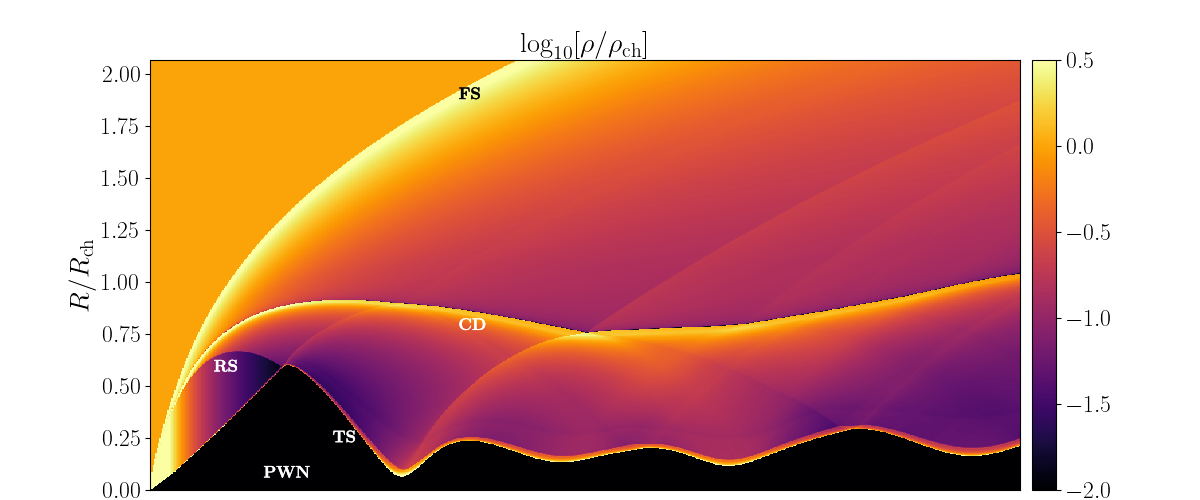}
     	\includegraphics[width=0.98\linewidth]{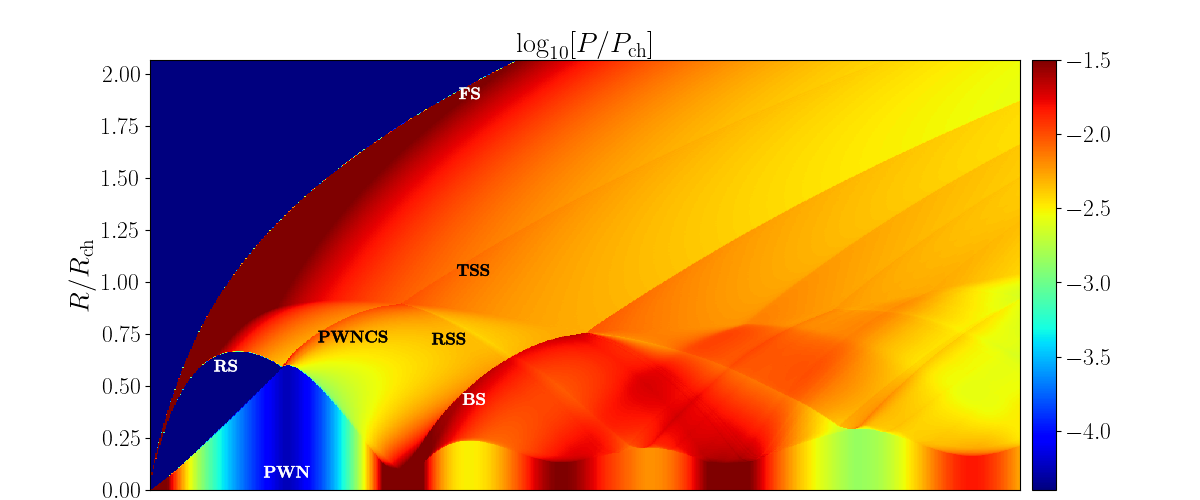}
     	\includegraphics[width=0.98\linewidth]{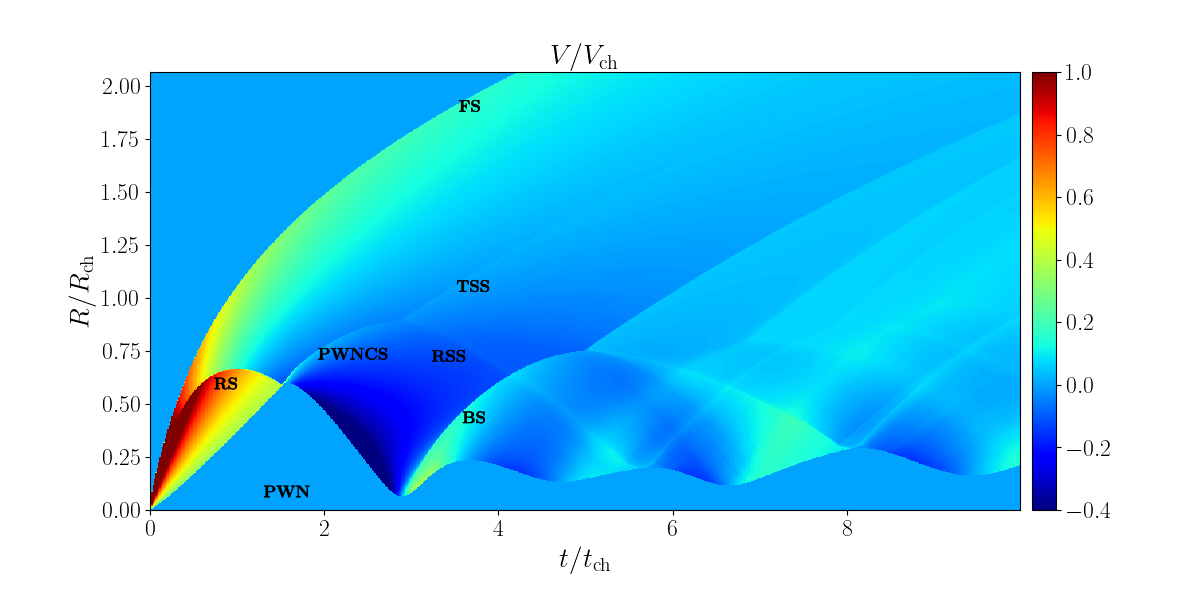}
    \caption{Evolution of the normalized density (\textit{Upper panel}), pressure (\textit{Center panel}), and velocity (\textit{Lower panel}) for the system \texttt{T01}.}  
    \label{high_cf}
\end{figure}
\begin{figure}[h!]
    \centering 
     	\includegraphics[width=0.98\linewidth]{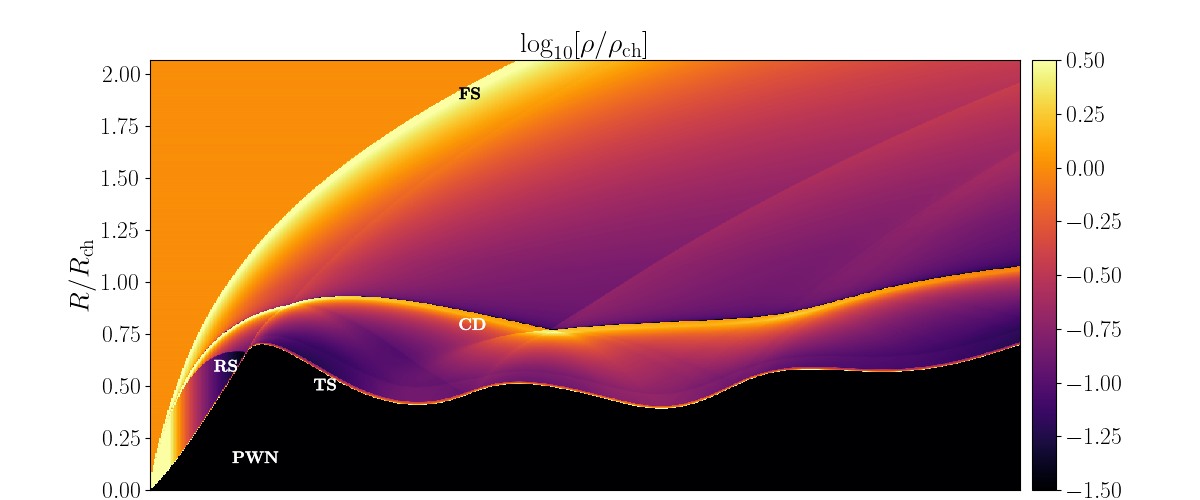}
     	\includegraphics[width=0.98\linewidth]{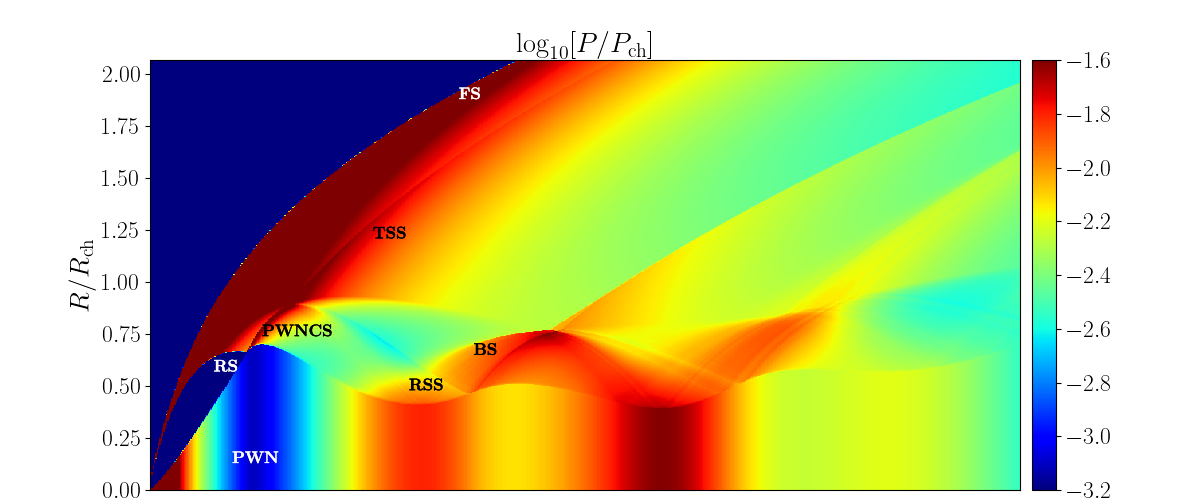}
     	\includegraphics[width=0.98\linewidth]{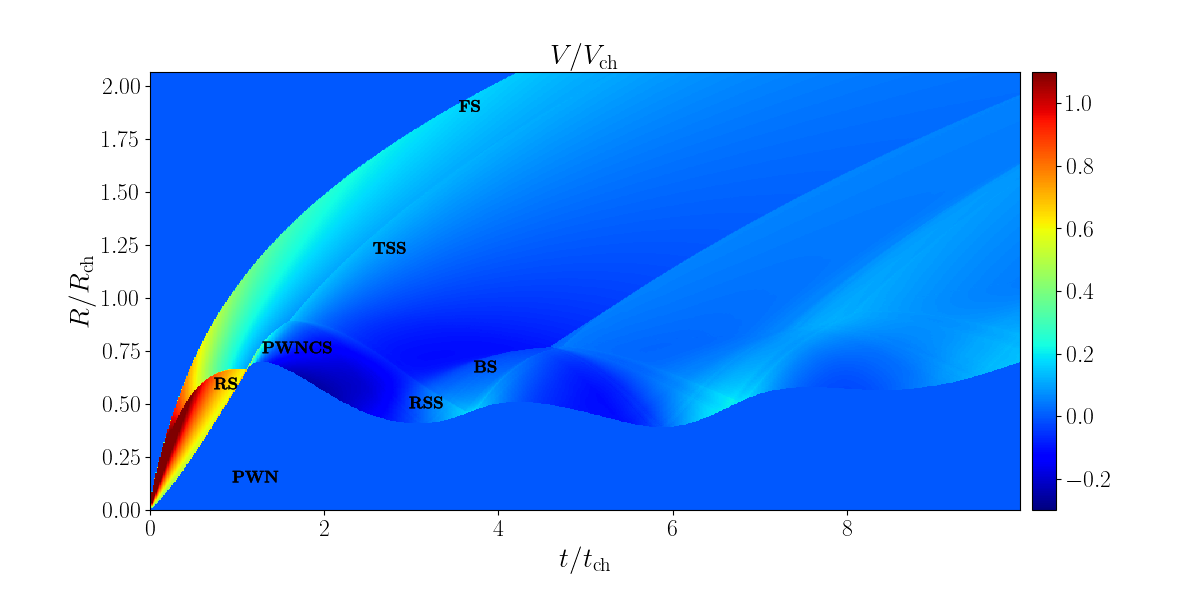}
    \caption{Evolution of the normalized density (\textit{Upper panel}), pressure (\textit{Center panel}), and velocity (\textit{Lower panel}) for the system \texttt{L05}.}  
    \label{low_cf}
\end{figure}
\noindent
The two systems exhibit extremely diverse evolutionary paths, characterized by hydrodynamic features that differ significantly in both profile and onset time. However, the general structure is preserved: the same structures are generated in all cases, spanning the full range from high to low compression.  
%
%
%
%
%
%
%
%
%
%
%
%
%
%
%
%
%
%
%
%
%
%
%
%
\chapter*{Acronyms} 
\markboth{Acronyms}{Acronyms} 
\addcontentsline{toc}{chapter}{Acronyms}
\markright{Acronyms}
\begin{description}[font=\bfseries, labelwidth=2cm, leftmargin=4cm, labelsep=0.5cm]
\item[AGN] Active Galactic Nuclei
\item[AMR] Adaptive Mesh Refinement (Module)
\item[AS$\gamma$] Air Shower Gamma
\item[ASTRI] Astronomia a Specchi a Tecnologia Replicante Italiana
\item[BS] Bounce Shock
\item[BH] Black Hole
\item[BSG] Blue Supergiants
\item[BSPWNe] Bow Shock Pulsar Wind Nebulae
\item[CF] Compression Factor
\item[CF(\texttt{NR})] Compression Factor for a non-relativistic nebula
\item[CFL] Courant-Friedrichs-Lewy (Condition)
\item[CCO] Central Compact Object
\item[CCSN] Core Collapse Supernova
\item[CD] Contact Discontinuity
\item[CDF] Cumulative Distribution Function
\item[CMB] Cosmic Microwave Background (Radiation)
\item[CR] Cosmic Rays
\item[CSM] Circumstellar Medium
\item[CTA] Cherenkov Telescope Array
\item[EAS] Extensive Air Shower
\item[EOS] Equation Of State
\item[FS] Forward Shock
\item[GRB] Gamma-Ray Burst
\item[HAWC] High-Altitude Water Cherenkov (Observatory)
\item[HD] Hydrodynamics
\item[HESS] High Energy Stereoscopic System
\item[\texttt{\textit{HLLC}}] Harten - Lax - van Leer - Contact (Solver)
\item[HST] Hubble Space Telescope
\item[IACT] Imaging Atmospheric Cherenkov Telescopes
\item[ICS] Inverse Compton Scattering
\item[IMF] Initial Mass Function
\item[IR] Infrared (background radiation)
\item[ISM] Interstellar Medium
\item[LAT] Large Area Telescope
\item[LC] Light Cylinder
\item[LHAASO] Large High Altitude Air Shower Observatory
\item[MAGIC] Major Atmospheric Gamma Imaging Cherenkov (Telescopes)
\item[\texttt{\textit{MC}}] Monotonized Central (Limiter)
\item[MHD] Magneto-Hydrodynamics
\item[NS] Neutron Star
\item[PDF] Probability Distribution Function
\item[PSR] Pulsar
\item[PWN] Pulsar Wind Nebula
\item[PWNCS] Pulsar Wind Nebula Compressive Shock
\item[QED] Quantum Electrodynamics
\item[\texttt{\textit{RK2}}] Second-Order Runge-Kutta (Temporal Reconstruction)
\item[RHD] Relativistic Hydrodynamics
\item[RMHD] Relativistic Magneto-Hydrodynamics
\item[ROI] Region Of Interest
\item[RSG] Red Supergiants
\item[RS] Reverse Shock
\item[RSS] Reflected Secondary Shock
\item[SN] Supernova
\item[SNR] Supernova Remnant
\item[TSS] Transmitted Secondary Shock
\item[TS] Thin Shell
\item[\texttt{\textit{TVD}}] Total Variation Diminishing (Reconstruction)
\item[VERITAS] Very Energetic Radiation Imaging Telescope Array System
\item[VHE] Very-High-Energy
\item[VLA] Very Large Array (Telescopes)
\item[WD] White Dwarf
\item[WTS] Wind Termination Shock
\end{description}
\chapter*{Symbols}
\markboth{Symbols}{Symbols}
\addcontentsline{toc}{chapter}{Symbols}
\markright{Symbols}
\begin{description}[font=\bfseries, labelwidth=2cm, leftmargin=4cm, labelsep=0.5cm]
    \item[$A$] Characteristic constant of the ejecta density profile
    \item[$A_0$] Maximum radial extent of dipole magnetic field lines
    \item[$a$] Acceleration
    \item[$\alpha$] Index of the Salpeter IMF, generic power-law index for the ejecta
    \item[$\beta$] Power-law index for the density of the ejecta
    \item[$\alpha_{\rho}$] Parameter of the relation for the density of the ejecta
    \item[$\alpha_1, \alpha_2, \alpha_3, \alpha_4$] Parameters of the relation of the moment of inertia of a NS
    \item[$a_1, a_2, a_3, a_4$] Parameters of the relation defining the RS
    \item[$B_{\rm core}$] Strength of the magnetic field in the star core
    \item[$B_{\rm pwn}$] Magnetic field of the PWN
    \item[$B_{\rm ns}$] Surface magnetic field of the NS
    \item[$B_{\rm u}$] Magnetic field upstream of the TS
    \item[$\mathcal{C}$] CDF of a physical quantity
    \item[$C_{X}$] Normalization factor of a PDF of a physical quantity $X$
    \item[$C(\delta, \omega)$] Multiplicative constant of the nebular radius
    \item[$C_{\rm CFL}$] CFL factor
    \item[$c_s$] Sound speed of the fluid
    \item[$c$] Speed of light in vacuum
    \item[$\Gamma$] Adiabatic index of a fluid
    \item[$\chi$] Inclination angle between the rotation and magnetic axes
    \item[$\Delta l_{i}$] Extension of the cell \textit{i}
    \item[$\Delta t$] Generic interval of time
    \item[$\Delta\theta$] Half amplitude of angular domain
    \item[$\delta$] Density index of the inner region of the ejecta
    \item[$\delta R$] Resolution of the radial domain
    \item[$\delta \rho$] Perturbation of density in the thin swept-up shell
    \item[$\delta\theta$] Resolution on the polar domain
    \item[$E_{\rm ava}$] Spin-down energy of the PSR available to sustain the reverberation
    \item[$E_{\rm B}$] Magnetic energy in the PWN
    \item[$E_{\rm e}$] Energy of the electron
    \item[$E_{\rm inj}$] Spin-down energy transfered by the PSR to the SNR
    \item[$E_{\rm pwn}$] Total energy stored in the PWN
    \item[$E_{\rm rot}$] Rotational kinetic energy of the NS
    \item[$E_{\rm sd}$] Total spin-down energy of the PSR
    \item[$E_{\rm sh}$] Kinetic energy of the swept-up shell
    \item[$E_{\rm sn}$] Kinetic energy of the SN
    \item[$\epsilon$] Specific Energy per unity of mass
    \item[$\eta$] Perturbation amplitude of the mass density of the swept-up shell
    \item[$\eta_B$] Fraction of NS spin-down luminosity converted into magnetic energy
    \item[$\phi$] Azimuthal angle
    \item[$\Phi_{\rm core}$] Flux of the magnetic field at the core surface
    \item[$\Phi_{\rm ns}$] Flux of the magnetic field at the NS surface
    \item[$f$]Auxiliary distribution function
    \item[$\gamma$] Lorentz factor of the particles
    \item[$\boldsymbol{I}$] Metric Matrix
    \item[$I_{\rm ns}$] Moment of inertia of the NS
    \item[$J$] Density Current on the NS
    \item[$K, K_{\rm ns}$] Braking constants of NSs
    \item[$\lambda_{i}^{\rm max}$] Maximum physical velocity of a signal in the cell \textit{i}
    \item[$l$] Characteristic length of acceleration in the magnetosphere
    \item[$l_{\rm max}$] Maximum number of refinement levels 
    \item[$\log B$] $\log_{10}[B_{\rm ns}/G]$
    \item[$\log n_0$] $\log_{10}[n_{0}/\text{cm}^{-3}]$
    \item[$L_0$] Initial spin-down luminosity of the PSR
    \item[$L_{\rm ch}$] Characteristic luminosity in the SNR
    \item[$L_{\rm sd}$] Spin-down luminosity of the PSR
    \item[$M$] Mass of progenitor star
    \item[$M_{\rm Pre-SN}$] Mass of progenitor star just before the SN explosion
    \item[$M_{\rm ej}$] Total mass of the SN ejecta
    \item[$M_{\rm sh}$] Mass of the swept-up shell
    \item[$m_0$] Mean particle mass of the ambient medium
    \item[$m$] Lagrangian mass coordinate
    \item[$\mu$] Magnetic dipole moment, Parameter of PDFs
    \item[$N$] Particle energy distribution function
    \item[$n$] Braking Index of PSRs
    \item[$n_{\theta}$] wavenumber of the mass-density perturbation in the swept-up shell
    \item[$n_0$] Number density of the ambient medium
    \item[$\nu_{\rm curv}$] Frequency of the curvature radiation
    \item[$\nu_{\rm IC}$] Frequency of IC radiation
    \item[$\nu_{\rm sync}$] Frequency of synchrotron radiation
    \item[$\mathcal{P}$] PDF of a generic physical quantity
    \item[$P_0$] Initial rotation period of the PSR
    \item[$P_{\rm ch}$] Characteristic SNR pressure
    \item[$P_{\rm ej}$] Pressure of the ejecta
    \item[$P_{\rm mag}$] Magnetic pressure inside the nebula
    \item[$P_{\rm ns}$] Rotation period of the NS in the present epoch
    \item[$P_{\rm pwn}$] Total pressure inside the nebula
    \item[$P_{\rm SNR}$] Pressure in the centre of the SNR
    \item[$q_{\rm e}$] Charge density in the magnetosphere of NSs
    \item[$Q$] Artificial Viscosity term
    \item[$Q_{\rm inj}$] Rate of energy or particle injection at the TS
    \item[$r, \ R$] Radial coordinate
    \item[$Q_{\rm rad}$] Total radiative energy losses
    \item[$R_{\rm ch}$] Characteristic SNR radius
    \item[$R_{\rm c}$] Radius of the core of the ejecta
    \item[$R_{\rm core}$] Radius of the progenitor star core; Radius of the core of the ejecta
    \item[$R_{\rm env}$] Radius of the envelope of the ejecta
    \item[$r_{\rm LC}$] Surface of the Light Cylinder
    \item[$R_{\rm LC}$] Radius of the Light Cylinder
    \item[$R_{\rm max}$] Maximum radius of the RS, Outer extension of the radial domain
    \item[$R_{\rm min}$] Minimal extension of the radial domain
    \item[$R_{\rm ns}$] Radius of the NS
    \item[$R_{\rm psr}$] Distance covered by the PSR over its life
    \item[$R_{\rm pwn}$] Radius of the PWN
    \item[$R^{(\rm app)}_{\rm pwn}$] Apparent Radius of the PWN
    \item[$R^{(\rm eff)}_{\rm pwn}$] Effective Radius of the PWN
    \item[$R_{\rm sh}^{(\rm out)}$] Outer boundary of the swept-up shell
    \item[$R_{\rm sh}^{(\rm in)}$] Inner boundary of the swept-up shell
    \item[$R_{\rm WTS}$] Radius of the WTS
    \item[$\rho_0$] Density of the medium surrounding SNRs, Density of the unperturbed swept-up shell
    \item[$\rho_{\rm ch}$] Characteristic SNR mass density
    \item[$\rho_{\rm ej}$] Ejecta density profile
    \item[$\rho_{\rm pwn}$] Mass density of the PWN
    \item[$\rho_{\rm u}$] Density of the fluid upstream of the TS
    \item[$\sigma$] Wind magnetization parameter, Parameter of PDFs
    \item[$t_{\rm ch}$] Characteristic time-scale of SNRs
    \item[$t_{\rm esc}$] Time-scale for a NS to escape from the nebula
    \item[$t_{\rm max}$] Time-scale where the RS reaches its maximum extension
    \item[$t_{\rm min}$] Time-scale where the RS reaches the centre of the SNR
    \item[$t_{\rm rev}$] Instant setting the beginning of the reverberation phase
    \item[$t_{\rm pwn}$] Age of a PWN
    \item[$\tau_0$] Initial spin-down time of the PSR
    \item[$\tau_{\rm c}$] Characteristic dipole age of the NS
    \item[$\tau_{\rm IC}$] Cooling time-scale for IC emission
    \item[$\tau_{\rm ns}$] Age of the NS
    \item[$\tau_{\rm sync}$] Cooling time-scale for synchrotron emission
    \item[$\theta$] Polar angle
    \item[$\theta_{\rm cap}$] Angle of the polar cap
    \item[$\theta_{\rm max}$] Maximum extension of the angular domain
    \item[$\theta_{\rm min}$] Minimum extension of the angular domain
    \item[$U_{\gamma}$] Energy density of a gas of photons
    \item[$U_{\rm cmb}$] Energy density of the CMB
    \item[$V$] Spatial volume
    \item[$v_{\rm c}$] Velocity of the core of the SN ejecta
    \item[$v_{\rm d}$] Velocity of the fluid in the downstream of the WTS
    \item[$v_{\rm ej}$] Velocity of the SN ejecta
    \item[$V_{\rm ch}$] Characteristic expansion velocity of SNRs
    \item[$V_{\rm out}$] Velocity of the outer ejecta
    \item[$V_{\rm pwn}$] Expansion velocity of the PWN
    \item[$V_{\rm psr}$] Kick velocity of the PSR
    \item[$v_{\rm u}$] Velocity of the fluid in the upstream of the TS
    \item[$\xi$] Compactness in geometrized units, Power-law index for the density of the ejecta
    \item[$\xi_1, \xi_2$] Parameters of the relation defining the RS
    \item[$\omega$] Density index for the outer region of the ejecta
    \item[$\Omega_{\rm ns}$] Angular rate rotation of the NS
\end{description}
%
%
%
%
%
%
%
%
%
%
%
%
%
%
%
%
%
%
%
%
%
%
%
%
%
%
%
%
%
%
%
%
%
%
%
%
%
%
%
%
%
%
%
%
%
%
%
%

%
%
%
%
%
%
%
%
%
%
%
%
%
%
%
%
%
%
%
%
%
%
%
%
%
%
%
%
%
%
%
%
%
%
%
%
%
%
%
%
%
%
%
%
%
%
%
%
%
%
%
%
%
%
%
%
%
%
%
%
%
%
%
%
%
%
%
%
%
%
\backmatter 
\cleardoublepage 
\phantomsection 
\addcontentsline{toc}{chapter}{Bibliography} 

\end{document}